\documentclass[letterpaper,notitlepage,11pt]{article}
\usepackage[colorlinks=true,urlcolor=blue,citecolor=magenta]{hyperref}
\usepackage{amssymb,amsmath,amsfonts}
\usepackage{epsfig}
\usepackage{graphicx}
\usepackage{epstopdf}
\usepackage{caption}
\usepackage{subcaption}
\usepackage{amscd}
\usepackage{stmaryrd}
\usepackage{amsthm}
\usepackage{latexsym}
\usepackage{amsbsy}
\usepackage[english]{babel}
\usepackage{psfrag}
\usepackage{tabularx}

\usepackage[acronym,nohypertypes={acronym,notation}]{glossaries}
\makeglossaries
\renewcommand{\glossarysection}[2][]{}
\newcommand{\regls}[1]{\glsentrylong{#1} (\gls{#1})}
\newcommand{\reglspl}[1]{\glsentrylongpl{#1} (\glspl{#1})}
\newcommand{\reGls}[1]{\Glsentrylong{#1} (\Gls{#1})}
\newcommand{\reGlspl}[1]{\Glsentrylongpl{#1} (\Glspl{#1})}

\usepackage{amssymb}
\usepackage{booktabs}
\usepackage{cancel}
\usepackage{setspace}
\usepackage{authblk}
\usepackage{delimset}
\usepackage{accents}
\usepackage{tikz}
\usetikzlibrary{matrix}
\numberwithin{equation}{section}

\newcommand{\order}[1]{\mathcal{O}\brk{#1}}
\renewcommand{\d}{\mathrm{d}}
\newcommand{\os}[2]{\accentset{\scriptscriptstyle{\brk{\hspace{-0.05em}#2\hspace{-0.05em}}}}{#1}{}}

\newcommand{\oss}[3]{{\os{#1}{#2}}_{\mathrm{#3}}}

\newacronym[longplural={Ward identities}]{wi}{WI}{Ward identity}
\newacronym{nc}{NC}{Newton--Cartan}
\newacronym{ncg}{NCG}{Newton--Cartan gravity}
\newacronym[description={Torsional Newton--Cartan, i.e. $\tau\wedge\d\tau\neq0$}]{tnc}{TNC}{torsional Newton--Cartan}
\newacronym{tov}{TOV}{Tolman--Oppenheimer--Volkoff}
\newacronym{gr}{GR}{general relativity}
\newacronym{nr}{NR}{non-relativistic}
\newacronym{cs}{CS}{Chern--Simons}
\newacronym{ads}{AdS}{anti-de Sitter}
\newacronym{flrw}{FLRW}{Friedmann--Lema\^itre--Robertson--Walker}
\newacronym{nrg}{NRG}{non-relativistic gravity}
\newacronym[description={Twistless torsional Newton--Cartan, i.e. $\tau\wedge\d\tau=0$}]{ttnc}{TTNC}{twistless torsional Newton--Cartan}
\newacronym[description={next-to-next-to-leading (or subsubleading) order}]{nnlo}{NNLO}{next-to-next-to-leading order}
\newacronym[description={next-to-leading (or subleading) order}]{nlo}{NLO}{next-to-leading order}
\newacronym{lo}{LO}{leading order}
\newacronym{eh}{EH}{Einstein--Hilbert}
\newacronym[longplural={equations of motion}]{eom}{EOM}{equation of motion}
\newacronym{snc}{SNC}{string Newton--Cartan}

\newacronym{ged}{GED}{Galilean electrodynamics}
\newacronym{pn}{PN}{post-Newtonian}

\usepackage{overpic}
\usepackage{array}
\usepackage{rotating}
\usepackage{color}

\def\be{\begin{equation}}
\def\ee{\end{equation}}
\def\bea{\begin{eqnarray}}
\def\eea{\end{eqnarray}}

\allowdisplaybreaks

\def\clock{{\count0=\time
           \divide\count0 60
           \ifnum\count0<10 0\fi\the\count0
           \multiply\count0 -60 \advance\count0 \time
           :\ifnum\count0<10 0\fi \the\count0
         }}
\newcommand{\timestamp}{{\small\vbox{\hbox{\tt\jobname.tex}
\hbox{\the\day/\the\month/\the\year, \clock}}}}

\title{Non-Relativistic Gravity and its Coupling to Matter}

\author[1]{\small Dennis Hansen}
\author[2]{\small Jelle Hartong}
\author[3,4]{\small Niels A. Obers}

\affil[1]{\footnotesize Institut f{\"u}r Theoretische Physik, Eidgen{\"o}ssische Technische Hochschule Z{\"u}rich,

Wolfgang-Pauli-Strasse 27, 8093 Z{\"u}rich, Switzerland.}
\affil[2]{\footnotesize School of Mathematics and Maxwell Institute for Mathematical Sciences,

University of Edinburgh, Peter Guthrie Tait road, Edinburgh EH9 3FD, UK.}

\affil[3]{\footnotesize Nordita, KTH Royal Institute of Technology and Stockholm University, 

Roslagstullsbacken 23, SE-106 91 Stockholm, Sweden.}

\affil[4]{\footnotesize The Niels Bohr Institute, Copenhagen University,

Blegdamsvej 17, DK-2100 Copenhagen \O , Denmark.}
\affil[ ]{\textit{dehansen@phys.ethz.ch, j.hartong@ed.ac.uk, obers@nbi.ku.dk}}

\begin{document}
\maketitle

\begin{abstract}
We study the non-relativistic expansion of general relativity coupled to matter.
This is done by expanding the metric and matter fields analytically in powers of $1/c^2$ where $c$ is the speed of light.
In order to perform this expansion it is shown to be very convenient to rewrite general relativity in terms of a timelike vielbein and a spatial metric.
This expansion can be performed covariantly and off shell.
We study the expansion of the Einstein--Hilbert action up to next-to-next-to-leading order.
We couple this to different forms of matter: point particles, perfect fluids, scalar fields (including an off-shell derivation of the Schr\"odinger--Newton equation) and electrodynamics (both its electric and magnetic limits).
We find that the role of matter is crucial in order to understand the properties of the Newton--Cartan geometry that emerges from the expansion of the metric.
It turns out to be the matter that decides what type of clock form is allowed, i.e. whether we have absolute time or a global foliation of constant time hypersurfaces. We end by studying a variety of solutions of non-relativistic gravity coupled to perfect fluids.
This includes the Schwarzschild geometry, the Tolman--Oppenheimer--Volkoff solution for a fluid star, the FLRW cosmological solutions and anti-de Sitter spacetimes. 
\end{abstract}

\newpage
\begingroup\setstretch{1.1}
\tableofcontents
\endgroup
\newpage

\section{Introduction}
Nature is relativistic at a fundamental level but nonetheless it often effectively appears to us as \gls{nr}.
This typically happens in many-body or condensed matter type systems but it can also be true for gravitational phenomena.
\Gls{gr} can often be well-approximated by a theory of non-relativistic gravity such as the \gls{pn} approximation.
Hence, uncovering the mathematical description of non-relativistic geometries, their dynamics and interaction with matter (classical and quantum) is a very relevant subject. In this paper, building on earlier work by \cite{Dautcourt:1996pm,Tichy:2011te,VandenBleeken:2017rij,Hansen:2018ofj}, we try to systematically build up a geometrical language that allows us to formulate certain gravitational problems in a manifestly non-relativistic manner. 

This work is mainly foundational and therefore not so much concerned with applications. Nevertheless we will show that the Friedmann equations, the \gls{tov} fluid star and the usual effects due to the Schwarzschild geometry can all be captured by the theory of non-relativistic gravity described here. More generally we expect this approach to be relevant whenever the gravitational interaction can effectively be treated as instantaneous.

\subsection{Background and motivation} 

Recent years have seen a revival in the study of non-relativistic gravity and its formulation in terms of \gls{nc} type geometries. NC geometry was originally introduced by Cartan in 1923 \cite{Cartan1,Cartan2} 
(see also e.g.  \cite{Trautman63,Havas:1964zza})  to geometrise Newton's laws of gravitation, 
following the successful use of pseudo-Riemannian geometry in the formulation of Einstein's theory of general relativity.
The recent developments in non-relativistic gravity have been spurred in part by modern advances leading towards a more general understanding of non-relativistic geometry. This includes in particular the discovery of a torsionful generalisation
of NC geometry, which allows for a non-closed clock form and was first observed as the boundary geometry in the context of Lifshitz holography \cite{Christensen:2013lma,Christensen:2013rfa,Hartong:2014pma}. Besides being
relevant in various non-relativistic gravity theories (see e.g. 
\cite{Hartong:2015zia,Bergshoeff:2015uaa,Afshar:2015aku,
Bergshoeff:2016lwr,Hartong:2016yrf,Bergshoeff:2017dqq,Hartong:2017bwq})
this geometry, called (type I) \gls{tnc} geometry in \cite{Hansen:2018ofj}, plays an important role as the background geometry to which non-relativistic field theories naturally couple 
\cite{Son:2013rqa,Jensen:2014aia,Hartong:2014oma,Geracie:2014nka}. It furthermore appears in the context of non-relativistic string theory 
\cite{Harmark:2017rpg,Kluson:2018egd,Harmark:2018cdl,Gallegos:2019icg,Harmark:2019upf} (see also references 
\cite{ Andringa:2012uz,Bergshoeff:2018yvt,Bergshoeff:2018vfn,Gomis:2019zyu,Bergshoeff:2019pij} for related theories
involving \gls{snc} geometry). 

Importantly, a novel version of TNC geometry (denoted as type II) was uncovered in \cite{Hansen:2018ofj} and shown to arise directly from a careful analysis of the large speed of light expansion of pseudo-Riemannian geometry, as
considered also in \cite{VandenBleeken:2017rij} following earlier work \cite{Dautcourt:1996pm,Tichy:2011te}.  
Correspondingly, it was found that type II TNC geometry is the correct framework to describe General Relativity (GR)
in the non-relativistic limit. In particular, this geometry allows us to formulate a non-relativistic gravity action
 \cite{Hansen:2018ofj} (see also \cite{Hansen:2019vqf,Hansen:2019svu}) in any spacetime dimension\footnote{See also \cite{Cariglia:2018hyr} for a non-relativistic action using first-order formalism, 
 and \cite{VandenBleeken:2019gqa} for related perspectives.}.
 This action has the property that the \glspl{eom} contain the Poisson equation of Newtonian gravity, thus providing for the first time an action principle for Newton's laws of gravity. Moreover, it generalises the latter by allowing for the effects of gravitational time dilation due to strong gravitational fields. The connection with Newtonian gravity follows since type II geometry (just like type I) reduces to standard NC geometry when time is absolute.  It was furthermore shown in \cite{Hansen:2019vqf} that the three classical tests of GR, namely perihelion precession, deflection of light and gravitational red-shift, are passed perfectly by this extension of Newtonian gravity since it includes gravitational time dilation effects even though retaining a non-relativistic causal structure. 
 
There are several motivations to study non-relativistic gravity as the dynamical theory of non-relativistic geometry.
First and foremost, as already mentioned above, it appears in a $1/c^2$ expansion of GR  and is
thus a relevant limit of a celebrated and well-tested theory. Moreover, it paves the way for a covariant formulation of the post-Newtonian expansion, to any order in principle. Even more so, it generalises this expansion since to a given
order in $1/c^2$ the theory retains all-order effects in Newton's constant $G_N$. Central to all this, 
is the appearance of a symmetry principle \cite{Hansen:2018ofj,Hansen:2019vqf}, which naturally arises from the $1/c^2$ expansion of the Poincar\'e algebra. The mathematical framework is that of Lie algebra expansions%
\footnote{See for example references \cite{deAzcarraga:2002xi,Izaurieta:2006zz,Khasanov:2011jr}
and also the recent applications \cite{Hansen:2019vqf,Bergshoeff:2019ctr,Gomis:2019fdh,deAzcarraga:2019mdn}.} which precisely
determines the local symmetry algebra of the ``effective geometry" at any given order in  $1/c^2$. In this way,
type II TNC geometry arises from gauging a novel non-relativistic algebra that differs from the Bargmann algebra,
while gauging the Bargmann algebra yields standard NC geometry \cite{Andringa:2010it}. 
For clarity, we emphasise here that there are other non-relativistic gravity theories than the one considered in this paper, 
arising from gauging type I or related avatars of NC geometry which, though not directly connected to General Relativity, are interesting in their own right from a more general perspective. 

Going beyond the classical level, there are even more fundamental reasons to pursue a deeper insight into non-relativistic  gravity theories, including the specific one considered in this paper, which arises from GR. One question is whether these theories have their own UV completion in terms of a non-relativistic quantum gravity theory. This in turn is interesting  since the construction of such non-relativistic quantum gravity theories could provide
an alternate route towards  (relativistic) quantum gravity, as opposed to approaching the latter from either the classical GR or quantum field theory perspective. In fact, the question of a UV completion of non-relativistic gravity theories
in terms of non-relativistic string theory has recently received a lot of attention
\cite{Andringa:2012uz,Harmark:2017rpg,Kluson:2018egd,Bergshoeff:2018yvt,Kluson:2018grx,Harmark:2018cdl,Kluson:2019ifd,Gomis:2019zyu,Gallegos:2019icg,Harmark:2019upf,Bergshoeff:2019pij,Blair:2019qwi,Kluson:2019xuo}
\footnote{These non-relativistic string theories can be considered to be the generalisation of the Gomis--Ooguri non-relativistic action~\cite{Gomis:2000bd} (see also \cite{Danielsson:2000gi}) to arbitrary backgrounds.} (see also 
\cite{Morand:2017fnv,Berman:2019izh,Blair:2019qwi,Cho:2019ofr,Cho:2019npq} for a relation to double field theory).
Related to this, another relevant application is that classical non-relativistic gravity may have an important role to play in novel types of holographic dualities (see e.g. \cite{Hartong:2016yrf,Hartong:2017bwq}).  Finally, it is relevant to mention that for fixed backgrounds, non-relativistic geometry has proven to be useful for understanding aspects such as energy-momentum tensors, Ward identities,  hydrodynamics and anomalies in the context of non-relativistic field theories, which are ubiquitous in condensed matter and biological systems (see e.g.
 \cite{Jensen:2014ama,Geracie:2015xfa,Hartong:2016nyx,deBoer:2017ing,deBoer:2017abi,Armas:2019gnb}).

\subsection{Outline and summary of the main results}
In this paper we give a comprehensive treatment of non-relativistic gravity and its coupling to matter as it appears from the large speed of light expansion of GR.
In particular, we first provide in Section \ref{sec:expgen} a detailed treatment of various aspects of the $1/c^2$ expansion of GR.
We will do this both from the geometric point of view, involving expanding the Lorentzian metric, as well as the algebraic perspective, which makes use of a Lie algebra expansion of the Poincar\'e algebra.
From the geometric side, it will be convenient to first write the metric $g_{\mu\nu}$ in a certain ``pre-non-relativistic'' parameterisation as $g_{\mu\nu}=-c^2 T_\mu T_\nu+\Pi_{\mu\nu}$ where $T_\mu$ is the timelike vielbein and $\Pi_{\mu\nu}$ a spatial tensor, i.e. a symmetric tensor with signature $(0,1,\ldots,1)$.
One finds that type II TNC geometry arises from the \gls{lo} and the \gls{nlo} fields in the $1/c^2$ expansion of $T_\mu$ and $\Pi_{\mu\nu}$.
In particular, the LO fields are the timelike vielbein  $\tau_\mu$ and symmetric spatial tensor $h_{\mu \nu}$ with signature $(0,1,\ldots,1)$ familiar from standard NC geometry. These are then accompanied by two further gauge fields, $m_\mu$ and $\Phi_{\mu \nu}$, respectively, which appear at NLO in the $1/c^2$ expansion of $T_\mu$ and $\Pi_{\mu\nu}$.

This set of four spacetime tensors, together with their gauge transformation
properties defines type II TNC.
The gauge transformations consist of the $1/c^2$ expansions of the diffeomorphisms of GR as well as the local Lorentz transformation that acts on $T_\mu$ and $\Pi_{\mu\nu}$.
The latter lead to local Galilean boosts and their subleading counterparts.
The former lead to diffeomorphisms plus gauge transformations (originating from the NLO terms in the $1/c^2$ expansion of the diffeomorphisms of GR) acting on the NLO fields $m_\mu$ and $\Phi_{\mu \nu}$.
We note that in NC geometry the torsion is determined by the properties of $\tau_\mu$:
$\d\tau =0$ corresponds to zero torsion (absolute time), $\tau \wedge\d\tau =0$ twistless torsion (i.e. \gls{ttnc} geometry with a foliation in terms of equal time hypersurfaces) and no condition on $\tau$ having arbitrary torsion (full \gls{tnc} geometry).

Just as in \cite{Hansen:2018ofj} we distinguish between type I and type II TNC geometry.
Type I is reserved to refer to the more familiar torsional Newton--Cartan geometry that can be viewed as originating from the gauging of the Bargmann algebra \cite{Andringa:2010it} while type II is reserved for the version of Newton--Cartan geometry that originates from the gauging of the expansion of the Poincar\'e algebra as discussed in \cite{Hansen:2018ofj} as well as the present paper.
The difference between the two consists of a) the gauge transformation properties of $m_\mu$ and b) the fact that in type II there is an extra field, namely $\Phi_{\mu\nu}$ which is not present in type I TNC geometry.
We note that when the clock 1-form $\tau_\mu$ is closed the type I and type II gauge transformations of $m_\mu$ agree and that furthermore the $\Phi_{\mu\nu}$ field decouples on shell from the equations of motion of Newton--Cartan gravity.
This explains why the type II structure was not manifestly present in older approaches to TNC gravity.
Nevertheless it is important to understand the structure of the Bianchi identities of TNC gravity as we discuss further in appendix \ref{sec:newtonian_gravity}. 
Type II TNC geometry arises from gauging a non-relativistic algebra that originates from a Lie algebra expansion, which can be viewed as tensoring the Poincar\'e algebra with the polynomial ring in $\sigma^n$, with $\sigma \equiv c^{-2}$, truncated  at NLO order (quotienting the algebra by removing all levels strictly higher than level 1).
The resulting algebra has twice as many generators as the Poincar\'e algebra and has the Galilei algebra as a subalgebra, corresponding to the LO algebra (level zero).
In this algebra the mass generator $N$ (which is the level 1 Hamiltonian) is not central anymore, see equation \eqref{eq:NRG_algebra} and so the Bargmann algebra is not a subalgebra.

We also display the structure that follows from expanding relativistic Lagrangians in $\sigma$, where the Lagrangians can be the \gls{eh} Lagrangian itself or that of matter coupled to GR. The resulting equations of motion  exhibit a cascading structure, such that at any given order the equations of motion include those of the previous order plus a set of new equations for the extra fields that appear at that given order. This is an important feature as it implies that there is a unique Lagrangian describing the geometric fields that arise at a given order in the
$1/c^2$ expansion and that this Lagrangian can be obtained by computing the corresponding order in the expansion of the relativistic Lagrangian one is interested in.

In order to perform the $1/c^2$ expansion of the Einstein--Hilbert Lagrangian it is very useful to express it first in terms of the fields $T_\mu$ and $\Pi_{\mu\nu}$.
This rewriting involves the choice of a connection that is different from the Levi--Civita connection.
This new connection has the property that at leading order in the $1/c^2$ expansion it provides us with a useful Newton--Cartan connection.
In this way we can rewrite the EH Lagrangian in a form that makes it substantially easier to perform a large speed of light expansion. 
The general structure and properties at any given order in the $1/c^2$ expansion is then studied in Section \ref{sec:NRG}.
We work out in detail the expansion of the EH Lagrangian up to \gls{nnlo}. The leading order action has the property that its equations of motion restrict the clock 1-form to be hypersurface orthogonal. This means that the geometry admits a foliation in terms of equal time hypersurfaces, with Riemannian geometry on these spatial slices, a case that is known
in the literature as \gls{ttnc} geometry \cite{Christensen:2013lma}. This result  is interesting by itself as it
shows that the dynamics restricts on shell to non-relativistic geometries that are causally well-defined (at least locally) \cite{Geracie:2015dea}. One of the main results of the paper is the derivation of the NNLO Lagrangian, which is the Lagrangian that involves the type II TNC fields described above as well as NNLO fields. However we show that, in case we truncate the expansion after the NLO, we are allowed, without loss of generality, to impose the TTNC condition off shell via a Lagrange multiplier. We refer to the resulting Lagrangian as \gls{nrg}. It is presented in equation  \eqref{eq:action_NRG}. We also present the equations of motion that result from this action.

In our earlier paper a Lagrangian on these type II TNC fields was obtained via another method, which we review here including many details that were not given in \cite{Hansen:2018ofj}. This method employs the type II TNC gauge symmetries and constructs the unique two-derivative action respecting this symmetry, starting with the correct kinetic term required for Newton's law of gravitation and then completing the full action. The result is presented in equation \eqref{eq:action_NRG_prime}. We also give the resulting form of the equations of motion that follow from that action. We furthermore show that, as expected,  the two Lagrangians \eqref{eq:action_NRG} and \eqref{eq:action_NRG_prime} are identical. The difference between the two non-relativistic Lagrangians stems from the fact that slightly different geometric variables are used. Depending on taste and type of application, one can work with either one of them.

In Section \ref{sec:coupling_to_matter} we discuss the general properties of matter coupled to type II TNC geometry. We work out the $1/c^2$ expansion of the energy-momentum tensor and relate this to the responses, with respect to variations of the fields of the type II TNC geometry, of the Lagrangians one obtains by expanding some matter Lagrangian order by order. We use this to derive the general matter coupled equations of motion for non-relativistic gravity. We also derive the form of the Ward identities resulting from various gauge invariances. These provide the analogue of the conservation of the relativistic energy-momentum tensor in the non-relativistic regime. We end Section \ref{sec:coupling_to_matter} by specifying what happens in the simpler case when the clock 1-form $\tau_\mu$ is closed which leads to NC gravity. We reproduce the well-known equations of motion of Newtonian gravity in equation \eqref{eq:mattercoupledNCgravity}.

We then proceed with a detailed analysis of the non-relativistic expansion and coupling of various types of well-known relativistic matter systems in Section \ref{sec:matter_examples}. In this section it will become clear that it is the matter sector that decides whether the geometry must have a closed clock 1-form, i.e. no torsion or whether TTNC geometry is allowed. We start with the simplest case, namely
that of a point particle. Already here the expansion exhibits a rich structure, revealing that there are two distinct cases depending on whether one has absolute time or one considers the more general case of TTNC geometry. For both cases the geodesic equation is obtained. We briefly consider the case of adding an electric charge and the role of the Lorentz force. As an illustration of the difference between Lorentzian and Newton--Cartan geometries and the role of geodesics we also
include a brief analysis of two-dimensional Rindler spacetime. Here, we will also see an example of the fact that
because of the analytic structure of the $1/c^2$ expansion, given a relativistic spacetime and two different charts that are related by a diffeomorphism that is not analytic in $c$, the $1/c^2$ expansion of the two charts will give rise to distinct, i.e. non-gauge equivalent charts of two non-relativistic spacetimes.

After this we treat the expansion of perfect fluids and show that there are different regimes depending on how we expand the energy and pressure as a function of $1/c^2$. We then turn to various important field theory examples, namely the case of a complex and a real scalar field as well as electrodynamics. In the case of a complex scalar field we show how we can expand this in $1/c^2$ such that we end up with the Lagrangian for the Schr\"odinger--Newton equation. This novel off shell description of the Schr\"odinger--Newton system includes fields whose equations of motion tell us that the clock 1-form must be closed. In the case of electrodynamics it is well-known that there are two limits, a magnetic and an electric limit depending on how we expand the gauge connection. We discuss the Lagrangian descriptions of both the magnetic and the electric expansion of Maxwell's theory.

A detailed and more extensive study of solutions is left for the future, but we conclude the paper by presenting some of the simplest solutions of the non-relativistic gravity action in Section \ref{sec:solutions}. 
In our list of solutions we first discuss two different expansions of the Schwarzschild solution depending on whether we treat the mass parameter as constant or as being of order $c^2$ when we expand in $1/c^2$, following \cite{VandenBleeken:2017rij}.
In the former case we obtain the well-known NC solution for a massive point particle while in the latter case we find a non-trivial TTNC geometry with spherical symmetry. We then proceed by studying the geodesics in this spherically symmetric TTNC background and we observe that the geodesic equations of motion for orbital motion around a centre are the same as in GR. This can be used to show that we can describe the three classical tests of GR using this non-relativistic perspective \cite{Hansen:2019vqf}. Next we obtain the non-relativistic analogue of the TOV equation. The main result here is that the resulting equations are again the same as in GR. Thus, the physical structure of fluid stars can be correctly described by non-relativistic gravity. We conclude by analysing cosmological \gls{flrw} metric solutions for which it is shown that, in parallel with the results above for massive objects, non-relativistic gravity yields the same Friedmann equations as one would obtain from GR. We end the section with some comments about the NR expansion of \gls{ads} spacetime in various coordinates. 

The paper is concluded with a discussion and outlook in Section \ref{sec:discussion}.

Many of the more technical details are collected in various appendices. In Appendix \ref{sec:notation_conventions} we discuss our notation and conventions.
Appendix \ref{sec:NC review} provides a review of the main elements of type I TNC geometry.
Newtonian gravity is presented in \ref{sec:null-reduction}. This appendix also includes a discussion about the null reduction of GR (which is invariant under type I NC gauge transformations) to contrast it with Newtonian gravity. We present an argument showing that Newtonian gravity cannot originate from a type I NC gauge invariant theory.
Appendix \ref{sec:NC_identities} contains a collection of many useful identities in TNC geometry without restriction on the type of clock 1-form we use.
A large number of variational identities, necessary for obtaining the equations of motion of the non-relativistic Lagrangians discussed in the main text are also collected in Appendix \ref{sec:NC_identities}. The final appendix
\ref{sec:TTNC_identities_special} contains many useful results that apply to TTNC geometries.

\section{Expansion generalities \label{sec:expgen} }
In this first section we set up the framework for working with $1/c^2$ expansions of field theories and geometry in a systematic way.
A very useful so-called ``pre-non-relativistic" parameterisation of Lorentzian geometry is defined.
This gives a convenient starting point for studying \regls{nr} expansions of the geometry in both vielbein and metric formalisms.
We then continue to show that a $1/c^2$ expansion of a Cartan connection taking values in the Poincaré algebra reveals the underlying NR local symmetry algebra of \regls{nrg}.
With the tools needed to study $1/c^2$ expansions of general Lagrangians then fully developed, we are prepared for tackling the \regls{eh} Lagrangian and \regls{gr}.
Finally, we also study \glspl{wi} and \reglspl{eom} in ``pre-non-relativistic'' parameterisation.

\subsection{Non-relativistic expansions}\label{sec:expansion_causal_structure}
The distinguishing feature of \regls{nr} physics is that in tangent space the light-cone is flattened out completely because of the causal structure of spacetime.
In Lorentzian geometry the slope of the light-cone is $1/c$, with $c$ denoting the speed of light. This means that in order to relate this to non-relativistic physics we need to perform an expansion around $c=\infty$.
With $c$ being dimensionful, what is meant more precisely by this statement is that we set $c=\hat c/\sqrt{\sigma}$, with $\sigma$ a small dimensionless parameter which in the non-relativistic expansion is expanded around $0$.
For convenience we choose units in which $\hat c=1$.
We will consider in this paper the most conventional case of expanding in even powers of $1/c$, i.e. in $\sigma= 1/c^2$.
Thus, up to an overall factor of $c$ to some power, actions and equations of motion of relativistic gravity and matter theories are studied in a $1/c^2$ expansion
\footnote{See \cite{Ergen:2020yop} for a study that also includes odd powers in $1/c$.}. 

We begin by discussing some general considerations related to the expansions of the fields \cite{Hansen:2019svu}.
Our starting assumption is that, up to an overall power of $c$ which will be factored out, any field $\phi^I(x;\sigma)$ (with $I$ a shorthand for all spacetime and/or internal indices) is analytic in $\sigma$ such that 
it enjoys a Taylor expansion around $\sigma=0$,
\begin{equation}\label{eq:expansion_field_general}
\phi^I(x;\sigma)=\phi^I_{(0)}(x)+\sigma\phi^I_{(2)}(x)+\sigma^2\phi^I_{(4)}(x)+\order{\sigma^3}\,,
\end{equation}
where $\phi^I_{(n)}(x)$ indicates that this is the coefficient of $c^{-n}$.
We will apply this expansion to both  the spacetime fields of relativistic gravity as well as other types of relativistic (bosonic) fields that couple to relativistic geometry.

The first main interest of this paper is to consider the expansion of \regls{gr} itself, and hence we first turn to applying it to the fields characterising a $(d+1)$-dimensional Lorentzian manifold, which are taken here to be the relativistic  vielbeine $E^A_\mu$, with spacetime indices $\mu =  0 \ldots d$ and tangent space indices $A=0 \ldots d$. Importantly, we need to explicitly choose the
overall factors of $c$ in these, such that the remaining fields have
the expansion \eqref{eq:expansion_field_general} starting at order $\sigma^0$.
The light-cone structure of the spacetime implies that the timelike vielbeine should scale different with $c$ as compared to the spacelike ones. Thus we write the vielbeine and their inverses as
\begin{eqnarray}
E^A_\mu&=&cT_\mu\delta^A_0+\mathcal{E}^a_\mu\delta^A_a\,,\label{eq:Viel1}\\
E_A^\mu&=&-c^{-1}T^\mu\delta_A^0+\mathcal{E}_a^\mu\delta_A^a\,,\label{eq:Viel2}
\end{eqnarray}
where the flat metric is $\eta_{AB}=\mathrm{diag}\left(-1,1,\ldots,1\right)$. Hence the spatial tangent space indices $a,b$ are raised and lowered with the Kronecker delta. We will denote this way
of parameterising the vielbeine as the pre-non-relativistic parameterisation of Lorentzian geometry. The fields $T_\mu,\,T^\mu,\,\mathcal{E}^a_\mu,\,\mathcal{E}_a^\mu$ are assumed to be analytic in $\sigma=1/c^2$ and exhibit the Taylor expansion  \eqref{eq:expansion_field_general}.
Since the relativistic vielbeine satisfy the completeness relations $E^{\mu}_A E_{\nu}^A=\delta_\nu^\mu$ and $E^{\mu}_B E_{\mu}^A=\delta_B^A$ we have the relations 
\begin{equation}\label{eq:completeness_relations_vielbeine_causal}
T_\mu\mathcal{E}^{\mu}_a=0\,,\qquad T^\mu\mathcal{E}_{\mu}^a=0\,,\qquad T_\mu T^\mu=-1\,,\qquad \mathcal{E}^{\mu}_a\mathcal{E}_{\mu}^b=\delta^b_a\,,\qquad \mathcal{E}^{\mu}_a\mathcal{E}_{\nu}^a=\delta_\nu^\mu+T^\mu T_\nu \,.
\end{equation}

The vielbeine transform under the gauge transformations of general relativity as $\delta E_{\mu}^A =\mathcal{L}_\Xi E_{\mu}^A+{\Lambda^A}_B E_{\mu}^B$, where $\Xi^\mu$ is a vector field generating the diffeomorphisms and ${\Lambda^A}_B=c^{-1}{\Lambda}_b \delta^A_0 \delta^b_B+c^{-1}{\Lambda^a} \delta^A_a \delta^0_B+{\Lambda^a}_b \delta^A_a \delta^b_B$ corresponds to the generator of infinitesimal local Lorentz transformations, where ${\Lambda}_{AB}=-{\Lambda}_{BA}$ and where we defined $\Lambda^0{}_b=c^{-1}\Lambda_b$. The factors of $c^{-1}$ follow from demanding that the local Lorentz transformations respect the appearance of $c$ in \eqref{eq:Viel1} and \eqref{eq:Viel2}. The inverse vielbeine transform as
$\delta E^{\mu}_A =\mathcal{L}_\Xi E^{\mu}_A-{\Lambda^B}_A E^{\mu}_B$. We thus find
\begin{eqnarray}
\delta T_\mu &=& \mathcal{L}_\Xi T_\mu  + c^{-2}{\Lambda}_b \mathcal{E}_\mu^b\,,\label{eq:transform_vielbeine_causal_expansion}\\
\delta \mathcal{E}^a_\mu &=& \mathcal{L}_\Xi \mathcal{E}^a_\mu + {\Lambda^a} T_\mu +{\Lambda^a}_b \mathcal{E}_\mu^b \,,\label{eq:transform_vielbeine_causal_expansion1}\\
\delta T^\mu &=& \mathcal{L}_\Xi T^\mu  + {\Lambda^b} \mathcal{E}^\mu_b\,,\label{eq:transform_vielbeine_causal_expansion_inv}\\
\delta \mathcal{E}_a^\mu &=& \mathcal{L}_\Xi \mathcal{E}_a^\mu -{\Lambda^b}_a \mathcal{E}^\mu_b + c^{-2} {\Lambda}_a T^\mu \,.\label{eq:transform_vielbeine_causal_expansion1_inv}
\end{eqnarray}

In terms of the geometric fields defined in \eqref{eq:Viel1}, \eqref{eq:Viel2}, the Lorentzian metric and its inverse take the form 
\begin{eqnarray}\label{eq:metrics_explicit_c_factors_no_expansion}
g_{\mu\nu} &\equiv& \eta_{AB} E^A_\mu E^B_\nu =- c^2 T_\mu T_\nu+\delta_{ab}\mathcal{E}^a_\mu\mathcal{E}^b_\nu\,,\\
g^{\mu\nu} &\equiv& \eta^{AB} E_A^\mu E_B^\nu =-\frac{1}{c^2} T^\mu T^\nu+\delta^{ab}\mathcal{E}_a^\mu\mathcal{E}_b^\nu\,.
\end{eqnarray}
We will define the spatial part of the metric and its inverse as 
\begin{equation}\label{eq:defPi}
\Pi^{\mu\nu}\equiv\delta^{ab}\mathcal{E}_a^\mu\mathcal{E}_b^\nu\,,\qquad \Pi_{\mu\nu}\equiv\delta_{ab}\mathcal{E}^a_\mu\mathcal{E}^b_\nu\,.
\end{equation}
These transform under the gauge transformations as
\begin{eqnarray}
\delta \Pi_{\mu\nu} &=& \mathcal{L}_\Xi \Pi_{\mu\nu} + \Lambda_a T_\mu \mathcal{E}^a_\nu+ \Lambda_a T_\nu \mathcal{E}^a_\mu\,,\label{eq:Pi_trafo1}\\
\delta \Pi^{\mu\nu} &=& \mathcal{L}_\Xi \Pi^{\mu\nu} + c^{-2}\Lambda^a T^\mu \mathcal{E}_a^\nu + c^{-2}\Lambda^a T^\nu \mathcal{E}_a^\mu\label{eq:Pi_trafo2}\,.
\end{eqnarray}
Note that we have the relations 
\begin{equation}\label{eq:completenessTandPi}
T_\mu\Pi^{\mu\nu}=0\,,\qquad T^\mu\Pi_{\mu\nu}=0\,,\qquad T_\mu T^\mu=-1\,,\qquad\Pi_{\mu\rho}\Pi^{\rho\nu}=\delta^\nu_\mu+T^\nu T_\mu\,.
\end{equation}
We will see in Section \ref{sec:causal_expansion_EH} that the pre-non-relativistic form of the metric in \eqref{eq:metrics_explicit_c_factors_no_expansion} enables to recast GR in a form that significantly simplifies its  non-relativistic expansion.

The main goal at this stage is thus to rewrite GR in terms of fields whose $1/c$ expansion starts at order $c^0$ and whose leading order fields are unconstrained. Note that the metric \eqref{eq:metrics_explicit_c_factors_no_expansion} does not satisfy these criteria because $g_{\mu\nu}$ starts at order $c^2$ with a leading order term that is constrained to be a product of two 1-forms. We thus need to write the \regls{eh} Lagrangian in terms of the fields $T_\mu$ and $\Pi_{\mu\nu}$. To this end it will prove convenient to work with a different connection than the usual Levi--Civita connection and consider what happens to the curvature of the spacetime with explicit factors of $c$ appearing due to \eqref{eq:metrics_explicit_c_factors_no_expansion}.
In the following we will denote the power in $1/c$ with an overscript and leave the fields $T_\mu$ and $\mathcal{E}^a_\mu$ unexpanded 
(see also appendix \ref{sec:notation_conventions} for notation conventions).

We can write the Christoffel connection $\Gamma^\rho_{\mu\nu} = \frac{1}{2} g^{\rho\lambda}\left(\partial_\mu g_{\nu\lambda}+\partial_\nu g_{\mu\lambda}-\partial_\lambda g_{\mu\nu}\right)$ as
\begin{equation}\label{eq:LC_connection_explicit_factors1c2}
\Gamma^\rho_{\mu\nu}=c^2\;\os{C}{-2}^{\rho}_{\mu\nu}+\os{C}{0}^{\rho}_{\mu\nu}+c^{-2}\os{C}{2}^{\rho}_{\mu\nu}\,,
\end{equation}
Here we define 
\begin{equation}
\os{C}{0}^{\rho}_{\mu\nu}\equiv C^\rho_{\mu\nu}+S^\rho_{\mu\nu}\,,
\end{equation}
with $C^\rho_{\mu\nu}$ the ``pre-non-relativistic'' connection 
\begin{equation}\label{eq:C_connection_notexpanded}
    C^{\rho}_{\mu\nu} = -T^\rho\partial_\mu T_\nu+\frac{1}{2}\Pi^{\rho\sigma}\left(\partial_\mu\Pi_{\nu\sigma}+\partial_\nu\Pi_{\mu\sigma}-\partial_\sigma\Pi_{\mu\nu}\right)\,,
\end{equation}
and the remaining terms given by 
\begin{eqnarray}
\os{C}{-2}^{\rho}_{\mu\nu} & = & \frac{1}{2}T_\mu\Pi^{\rho\sigma}\left(\partial_\sigma T_\nu-\partial_\nu T_\sigma\right)+\frac{1}{2}T_\nu\Pi^{\rho\sigma}\left(\partial_\sigma T_\mu-\partial_\mu T_\sigma\right)\,,\\
S^{\rho}_{\mu\nu} & = & \frac{1}{2}T^\rho\left(\partial_\mu T_\nu-\partial_\nu T_\mu-T_\mu\mathcal{L}_T T_\nu-T_\nu\mathcal{L}_T T_\mu\right)\,,  \label{Stensor}
\\
\os{C}{2}^{\rho}_{\mu\nu} & = & \frac{1}{2}T^\rho\mathcal{L}_T\Pi_{\mu\nu}\,,
\end{eqnarray}
where all the Lie derivatives are with respect to $T^\mu$. We emphasise that there is still the implicit $c$ dependence of the background fields, which will be dealt with later. 

It will be useful in the following to use the torsionful connection $C^{\rho}_{\mu\nu}$ in \eqref{eq:C_connection_notexpanded} for calculating covariant derivatives. We denote  this by $\os{\nabla}{C}_\mu$ and the associated Riemann tensor $\os{R}{C}_{\mu\nu\rho}{}^\sigma$ is defined in the usual way. The covariant derivative $\os{\nabla}{C}_\mu$ obeys
\begin{equation}
\os{\nabla}{C}_\mu T_\nu=0\,,\qquad \os{\nabla}{C}_\mu\Pi^{\nu\rho}=0\,,\qquad \os{\nabla}{C}_\mu T^\nu=\frac{1}{2}\Pi^{\nu\rho}\mathcal{L}_T\Pi_{\rho\mu}\,,\qquad\os{\nabla}{C}_\mu \Pi_{\nu\rho}=T_{(\nu}\mathcal{L}_T\Pi_{\rho)\mu}\,.
\end{equation}

The Ricci tensor $R_{\mu\nu}$ associated with the Levi--Civita connection is $1/c^2$-expanded as
\begin{equation}
R_{\mu\nu}=c^4\;\os{R}{-4}_{\mu\nu}+c^2\; \os{R}{-2}_{\mu\nu}+\os{R}{0}_{\mu\nu}+c^{-2}\os{R}{2}_{\mu\nu}\,,
\end{equation}
where
\begin{eqnarray}
\os{R}{-4}_{\mu\nu} &=& \frac{1}{4}T_{\mu}T_\nu\Pi^{\alpha\beta}\Pi^{\rho\sigma}T_{\alpha\rho}T_{\beta\sigma}\,,\\
\os{R}{-2}_{\mu\sigma}&=&\os{\nabla}{C}_\nu \os{C}{-2}^\nu_{\mu\sigma}+\os{C}{-2}^\lambda_{\mu\sigma}S^\nu_{\nu\lambda}-\os{C}{-2}^{\nu}_{\mu\lambda}S^{\lambda}_{\nu\sigma}-\os{C}{-2}^{\lambda}_{\nu\sigma}S^{\nu}_{\mu\lambda}-2C^\lambda_{[\mu\nu]}\os{C}{-2}^\nu_{\lambda\sigma}\,,\\
\os{R}{0}_{\mu\sigma}&=&\os{R}{C}_{\mu\sigma}-\os{C}{-2}^{\nu}_{\mu\lambda}\os{C}{2}^{\lambda}_{\nu\sigma}-\os{C}{-2}^{\nu}_{\sigma\lambda}\os{C}{2}^{\lambda}_{\nu\mu}-\os{\nabla}{C}_\mu S^\nu_{\nu\sigma}+\os{\nabla}{C}_\nu S^\nu_{\mu\sigma}-2C^\lambda_{[\mu\nu]}S^\nu_{\lambda\sigma}\,,\\
\os{R}{2}_{\mu\sigma}&=&\os{\nabla}{C}_\nu \os{C}{2}^\nu_{\mu\sigma}\,.
\end{eqnarray}
It is useful to note from \eqref{Stensor} that $S^\rho_{\mu\nu}T^\nu=0$
and that the $C$-connection satisfies 
\begin{equation}
C^\rho_{\rho\nu}=-\mathcal{L}_T T_\nu+\partial_\nu\log\sqrt{-\text{det}\,(-T_\alpha T_\beta+\Pi_{\alpha\beta})}\,.
\end{equation}
One can furthermore derive that the components of the Ricci tensor obey
\begin{eqnarray}
\Pi^{\mu\nu}\os{R}{-2}_{\mu\nu} & = &  \frac{1}{2}\Pi^{\alpha\beta}\Pi^{\rho\sigma}T_{\alpha\rho}T_{\beta\sigma}\,,\\
T^\mu T^\nu\os{R}{-2}_{\mu\nu} & = & 0\,,\\
\Pi^{\mu\nu}\os{R}{0}_{\mu\nu} & = & \Pi^{\mu\nu}\os{R}{C}_{\mu\nu}\,,\\
T^\mu T^\nu\os{R}{0}_{\mu\nu} & = & T^\mu T^\nu\os{R}{C}_{\mu\nu}\,,\\
\Pi^{\mu\nu}\os{R}{2}_{\mu\nu} & = & 0\,,\\
T^\mu T^\nu\os{R}{2}_{\mu\nu} & = & 0\,,
\end{eqnarray}
where we defined
\begin{equation}
T_{\mu\nu}\equiv\partial_\mu T_\nu-\partial_\nu T_\mu\,.
\end{equation}
Some of the identities above are only true up to a total derivative and we point out that integration by parts gives an extra term proportional to torsion:
\begin{equation}
E^{-1}\partial_\mu \left(EX^\mu\right) = \os{\nabla}{C}_\mu X^\mu +T_{\mu\nu} T^\mu X^\nu\,,
\end{equation}
for any vector field $X^\mu$ and where the integration measure is given by
\begin{equation}\label{eq:measure_unexpanded}
    cE \equiv \sqrt{-g}=c\sqrt{-\text{det}\,(-T_\alpha T_\beta+\Pi_{\alpha\beta})}\,.
\end{equation}
It follows that the Ricci scalar associated with the Levi--Civita connection 
takes the following form when expressed in terms of pre-non-relativistic fields and the $C$-connection
\begin{equation}\label{eq:RicciscalarTPi}
    R=\frac{c^2}{4}\Pi^{\mu\nu}\Pi^{\rho\sigma}T_{\mu\rho}T_{\nu\sigma}+\Pi^{\mu\nu}\os{R}{C}_{\mu\nu}-\frac{1}{c^2}T^\mu T^\nu\os{R}{C}_{\mu\nu}\,.
\end{equation}
This result will be useful when we expand the Einstein--Hilbert Lagrangian in Section \ref{sec:causal_expansion_EH}.

\subsection{Vielbeine}\label{sec:expansion:bckg_fields}

By assumption the fields are taken to be analytic in $\sigma=1/c^2$ and they thus admit a Taylor expansion. For the vielbeine this means that they can be expanded to subleading orders as
\begin{eqnarray}
T_\mu & = & \tau_\mu+c^{-2}m_\mu+c^{-4}B_\mu+\order{c^{-6}}\,,\label{eq:Tmu}\\
\mathcal{E}^a_\mu & = & e^a_\mu+c^{-2}\pi^a_\mu+\order{c^{-4}}\,. \label{eq:Emu}
\end{eqnarray}
For the inverse vielbeine we have 
\begin{eqnarray}
T^\mu & = & v^\mu+c^{-2}\left(v^\mu v^\rho m_\rho-e^\mu_b v^\rho\pi_\rho^b\right)+\order{c^{-4}}\,,\label{eq:Tmu-up}\\
\mathcal{E}^\mu_a & = & e^\mu_a + c^{-2}\left(v^\mu m_a-e^\mu_c e^\rho_a\pi^c_\rho\right)+\order{c^{-4}}\,,
\end{eqnarray}
where $m_a \equiv e^\rho_a m_\rho$ as well as the $1/c^2$ expansion of the completeness relations \eqref{eq:completeness_relations_vielbeine_causal}.
The leading order fields in \eqref{eq:Tmu}, \eqref{eq:Emu} satisfy the completeness relations
\begin{equation}
\tau_\mu e^{\mu}_a=0\,,\qquad v^\mu e_{\mu}^a=0\,,\qquad \tau_\mu v^\mu=-1\,,\qquad e^{\mu}_a e_{\mu}^b=\delta^b_a\,,\qquad e^{\mu}_a e_{\nu}^a=\delta_\nu^\mu+v^\mu \tau_\nu \,,
\end{equation}
which are simply the leading order counterparts of \eqref{eq:completenessTandPi}.

Let us now turn our attention towards the gauge transformation of the various fields.
The components of the infinitesimal local Lorentz transformation $\Lambda^A{}_B$ and diffeomorphism generating vector $\Xi^\mu$ are expanded as
\begin{eqnarray}
\Xi^\mu & = & \xi^\mu+c^{-2}\zeta^\mu+\order{c^{-4}}\,,\label{eq:diffeo_expansion}\\
\Lambda^a & = & \lambda^a+c^{-2}\eta^a+\order{c^{-4}}\,,\\
\Lambda^a{}_b & = & \lambda^a{}_b+c^{-2}\sigma^a{}_b+\order{c^{-4}}\,,\label{eq:Lambdaexp}
\end{eqnarray}
where $\Lambda^a{}_0=c^{-1}\Lambda^a$. This ensures that the gauge transformations respect the $1/c^2$ expansion of the fields. The interpretation is that $\xi^\mu$ is a diffeomorphism generating vector field, while $\zeta^\mu$ (being subleading) generates gauge transformations that act on the subleading fields $m_\mu$ and $\pi_\mu^a$. They will be studied in more detail below.
In particular, we will see that $\lambda^a$ is a local Galilean (Milne) boost and $\eta^a$ its  subleading version, while the parameter
$\lambda^a{}_b$ corresponds to a local spatial rotation and $\sigma^a{}_b$ its
subleading version. 

It is important to realise that because of the analytic structure of the $1/c^2$ expansion, given a relativistic spacetime and two different charts that are related by a diffeomorphism that is not analytic in $c$, the $1/c^2$ expansion of the two charts will give rise to distinct, i.e. non-gauge equivalent charts of two non-relativistic spacetimes.
We shall give two examples of this in Sections \ref{subsubsec:Rindler} and \ref{sec:AdS_spacetime_1c} for the $1/c^2$ expansion of 2-dimensional Rindler spacetime and of \regls{ads} spacetime in any dimension.

Expanding the parameters and fields in \eqref{eq:transform_vielbeine_causal_expansion}-\eqref{eq:transform_vielbeine_causal_expansion1} leads to the following transformations deduced from collecting terms order by order in $1/c^2$:
\begin{eqnarray}
\delta\tau_\mu & = & \mathcal{L}_\xi\tau_\mu\,,\label{eq:trafo1}\\
\delta m_\mu & = & \mathcal{L}_\xi m_\mu+\mathcal{L}_\zeta\tau_\mu+\lambda_a e^a_\mu\,,\label{eq:trafo2}\\
\delta e_\mu^a & = & \mathcal{L}_\xi e_\mu^a+\lambda^a\tau_\mu+\lambda^a{}_b e_\mu^b\,,\label{eq:trafo3}\\
\delta\pi_\mu^a & = & \mathcal{L}_\xi\pi_\mu^a+\mathcal{L}_\zeta e_\mu^a+\lambda^a m_\mu+\eta^a\tau_\mu+\lambda^a{}_b\pi_\mu^b+\sigma^a{}_b e^b_\mu\,.\label{eq:trafo4}
\end{eqnarray}
For the inverse vielbeine the leading order terms transform as
\begin{eqnarray}
\delta v^\mu & = & \mathcal{L}_\xi v^\mu+\lambda^a e^\mu_a\,,\label{eq:trafov}\\
\delta e^\mu_a & = & \mathcal{L}_\xi e^\mu_a+\lambda_a{}^b e^\mu_b\,.
\end{eqnarray}

In appendix \ref{sec:NC review} we give a review of (torsional) \regls{nc} geometry\footnote{The relation between the vielbeine $e_\mu^a$ and $h_{\mu\nu}$ used in appendix \ref{sec:NC review} is given in the equation \eqref{eq:defh}. Equation \eqref{eq:definvh} gives a similar result for the inverses}. By contrasting this with the $1/c^2$ expansion of a Lorentzian geometry we see from \eqref{eq:trafo1}-\eqref{eq:trafo4} that there are two noticeable differences. The first one is the transformation rule for $m_\mu$. This agrees with equation \eqref{eq:typeItrafom} only when $\tau_\mu$ is exact. Secondly the $1/c^2$ expansion gives us a new field, namely $\pi_\mu^a$. For these reasons we will refer to the geometry originating from the $1/c^2$ expansion as {\it{type II}} NC geometry and the version of NC geometry reviewed in appendix \ref{sec:NC review} as {\it{type I}} NC geometry.

\subsection{Metric}\label{sec:expansion_metric}

In Section \ref{sec:expansion:bckg_fields} we found that $T_\mu,\,T^\mu$ are expanded as in equations \eqref{eq:Tmu} and \eqref{eq:Tmu-up}, respectively. The fields $\Pi_{\mu\nu},\,\Pi^{\mu\nu}$, defined in \eqref{eq:defPi}, admit the following $1/c^2$ expansions
\begin{eqnarray}
\Pi_{\mu\nu} & = & h_{\mu\nu}+c^{-2}\Phi_{\mu\nu}+\order{c^{-4}}\,,\\
\Pi^{\mu\nu} & = & h^{\mu\nu}+c^{-2}\left(2h^{\rho(\mu}v^{\nu)}m_\rho - h^{\mu\rho}h^{\nu\sigma}\Phi_{\rho\sigma} \right) +\order{c^{-4}}\,.
\end{eqnarray}
where we defined, in terms of the vielbeine of Section \ref{sec:expansion:bckg_fields}, the spacetime tensors
\begin{eqnarray}
h_{\mu\nu} & = & \delta_{ab}e_\mu^a e_\nu^b\,,\label{eq:defh}\\
h^{\mu\nu} & = & \delta^{ab}e^\mu_a e^\nu_b\,,\label{eq:definvh}\\
\Phi_{\mu\nu} & = & \delta_{ab}\left(e_\mu^a \pi_\nu^b+e_\nu^a\pi_\mu^b \right)\,.
\end{eqnarray}

With these conventions the expansion of the metric \eqref{eq:metrics_explicit_c_factors_no_expansion} can be written concisely as
\begin{eqnarray}
g_{\mu\nu} &=&-c^2\tau_\mu \tau_\nu+\bar h_{\mu\nu}+c^{-2}\bar \Phi_{\mu\nu}+\order{c^{-4}}\,,\label{eq:metric_expansion1}\\
g^{\mu\nu} &=& h^{\mu\nu}-c^{-2}\left(\hat v^\mu \hat v^\nu + h^{\mu\rho}h^{\nu\sigma}\bar \Phi_{\rho\sigma} \right) +c^{-4}\left(2\hat v^\mu \hat v^\nu \hat \Phi+Y^{\mu\nu}\right) +\order{c^{-6}}\label{eq:metric_expansion2}\,,
\end{eqnarray}
where we defined the tensors
\begin{eqnarray}
\bar h_{\mu\nu} & \equiv & h_{\mu\nu}-2\tau_{(\mu}m_{\nu)}\,,\label{eq:def_barh}\\
\hat v^\mu & \equiv & v^\mu - h^{\mu\rho}m_\rho\,,\label{eq:def_hatv}\\
\hat \Phi  & \equiv & -v^{\mu}m_{\mu}+\frac{1}{2}h^{\mu\nu}m_{\mu}m_{\nu}\,,\label{eq:def_hatPhi}\\
\bar \Phi_{\mu\nu} & \equiv &\Phi_{\mu\nu}-m_\mu m_\nu-2B_{(\mu}\tau_{\nu)}\,,\label{eq:def_barPhi}
\end{eqnarray}
and we have 
\begin{equation}
    \tau_\mu\tau_\nu Y^{\mu\nu} = 0\,.
\end{equation}
We will not need any other components of $Y^{\mu\nu}$. Notice that the contribution from the $\order{c^{-4}}$ term in $T_\mu$, the \regls{nnlo} field $B_\mu$, enters in $\bar \Phi_{\mu\nu}$.

The \regls{lo} terms in the expansion of the metric and its inverse define the metric structures of the non-relativistic geometry. This is thus given by the clock form via $\tau_\mu\tau_\nu$ and the inverse spatial metric $h^{\mu\nu}$ which has rank $d$ and thus one zero eigenvalue. Its kernel is spanned by $\tau_\mu$. The objects $v^\mu$ and $h_{\mu\nu}$ are not metric tensors because they transform under the Milne boosts with parameter $\lambda^a$ as can be seen from equations \eqref{eq:trafo3} and \eqref{eq:trafov}. We have the Milne boost invariant relations 
\begin{equation}
\tau_\mu h^{\mu\nu}=0\,,\qquad v^\mu h_{\mu\nu}=0\,,\qquad \tau_\mu v^\mu=-1\,,\qquad h_{\mu\rho} h^{\rho\nu}=\delta^\nu_\mu+v^\nu \tau_\mu\,.
\end{equation}
The \regls{nlo} fields $m_\mu,\,\Phi_{\mu\nu}$ are to be thought of as gauge fields in this context.
The 1-form $m_\mu$ is related to the Newtonian potential
\begin{equation}
\Phi \equiv -v^\mu m_\mu\,.    
\end{equation}
The NLO field $\Phi_{\mu\nu}$ is less well-known but also appeared in our previous work \cite{Hansen:2018ofj}.

Under a diffeomorphism generated by $\Xi^\mu$ the metric transforms as $\delta g_{\mu\nu}=\mathcal{L}_\Xi g_{\mu\nu}$. Expanding the transformation of the metric using $\Xi^\mu$ as given by \eqref{eq:diffeo_expansion} we find that the fields in the expansions \eqref{eq:metric_expansion1}-\eqref{eq:metric_expansion2} transform as
\begin{eqnarray}
\delta\tau_\mu & = & \mathcal{L}_\xi\tau_\mu\,,\label{eq:trafo5a}\\
\delta\bar h_{\mu\nu} & = & \mathcal{L}_\xi\bar h_{\mu\nu}-\tau_\mu\mathcal{L}_\zeta \tau_\nu-\tau_\nu\mathcal{L}_\zeta \tau_\mu\,,\label{eq:trafo7}\\
\delta h^{\mu\nu} & = & \mathcal{L}_\xi h^{\mu\nu}\,,\label{eq:trafo6}\\
\delta \hat v^\mu & = &  \mathcal{L}_\xi \hat v^\mu  - h^{\mu\rho}\mathcal{L}_\zeta \tau_\rho\,,\label{eq:trafo8}\\
\delta\hat\Phi & = & \mathcal{L}_\xi\hat\Phi-\hat v^\mu\mathcal{L}_\zeta\tau_\mu\,,\label{eq:trafo10}\\
\delta \bar\Phi_{\mu\nu} & = & \mathcal{L}_\xi \bar\Phi_{\mu\nu}+\mathcal{L}_\zeta \bar h_{\mu\nu}\label{eq:trafo9}\,,
\end{eqnarray}
Some of the fields appearing in the expansion of the vielbeine transform under Milne boosts with parameter $\lambda^a$. The combinations $\bar h_{\mu\nu}$, $\hat v^\mu$ and $\bar\Phi_{\mu\nu}$ that appear in the metric are all Milne boost invariant. In fact they are also invariant under the spatial rotations $\lambda^a{}_b$ and the subleading transformations with parameters $\eta^a$ and $\sigma^a{}_b$ that appear in the transformation of $\pi_\mu^a$.

Depending on the setting we will either work with the fields that appear in the expansions of $T_\mu$ and $\Pi_{\mu\nu}$, i.e. $\tau_\mu$, $m_\mu$, $h_{\mu\nu}$ and $\Phi_{\mu\nu}$ or we will work with the fields that appear in the expansion of the metric, i.e. $\tau_\mu$, $\bar h_{\mu\nu}$ and $\bar\Phi_{\mu\nu}$. 
The transformations of the first set of fields can be readily found from the results in this and the previous section and we have
\begin{eqnarray}
\delta h_{\mu\nu} & = & \mathcal{L}_\xi h_{\mu\nu}+\tau_\mu\lambda_\nu+\tau_\nu\lambda_\mu\,,\label{eq:trafo5}\\
\delta m_\mu & = & \mathcal{L}_\xi m_\mu+\lambda_\mu+\partial_\mu \Lambda -\Lambda a_\mu+h^{\rho\sigma}\zeta_\sigma\left(\partial_\rho\tau_\mu-\partial_\mu\tau_\rho\right)\,,\label{eq:torsional_u(1)_1}\\
\delta \Phi_{\mu\nu} & = & \mathcal{L}_\xi \Phi_{\mu\nu}+2\lambda_a\left(\tau_{(\mu}\pi^a_{\nu)}+m_{(\mu}e^a_{\nu)}\right)+2\eta_a e^a_{(\mu}\tau_{\nu)}+2\Lambda K_{\mu\nu}+\check{\nabla}_\mu\zeta_\nu+\check{\nabla}_\nu\zeta_\mu\,,\label{eq:torsional_u(1)_2}
\end{eqnarray}
where $\lambda_\mu\equiv e_\mu^a \lambda_a$ is the Galilean boost parameter which obeys $v^\mu\lambda_\mu=0$, $\eta_a$ is the parameter for subleading boosts and where we wrote the subleading diffeomorphisms $\zeta^\mu$ as
\begin{equation}
\zeta^\mu = -\Lambda v^\mu+h^{\mu\nu}\zeta_\nu\,.\label{eq:torsionalU1_subleading_diffeo}
\end{equation}
We furthermore defined the torsion vector
\begin{equation}
a_\mu \equiv \mathcal{L}_v \tau_\mu\,,
\end{equation}
and the extrinsic curvature
\begin{equation}
K_{\mu\nu} \equiv -\frac{1}{2}\mathcal{L}_v h_{\mu\nu}\,.
\end{equation}
It is convenient to define a covariant derivative with respect to the torsionful connection $\check\Gamma^\rho_{\mu\nu}$ defined as the leading order of $C^\rho_{\mu\nu}$ \eqref{eq:C_connection_notexpanded}:
\begin{equation}
\check\Gamma^\rho_{\mu\nu} \equiv \left.C^\rho_{\mu\nu}\right|_{\sigma=0} = -v^\rho\partial_\mu\tau_\nu+\frac{1}{2}h^{\rho\sigma}\left(\partial_\mu h_{\nu\sigma}+\partial_\nu h_{\mu\sigma}-\partial_\sigma h_{\mu\nu}\right)\,.
\end{equation}
This is the minimal collection of terms that transforms as an affine connection under diffeomorphisms, the remaining terms in the expansion of the Levi--Civita connection are tensorial.
$\check\Gamma^\rho_{\mu\nu}$ is a Newton--Cartan metric compatible connection satisfying the properties \eqref{eq:Cov_Der_check_metrics_1}-\eqref{eq:Cov_Der_check_metrics_4}, but it does transform under local Galilean boosts.

\subsection{Poincar\'e algebra}\label{sec:Poincare_algebra_expansion}

It is well-known that the properties of Lorentzian geometry can be understood by starting from a Cartan connection that takes values in the Poincar\'e algebra. It is therefore natural to study the $1/c^2$ expansion from this algebraic point of view using the method of Lie algebra expansions.
The latter has been considered e.g. in \cite{deAzcarraga:2002xi,Izaurieta:2006zz,Khasanov:2011jr} and 
recently been applied to the $1/c^2$ expansion of the Poincar\'e algebra in 
\cite{Hansen:2019vqf} and subsequently in \cite{Bergshoeff:2019ctr,Gomis:2019fdh,Gomis:2019sqv}. 

Writing the Poincar\'e generators as $T_I=\{H,P_a,B_a,J_{ab}\}$, where $H$ is the Hamiltonian, $P_a$ the spatial momenta, $B_a$ the Lorentz boost and $J_{ab}$ the spatial rotations and re-instating all factors of $\sigma$ in the structure constants, the Poincaré algebra becomes
\begin{eqnarray}\label{eq:Poincare_algebra_explicit_c}
\left[H,B_a\right]&=&P_a\,,\quad\left[P_a,B_b\right]=\sigma\delta_{ab}H\,\quad\left[B_a,B_b\right]=-\sigma J_{ab}\,,\nonumber\\
\left[J_{ab},P_c\right]&=&\delta_{ac}P_b-\delta_{bc}P_a\,,\quad\left[J_{ab},B_c\right]=\delta_{ac}B_b-\delta_{bc}B_a\,,\nonumber\\ \left[J_{ab},J_{cd}\right]&=&\delta_{ac}J_{bd}-\delta_{bc}J_{ad}-\delta_{ad}J_{bc}+\delta_{bd}J_{ac}\,.
\end{eqnarray}

The Cartan connection is 
\begin{equation}
    A_\mu=H T_\mu+P_a\mathcal{E}_\mu^a+B_a\Omega_\mu{}^a+\frac{1}{2}J_{ab}\Omega_\mu{}^{ab}\,,
\end{equation}
where the boost connection $\Omega_\mu{}^a$ and rotation connection $\Omega_\mu{}^{ab}$ together form the usual Lorentz connection. Let us schematically write this as $A_\mu=T_I\mathcal{A}_\mu^I$. If we now expand the gauge connections $\mathcal{A}_\mu^I=\sum_{n=0}^\infty \sigma^n\os{\mathcal{A}}{2n}_\mu^I$ then we obtain the new generators $T^{(n)}_I\equiv T_I \otimes \sigma^n$, where $n\ge 0$ will be referred to as the level.
One then obtains an algebra by expanding in the basis of generators $T^{(n)}$, with nonzero commutation relations of the form \cite{Hansen:2019vqf}
\begin{eqnarray}\label{eq:commutation_relations_ring}
\left[H^{(m)},B^{(n)}_a\right]&=&P^{(m+n)}_a\,,\quad\left[P^{(m)}_a,B^{(n)}_b\right]=\delta_{ab}H^{(m+n+1)}\,\quad\left[B^{(m)}_a,B^{(n)}_b\right]=-J^{(m+n+1)}_{ab}\,,\nonumber\\
\left[J^{(m)}_{ab},P^{(n)}_c\right]&=&\delta_{ac}P_b^{(m+n)}-\delta_{bc}P_a^{(m+n)}\,,\quad\left[J^{(m)}_{ab},B^{(n)}_c\right]=\delta_{ac}B_b^{(m+n)}-\delta_{bc}B_a^{(m+n)}\,,\nonumber\\ \left[J_{ab}^{(m)},J_{cd}^{(n)}\right]&=&\delta_{ac}J_{bd}^{(m+n)}-\delta_{bc}J_{ad}^{(m+n)}-\delta_{ad}J_{bc}^{(m+n)}+\delta_{bd}J_{ac}^{(m+n)}\,.
\end{eqnarray}

We can quotient out all generators with level $n>L$ for some $L$ which amounts to truncating the $1/c^2$ expansion. At the lowest level level $L=0$ the algebra is isomorphic to the Galilean algebra when identifying $H\equiv H^{(0)},P_a\equiv P_a^{(0)},G_a\equiv B_a^{(0)},J_{ab}\equiv J_{ab}^{(0)}$, where $G_a$ is the Galilean boost generator.
At the next level $L=1$ we have furthermore the generators $N\equiv H^{(1)},\,T_a\equiv P_a^{(1)},\,B_a\equiv B_a^{(1)},\,S_{ab}\equiv J_{ab}^{(1)}$. Written out in detail the non-zero commutation relations of the algebra obtained by modding out all levels $n>1$ are
\begin{eqnarray}\label{eq:NRG_algebra}
&\left[H\,,G_a\right]=P_a\,,\qquad \left[P_a\,,G_b\right]=N\delta_{ab}\,,\nonumber\\
&\left[N\,,G_a\right]=T_a\,,\qquad\left[H\,,B_a\right]=T_a\,,\qquad\left[S_{ab}\,,P_c\right]=\delta_{ac}T_b-\delta_{bc}T_a\,,\nonumber\\
&\left[S_{ab}\,,G_c\right]=\delta_{ac}B_b-\delta_{bc}B_a\,,\qquad\left[G_a\,,G_b\right]=-S_{ab}\,,\nonumber\\
&\left[J_{ab}\,,X_c\right]=\delta_{ac}X_b-\delta_{bc}X_a\,,\label{eq:expansion_tangent_space_algebra}\\
&\left[J_{ab}\,,J_{cd}\right]=\delta_{ac}J_{bd}-\delta_{bc}J_{ad}-\delta_{ad}J_{bc}+\delta_{bd}J_{ac}\,,\nonumber\\
&\left[J_{ab}\,,S_{cd}\right]=\delta_{ac}S_{bd}-\delta_{bc}S_{ad}-\delta_{ad}S_{bc}+\delta_{bd}S_{ac}\,,\nonumber
\end{eqnarray}
where $X_a$ denotes $P_a$, $T_a$, $G_a$ and $B_a$.

In particular one finds that $N=H^{(1)}$ is not central and the Bargmann algebra is not a subalgebra. This algebra was determined in \cite{Hansen:2018ofj} to be the relevant local symmetry algebra of type II Newton--Cartan geometry. As pointed out in \cite{Hansen:2019vqf}, the fact that one does not get the Bargmann algebra, gives a group theoretical perspective on the difference between type I and type II Newton--Cartan geometry.

Let us briefly review how we can obtain type II NC geometry by gauging \eqref{eq:NRG_algebra}. This procedure has previously been studied in \cite{Andringa:2010it, Hartong:2015zia} for other local symmetry algebras and is a powerful way to construct relevant non-relativistic geometries\footnote{See also \cite{Bergshoeff:2014uea} on gauging of the Schr\"odinger algebra.}.
Consider  first a Cartan connection $A_\mu$ that takes values in the algebra \eqref{eq:expansion_tangent_space_algebra}:
\begin{equation}
A_\mu = H\tau_\mu+P_a e^a_\mu+Nm_\mu+T_a\pi_\mu^a+G_a\omega_\mu{}^a+B_a\Omega_\mu{}^a+\frac{1}{2}J_{ab}\omega_\mu{}^{ab}+\frac{1}{2}S_{ab}\Omega_\mu{}^{ab}\,,
\end{equation}
whose adjoint transformation is given by
\begin{equation}
\delta_{\text{Ad}}\,A_\mu=\partial_\mu\Lambda+\left[A_\mu\,,\Lambda\right]=H\delta_{\text{Ad}}\,\tau_\mu+P_a \delta_{\text{Ad}}\, e_\mu^a+N\delta_{\text{Ad}}\, m_\mu+T_a\delta_{\text{Ad}}\,\pi_\mu^a+\ldots\,,
\end{equation}
where $\delta_{\text{Ad}}$ denotes the adjoint transformations. Define a new set of transformations via
\begin{eqnarray}
\delta\tau_\mu & = & \delta_{\text{Ad}}\,\tau_\mu-\xi^\nu R_{\mu\nu}(H)\,, \label{eq:deltataugroup}\\
\delta e_\mu^a & = & \delta_{\text{Ad}}\, e_\mu^a-\xi^\nu R_{\mu\nu}{}^a(P)\,,\\
\delta m_\mu & = & \delta_{\text{Ad}}\, m_\mu-\xi^\nu R_{\mu\nu}(N)-\zeta^\nu R_{\mu\nu}(H)\,,\\
\delta\pi_\mu^a & = & \delta_{\text{Ad}}\,\pi_\mu^a-\xi^\nu R_{\mu\nu}{}^a(T)-\zeta^\nu R_{\mu\nu}{}^a(P)\,.\label{eq:deltapigroup}
\end{eqnarray}
In here $R_{\mu\nu}(X)$ denotes a curvature corresponding to generator $X$, defined by
\begin{eqnarray}
F_{\mu\nu} & = & \partial_\mu A_\nu-\partial_\nu A_\mu+\left[A_\mu\,, A_\nu\right]\nonumber\\
& = & HR_{\mu\nu}(H)+P_a R_{\mu\nu}{}^a(P)+NR_{\mu\nu}(N)+T_a R_{\mu\nu}{}^a(T)+\ldots\,,
\end{eqnarray}
from which the $R_{\mu\nu}(X)$'s can be determined. Without loss of generality we can take for $\Lambda$ (appearing in the adjoint transformation) the following gauge transformation
\begin{eqnarray}
\Lambda & = & H\xi^\nu\tau_\nu+P_a \xi^\nu e^a_\nu+N\left(\xi^\nu m_\nu+\zeta^\nu\tau_\nu\right)+T_a\left(\xi^\nu\pi_\nu^a+\zeta^\nu e_\nu^a\right)+G_a\left(\xi^\nu\omega_\nu{}^a+\lambda^a\right)\nonumber\\
&&+B_a\left(\xi^\nu\Omega_\nu{}^a+\zeta^\nu\omega_\nu{}^a+\eta^a\right)+\frac{1}{2}J_{ab}\left(\xi^\nu\omega_\nu{}^{ab}+\lambda^{ab}\right)\nonumber\\
&&+\frac{1}{2}S_{ab}\left(\xi^\nu\Omega_\nu{}^{ab}+\zeta^\nu\omega_\nu{}^{ab}+\sigma^{ab}\right)\,.
\end{eqnarray}
If we now compute $\delta\tau_\mu$ etc as defined in \eqref{eq:deltataugroup}-\eqref{eq:deltapigroup} we reproduce the transformations of $\tau_\mu$, $m_\mu$, $e_\mu^a$ and $\pi_\mu^a$ given in equations \eqref{eq:trafo1}-\eqref{eq:trafo4}. This shows that the $1/c^2$ expansion of Lorentzian geometry to subleading order can be viewed as the gauging of the level 1 expansion of the Poincaré algebra.
More generally, this procedure can be used to any order in the  
$1/c^2$ expansion to obtain the relevant geometric fields describing gravity to that particular order.

\subsection{Lagrangians}\label{sec:Lagrangians}

We now present the systematics of the $1/c^2$ expansion of a given theory at the Lagrangian level. 
Consider a Lagrangian that is a function of some field $\phi(x;\sigma)$ and its derivatives, i.e. $\mathcal{L}=\mathcal{L}(\sigma,\phi,\partial_\mu\phi)$ where we also allow for an explicit dependence on the speed of light.
We now want to expand the Lagrangian and $\phi$ according to \eqref{eq:expansion_field_general}. Further below we generalise this to a Lagrangian depending on multiple fields. The explicit $\sigma$ dependence can for example come from the expansion of the background metric or matter fields as well as from parameters appearing in the kinetic or potential terms.
Assuming the overall power of the Lagrangian is $\sigma^{-N/2}=c^N$, we define $\tilde{\mathcal{L}}(\sigma)=\sigma^{N/2}\mathcal{L}(\sigma,\phi,\partial_\mu\phi)$ such that $\tilde{\mathcal{L}}$ starts at order zero.
Now we can Taylor expand $\tilde{\mathcal{L}}(\sigma)$ around $\sigma=0$, i.e.
\begin{equation}
\tilde{\mathcal{L}}(\sigma)=\tilde{\mathcal{L}}(0)+\sigma \tilde{\mathcal{L}}'(0)+\frac{1}{2}\sigma^2\tilde{\mathcal{L}}''(0)+\order{\sigma^3}\,,
\end{equation}
where the prime denotes differentiation with respect to $\sigma$. We have
\begin{equation}
\frac{\d}{\d\sigma}=\frac{\partial}{\partial\sigma}+\frac{\partial\phi}{\partial\sigma}\frac{\partial}{\partial\phi}+\frac{\partial\partial_\mu\phi}{\partial\sigma}\frac{\partial}{\partial\partial_\mu\phi}\,,
\end{equation}
so that
\begin{eqnarray}
\tilde{\mathcal{L}}(\sigma) & = & \tilde{\mathcal{L}}(0)+\left.\sigma\left(\frac{\partial\tilde{\mathcal{L}}}{\partial\sigma}+\frac{\partial\phi}{\partial\sigma}\frac{\partial\tilde{\mathcal{L}}}{\partial\phi}+\frac{\partial\partial_\mu\phi}{\partial\sigma}\frac{\partial\tilde{\mathcal{L}}}{\partial\partial_\mu\phi}\right)\right|_{\sigma=0}+\cdots\\
& = & \tilde{\mathcal{L}}(0)+\sigma\left(\frac{\partial\tilde{\mathcal{L}}}{\partial\sigma}\vert_{\sigma=0}+\phi_{(2)}\frac{\partial\tilde{\mathcal{L}}(0)}{\partial\phi_{(0)}}+\partial_\mu\phi_{(2)}\frac{\partial\tilde{\mathcal{L}}(0)}{\partial\partial_\mu\phi_{(0)}}\right)+\cdots\,.
\end{eqnarray}
Hence we can write an expansion in $\sigma=1/c^2$ as
\begin{equation}\label{eq:expLagrangian}
\mathcal{L}(c^2,\phi,\partial_\mu\phi)=c^N\tilde{\mathcal{L}}(\sigma)=c^N\;\oss{\mathcal{L}}{-N}{LO}+c^{N-2}\;\;\oss{\mathcal{L}}{2-N}{NLO}+c^{N-4}\;\;\oss{\mathcal{L}}{4-N}{NNLO}+\order{c^{N-6}}\,,
\end{equation}
where all the $c$-dependence is in the prefactors. The task is then to determine the coefficients. These can be found to be given by 
\begin{eqnarray}
\oss{\mathcal{L}}{-N}{LO} & = & \tilde{\mathcal{L}}(0)=\oss{\mathcal{L}}{-N}{LO}(\phi_{(0)},\partial_\mu\phi_{(0)})\,,\label{eq:expLagrangian1}\\
\oss{\mathcal{L}}{2-N}{NLO} & = & \tilde{\mathcal{L}}'(0)=\left.\frac{\partial\tilde{\mathcal{L}}}{\partial\sigma}\right|_{\sigma=0}+\phi_{(2)}\frac{\partial\;\oss{\mathcal{L}}{-N}{LO}}{\partial\phi_{(0)}}+\partial_\mu\phi_{(2)}\frac{\partial\;\oss{\mathcal{L}}{-N}{LO}}{\partial\partial_\mu\phi_{(0)}}\nonumber\\
&=&\left.\frac{\partial\tilde{\mathcal{L}}}{\partial\sigma}\right|_{\sigma=0}+\phi_{(2)}\frac{\delta\;\oss{\mathcal{L}}{-N}{LO}}{\delta\phi_{(0)}}\,.\label{eq:expLagrangian2}
\end{eqnarray}
Hence we see that the \regls{eom} of the subleading field of the subleading action is the EOM of the leading field of the leading action. A very similar calculation gives for the NNLO Lagrangian
\begin{eqnarray}
\hspace{-1.5cm}\oss{\mathcal{L}}{4-N}{NNLO} & = & \frac{1}{2}\tilde{\mathcal{L}}''(0) = \left.\frac{1}{2}\frac{\partial^2\tilde{\mathcal{L}}}{\partial\sigma^2}\right|_{\sigma=0}+\phi_{(2)}\frac{\delta}{\delta\phi_{(0)}}\frac{\partial\tilde{\mathcal{L}}}{\partial\sigma}\vert_{\sigma=0}+\phi_{(4)}\frac{\delta\;\oss{\mathcal{L}}{-N}{LO}}{\delta\phi_{(0)}}\nonumber\\
&&\hspace{-1.5cm}+\frac{1}{2}\Bigg[\phi_{(2)}^2\frac{\partial^2\;\oss{\mathcal{L}}{-N}{LO}}{\partial\phi_{(0)}^2}
+2\phi_{(2)}\partial_\mu\phi_{(2)}\frac{\partial^2\;\oss{\mathcal{L}}{-N}{LO}}{\partial\phi_{(0)}\partial(\partial_\mu\phi_{(0)})}+\partial_\mu\phi_{(2)}\partial_\nu\phi_{(2)}\frac{\partial^2\;\oss{\mathcal{L}}{-N}{LO}}{\partial(\partial_\mu\phi_{(0)})\partial(\partial_\nu\phi_{(0)})}\Bigg]\,.\label{eq:NNLO}
\end{eqnarray}
The second line forms the second variation of the LO Lagrangian and is a quadratic form involving the Hessian of the LO Lagrangian. It can be shown that
\begin{equation}\label{eq:NNLO_Nvar_NLO_var}
\frac{\delta\;\;\oss{\mathcal{L}}{4-N}{NNLO}}{\delta\phi_{(2)}}=\frac{\delta\;\;\oss{\mathcal{L}}{2-N}{NLO}}{\delta\phi_{(0)}}\,.
\end{equation}
Combining this with the fact that the EOM of $\phi_{(4)}$ gives the EOM of the LO Lagrangian we see that the NNLO Lagrangian reproduces all of the EOM of the NLO Lagrangian. 

When there is more than one field $\phi$ then we also get mixed derivatives of the LO Lagrangian in the second line of \eqref{eq:NNLO}. We can generalise the result by adding an index $I$ to $\phi$ as follows at NNLO
\begin{eqnarray}
\oss{\mathcal{L}}{4-N}{NNLO} & = & \frac{1}{2}\tilde{\mathcal{L}}''(0) = \frac{1}{2}\frac{\partial^2\tilde{\mathcal{L}}}{\partial\sigma^2}\vert_{\sigma=0}+\phi^I_{(2)}\frac{\delta}{\delta\phi^I_{(0)}}\frac{\partial\tilde{\mathcal{L}}}{\partial\sigma}\vert_{\sigma=0}+\phi^I_{(4)}\frac{\delta\;\oss{\mathcal{L}}{-N}{LO}}{\delta\phi^I_{(0)}}+\frac{1}{2}\Bigg[\phi_{(2)}^I\phi_{(2)}^J\frac{\partial^2\;\oss{\mathcal{L}}{-N}{LO}}{\partial\phi_{(0)}^I\partial\phi_{(0)}^J}\nonumber\\
\hspace{-1.5cm}&&+2\phi^I_{(2)}\partial_\mu\phi^J_{(2)}\frac{\partial^2\;\oss{\mathcal{L}}{-N}{LO}}{\partial\phi^I_{(0)}\partial(\partial_\mu\phi^J_{(0)})}+\partial_\mu\phi^I_{(2)}\partial_\nu\phi^J_{(2)}\frac{\partial^2\;\oss{\mathcal{L}}{-N}{LO}}{\partial(\partial_\mu\phi^I_{(0)})\partial(\partial_\nu\phi^J_{(0)})}\Bigg]\,,\label{eq:NNLO2}
\end{eqnarray}
and similar at pre-leading orders.

The expansion can be straightforwardly extended to include higher orders in $\sigma$. Doing so we will find relations analogous to \eqref{eq:NNLO_Nvar_NLO_var}, i.e. lower-order EOM are reproduced when going to higher orders.

\subsection{Einstein--Hilbert Lagrangian}\label{sec:causal_expansion_EH}
As a first step towards a $1/c^2$ expansion of the \regls{eh} Lagrangian, we must discuss its dimensionful normalisation.
When we write $g_{\mu\nu}$ in terms of $T_\mu$ and $\Pi_{\mu\nu}$ the line element maintains its property that its dimension is $L^2$ with $L=\text{length}$. Thus $T_\mu \d x^\mu$ has the dimension of length (if we set $\hat c=1$) or time (if we keep $\hat c$) and $\Pi_{\mu\nu}\d x^\mu \d x^\nu$ has the dimension of $L^2$. 
The measure $E \d^dx \d t$ defined in \eqref{eq:measure_unexpanded} has dimensions $TL^d$. Since we have $\sqrt{-g}=cE$ we take the EH action to be
\begin{equation}
S=\frac{c^3}{16\pi G_N}\int \d^dx \d t\sqrt{-g}R\,.
\end{equation}
Using the results of Section \ref{sec:expansion_causal_structure}, in particular equation \eqref{eq:RicciscalarTPi}, we find that the EH Lagrangian can now be written as 
\begin{eqnarray}\label{eq:EH_Lagrangian_tildeL}
\mathcal{L}_{\text{EH}}=\frac{c^6}{16\pi G_N}\tilde{\mathcal{L}}(\sigma,T,\Pi,\partial)\,,
\end{eqnarray}
where $\tilde{\mathcal{L}}(\sigma,T,\Pi,\partial)$ only depends on fields analytic in $\sigma$. The prefactor of $c^6$ follows from the fact that the Ricci scalar is order $c^2$ and $\sqrt{-g}$ is order $c$.
In here $\tilde{\mathcal{L}}$ can be written as
\begin{equation}\label{eq:explicitsigma}
\tilde{\mathcal{L}} = E\left[\frac{1}{4}\Pi^{\mu\nu}\Pi^{\rho\sigma}T_{\mu\rho}T_{\nu\sigma}+\sigma\Pi^{\mu\nu}\os{R}{C}_{\mu\nu}-\sigma^2 T^\mu T^\nu\os{R}{C}_{\mu\nu}\right]\,.
\end{equation}
This is the form of the EH Lagrangian to which we can apply the results \eqref{eq:expLagrangian}-\eqref{eq:NNLO2}.

At this stage, without the expansion of the fields $T_\mu$ and $\Pi_{\mu\nu}$, equation \eqref{eq:explicitsigma} is completely equivalent to the Einstein--Hilbert Lagrangian.
This form of the Lagrangian is crucial as it establishes a starting point for a non-relativistic expansion of general relativity.
In principle one could expand it to any desired order, keeping a manifest non-relativistic symmetry structure at each order.
This is expected to be closely related to the usual post-Newtonian expansion of general relativity, except that we here work in a manifestly covariant framework and do not make the assumption that the fields are weak.

We define the variation of the Einstein--Hilbert action as
\begin{equation}\label{eq:EH_variation_explicit}
\delta\mathcal{L}_{\text{EH}}\equiv-\frac{c^6}{8\pi G_N}E\left(E^\mu_g\delta T_\mu+\frac{1}{2}E^{\mu\nu}_{g}\delta\Pi_{\mu\nu}\right)\,,
\end{equation}
and the coupling to matter through a given matter Lagrangian $\mathcal{L}_{\text{mat}}=\mathcal{L}_{\text{mat}}(\sigma,\phi,\partial_\mu\phi)$ starting at order $\order{c^N}$ with variations defined as
\begin{equation}\label{eq:rel_EM_tensors_matter_Lagrangian}
\delta\mathcal{L}_{\text{mat}}\equiv c^N E \left(E^\mu_{\text{mat}}\delta T_\mu+\frac{1}{2}E^{\mu\nu}_{\text{mat}}\delta\Pi_{\mu\nu}\right)\,,
\end{equation}
where we left out the variations of the matter fields. This gives the Einstein field equations of motion in the form
\begin{equation}\label{eq:Einstein_field_eqs}
E^\mu_g = 8\pi G_N c^{N-6}E^\mu_{\text{mat}}\,,\qquad E^{\mu\nu}_{g}=8\pi G_N c^{N-6}E^{\mu\nu}_{\text{mat}}\,.
\end{equation}
The equations of motion satisfy two Ward identities as a consequence of diffeomorphism invariance of the action as well as invariance under local Lorentz transformations acting on $T_\mu$ and $\Pi_{\mu\nu}$ as expressed in equation \eqref{eq:Pi_trafo1}.

\subsubsection{Energy--momentum conservation}
Diffeomorphism invariance implies the equivalent of the divergencelessness of the usual Einstein tensor, namely
\begin{equation}\label{eq:Bianchi_relativistic}
T_\rho\left(\os{\nabla}{C}_\mu+\mathcal{L}_T T_\mu\right) E_{g}^\mu  + \Pi_{\nu\rho}\left(\os{\nabla}{C}_\mu+\mathcal{L}_T T_\mu\right) E_{g}^{\mu\nu} 
+E_{g}^\mu T_{\mu\rho}+\frac{1}{2}T_\rho E_{g}^{\mu\nu}\mathcal{L}_T \Pi_{\mu\nu}=0\,.
\end{equation}
Under a local Lorentz transformation the equations transform into each other as
\begin{equation}\label{eq:LboostWI}
c^{-2} E^\mu_g\mathcal{E}^a_\mu + E^{\mu\nu}_{g} T_\mu \mathcal{E}^a_\nu=0\,.
\end{equation}
Likewise the matter Lagrangian must be invariant under diffeomorphisms and local Lorentz transformations that act simultaneously on the matter fields and on the geometric fields they couple to. On shell this leads to 
the relativistic conservation law for energy-momentum conservation as derived from diffeomorphism invariance of \eqref{eq:rel_EM_tensors_matter_Lagrangian} (where we leave out the diffeomorphisms acting on the matter fields which is justified on shell as these are proportional to the matter equations of motion),
\begin{equation}\label{eq:EM_conservation_relativistic}
T_\rho\left(\os{\nabla}{C}_\mu+\mathcal{L}_T T_\mu\right) E_{\text{mat}}^\mu  + \Pi_{\nu\rho}\left(\os{\nabla}{C}_\mu+\mathcal{L}_T T_\mu\right) E_{\text{mat}}^{\mu\nu} 
+E_{\text{mat}}^\mu T_{\mu\rho}+\frac{1}{2}T_\rho E_{\text{mat}}^{\mu\nu}\mathcal{L}_T \Pi_{\mu\nu}=0\,,
\end{equation}
with projections
\begin{eqnarray}
0&=&\left(\os{\nabla}{C}_\mu+2\mathcal{L}_T T_\mu\right) E_{\text{mat}}^\mu +\frac{1}{2} E_{\text{mat}}^{\mu\nu}\mathcal{L}_T \Pi_{\mu\nu} \,,\label{eq:conservation1}\\
0&=&\Pi^{\rho\sigma}\Pi_{\rho\nu}\left(\os{\nabla}{C}_\mu +\mathcal{L}_T T_\mu\right)E_{\text{mat}}^{\mu\nu}  + \Pi^{\rho\sigma}T_{\mu\rho}E^\mu_{\text{mat}}\,.\label{eq:conservation2}
\end{eqnarray}
The $1/c^2$ expansion of these results will be studied in Section \ref{sec:Ward_identities}.

Assuming that the matter fields are inert under the local Lorentz transformation we find that for the matter currents we have (off shell)
\begin{equation}\label{eq:LboostWImat}
0=c^{-2} E^\mu_{\text{mat}}\mathcal{E}^a_\mu + E^{\mu\nu}_{\text{mat}} T_\mu \mathcal{E}^a_\nu\,.
\end{equation}

The currents $E^\mu_{\text{mat}},\,E^{\mu\nu}_{\text{mat}}$ are related to the usual Hilbert energy-momentum tensor $T^{\mu\nu}$ via
\begin{eqnarray}
    \delta\mathcal{L}_{\text{mat}}&=&\frac{c^{-1}}{2}\sqrt{-g}T^{\mu\nu}\delta g_{\mu\nu}=\frac{1}{2}ET^{\mu\nu}\left(-2c^2 T_\nu\delta T_\mu+\delta\Pi_{\mu\nu}\right)\nonumber\\
    &=&c^N E\left(E^\mu_{\text{mat}}\delta T_\mu+\frac{1}{2}E^{\mu\nu}_{\text{mat}}\delta\Pi_{\mu\nu}\right)\,,\label{eq:varLmat}
\end{eqnarray}
so that 
\begin{eqnarray}
E^\mu_{\text{mat}} &=& -c^{-N+2} \ T^{\mu\nu}T_{\nu}\,,\label{eq:Matter_EH_tensor1}\\
\mathcal{E}_\mu^a E^{\mu\nu}_{\text{mat}} &=& c^{-N}\mathcal{E}_\mu^a  T^{\mu\nu}\label{eq:Matter_EH_tensor2}\,,
\end{eqnarray}
where the latter equation follows from writing $\Pi_{\mu\nu}$ in terms of the spatial vielbeine and varying those. This is clearly consistent with the Lorentz boost Ward identity \eqref{eq:LboostWImat} which is of course nothing other than the symmetry of the Hilbert energy-momentum tensor. The power of $c$ in the first equality of \eqref{eq:varLmat} is fixed by demanding that the Einstein equation reads $G_{\mu\nu}=\frac{8\pi G_N}{c^4}T_{\mu\nu}$. The conservation equation \eqref{eq:EM_conservation_relativistic} is equivalent to that of the Hilbert energy-momentum tensor 
\begin{equation}\label{eq:EM_conservation_Hilbert}
\nabla_\mu T^{\mu\nu}=0\,,
\end{equation}
where the covariant derivative is taken with respect to the Levi--Civita connection.

\subsubsection{The Pre-Poisson equation}\label{sec:pre_poisson_eqn}
An interesting combination of the equations of motion comes from a particular rescaling of the metric components:
\begin{equation}
\delta T_\mu=\alpha\omega T_\mu\,,\qquad\delta\Pi_{\mu\nu}=\beta\omega\Pi_{\mu\nu}\,,
\end{equation}
where $\alpha$ and $\beta$ are numbers.
We keep the scalar function $\omega$ arbitrary. To find the equation of motion of $\omega$ the following results are useful
\begin{eqnarray}
\hspace{-1.5cm}\delta_\omega\log E & = & \left(\alpha+\frac{d}{2}\beta\right)\omega\,,\\
\hspace{-1.5cm}\delta_\omega\left(\Pi^{\mu\nu}\Pi^{\rho\sigma}T_{\mu\rho}T_{\nu\sigma}\right) & = & 2(\alpha-\beta)\omega\Pi^{\mu\nu}\Pi^{\rho\sigma}T_{\mu\rho}T_{\nu\sigma}\,,\\
\hspace{-1.5cm}\delta_\omega C^\rho_{\mu\nu} & = & -\alpha T^\rho T_\nu\partial_\mu\omega+\beta\Pi^{\rho\sigma}\Pi_{\sigma(\mu}\partial_{\nu)}\omega-\frac{1}{2}\beta\Pi_{\mu\nu}\Pi^{\rho\sigma}\partial_\sigma\omega\,,\\
\hspace{-1.5cm}\delta_\omega\left(\Pi^{\mu\nu}\os{R}{C}_{\mu\nu}\right) & = & -\beta\omega\Pi^{\mu\nu}\os{R}{C}_{\mu\nu}-(d-1)\beta\omega\Pi^{\rho\sigma}\left(\os{\nabla}{C}_\rho+\mathcal{L}_T T_\rho\right)\mathcal{L}_T T_\sigma\,,\\
\hspace{-1.5cm}\delta_\omega\left(T^\mu T^\nu \os{R}{C}_{\mu\nu}\right) & = & -2\alpha\omega T^\mu T^\nu \os{R}{C}_{\mu\nu}-\left(\alpha+\frac{d-2}{2}\beta\right)\omega\os{\nabla}{C}_{\rho}\left(T^\rho\omega\os{\nabla}{C}_{\mu}T^\mu\right)\,.
\end{eqnarray}
If we now set
\begin{equation}
\alpha=-(d-2)\,,\qquad\beta=2\,,
\end{equation}
one obtains
\begin{multline}
    \delta_\omega{\mathcal{L}_\text{EH}}=-\frac{c^6}{8\pi G_N}(d-1)\omega E\\\times \left[\frac{1}{4}\Pi^{\mu\nu}\Pi^{\rho\sigma}T_{\mu\rho}T_{\nu\sigma}+\sigma \Pi^{\mu\nu}\left(\os{\nabla}{C}_\mu+\mathcal{L}_T T_\mu\right)\mathcal{L}_T T_\nu+\sigma^2 T^\mu T^\nu \os{R}{C}_{\mu\nu}\right]\,.
\end{multline}
Expressing the left hand side using the chain rule in terms of variations of the Lagrangian with respect to the $T_\mu$ and $\Pi_{\mu\nu}$ and using the Einstein equations \eqref{eq:Einstein_field_eqs} this this can be seen to be equivalent to
\begin{multline}\label{eq:scalingrel}
8\pi G_N c^{N-6}\left[-(d-2)E^\mu_{\text{mat}}T_\mu+E^{\mu\nu}_{\text{mat}}\Pi_{\mu\nu}\right]=\\-(d-1)\left(\frac{1}{4}\Pi^{\mu\nu}\Pi^{\rho\sigma}T_{\mu\rho}T_{\nu\sigma}+\sigma \Pi^{\mu\nu}\left(\os{\nabla}{C}_\mu+\mathcal{L}_T T_\mu\right)\mathcal{L}_T T_\nu+\sigma^2 T^\mu T^\nu \os{R}{C}_{\mu\nu}\right)\,.
\end{multline}
It will turn out that this equation, when expanded in $\sigma=c^{-2}$,  contains important equations like the Poisson equation and the sourcing of Newton--Cartan torsion. This will be shown in Section \ref{sec:NRG_EOMs}.

\section{Non-relativistic gravity \label{sec:NRG} }

In this section we will use the results of the previous section to obtain the action and \reglspl{eom} that result from the $1/c^2$ expansion of \regls{gr}. In particular we will focus on the theory that governs the dynamics of the \regls{lo} and \regls{nlo} fields in the expansion of the metric. For definiteness we refer to this as ``non-relativistic gravity'' (\gls{nrg}). As was shown already in \cite{Hansen:2018ofj} this includes Newtonian gravity but goes beyond it, as it also includes  geometries with gravitational time dilaton
\footnote{We emphasise that our methods can be used in principle to obtain the dynamics of the fields appearing to any order in the large speed of light expansion of GR.}. 
We discuss in this section two distinct methods to obtain non-relativistic gravity. We start with the direct approach which uses the $1/c^2$ expansion. Alternatively one can follow a symmetry-based route which uses gauge invariance, from which a unique two-derivative action can be obtained given a kinetic term that is required to include Newtonian gravity. Satisfyingly, we will show that the two methods lead to the same action, though in a slightly different form.

\subsection{Theory from \texorpdfstring{$1/c^2$}{1/c2} expansion}\label{sec:NRG_1c2_expansion}
\subsubsection{General structure}
\begin{figure}[hb!]
    \centering
\begin{tikzpicture}
\matrix (m) [matrix of math nodes,row sep=2em,column sep=2em,minimum width=2em]
{
    && \os{G}{-6}_{h}^{\mu\nu} & \os{G}{-6}_{\tau}^{\mu} &&\\
     &\os{G}{-4}_{\Phi}^{\mu\nu} & \os{G}{-4}_{h}^{\mu\nu} &  \os{G}{-4}_{\tau}^{\mu}  &  \os{G}{-4}_{m}^{\mu}  & \\
     \os{G}{-2}_{\psi}^{\mu\nu} &\os{G}{-2}_{\Phi}^{\mu\nu} & \os{G}{-2}_{h}^{\mu\nu} &  \os{G}{-2}_{\tau}^{\mu}  &  \os{G}{-2}_{m}^{\mu} &\os{G}{-2}_{B}^{\mu}\\};

\draw[-,white] (m-1-3) -- node [midway,sloped,black] {=} (m-2-2);
\draw[-,white] (m-2-2) -- node [midway,sloped,black] {=} (m-3-1);
\draw[-,white] (m-1-4) -- node [midway,sloped,black] {=} (m-2-5);
\draw[-,white] (m-2-5) -- node [midway,sloped,black] {=} (m-3-6);

\draw[-,white] (m-2-3) -- node [midway,sloped,black] {=} (m-3-2);
\draw[-,white] (m-2-4) -- node [midway,sloped,black] {=} (m-3-5);

\end{tikzpicture}
\caption{Structure of the equations of motion in the $1/c^2$ expansion, of which many will enter in the Lagrangian at subleading orders. Because of the way the EH Lagrangian is expanded and the property \eqref{eq:NNLO_Nvar_NLO_var} there will only be two new EOMs at each order to solve, the remaining ones being recursively equal to those of the previous order. Notice that when we impose TTNC off shell, all the outermost equations are zero since the LO EOMs are $\propto \tau\wedge \d\tau$ as explained in Section \ref{sec:Lagrangian_NNLO_1c2_expansion}.}
\label{fig:structure_EOMs_expansion}
\end{figure}

In this section we want to determine the Lagrangian that arises when we expand the fields in \eqref{eq:explicitsigma} using the methods of Section \ref{sec:Lagrangians}. This means that we will end up with a theory that is expressed in terms of the fields $\phi^I_{(0)}=\{\tau_\mu,\,h_{\mu\nu}\}$, $\phi^I_{(2)}=\{m_\mu,\,\Phi_{\mu\nu}\}$ and $\phi^I_{(4)}=\{B_\mu,\,\psi_{\mu\nu}\}$. In the next section we will rederive similar results in terms of the fields $\tau_\mu$, $\bar h_{\mu\nu}$, $\bar\Phi_{\mu\nu}$ that appear in the expansion of the metric.

The $1/c^2$ expansion of the \regls{eh} Lagrangian will take the form
\begin{equation}
    \mathcal{L}_{\text{EH}}=c^6\left(\oss{\mathcal{L}}{-6}{\text{LO}}+\sigma\;\oss{\mathcal{L}}{-4}{\text{NLO}}+\sigma^2\; \oss{\mathcal{L}}{-2}{\text{NNLO}}+\mathcal{O}(\sigma^3)\right)\,.
\end{equation}

We now define for $n\in \mathbb{N}$, including zero, the equations of motion of $h_{\alpha\beta}$ and $\tau_\alpha$ of the $\mathrm{N}^n\mathrm{LO}$ Lagrangian as
\begin{eqnarray}\label{eq:1c2_expansion_def_EOMs}
\frac{1}{16\pi G_N}\;\os{G}{2n-6}^{\alpha\beta}_{h}&\equiv& -e^{-1}\frac{\delta\;\;{\os{\mathcal{L}}{2n-6}}_{\mathrm{N}^n\mathrm{LO}}}{\delta h_{\alpha\beta}}\,,\\
\frac{1}{8\pi G_N}\;\os{G}{2n-6}^\alpha_{\tau} &\equiv& -e^{-1}\frac{\delta\;\;{\os{\mathcal{L}}{2n-6}}_{\mathrm{N}^n\mathrm{LO}}}{\delta \tau_{\alpha}}\,,
\end{eqnarray}
where the Galilean boost invariant integration measure of both type I and type II \regls{nc} geometry is given by
\begin{equation}
e  \equiv  \left(-\mathrm{det}\left(-\tau_\mu\tau_\nu+h_{\mu\nu}\right)\right)^{1/2}\,.
\end{equation}
These equations of motion will also appear in the $1/c^2$ expansion of $E^\mu_g$, $E^{\mu\nu}_{g}$, defined as the response to the variations of $T_\mu$ and $\Pi_{\mu\nu}$, respectively
(see \eqref{eq:EH_variation_explicit}). We will give the explicit relationship between the two in Section \ref{sec:NRG_EOMs}.

When we go beyond leading order (for which $n=0$) we encounter subleading fields in the Lagrangian. For example $\oss{\mathcal{L}}{-4}{\text{NLO}}$ depends on both LO fields $\tau_\mu$, $h_{\mu\nu}$ as well as on the NLO fields $m_\mu$ and $\Phi_{\mu\nu}$ etc. 
At order N$^n$LO there are $2(n+1)$  fields, each of which has its own equation of motion. However, the equations of motion that appear at order N$^{n-1}$LO are all reproduced at the $n$th order, see for example \eqref{eq:NNLO_Nvar_NLO_var}. The additional equations appearing at order $n$ that are not already present at order $n-1$ involve the $n$th order subleading fields. For the NLO fields we define in analogy with \eqref{eq:1c2_expansion_def_EOMs}
\begin{eqnarray}
\frac{1}{16\pi G_N}\;\os{G}{2n-6}^{\alpha\beta}_{\Phi}&\equiv&-e^{-1}\frac{\delta\;\;{\os{\mathcal{L}}{2n-6}}_{\mathrm{N}^n\mathrm{LO}}}{\delta \Phi_{\alpha\beta}}\,,\\
\frac{1}{8\pi G_N}\;\os{G}{2n-6}^\alpha_{m} &\equiv& -e^{-1}\frac{\delta\;\;{\os{\mathcal{L}}{2n-6}}_{\mathrm{N}^n\mathrm{LO}}}{\delta m_{\alpha}}\,,
\end{eqnarray}
where $\os{G}{-6}^{\alpha\beta}_{\Phi}=\os{G}{-6}^\alpha_{m}=0$ because $\Phi_{\mu\nu},\,m_\mu$ do not appear at the leading order.
In particular because of \eqref{eq:NNLO_Nvar_NLO_var} we have
\begin{eqnarray}
\os{G}{-2}^{\alpha\beta}_{\psi}&=& \os{G}{-4}^{\alpha\beta}_{\Phi}= \os{G}{-6}^{\alpha\beta}_{h}\,,\\
\os{G}{-2}^\alpha_{B} &=& \os{G}{-4}^\alpha_{m} = \os{G}{-6}^\alpha_{\tau}\,.
\end{eqnarray}
The structure of the expansion of the equations of motion is summarised in Figure \ref{fig:structure_EOMs_expansion}.

\subsubsection{NNLO Lagrangian: Non-relativistic gravity}\label{sec:Lagrangian_NNLO_1c2_expansion}
Using \eqref{eq:expLagrangian1} we find the leading order part of the EH Lagrangian ${\mathcal{L}_\text{EH}}$ given by \eqref{eq:EH_Lagrangian_tildeL} to be
\begin{equation}
\oss{\mathcal{L}}{-6}{LO}=\frac{E}{16\pi G_N}\left.\frac{1}{4}\Pi^{\mu\nu}\Pi^{\rho\sigma}T_{\mu\rho}T_{\nu\sigma}\right|_{\sigma=0}=\frac{e}{16\pi G_N}\frac{1}{4}h^{\mu\nu}h^{\rho\sigma}\tau_{\mu\rho}\tau_{\nu\sigma}\,,
\end{equation}
where we defined 
\begin{equation}
\tau_{\mu\nu} \equiv \partial_\mu\tau_\nu-\partial_\nu\tau_\mu\,.\\
\end{equation}

With the above conventions we can then  write the variation of the leading order Lagrangian:
\begin{equation}
\delta\oss{\mathcal{L}}{-6}{LO} =  -\frac{1}{8\pi G_N}e\left(\os{G}{-6}^\alpha_{\tau}\delta \tau_{\alpha}+\frac{1}{2}\os{G}{-6}^{\alpha\beta}_{h}\delta h_{\alpha\beta}\right)\,,
\end{equation}
where the leading order equations of motion are
\begin{eqnarray}
\os{G}{-6}^{\alpha\beta}_{h}&=& -\frac{1}{8}h^{\mu\nu}h^{\rho\sigma}\tau_{\mu\rho}\tau_{\nu\sigma}h^{\alpha\beta}+\frac{1}{2}h^{\mu\alpha}h^{\nu\beta}h^{\rho\sigma}\tau_{\mu\rho}\tau_{\nu\sigma}\,,\\
\os{G}{-6}^\alpha_{\tau} &=&\frac{1}{8}h^{\mu\nu}h^{\rho\sigma}\tau_{\mu\rho}\tau_{\nu\sigma}v^\alpha + \frac{1}{2}a_\mu h^{\mu\nu} h^{\rho\alpha}\tau_{\nu\rho}+\frac{1}{2}e^{-1}\partial_\mu\left(eh^{\mu\nu}h^{\rho\alpha}\tau_{\nu\rho}\right)\,.
\end{eqnarray}
Contracting $\os{G}{-6}^\mu_{\tau}$ with $\tau_\mu$ tells us that on shell
\begin{equation}\label{eq:HSO}
h^{\mu\nu}h^{\rho\sigma}\tau_{\mu\rho}\tau_{\nu\sigma}=0\,.
\end{equation}
As this is a sum of squares it implies that $h^{\mu\nu}h^{\rho\sigma}\tau_{\mu\rho}=0$ and thus that $\tau\wedge \d\tau=0$ on shell. This is the Frobenius integrability condition for the existenc of a foliation with normal 1-form $\tau=N \d T$ where $N$ and $T$ are scalars. 
This implies that there is a foliation of the Newton--Cartan spacetime in terms of hypersurfaces of absolute simultaneity, foliated by leaves of constant $T$. The on shell geometry arising from the expansion is thus a \regls{ttnc} geometry \cite{Christensen:2013lma} (albeit of type II but that distinction only affects the gauge fields defined on the geometry described by the LO fields $\tau_\mu$ and $h_{\mu\nu}$).

This completes the discussion of the LO Lagrangian. We will continue with the NLO Lagrangian which can be obtained from \eqref{eq:expLagrangian2} generalised to include multiple fields. From \eqref{eq:expLagrangian2}, \eqref{eq:EH_Lagrangian_tildeL} and \eqref{eq:explicitsigma} we can see that one first of all needs to compute the derivative of \eqref{eq:explicitsigma}, which gives
\begin{eqnarray}
\left.\frac{\partial\tilde{\mathcal{L}}}{\partial\sigma}\right|_{\sigma=0} & = & \left. E\Pi^{\mu\nu}\os{R}{C}_{\mu\nu}\right|_{\sigma=0}=eh^{\mu\nu}\check R_{\mu\nu}\,,
\end{eqnarray}
where we recall that $\check R_{\mu\nu}$ is defined in \eqref{eq:Ricci_tensor_LC}. 
Using \eqref{eq:expLagrangian2} this combines with the leading order EOMs contracted with the subleading fields so that we obtain
\begin{equation}\label{eq:NLOLag}
\oss{\mathcal{L}}{-4}{NLO}=-\frac{e}{8\pi G_N}\left(-\frac{1}{2}h^{\mu\nu}\check R_{\mu\nu}+\os{G}{-6}_\tau^\mu\, m_\mu+\frac{1}{2}\os{G}{-6}_h^{\mu\nu}\Phi_{\mu\nu}\right)\,.
\end{equation}
We will refer to this Lagrangian as the Lagrangian of Galilean gravity.
This theory was studied in \cite{Bergshoeff:2017btm} using a first-order formalism. Equation
\eqref{eq:NLOLag} can be related to the Lagrangian appearing in that work by a specific choice of the undetermined Lagrange multipliers that appears.
That theory also has the scaling properties described in Section \ref{sec:pre_poisson_eqn}.
All the leading order equations of motion are included in the NLO Lagrangian as they are obtained by varying with respect to the subleading fields. The $\tau_\mu$ and $h_{\mu\nu}$ equations of motion are
\begin{eqnarray}
\hspace{-1.3cm}\os{G}{-4}_{\tau}^\nu &=& \frac{1}{2}\Big[ 2\left(h^{\mu\rho}h^{\nu\sigma}-h^{\mu\nu}h^{\rho\sigma}\right)\check{\nabla}_\mu K_{\rho\sigma}
+v^\nu h^{\rho\sigma}\check R_{\rho\sigma}
+\left(\check{\nabla}_\mu+2a_\mu \right)h^{\mu\rho}h^{\nu\sigma}F_{\rho\sigma}\Big]+\ldots\,,\label{eq:NLO_EOM1}\\
\hspace{-1.3cm}\os{G}{-4}_{h}^{\rho\sigma}&=&h^{\mu\rho}h^{\nu\sigma}\left(\check R_{\mu\nu}
-\frac{1}{2}h_{\mu\nu}h^{\kappa\lambda}\check R_{\kappa\lambda}-\left(\check{\nabla}_\mu +a_\mu\right)a_\nu 
+h_{\mu\nu}h^{\kappa\lambda} \left(\check\nabla_{\kappa}+a_\kappa\right)a_\lambda\right)+\ldots\,,\label{eq:NLO_EOM2}
\end{eqnarray}
where the dots denote terms that vanish on shell upon using the $m_\mu$ and $\Phi_{\mu\nu}$ equations of motion.
These extra terms can easily be calculated from $m_\mu$ and $\Phi_{\mu\nu}$ variations of the NNLO Lagrangian \eqref{eq:LNNLO} below using $\os{G}{-4}^\alpha_{\tau} = \os{G}{-2}^\alpha_{m} $ and $\os{G}{-4}^{\alpha\beta}_{h} = \os{G}{-2}^{\alpha\beta}_{\Phi}$.

The NNLO Lagrangian is found from \eqref{eq:NNLO}.
For that we need the second order derivative of $\tilde{\mathcal{L}}$ which reads
\begin{eqnarray}
\left.\frac{\partial^2\tilde{\mathcal{L}}}{\partial\sigma^2}\right|_{\sigma=0} & = & \left.-2E T^\mu T^\nu\os{R}{C}_{\mu\nu}\right|_{\sigma=0}=-2e v^{\mu}v^{\nu}\check R_{\mu\nu}\,.
\end{eqnarray}
Consider the general form of the Lagrangian \eqref{eq:NNLO}. Adapted to the case of the EH Lagrangian we have the general result
\begin{eqnarray}\label{eq:LNNLO}
\oss{\mathcal{L}}{-2}{NNLO} & = & \frac{1}{16\pi G_N}\left( -e v^\mu v^\nu\check R_{\mu\nu}+m_\alpha\frac{\delta}{\delta\tau_\alpha}\left(eh^{\mu\nu}\check R_{\mu\nu}\right)+\Phi_{\alpha\beta}\frac{\delta}{\delta h_{\alpha\beta}}\left(eh^{\mu\nu}\check R_{\mu\nu}\right)\right)\nonumber\\
&&+B_\mu\frac{\delta\oss{\mathcal{L}}{-6}{LO}}{\delta\tau_\mu}+\psi_{\mu\nu}\frac{\delta\oss{\mathcal{L}}{-6}{LO}}{\delta h_{\mu\nu}}+\left[\frac{1}{2}m_\mu m_\nu\frac{\partial^2\oss{\mathcal{L}}{-6}{LO}}{\partial\tau_\mu\partial\tau_\nu}+m_\mu\Phi_{\nu\rho}\frac{\partial^2\oss{\mathcal{L}}{-6}{LO}}{\partial\tau_\mu\partial h_{\nu\rho}}\right.\nonumber\\
&&+\frac{1}{2}\Phi_{\mu\nu}\Phi_{\rho\sigma}\frac{\partial^2\oss{\mathcal{L}}{-6}{LO}}{\partial h_{\mu\nu}\partial h_{\rho\sigma}}+\Phi_{\rho\sigma}\partial_\mu m_\nu\frac{\partial^2\oss{\mathcal{L}}{-6}{LO}}{\partial h_{\rho\sigma}\partial(\partial_\mu\tau_\nu)}+m_\rho\partial_\mu m_\nu \frac{\partial^2\oss{\mathcal{L}}{-6}{LO}}{\partial\tau_\rho\partial(\partial_\mu\tau_\nu)}\nonumber\\
&&\left.+\frac{1}{2}\partial_\mu m_\rho\partial_\nu m_\sigma \frac{\partial^2\oss{\mathcal{L}}{-6}{LO}}{\partial(\partial_\mu\tau_\rho)\partial(\partial_\nu\tau_\sigma)}\right]\,.
\end{eqnarray}
The term in square brackets is the second variation of the LO Lagrangian and we used that $\oss{\mathcal{L}}{-6}{LO}$ does not depend on derivatives of $h_{\mu\nu}$. The field $\psi_{\mu\nu}$ is the NNLO field in the expansion of $\Pi_{\mu\nu}$. The term in square brackets is given by
\begin{eqnarray}
[\ldots] & = & \frac{e}{16\pi G_N}\left(\frac{1}{4}h^{\mu\rho}h^{\nu\sigma}F_{\mu\nu}F_{\rho\sigma}+\frac{1}{2}\Phi h^{\mu\rho}h^{\nu\sigma}\tau_{\mu\nu}F_{\rho\sigma}+\frac{1}{4}h^{\kappa\lambda}\Phi_{\kappa\lambda}h^{\mu\rho}h^{\nu\sigma}\tau_{\mu\nu}F_{\rho\sigma}\right.\nonumber\\
&&\left.-h^{\mu\alpha}h^{\nu\gamma}h^{\beta\delta}\Phi_{\gamma\delta}\tau_{\mu\nu}F_{\alpha\beta}+h^{\mu\alpha}m_\alpha v^\beta F_{\beta\gamma}h^{\gamma\nu}\tau_{\mu\nu}+\frac{1}{4}m_\alpha h^{\alpha\beta}v^\gamma\Phi_{\beta\gamma}h^{\mu\rho}h^{\nu\sigma}\tau_{\mu\nu}\tau_{\rho\sigma}\right.\nonumber\\
&&\left.-m_\alpha h^{\alpha\mu}h^{\nu\sigma}h^{\rho\beta}v^\gamma\Phi_{\beta\gamma}\tau_{\rho\sigma}\tau_{\mu\nu}+\frac{1}{8}\Phi h^{\beta\gamma}\Phi_{\beta\gamma}h^{\mu\rho}h^{\nu\sigma}\tau_{\mu\nu}\tau_{\rho\sigma}-\frac{1}{2}\Phi\Phi_{\beta\gamma}h^{\mu\beta}h^{\rho\gamma}h^{\nu\sigma}\tau_{\mu\nu}\tau_{\rho\sigma}\right.\nonumber\\
&&\left.+\frac{1}{32}\left(h^{\alpha\beta}\Phi_{\alpha\beta}\right)^2h^{\mu\rho}h^{\nu\sigma}\tau_{\mu\nu}\tau_{\rho\sigma}-\frac{1}{16}h^{\alpha\gamma}h^{\beta\delta}\Phi_{\alpha\beta}\Phi_{\gamma\delta}h^{\mu\rho}h^{\nu\sigma}\tau_{\mu\nu}\tau_{\rho\sigma}\right.\nonumber\\
&&\left.-\frac{3}{8}h^{\gamma\delta}\Phi_{\gamma\delta}h^{\mu\alpha}h^{\rho\beta}\Phi_{\alpha\beta}h^{\nu\sigma}\tau_{\mu\nu}\tau_{\rho\sigma}+\frac{1}{2}h^{\mu\gamma}h^{\alpha\delta}\Phi_{\gamma\delta}\Phi_{\alpha\beta}h^{\rho\beta}h^{\nu\sigma}\tau_{\mu\nu}\tau_{\rho\sigma}\right.\nonumber\\
&&\left.+\frac{1}{4}h^{\mu\alpha}h^{\rho\beta}\Phi_{\alpha\beta}h^{\nu\gamma}h^{\sigma\delta}\Phi_{\gamma\delta}\tau_{\mu\nu}\tau_{\rho\sigma}\right)\,,\label{eq:NRG_nnlo_terms_full_torsion}
\end{eqnarray}
where
\begin{equation}
F_{\mu\nu} \equiv \partial_\mu m_\nu - \partial_\nu m_\mu - a_\mu m_\nu + a_\nu m_\mu\,,
\end{equation}
which can be thought of as the field strength of $m_\mu$ (for the torsional $U(1)$ transformation \eqref{eq:torsional_u(1)_1}). 
The terms in \eqref{eq:NRG_nnlo_terms_full_torsion} can be written as
\begin{equation}
    [\ldots]=\frac{e}{16\pi G_N}\left(\frac{1}{4}h^{\mu\rho}h^{\nu\sigma}F_{\mu\nu}F_{\rho\sigma}+h^{\mu\rho}h^{\nu\sigma}\tau_{\mu\nu}X_{\rho\sigma}\right)\,,
\end{equation}
where $X_{\rho\sigma}$ is some tensor.
With the help of the results of appendix \ref{sec:curvature_variations} it can be shown that (for general $\tau_\mu$)
\begin{eqnarray}
    \frac{\delta}{\delta\tau_\alpha}\left(eh^{\mu\nu}\check R_{\mu\nu}\right) & = & e\left(-v^\alpha h^{\mu\nu}\check R_{\mu\nu}-2\left(h^{\mu\rho}h^{\sigma\alpha}-h^{\rho\sigma}h^{\mu\alpha}\right)\check\nabla_\mu K_{\rho\sigma}\right)\,,\label{eq:varRic_tau}\\
    \frac{\delta}{\delta h_{\alpha\beta}}\left(eh^{\mu\nu}\check R_{\mu\nu}\right) & = & eh^{\mu\alpha}h^{\nu\beta}\bigg[-\check R_{\mu\nu}+\frac{1}{2}h_{\mu\nu}h^{\rho\sigma}\check R_{\rho\sigma}+\left(\check\nabla_{\mu}+a_{\mu}\right)a_{\nu}-h_{\mu\nu}h^{\rho\sigma}\left(\check\nabla_\rho+a_\rho\right)a_\sigma\nonumber\\
    &&-\frac{1}{2}\check\nabla_\rho\left(v^\rho\tau_{\mu\nu}\right)\bigg]+e\left(\check\nabla_\mu+2a_\mu\right)\left(h^{\mu\sigma}h^{\rho(\alpha}v^{\beta)}\tau_{\rho\sigma}\right)\,,\label{eq:varRic_h}
\end{eqnarray}
where we used that \eqref{eqn:variation_connection_check} yields
\begin{eqnarray}
    \delta_\tau\check\Gamma^\rho_{\mu\nu} & = & -v^\rho\check\nabla_\mu\delta\tau_\nu+h^{\rho\alpha}K_{\mu\nu}\delta\tau_\alpha\,,\\
    \delta_h\check\Gamma^\rho_{\mu\nu} & = & \frac{1}{2} v^\beta h^{\rho\alpha}\tau_{\mu\nu}\delta h_{\alpha\beta}-\frac{1}{2}v^\lambda h^{\rho\sigma}\tau_{\nu\sigma}\delta h_{\mu\lambda}-\frac{1}{2}v^\lambda h^{\rho\sigma}\tau_{\mu\sigma}\delta h_{\nu\lambda}\nonumber\\
    &&+\frac{1}{2}h^{\rho\sigma}\left(\check\nabla_\mu\delta h_{\nu\sigma}+\check\nabla_\nu\delta h_{\mu\sigma}-\check\nabla_\sigma\delta h_{\mu\nu}\right)\,.
\end{eqnarray}
We note that the term in square brackets in \eqref{eq:varRic_h} is symmetric in $\alpha\beta$ due to the fact that the antisymmetric part of the Ricci tensor is
\begin{equation}
    \check R_{[\mu\nu]}=\check\nabla_{[\mu} a_{\nu]}-\frac{1}{2}\check\nabla_\rho\left(v^\rho\tau_{\mu\nu}\right)\,.
\end{equation}

The reason we do not need to use $X_{\rho\sigma}$ is as follows. The terms involving the NNLO fields $B_\mu$ and $\psi_{\mu\nu}$ are both also of the form 
$e h^{\mu\rho}h^{\nu\sigma}\tau_{\mu\nu}X_{\rho\sigma}$. Furthermore the EOM of the NNLO fields lead to the familiar condition $h^{\mu\rho}h^{\nu\sigma}\tau_{\mu\nu}=0$. This means that any variation of $e h^{\mu\rho}h^{\nu\sigma}\tau_{\mu\nu}X_{\rho\sigma}$ that is proportional to $h^{\mu\rho}h^{\nu\sigma}\tau_{\mu\nu}=0$ does not contribute on shell. It turns out that the only variation of $e h^{\mu\rho}h^{\nu\sigma}\tau_{\mu\nu}X_{\rho\sigma}$ that contributes on shell is the variation of $\tau_\mu$ except for the special case $\delta\tau_\mu=\Omega\tau_\mu$ for arbitrary $\Omega$.
Put another way, terms of the kind $h^{\mu\rho}h^{\nu\sigma}\tau_{\mu\nu}X_{\rho\sigma}$ can be ignored except for variations of the type $h^{\mu\nu}\delta\tau_\mu$.
These variations give us an equation for $B_\mu$ and so one arrives  at the important conclusion that if we only care about EOM for the NLO fields we can ignore the term $e h^{\mu\rho}h^{\nu\sigma}\tau_{\mu\nu}X_{\rho\sigma}$ in the NNLO Lagrangian as well as the terms involving the NNLO fields. Effectively we can set $h^{\mu\rho}h^{\nu\sigma}\tau_{\mu\nu}=0$ i.e. impose the TTNC condition $\tau\wedge\d\tau=0$ off shell. If we do this we are only allowed to vary $\tau_\mu$ as $\delta\tau_\mu=\Omega\tau_\mu$.

This procedure gives us what we call the \regls{nrg} Lagrangian:
\begin{eqnarray}\label{eq:action_NRG}
\hspace{-1.5cm}\mathcal{L}_\mathrm{NRG} & \equiv & \left.\oss{\mathcal{L}}{-2}{NNLO}\right|_{\tau\wedge \d\tau=0}+
\frac{e}{16\pi G_N}\frac{1}{2}\zeta_{\rho\sigma}h^{\mu\rho}h^{\nu\sigma}(\partial_\mu\tau_\nu-\partial_\nu\tau_\mu)\nonumber\\
\hspace{-1.5cm}& = & \frac{e}{16\pi G_N}\Bigg[h^{\mu\rho}h^{\nu\sigma}K_{\mu\nu}K_{\rho\sigma}-\left(h^{\mu\nu}K_{\mu\nu}\right)^2
-2m_\nu\left(h^{\mu\rho}h^{\nu\sigma}-h^{\mu\nu}h^{\rho\sigma}\right)\check{\nabla}_\mu K_{\rho\sigma}
\nonumber\\
\hspace{-1.5cm}&&+\Phi h^{\mu\nu}\check R_{\mu\nu}
+\frac{1}{4}h^{\mu\rho}h^{\nu\sigma}F_{\mu\nu}F_{\rho\sigma}+\frac{1}{2}\zeta_{\rho\sigma}h^{\mu\rho}h^{\nu\sigma}(\partial_\mu\tau_\nu-\partial_\nu\tau_\mu)\nonumber\\
\hspace{-1.5cm}&&
-\Phi_{\rho\sigma}h^{\mu\rho}h^{\nu\sigma}\left(\check R_{\mu\nu}-\check{\nabla}_\mu a_\nu - a_\mu a_\nu
-\frac{1}{2}h_{\mu\nu}h^{\kappa\lambda}\check R_{\kappa\lambda}
+h_{\mu\nu} e^{-1} \partial_{\kappa}\left(e h^{\kappa\lambda} a_\lambda\right)\right)\Bigg]\,,
\end{eqnarray}
where $\Phi \equiv -v^\mu m_\mu$ is the Newtonian potential. We added a Lagrange multiplier term to enforce the TTNC condition and also used the identity
\begin{equation}
v^\mu v^\nu\check R_{\mu\nu}=\left(h^{\mu\nu}K_{\mu\nu}\right)^2-h^{\mu\rho}h^{\nu\sigma}K_{\mu\nu}K_{\rho\sigma}+\check\nabla_\rho\left(v^\rho h^{\mu\nu}K_{\mu\nu}\right)\,.
\end{equation}

In order to obtain \eqref{eq:action_NRG}, which is one of the central results of this paper, 
we worked out the variations of the first line of \eqref{eq:LNNLO} using the TTNC condition. In the next subsection it will be shown that this Lagrangian is equivalent to the one given in our previous work \cite{Hansen:2018ofj} which was obtained from gauge symmetry principles.
It is clear that the expansion of the Einstein--Hilbert Lagrangian in $\sigma=1/c^2$ can be done systematically using the above framework.
There is nothing, except computational complexity, that prevents an expansion to any given order in $\sigma$,
yielding more and more relativistic corrections to non-relativistic gravity. We will discuss applications of this approach further in Section \ref{sec:discussion}.

\subsubsection{Equations of motion}\label{sec:NRG_EOMs}
We derive here the equations of motion based on the variations of $\mathcal{L}_\mathrm{NRG}$ with respect to $\tau_\mu$, $h_{\mu\nu}$, $m_\mu$ and $\Phi_{\mu\nu}$ where for the $\tau_\mu$ variation we only consider $\delta\tau_\mu=\Omega\tau_\mu$ with $\Omega$ an arbitrary function.

Because TTNC is imposed off shell a number of identities that applies to TTNC geometry has to be used in the process, which we have collected in appendix \ref{sec:TTNC_identities_special}.
With these identities at hand, the variations that need to be done are straightforward albeit slightly tedious.

The variation of NRG in terms of the LO and NLO fields  can be written as

\begin{equation}\label{eq:EOMs_NRG_variation}
\delta\mathcal{L_{\mathrm{NRG}}} \equiv -\frac{e}{8\pi G_N} \left(\mathcal{G}_{\tau}^\mu \delta \tau_\mu + \mathcal{G}_{m}^\mu \delta m_\mu
+\frac{1}{2}\mathcal{G}_{h}^{\mu\nu}\delta h_{\mu\nu}+\frac{1}{2}\mathcal{G}_{\Phi}^{\mu\nu}\delta{\Phi}_{\mu\nu}\right)\,.
\end{equation}
We denoted the responses with caligraphic symbols to distinguish them from the variations of the full NNLO Lagrangian in which the TTNC condition has not been imposed, i.e. $\mathcal{G}_\phi\equiv \os{G}{-2}_{\phi}\vert_{\tau\wedge \d\tau=0}$ for the field $\phi$.

Ignoring diffeomorphisms and using TTNC, it follows from \eqref{eq:trafo5a}-\eqref{eq:trafo9} that $h_{\mu\nu}$ and the subleading fields transform as 
\begin{eqnarray}
\delta h_{\mu\nu} & = & \lambda_\mu\tau_\nu+\lambda_\nu\tau_\mu\,,\\
\delta m_\mu & = & \lambda_\mu+\partial_\mu\Lambda -\Lambda a_\mu+h^{\rho\sigma}\zeta_\sigma a_\rho\tau_\mu\,,\\
h^{\mu\rho}h^{\nu\sigma}\delta\Phi_{\mu\nu} & = & h^{\mu\rho}h^{\nu\sigma}\left(\lambda_\mu m_\nu+\lambda_\nu m_\mu+2\Lambda K_{\mu\nu}+\check\nabla_\mu\zeta_\nu+\check\nabla_\nu\zeta_\mu\right)\,,
\end{eqnarray}
where $v^\mu\lambda_\mu=0$. We define the spatial projector $P_\mu^\nu$ as
\begin{equation}
    P_\mu^\nu=\delta^\nu_\mu+v^\nu\tau_\mu\,.
\end{equation}
The Ward identity for Galilean boost invariance with parameter $\lambda_\mu$ is given by
\begin{equation}\label{eq:WIGalboosts}
    P_\mu^\rho\left(\mathcal{G}_m^\mu+\tau_\nu\mathcal{G}_h^{\mu\nu}+m_\nu\mathcal{G}_\Phi^{\mu\nu}\right)=0\,.
\end{equation}
We can use this to simplify the process of varying $h_{\mu\nu}$ which is by far the most laborious variation. We can write 
\begin{equation}
    \frac{1}{2}\mathcal{G}_h^{\mu\nu}\delta h_{\mu\nu}=\frac{1}{2}\mathcal{G}_h^{\rho\sigma}P_\rho^\mu P_\sigma^\nu\delta h_{\mu\nu}- \mathcal{G}_h^{\rho\sigma}\tau_\rho P_\sigma^\nu v^\mu\delta h_{\mu\nu}\,.
\end{equation}
We see that the part in front of the $v^\mu\delta h_{\mu\nu}$ variation is fixed by the Ward identity \eqref{eq:WIGalboosts}. We can thus without any loss of generality restrict ourselves to the $P$-projected variation
\begin{equation}
    \delta_P h_{\rho\sigma}=P_\rho^\mu P_\sigma^\nu\delta h_{\mu\nu}\,.
\end{equation}

One finds after a bit of work that the equations of motion are given by
\begin{eqnarray}
\tau_\mu \mathcal{G}_{\tau}^\mu &=& -\frac{1}{2}\Big[\left(h^{\mu\nu}K_{\mu\nu}\right)^2-h^{\mu\rho}h^{\nu\sigma}K_{\mu\nu}K_{\rho\sigma}+\frac{3}{4}h^{\mu\rho}h^{\nu\sigma}F_{\mu\nu}F_{\rho\sigma}\nonumber \\
&&+m_\nu\left[\left(\check\nabla_\mu+2a_\mu\right)h^{\mu\rho}h^{\nu\sigma}F_{\rho\sigma}-2\left(h^{\mu\rho}h^{\nu\sigma}-h^{\mu\nu}h^{\rho\sigma}\right)\check\nabla_\mu K_{\rho\sigma}\right]\nonumber\\
&&-2\left(h^{\mu\rho}h^{\nu\sigma}-h^{\mu\nu}h^{\rho\sigma}\right)K_{\rho\sigma}\left(\check\nabla_\mu+a_\mu\right)m_\nu
+\left(h^{\mu\rho}h^{\nu\sigma}-h^{\mu\nu}h^{\rho\sigma}\right)\check{\nabla}_\mu \check{\nabla}_\nu\Phi_{\rho\sigma}\nonumber \\
&&-\Phi_{\rho\sigma}h^{\mu\rho}h^{\nu\sigma}\left(\check R_{\mu\nu}-\frac{1}{2}h_{\mu\nu}h^{\kappa\lambda}\check R_{\kappa\lambda}\right)\Big]\,,\label{eq:NRG_EOM1}\\
\mathcal{G}_{m}^\nu &=& \frac{1}{2}\Big[ 2\left(h^{\mu\rho}h^{\nu\sigma}-h^{\mu\nu}h^{\rho\sigma}\right)\check{\nabla}_\mu K_{\rho\sigma}
+v^\nu h^{\mu\nu}\check R_{\mu\nu}
+\left(\check{\nabla}_\mu+2a_\mu \right)h^{\mu\rho}h^{\nu\sigma}F_{\rho\sigma}\Big]\,,\label{eq:NRG_EOM2}\\
\mathcal{G}_{\Phi}^{\rho\sigma}&=&h^{\mu\rho}h^{\nu\sigma}\left(\check R_{\mu\nu}
-\frac{1}{2}h_{\mu\nu}h^{\kappa\lambda}\check R_{\kappa\lambda}-\left(\check{\nabla}_\mu +a_\mu\right)a_\nu 
+h_{\mu\nu}h^{\kappa\lambda} \left(\check\nabla_{\kappa}+a_\kappa\right)a_\lambda\right)\,,\label{eq:NRG_EOM3}\\
\mathcal{G}_{h}^{\mu\nu}P^\alpha_\mu P^\beta_\nu &=& -\frac{1}{2}h^{\alpha\beta}\left(h^{\mu\rho}h^{\nu\sigma}-h^{\mu\nu}h^{\rho\sigma}\right)K_{\mu\nu}K_{\rho\sigma}-\check\nabla_\lambda\left(v^\lambda\left(h^{\mu\alpha}h^{\nu\beta}-h^{\alpha\beta}h^{\mu\nu}\right)K_{\mu\nu}\right)\nonumber\\
&&-h^{\alpha\beta}h^{\mu\rho}h^{\nu\sigma}K_{\rho\sigma}\left(\check\nabla_\mu+a_\mu\right)m_\nu+h^{\alpha\beta}\check\nabla_\lambda\left(v^\lambda h^{\mu\nu}\left(\check\nabla_\mu+a_\mu\right)m_\nu\right)\nonumber\\
&&-\frac{1}{2}\check\nabla_\lambda\left(v^\lambda\left(h^{\mu\alpha}h^{\nu\beta}+h^{\nu\alpha}h^{\mu\beta}\right)\left(\check\nabla_\mu+a_\mu\right)m_\nu\right)\nonumber\\
&&+h^{\alpha\rho}h^{\beta\sigma}K_{\rho\sigma}h^{\mu\nu}\left(\check\nabla_\mu+a_\mu\right)m_\nu-\left(h^{\alpha\beta}h^{\rho\sigma}-h^{\alpha\rho}h^{\beta\sigma}\right)h^{\mu\nu}m_\mu\check\nabla_\nu K_{\rho\sigma}\nonumber\\
&&-h^{\mu\sigma}\left(h^{\alpha\rho}h^{\beta\nu}+h^{\beta\rho}h^{\alpha\nu}\right)m_\nu\left(\check\nabla_\mu K_{\rho\sigma}-\check\nabla_\rho K_{\mu\sigma}\right)\nonumber\\
&&-\frac{1}{2}\left(h^{\alpha\rho}h^{\beta\nu}+h^{\beta\rho}h^{\alpha\nu}\right)h^{\mu\sigma}K_{\rho\sigma}\left(F_{\mu\nu}+2a_\mu m_\nu-2a_\nu m_\mu\right)\nonumber\\
&&+\frac{1}{2}h^{\mu\alpha}h^{\rho\beta}h^{\nu\sigma}F_{\mu\nu}F_{\rho\sigma}-\frac{1}{8}h^{\alpha\beta}h^{\mu\rho}h^{\nu\sigma}F_{\mu\nu}F_{\rho\sigma}\nonumber\\
&&+\Phi h^{\mu\alpha}h^{\nu\beta}\left(\check R_{\mu\nu}-\frac{1}{2}h_{\mu\nu}h^{\rho\sigma}\check R_{\rho\sigma}\right)+h^{\alpha\beta}h^{\rho\sigma}\left(\check\nabla_\rho+a_\rho\right)\left(\check\nabla_\sigma+a_\sigma\right)\Phi\nonumber\\
&&-\frac{1}{2}\left(h^{\mu\alpha}h^{\rho\beta}+h^{\mu\beta}h^{\rho\alpha}\right)\left(\check\nabla_\mu+a_\mu\right)\left(\check\nabla_\rho+a_\rho\right)\Phi\nonumber\\
&&-\frac{1}{2}\left(h^{\alpha\beta}h^{\mu\rho}h^{\nu\sigma}+h^{\mu\nu}h^{\alpha\rho}h^{\beta\sigma}-h^{\nu\sigma}\left(h^{\mu\alpha}h^{\rho\beta}+h^{\mu\beta}h^{\rho\alpha}\right)\right)\Phi_{\rho\sigma}\left(\check\nabla_\mu+a_\mu\right)a_\nu\nonumber\\
&&-\frac{1}{2}h^{\mu\rho}\left(h^{\nu\beta}h^{\sigma\alpha}+h^{\nu\alpha}h^{\sigma\beta}\right)a_\nu\check\nabla_\mu\Phi_{\rho\sigma}+\frac{1}{2}h^{\rho\sigma}\left(h^{\mu\alpha}h^{\nu\beta}+h^{\mu\beta}h^{\nu\alpha}\right)a_\nu\check\nabla_\mu\Phi_{\rho\sigma}\nonumber\\
&&+\frac{1}{2}h^{\mu\nu}\left(h^{\alpha\rho}h^{\beta\sigma}-h^{\alpha\beta}h^{\rho\sigma}\right)a_\nu\check\nabla_\mu\Phi_{\rho\sigma}+\frac{1}{2}h^{\kappa\lambda}\Phi_{\kappa\lambda}h^{\mu\alpha}h^{\nu\beta}\left(\check R_{\mu\nu}-\frac{1}{2}h_{\mu\nu}h^{\rho\sigma}\check R_{\rho\sigma}\right)\nonumber\\
&&-\frac{1}{4}h^{\kappa\lambda}\left(h^{\mu\alpha}h^{\rho\beta}+h^{\mu\beta}h^{\rho\alpha}\right)\left(\check\nabla_\mu+a_\mu\right)\left(\check\nabla_\rho+a_\rho\right)\Phi_{\kappa\lambda}\nonumber\\
&&+\frac{1}{2}h^{\rho\alpha}h^{\sigma\beta}\Phi_{\rho\sigma}h^{\mu\nu}\check R_{\mu\nu}+\frac{1}{2}h^{\kappa\lambda}h^{\alpha\beta}h^{\rho\sigma}\left(\check\nabla_\rho+a_\rho\right)\left(\check\nabla_\sigma+a_\sigma\right)\Phi_{\kappa\lambda}\nonumber\\
&&-\left(h^{\mu\alpha}h^{\rho\beta}h^{\nu\sigma}+h^{\mu\rho}h^{\nu\alpha}h^{\sigma\beta}\right)\Phi_{\rho\sigma}\check R_{\mu\nu}+\frac{1}{2}h^{\alpha\beta}h^{\mu\rho}h^{\nu\sigma}\Phi_{\rho\sigma}\check R_{\mu\nu}\nonumber\\
&&+\frac{1}{2}h^{\mu\rho}\left(h^{\alpha\sigma}h^{\beta\nu}+h^{\beta\sigma}h^{\alpha\nu}\right)\left(\check\nabla_\mu+a_\mu\right)\left(\check\nabla_\nu+a_\nu\right)\Phi_{\rho\sigma}\nonumber\\
&&+\left(h^{\alpha\rho}h^{\beta\sigma}h^{\mu\nu}+h^{\nu\sigma}h^{\mu\rho}h^{\alpha\beta}\right)\left(\check\nabla_\mu+a_\mu\right)\left(\check\nabla_\nu+a_\nu\right)\Phi_{\rho\sigma}\label{eq:NRG_EOM4}\,.
\end{eqnarray}
These equations are somewhat lengthy and perhaps they would be easier to handle in a first-order formalism at the price of working with more fields. This will be studied further in upcoming work \cite{Hansen:2020a}. We will see in Section \ref{sec:NRGprime_EOMs} that the equations of motion acquire a more compact form if one 
uses boost-invariant fields. Finally, it can be shown that when  $\d\tau=0$ it follows that $\Phi_{\mu\nu}$ decouples and that there are drastic simplifications as we will be studied further in Section \ref{sec:torsionless_Newtonian_gravity}.

There are three \reGlspl{wi} that result from invariances under the LO and NLO diffeomorhisms. The latter are the gauge transformations with parameters $\Lambda$ and $\zeta_\mu$. For the $\Lambda$-transformation we find
\begin{equation}\label{eq:LambdaBianchi}
    \left(\check\nabla_\mu+2a_\mu\right)\mathcal{G}_m^\mu-K_{\mu\nu}\mathcal{G}_\Phi^{\mu\nu}=0\,,
\end{equation}
while for $\zeta_\mu$ we obtain
\begin{equation}\label{eq:zetaBianchi}
    \left(\check\nabla_\nu+a_\nu\right)\mathcal{G}_\Phi^{\mu\nu}-\mathcal{G}_m^\nu\tau_\nu h^{\mu\rho}a_\rho=0\,.
\end{equation}

An important role will be played by two equations that can be obtained from various contractions of the above equations of motion. The first of these is the combination 
\begin{equation}\label{eq:TTNCsourcing}
    -(d-2)\tau_\mu\mathcal{G}_m^\mu+h_{\mu\nu}\mathcal{G}_\Phi^{\mu\nu}=(d-1)h^{\mu\nu}\left(\check\nabla_\mu+a_\mu\right)a_\nu=(d-1)e^{-1}\partial_\mu\left(eh^{\mu\nu}a_\nu\right)\,.
\end{equation}
We note that TTNC implies $\tau=N \d T$ where $N$ is like a non-relativistic lapse function and that this in turn implies $h^{\mu\nu}a_\nu=h^{\mu\nu}N^{-1}\partial_\nu N$ so that the right hand side of the above equation is the Laplacian of the non-relativistic lapse function. When we study matter couplings in the next section this will tell us something about what type of matter sources TTNC torsion. The second equation follows from the following combination
\begin{eqnarray}
&&-(d-2)\tau_\mu \mathcal{G}_{\tau}^\mu-(d-2)m_\mu \mathcal{G}_{m}^\mu+h_{\mu\nu}\mathcal{G}_{h}^{\mu\nu}+\Phi_{\mu\nu}\mathcal{G}_{\Phi}^{\mu\nu}=(d-1)\left[v^\mu v^\nu\check R_{\mu\nu}\right.\nonumber\\
&&\left.+\check\nabla_\rho\left(v^\rho h^{\mu\nu}\left(\check\nabla_\mu+a_\mu\right)m_\nu\right)+\frac{1}{4}h^{\mu\rho}h^{\nu\sigma}F_{\mu\nu}F_{\rho\sigma}+h^{\mu\nu}\left(\check\nabla_\mu+a_\mu\right)\left(\check\nabla_\nu+a_\nu\right)\Phi\right.\nonumber\\
&&\left.-\left(h^{\mu\rho}h^{\nu\sigma}-\frac{1}{2}h^{\mu\nu}h^{\rho\sigma}\right)\left(\check\nabla_\mu+a_\mu\right)\left(a_\nu\Phi_{\rho\sigma}+2m_\nu K_{\rho\sigma}\right)\right]\,.\label{eq:scaleeq}
\end{eqnarray}
Equation \eqref{eq:scaleeq} contains the Laplacian of the Newtonian potential $\Phi$. When $\d\tau=0$ and hence 
$a_\mu=0$ the field $\Phi_{\mu\nu}$ decouples from this equation. When coupling this equation to matter we will be able to identify the sources of Newtonian gravity.

Given the simplicity of \eqref{eq:scaleeq} in comparison to the $h_{\mu\nu}$ equation of motion one might wonder if there exists a simpler way to obtain this result. The answer is affirmative and this conclusion can be obtained by going back to \eqref{eq:scalingrel}. To show this we need to expand the Einstein equation in powers of $1/c^2$.
We can relate the $1/c^2$ expansion of the Einstein equations $E^{\mu\nu}_{g},\,E^{\mu}_{g}$ defined through \eqref{eq:Einstein_field_eqs} to the EOMs defined through \eqref{eq:1c2_expansion_def_EOMs}.
When TTNC is imposed off shell we find the non-vanishing orders to give
\begin{eqnarray}
\os{G}{-4}^{\mu\nu}_{h}=\os{G}{-2}^{\mu\nu}_{\Phi}&=& \os{E}{-4}^{\mu\nu}_{g}\,,\\
\os{G}{-4}^\mu_{\tau}=\os{G}{-2}^\mu_{m} &=& \os{E}{-4}^\mu_{g}\,,\\
\os{G}{-2}^{\mu\nu}_{h}&=& \os{E}{-2}^{\mu\nu}_{g}+\left(\Phi+\frac{1}{2}h^{\alpha\beta}\Phi_{\alpha\beta}\right)\os{E}{-4}^{\mu\nu}_{g}\,,\\
\os{G}{-2}^\mu_{\tau} &=& \os{E}{-2}^\mu_{g}+\left(\Phi+\frac{1}{2}h^{\alpha\beta}\Phi_{\alpha\beta}\right)\os{E}{-4}^\mu_{g}\,,
\end{eqnarray}
where the $\Phi+\frac{1}{2}h^{\alpha\beta}\Phi_{\alpha\beta}$ terms originate from expanding $\sqrt{-g}$. In deriving this result we expanded $E^\mu_g$ and $E^{\mu\nu}_g$ as 
\begin{equation}
E^\mu_g=\os{E}{-6}_{g}^\mu+\sigma\os{E}{-4}_{g}^\mu+\sigma^2\os{E}{-2}_{g}^\mu+\order{\sigma^3}\,.
\end{equation}
The notation follows from \eqref{eq:EH_variation_explicit} which has an overall factor of $c^6$ and this is why we start with $\os{E}{-6}_{g}^\mu$.
Since we work here with off shell TTNC it follows that $\os{E}{-6}_{g}^\mu=0$.
Similar statements apply to $E^{\mu\nu}_g$ mutatis mutandis.

When expanding the anisotropic scaling equation \eqref{eq:scalingrel} in GR in Section \ref{sec:causal_expansion_EH} we see that at LO it reproduces the TTNC condition while at NLO it gives rise to \eqref{eq:TTNCsourcing}. Finally at the NNLO it reproduces \eqref{eq:scaleeq}.

\subsection{Theory from gauge invariances}\label{sec:action_from_torsional_u1_invariance}
\subsubsection{Lagrangian}
In the previous section we derived the Lagrangian of non-relativistic gravity from the $1/c^2$ expansion of the Einstein--Hilbert Lagrangian. In this section we will derive the same result using a different method. Starting with the field content that originates from the $1/c^2$ expansion of the metric  we will derive a (two-derivative) Lagrangian that has all the gauge invariances associated with the type II NC gauge transformations of these fields. 

We will work with manifestly Galilean boost invariant quantities, i.e. $\tau_\mu$, $\bar h_{\mu\nu}$, $\hat\Phi$ (and their inverses $\hat v^\mu$ and $h^{\mu\nu}$) as well as $\bar\Phi_{\mu\nu}$ (see equations \eqref{eq:def_barh}-\eqref{eq:def_barPhi} for their definitions). These fields transform as in \eqref{eq:trafo5a}-\eqref{eq:trafo9}. In this section we will mostly be concerned with the $\Lambda=\tau_\mu\zeta^\mu$ part of the these transformations which can be rewritten as 
\begin{eqnarray}
\delta_\Lambda m_\mu & = &\partial_\mu \Lambda -\Lambda \hat a_\mu\,,\label{eq:torsional_U1_prime1}\\
h^{\mu\rho}h^{\nu\sigma}\delta_\Lambda \bar{\Phi}_{\rho\sigma} & = &2\Lambda h^{\mu\rho}h^{\nu\sigma} \bar K_{\rho\sigma}\,,\label{eq:torsional_U1_prime2}
\end{eqnarray}
where we defined the boost invariant torsion vector and the extrinsic curvature\footnote{We have in previous work defined `$a_\mu$' as what we here call $\hat a_\mu$. However, since we usually encounter spatial contractions of it, for which $h^{\mu\nu}\hat a_\mu=h^{\mu\nu}a_\mu$, we can be relaxed about their difference. A conversion table between various notations can be found in appendix \ref{sec:conventions_change}.}
\begin{eqnarray}
\hat a_\mu &\equiv&\mathcal{L}_{\hat v} \tau_\mu\,,\\
\bar K_{\mu\nu} &\equiv& -\frac{1}{2}\mathcal{L}_{\hat v} \bar h_{\mu\nu}\,.
\end{eqnarray}
Furthermore we will work with the Galilean boost invariant connection $\bar\Gamma^\rho_{\mu\nu}$ defined in equation \eqref{eq:Special TNC connection}. We will refer to this as the torsional $U(1)$ gauge transformation to contrast it with the type I Bargmann $U(1)$ gauge transformation with parameter $\sigma$.

We will assume that the Lagrangian is at most second order in derivatives and that it has at least the kinetic term $\hat v^\mu\hat v^\nu\bar R_{\mu\nu}$. We will simply add terms until we obtain invariance under all the desired non-relativistic symmetries \eqref{eq:torsional_u(1)_1}-\eqref{eq:torsional_u(1)_2} with off shell TTNC condition $\tau\wedge \d\tau=0$ imposed.
This was the original approach used to find the action presented in our previous work \cite{Hansen:2018ofj}.

If we ignore the field $\bar\Phi_{\mu\nu}$ the only difference between type I and type II NC geometry is the type II transformation of $m_\mu$ under the $\Lambda$-transformation which should be contrasted with the type I $\sigma$ transformation given in \eqref{eq:typeItrafom}. Note that for zero torsion the type II transformation of $m_\mu$
reduces to the Bargmann $U(1)$ gauge transformation.

Type I NC geometry can be obtained by null reduction as detailed in appendix \ref{subsec:null-redGR}. Since from the point of view of the fields $\tau_\mu$, $m_\mu$ and $h_{\mu\nu}$ the difference between type I and type II is just one term in the transformation rule of $m_\mu$ we will work with the reasonable assumption that to build a type II invariant action the non-relativistic gravity Lagrangian should contain the term $\hat G^{uu}$. This is the $uu$ component of the `null uplifted' Einstein tensor $\hat G^{MN}$ (see appendix \ref{subsec:null-redGR} for more details).
This term indeed contains the kinetic term $\hat v^\mu\hat v^\nu\bar R_{\mu\nu}$ we would like there to be. The other terms in the Lagrangian can be found by demanding that they cancel the non-invariance of $\hat G^{uu}$ under the type II $\Lambda$-transformation.

To this end let us consider the transformations of $\hat G^{uu}$ under the variation of $m_\mu$. We start with the transformation of the connection $\bar\Gamma^\rho_{\mu\nu}$ defined in \eqref{eq:Special TNC connection}.
Using TTNC throughout we have
\begin{eqnarray}
\delta_m\bar\Gamma^\rho_{\mu\nu} & = & \frac{1}{2}\left(\hat a_\mu\tau_\nu+\hat a_\nu\tau_\mu\right)h^{\rho\sigma}\delta m_\sigma-\tau_\mu\tau_\nu h^{\rho\sigma}\hat a_\sigma\hat v^\lambda\delta m_\lambda\nonumber\\
&&+\frac{1}{2}\tau_\mu h^{\rho\sigma}\left[\left(\bar\nabla_\sigma+\hat a_\sigma\right)\delta m_\nu-\left(\bar\nabla_\nu+\hat a_\nu\right)\delta m_\sigma\right]\nonumber\\
&&+\frac{1}{2}\tau_\mu h^{\rho\sigma}\left[\left(\bar\nabla_\sigma+\hat a_\sigma\right)\delta m_\nu-\left(\bar\nabla_\nu+\hat a_\nu\right)\delta m_\sigma\right]\,.
\end{eqnarray}
Using this result as well as equation \eqref{eq:nullredmunu} it follows that for the $\mu\nu$ component of the 
$(d+2)$-dimensional Ricci tensor $\hat R_{MN}$ we have
\begin{eqnarray}
\delta_m\hat R_{\mu\nu}&=&\left(\bar\nabla_\rho+\hat a_\rho\right)\delta_m\bar\Gamma^\rho_{(\mu\nu)}-\tau_\mu\tau_\nu h^{\rho\sigma}\hat a_\rho \hat a_\sigma\hat v^\lambda\delta m_\lambda\nonumber\\
&&-\frac{1}{2}\tau_\mu h^{\rho\sigma}\hat a_\rho \bar\nabla_\nu\delta m_\sigma-\frac{1}{2}\tau_\nu h^{\rho\sigma}\hat a_\rho\bar\nabla_\mu\delta m_\sigma\,.
\end{eqnarray}
This implies that 
\begin{equation}\label{eq:nom}
h^{\mu\rho}h^{\nu\sigma}\delta_m\hat R_{\rho\sigma}=0\,,
\end{equation}
a result that will be useful later. It also implies (up to a total derivative) that when taking $m_\mu$ to transform under the torsional $U(1)$ one has 
\begin{eqnarray}
\hat v^\mu\hat v^\nu\delta_\Lambda\hat R_{\mu\nu} & = & e^{-1}\partial_\mu\left(eh^{\mu\nu}\hat a_\nu\right)\delta_\Lambda\hat\Phi+\frac{1}{2}h^{\mu\nu}\hat a_\mu \hat a_\nu\delta_\Lambda\hat\Phi\nonumber\\
&&+\Lambda \bar K_{\mu\nu}\left(h^{\mu\rho}h^{\nu\sigma}\hat a_\rho \hat a_\sigma-\frac{1}{2}h^{\mu\nu}h^{\rho\sigma}\hat a_\rho \hat a_\sigma\right)\,.
\end{eqnarray}
Using 
\begin{equation}
\hat R^{uu}=g^{uM}g^{uN}\hat R_{MN}=\hat v^\mu\hat v^\nu\hat R_{\mu\nu}-2\hat\Phi e^{-1}\partial_\mu\left(eh^{\mu\nu}\hat a_\nu\right)\,,
\end{equation}
where in the second equality we used \eqref{eq:hatRmuu}, we find
\begin{equation}
\delta_\Lambda\hat R^{uu}=2\hat R^{\nu u}\left(\partial_\nu\Lambda-\hat a_\nu\Lambda\right)+\frac{1}{2}h^{\mu\nu}\hat a_\mu \hat a_\nu\delta_\Lambda\hat\Phi+\Lambda \bar K_{\mu\nu}\left(h^{\mu\rho}h^{\nu\sigma}-\frac{1}{2}h^{\mu\nu}h^{\rho\sigma}\right)\hat a_\rho \hat a_\sigma\,.
\end{equation}
In other words since $h^{\mu\nu}\delta_m \hat a_\nu=0$ we can write this as
\begin{equation}\label{eq:intermediate}
\delta_\Lambda\left[\hat R^{uu}-\frac{1}{2}\hat\Phi h^{\mu\nu}\hat a_\mu \hat a_\nu-\frac{1}{2}\bar{\Phi}_{\mu\nu}\left(h^{\mu\rho}h^{\nu\sigma}-\frac{1}{2}h^{\mu\nu}h^{\rho\sigma}\right)\hat a_\rho \hat a_\sigma\right]=2\hat R^{\mu u}\left(\partial_\mu\Lambda-\hat a_\mu\Lambda\right)\,.
\end{equation}

We can straightforwardly replace $\hat R^{uu}$ in \eqref{eq:intermediate} by $\hat G^{uu}$ because for TTNC $\hat R$ is independent of $m_\mu$ (for TNC it would vary into $\hat R^\mu{}_u\delta m_\mu$ but for TTNC $\hat R^\mu{}_u=0$). Finally, since for TTNC we also have that $\delta_m\hat G^{\mu\nu}=0$, which follows essentially from \eqref{eq:nom}, we can subtract $\delta_\Lambda\bar{\Phi}_{\mu\nu}\hat G^{\mu\nu}$ from both sides of \eqref{eq:intermediate}. Putting it all together we obtain the transformation rule with respect to the type II $\Lambda$-transformation,
\begin{eqnarray}
&&\delta_\Lambda\left[\hat G^{uu}-\frac{1}{2}\hat\Phi h^{\mu\nu}\hat a_\mu \hat a_\nu+\bar{\Phi}_{\mu\nu}\hat G^{\mu\nu}-\frac{1}{2}\bar{\Phi}_{\mu\nu}\left(h^{\mu\rho}h^{\nu\sigma}\hat a_\rho \hat a_\sigma-\frac{1}{2}h^{\mu\nu}h^{\rho\sigma}\hat a_\rho \hat a_\sigma\right)\right]\nonumber\\
&&=2\hat G^{\mu u}\left(\partial_\mu\Lambda-\hat a_\mu\Lambda\right)+2\Lambda\bar K_{\mu\nu}\hat G^{\mu\nu}\,.
\end{eqnarray}
It then follows that the Lagrangian given by
\begin{equation}\label{eq:action}
\mathcal{L}=e\left[-\hat G^{uu}+\frac{1}{2}\hat\Phi h^{\mu\nu}\hat a_\mu \hat a_\nu-\bar{\Phi}_{\mu\nu}\left(\hat G^{\mu\nu}-\frac{1}{2}h^{\mu\rho}h^{\nu\sigma}\hat a_\rho \hat a_\sigma+\frac{1}{4}h^{\mu\nu}h^{\rho\sigma}\hat a_\rho \hat a_\sigma\right)\right]\,,
\end{equation}
is invariant under the torsional $U(1)$ transformation  \eqref{eq:torsional_U1_prime1}-\eqref{eq:torsional_U1_prime2} (after partial integration) by virtue of the Bianchi identity \eqref{eq:nullredBI2}. We can then write
\begin{equation}
\hat G^{\mu\nu}=h^{\mu\rho}h^{\nu\sigma}\hat G_{\rho\sigma}=h^{\mu\rho}h^{\nu\sigma}\hat R_{\rho\sigma}-\frac{1}{2}h^{\mu\nu}\hat R\,,
\end{equation} 
where $\hat R_{\mu\nu}$ and $\hat R$ are given in \eqref{eq:nullredmunu} and \eqref{eq:hatRicci}, respectively. In terms of more intrinsically defined objects this can be rewritten using (up to total derivatives)
\begin{equation}
\hat G^{uu}-\frac{1}{2}\hat\Phi h^{\mu\nu}a_\mu a_\nu=\hat v^\mu\hat v^\nu\bar R_{\mu\nu}-\hat\Phi h^{\mu\nu}\bar R_{\mu\nu}=\left(h^{\mu\nu}\bar K_{\mu\nu}\right)^2-h^{\mu\rho}h^{\nu\sigma}\bar K_{\mu\nu}\bar K_{\rho\sigma}-\hat\Phi h^{\mu\nu}\bar R_{\mu\nu}\,,
\end{equation}
as follows from \eqref{eq:Guu} and \eqref{eq:kinetic}.

The Lagrangian \eqref{eq:action} (with a restored prefactor $1/16\pi G_N$) can finally be written as what we will call the (primed) Non-Relativistic Gravity Lagrangian.
\begin{eqnarray}\label{eq:action_NRG_prime}
\mathcal{L}'_{\mathrm{NRG}}&=&\frac{e}{16\pi G_N}\Bigg[-\hat v^\mu\hat v^\nu\bar R_{\mu\nu}+\hat\Phi h^{\mu\nu}\bar R_{\mu\nu}
-\bar\Phi_{\rho\sigma}h^{\mu\rho}h^{\nu\sigma}\Big(\bar R_{\mu\nu}-\bar\nabla_\mu \hat a_\nu-\hat a_\mu \hat a_\nu \nonumber\\
&&\qquad\qquad\,-\frac{1}{2}h_{\mu\nu}h^{\kappa\lambda}\bar R_{\kappa\lambda}+h_{\mu\nu}e^{-1}\partial_\kappa\left(eh^{\kappa\lambda}\hat a_\lambda\right)\Big)\Bigg]\,.
\end{eqnarray}

The Lagrangian \eqref{eq:action_NRG_prime} has an additional gauge symmetry which reads
\begin{equation}\label{eq:zeta_gauge_transform_prime}
h^{\mu\rho}h^{\nu\sigma}\delta_{\zeta}\bar{\Phi}_{\rho\sigma}=h^{\mu\rho}h^{\nu\sigma}\left(\bar\nabla_\rho\zeta_\sigma+\bar\nabla_\sigma\zeta_\rho\right)
\,,\qquad\delta_{\zeta}\hat\Phi=h^{\mu\nu}\hat a_\mu\zeta_\nu\,.
\end{equation}
In order to show this we need to use two identities.
The first one is the $N=\nu$ component of the Bianchi identity $\hat\nabla_{M}\hat G^{M N}=0$, which can be written as
\begin{equation}
\left(\bar\nabla_\mu+\hat a_\mu\right)\hat G^{\mu\nu}-\frac{1}{2}h^{\nu\rho}\hat a_\rho e^{-1}\partial_\mu\left(eh^{\mu\sigma}\hat a_\sigma\right)+\frac{1}{4}h^{\nu\rho}\hat a_\rho h^{\mu\sigma}\hat a_\mu \hat a_\sigma+\frac{1}{2}h^{\nu\rho}\hat a_\rho h^{\mu\sigma}\bar R_{\mu\sigma}=0\,.
\end{equation}
The second one is the identity
\begin{equation}
\left(\bar\nabla_\mu+\hat a_\mu\right)\left(h^{\mu\rho}h^{\nu\sigma}\hat a_\rho \hat a_\sigma-\frac{1}{2}h^{\mu\nu}h^{\rho\sigma}\hat a_\rho \hat a_\sigma\right)=h^{\nu\rho}\hat a_\rho e^{-1}\partial_\mu\left(eh^{\mu\sigma}\hat a_\sigma\right)-\frac{1}{2}h^{\nu\rho}\hat a_\rho h^{\mu\sigma}\hat a_\mu \hat a_\sigma\,.
\end{equation}
The $\zeta$ transformations together with the $\Lambda$-transformation make the Lagrangian invariant under \eqref{eq:trafo5a}-\eqref{eq:trafo9}. It is interesting to note that, as shown above, the Lagrangian is fixed already by the $\Lambda$-invariance, with the $\zeta$ transformations appearing as an extra gauge symmetry.
This completes the construction of a Lagrangian that is invariant under the type II gauge transformations.

\subsubsection{Equations of motion}\label{sec:NRGprime_EOMs}
The details of the variational calculus of the action \eqref{eq:action_NRG_prime} can be found in appendix \ref{sec:TTNC_identities_special}.
A number of identities that applies to TTNC geometry has to be applied in the process.
If we define 
\begin{equation}\label{eq:EOMs_prime_variation}
\delta\mathcal{L}'_{\mathrm{NRG}} \equiv \frac{e}{8\pi G_N} \left(\mathcal G_{\hat\Phi}\delta\hat\Phi-\mathcal G^{\hat v}_\mu\delta\hat v^\mu+\frac{1}{2}\mathcal G^h_{\mu\nu}\delta h^{\mu\nu}-\frac{1}{2}\mathcal G^{\mu\nu}_{\bar \Phi}\delta\bar{\Phi}_{\mu\nu}\right)\,,
\end{equation}
one finds after a bit of work that the EOMs are given by
\begin{eqnarray}
\mathcal G_{\hat\Phi} & = & \frac{1}{2}h^{\mu\nu}\bar R_{\mu\nu}\nonumber\\
\mathcal G^{\mu\nu}_{\bar \Phi} &=& h^{\mu\rho}h^{\nu\sigma}\left(\bar R_{\rho\sigma}-\hat a_\rho \hat a_\sigma-\bar\nabla_{\rho}\hat a_{\sigma}\right)-\frac{1}{2}h^{\mu\nu}\left(h^{\rho\sigma}\bar R_{\rho\sigma}-2e^{-1}\partial_\rho\left(eh^{\rho\sigma}\hat a_\sigma\right)\right)\,,\nonumber\\
h^{\rho\mu}\mathcal G^{\hat v}_\mu & = & h^{\rho \mu} \hat v^\nu \bar R_{\mu\nu} \,,\nonumber\\
\hat v^\mu \mathcal G^{\hat v}_\mu & = & \hat\Phi \mathcal G_{\hat\Phi}-\frac{1}{2}\bar{\Phi}_{\mu\nu}\mathcal G_{\bar \Phi}^{\mu\nu}+\frac{1}{2}h^{\mu\nu}\bar{\Phi}_{\mu\nu}e^{-1}\partial_\rho\left(eh^{\rho\sigma}\hat a_\sigma\right)-\frac{1}{2}h^{\mu\rho}h^{\nu\sigma}\bar{\Phi}_{\mu\nu}\left(\bar\nabla_\rho \hat a_\sigma+\hat a_\rho \hat a_\sigma\right)\nonumber\\
&&+\frac{1}{2}\left(h^{\rho\sigma}\bar K_{\rho\sigma}\right)^2-\frac{1}{2}h^{\rho\sigma}h^{\kappa\lambda}\bar K_{\rho\kappa}\bar K_{\sigma\lambda}-\frac{1}{2}\bar\nabla_\mu\left[h^{\mu\rho}h^{\nu\sigma}\left(\bar\nabla_\rho\bar{\Phi}_{\nu\sigma}-\bar\nabla_\nu\bar{\Phi}_{\rho\sigma}\right)\right]\,,\nonumber\\
\mathcal G_h^{\alpha\beta} & = & 
\left(h^{\mu\alpha}h^{\nu\beta}\bar{\Phi}_{\mu\nu}-\frac{1}{2}h^{\alpha\beta}h^{\mu\nu}\bar{\Phi}_{\mu\nu}\right)\left(e^{-1}\partial_\rho\left(eh^{\rho\sigma}\hat a_\sigma\right)-\mathcal G_{\hat\Phi}\right)\nonumber\\
&&\frac{1}{2}h^{\alpha\beta}\bar{\Phi}_{\mu\nu}\mathcal G_{\bar \Phi}^{\mu\nu}-h^{\mu\alpha}\bar{\Phi}_{\mu\rho}\mathcal G_{\bar \Phi}^{\rho\beta}-h^{\mu\beta}\bar{\Phi}_{\mu\rho}\mathcal G_{\bar \Phi}^{\rho\alpha}+\frac{1}{2}h^{\rho\sigma}\bar{\Phi}_{\rho\sigma}\mathcal G_{\bar \Phi}^{\alpha\beta}+\hat\Phi \mathcal G_{\bar \Phi}^{\alpha\beta}\nonumber\\
&&+\frac{1}{2}h^{\alpha\beta}\left[\left(h^{\mu\nu}\bar K_{\mu\nu}\right)^2-h^{\mu\rho}h^{\nu\sigma}\bar K_{\mu\nu}\bar K_{\rho\sigma}\right]-\bar\nabla_\rho\left[\hat v^\rho h^{\mu\alpha}h^{\nu\beta}\bar K_{\mu\nu}-\hat v^\rho h^{\alpha\beta}h^{\mu\nu}\bar K_{\mu\nu}\right]\nonumber\\
&&-h^{\mu\alpha}h^{\nu\beta}\bar\nabla_\mu\partial_\nu\hat\Phi-h^{\mu\alpha}h^{\nu\beta}\left(\hat a_\mu\partial_\nu\hat\Phi+\hat a_\nu\partial_\mu\hat\Phi\right)+h^{\alpha\beta}h^{\mu\nu}\bar\nabla_\mu\partial_\nu\hat\Phi\nonumber\\
&&+2h^{\alpha\beta}h^{\mu\nu}\hat a_\mu\partial_\nu\hat\Phi-\frac{1}{2}h^{\alpha\beta}h^{\mu\nu}h^{\rho\sigma}\left(\bar\nabla_\mu+\hat a_\mu\right)\left(\bar\nabla_\rho+\hat a_\rho\right)\bar{\Phi}_{\nu\sigma}\nonumber\\
&&+h^{\mu\alpha}h^{\nu\beta}h^{\rho\sigma}\left(\bar\nabla_\rho+\hat a_\rho\right)\left(\bar\nabla_{(\mu}\bar{\Phi}_{\nu)\sigma}-\frac{1}{2}\bar\nabla_\sigma\bar{\Phi}_{\mu\nu}\right)\nonumber\\
&&+\frac{1}{2}h^{\alpha\beta}h^{\mu\nu}h^{\rho\sigma}\left(\bar\nabla_\mu+\hat a_\mu\right)\bar\nabla_\nu\bar{\Phi}_{\rho\sigma}-\frac{1}{2}h^{\mu\alpha}h^{\nu\beta}h^{\rho\sigma}\bar\nabla_\mu\bar\nabla_\nu\bar{\Phi}_{\rho\sigma}\label{eq:EOMS_TTNC_action}\,,
\end{eqnarray}
where we found it convenient to split the $\hat v^\mu$ variation in two terms $h^{\rho\mu}\mathcal G^{\hat v}_\mu$ and $\hat v^\mu \mathcal G^{\hat v}_\mu$. We only need to consider the variation $P^\alpha_\mu P^\beta_\nu\delta h^{\mu\nu}$ because this projection is the one where the NNLO fields decouple. This is taken care of by contracting the $\mathcal G^h_{\mu\nu}\delta h^{\mu\nu}\in\delta \mathcal{L}'_{\mathrm{NRG}}$ variation with the inverse spatial metrics $\mathcal G_h^{\alpha\beta} \equiv h^{\mu\alpha}h^{\nu\beta} \mathcal G^h_{\mu\nu}$.
By taking the trace of $\mathcal G_{\bar \Phi}^{\alpha\beta}$ and using $\hat v^\mu \mathcal G^{\hat v}_\mu$ we find the equation analogous to \eqref{eq:scaleeq},
\begin{eqnarray}
(d-2)\hat v^\mu \mathcal G^{\hat v}_\mu+h_{\mu\nu}\mathcal G_h^{\mu\nu} + \bar{\Phi}_{\mu\nu}\mathcal G_{\bar \Phi}^{\mu\nu}& = &\nonumber\\
&&\hspace{-6cm} (d-1)\left[ \hat v^\mu\hat v^\nu\bar R_{\mu\nu}- \left(\bar\nabla_\mu+\hat a_\mu\right)\left(h^{\mu\nu}\hat a_\nu\left(\hat\Phi-\frac{1}{2}h^{\rho\sigma}\bar{\Phi}_{\rho\sigma}\right)+h^{\mu\nu}h^{\rho\sigma}\hat a_\rho\bar{\Phi}_{\nu\sigma}\right)\right] \,.\label{eq:identity_EOMs_NRGprime}
\end{eqnarray}

\subsection{Equality of \texorpdfstring{$\mathcal{L}_{\mathrm{NRG}}$}{LNRG} and \texorpdfstring{$\mathcal{L}'_{\mathrm{NRG}}$}{LNRG prime}}\label{sec:NRG_theories_equivalence}
The two actions \eqref{eq:action_NRG} and \eqref{eq:action_NRG_prime} are equivalent.
To see this one can express all Galilean boost invariant fields $\tau_\mu$, $h^{\mu\nu}$, $\hat v^\mu$, $\bar h_{\mu\nu}$, $\hat \Phi$, $\bar \Phi_{\mu\nu}$ in terms of the fields $\tau_\mu$, $h^{\mu\nu}$, $v^\mu$, $h_{\mu\nu}$, $m_\mu$, $\Phi_{\mu\nu}$ and $B_\mu$, although the latter will drop out when we use TTNC off shell. To make the comparison we also have to change the connection and the Ricci tensor associated to it. The difference between the two connections is (for TTNC geometries)
\begin{eqnarray}
\bar{\Gamma}_{\mu\nu}^{\lambda}-\check{\Gamma}_{\mu\nu}^{\lambda}&=&-\frac{1}{2}h^{\lambda\sigma}\tau_\mu\left(\partial_\nu m_\sigma-\partial_\sigma m_\nu+a_\nu m_\sigma-a_\sigma m_\nu\right)\nonumber\\
&&-\frac{1}{2}h^{\lambda\sigma}\tau_\nu\left(\partial_\mu m_\sigma-\partial_\sigma m_\mu+a_\mu m_\sigma-a_\sigma m_\mu\right)\,.
\end{eqnarray}
Assuming $\tau\wedge \d \tau =0$ we can use some of the results derived in appendix  \ref{sec:TTNC_identities_special}. In particular the following relations are useful
\begin{eqnarray}
h^{\mu\rho}h^{\nu\sigma}\bar K_{\mu\nu} & = & h^{\mu\rho}h^{\nu\sigma}\left(K_{\mu\nu}+\frac{1}{2}(\check\nabla_\mu+a_\mu) m_\nu+\frac{1}{2}(\check\nabla_\nu+ a_\nu) m_\mu\right)\,,\label{eq:relbarKtoK}\\
h^{\mu\rho}h^{\nu\sigma}\bar R_{\mu\nu} & = & h^{\mu\rho}h^{\nu\sigma}\check R_{\mu\nu}\,,\\
h^{\mu\nu}\hat a_\mu & = & h^{\mu\nu}a_\mu\,.
\end{eqnarray}
A straightforward calculation then shows that the Lagrangian \eqref{eq:action_NRG_prime} is equal to  \eqref{eq:action_NRG}.
In deriving the actions we see that they arise from two different (but of course closely connected) approaches: The $1/c^2$ expansion makes it obvious how the NRG theory is related to Einstein's theory of general relativity. On the other hand, from the perspective of gauging the algebra, the second approach makes it clear what the role of the local symmetry algebra is.

We can also relate the equations of motion of the two Lagrangians to each other by changing the basis of the variational calculus as:
\begin{eqnarray}
\delta \hat \Phi &=& h^{\rho\mu}m_\mu\left(v^\sigma-\frac{1}{2}h^{\sigma\nu}m_\nu\right)\delta h_{\rho\sigma}+\Phi \hat v^\rho \delta \tau_\rho-\hat v^\rho \delta m_\rho\,,\label{eq:variation_boostinv_var1}\\
\delta \hat v^\mu &=& -\hat v^\rho h^{\sigma\mu}\delta h_{\rho\sigma}+\left(\Phi h^{\mu\rho}+v^\mu \hat v^\rho \right) \delta \tau_\rho-h^{\mu\rho} \delta m_\rho\,,\label{eq:variation_boostinv_var2}\\
\delta h^{\mu\nu} &=&-h^{\rho\mu} h^{\sigma\nu} \delta h_{\rho\sigma}+\left(v^\mu h^{\nu \rho}+v^\nu h^{\mu \rho}\right)\delta \tau_\rho\,,\label{eq:variation_boostinv_var3}\\
\delta \bar\Phi_{\mu\nu} &=& \delta \Phi_{\mu\nu}-2m_{(\mu}\delta m_{\nu)}-2\tau_{(\mu}\delta B_{\nu)}-2B_{(\mu}\delta \tau_{\nu)}\,\label{eq:variation_boostinv_var4}.
\end{eqnarray}
When TTNC is imposed, $B_\mu$ and its variation decouple in the projections we are interested in.
Inserting these variations in \eqref{eq:EOMs_prime_variation} and equating with \eqref{eq:EOMs_NRG_variation}, we can read off the following relation between the EOMs
\begin{eqnarray}
\mathcal{G}_{\tau}^\rho\tau_\rho &=& \mathcal G_{\hat\Phi} \Phi - \mathcal G_\mu^{\hat v}v^\mu \,,\\
\mathcal{G}_{m}^\rho &=& \mathcal G_{\hat\Phi} \hat v^\rho - \mathcal G_\mu^{\hat v} h^{\mu\rho} - \mathcal G_{\bar\Phi}^{\rho\sigma}m_\sigma\,,\\
\mathcal{G}_{h}^{\rho\sigma} &=& -2\mathcal G_{\hat\Phi} h^{\mu(\rho} m_\mu \left(v^{\sigma)}-\frac{1}{2}h^{\sigma)\nu}m_\nu\right) - 2\mathcal G_\mu^{\hat v} \hat v^{(\rho} h^{\sigma)\mu} + \mathcal G^{h}_{\rho\sigma}h^{\mu\rho}h^{\nu\sigma}\,,\\
\mathcal{G}_{\Phi}^{\rho\sigma} &=& \mathcal G_{\bar\Phi}^{\rho\sigma}\,.
\end{eqnarray}
It can be checked that these relation obey the Galilean boost Ward identity \eqref{eq:WIGalboosts}.

\subsection{Comments on imposing \texorpdfstring{$\tau\wedge \d\tau=0$}{HSO} and \texorpdfstring{$\d\tau=0$}{torsionlessness} with a Lagrange multiplier}\label{sec:3D_CS_TTNC}

We have seen that the TTNC condition is imposed via the NNLO fields. Alternatively we can enforce this condition with a Lagrange multiplier by adding the term 
\begin{equation}
\mathcal{L}_{\mathrm{LM}}=\frac{e}{16\pi G_N}\frac{1}{2}\zeta_{\rho\sigma}h^{\mu\rho}h^{\nu\sigma}(\partial_\mu\tau_\nu-\partial_\nu\tau_\mu)\,,
\end{equation}
to the Lagrangian where $\zeta_{\mu\nu}=-\zeta_{\nu\mu}$. When varying the measure or $\tau_\mu$ (in the direction along $\tau_\mu$) in this expression we do not find any new contributions to the on shell equations of motion because of the TTNC condition enforced by $\zeta_{\mu\nu}$. On the other hand the variation $h^{\mu\nu}\delta\tau_\mu$ leads to an equation of motion for $\zeta_{\mu\nu}$ which is decoupled from all the other equations.

If we take instead  
\begin{equation}
\mathcal{L}'_{\mathrm{LM}}=\frac{e}{16\pi G_N}\frac{1}{2}\zeta^{\mu\nu}(\partial_\mu\tau_\nu-\partial_\nu\tau_\mu)\,,
\end{equation}
with $\zeta^{\mu\nu}=-\zeta^{\nu\mu}$ unconstrained then the equation of motion of $\zeta^{\mu\nu}$ enforces a NC geometry with $\d\tau=0$. However the field $\zeta^{\mu\nu}$ does not decouple from the equations of motion. This is because it now also appears in the equation of motion for $\Omega$ defined as $\delta\tau_\mu=\Omega\tau_\mu$. This is what happens in the 3D \gls{cs} actions for extended Bargmann algebras where $\zeta^{\mu\nu}=\epsilon^{\mu\nu\rho}\zeta_\rho$ in which $\zeta_\rho$ is associated with the central extension of the 3D Bargmann algebra \cite{Papageorgiou:2009zc,Bergshoeff:2016lwr,Hartong:2016yrf, Hartong:2017bwq}.
See also references \cite{Ozdemir:2019orp,Penafiel:2019czp,Concha:2019dqs,Gomis:2019nih} for recent work on related CS theories.

We conclude that setting $\d\tau=0$ with a Lagrange multiplier does not lead to the equations of motion that are obtained from the on shell $1/c^2$ expansion of GR with $\d\tau$ put to zero by hand. However, it does provide us with an alternative theory obtained by adding $\mathcal{L}'_{\mathrm{LM}}$ to \eqref{eq:action} for an unconstrained $\zeta^{\mu\nu}$. Since in this theory on shell $\d\tau=0$ we can remove all terms with $\hat a_\mu$ since they can be absorbed into the $\zeta^{\mu\nu}$ term. Doing so leads to the following Lagrangian
\begin{equation}\label{eq:action2}
\mathcal{L}=\frac{e}{16\pi G_N}\left[h^{\mu\rho}h^{\nu\sigma}\bar K_{\mu\nu}\bar K_{\rho\sigma}-\left(h^{\mu\nu}\bar K_{\mu\nu}\right)^2+\hat\Phi h^{\mu\nu}\bar R_{\mu\nu}-\bar{\Phi}_{\mu\nu}\hat G^{\mu\nu}+\frac{1}{2}\zeta^{\mu\nu}\tau_{\mu\nu}\right]\,.
\end{equation}
The interesting feature of this theory is that it is Bargmann $U(1)$ invariant since all the non-invariance is proportional to $\d\tau$ which can be compensated for by an appropriate transformation of $\zeta^{\mu\nu}$. The $\bar\Phi_{\mu\nu}$ term can be rewritten using \eqref{eq:nullredmunu} and $\hat a_\mu=0$ to
\begin{equation}
\bar{\Phi}_{\mu\nu}\hat G^{\mu\nu}=\bar{\Phi}_{\mu\nu}h^{\mu\rho}h^{\nu\sigma}\left(\bar R_{\rho\sigma}-\frac{1}{2}\bar h_{\rho\sigma}h^{\lambda\kappa}\bar R_{\lambda\kappa}\right)\,.
\end{equation}
The term in parenthesis is the Einstein tensor of the Riemannian geometry of the constant time slices. If we are in 2+1 dimensions then this vanishes identically and by writing $\zeta^{\mu\nu}=\epsilon^{\mu\nu\rho}\zeta_\rho$ we recover the \gls{cs} model for the extended Bargmann algebra. Hence \eqref{eq:action2} can be thought of as a novel higher-dimensional generalisation of the CS model. 

We have thus found two different classes of theories. As it turns out only the one based on the torsional $U(1)$ symmetry has a pure GR origin, while the Bargmann invariant case requires already in 3D to consider GR coupled to a pair of $U(1)$ gauge fields \cite{Bergshoeff:2016lwr}. In 4D the $\zeta^{\mu\nu}$ can be dualised to another 2-form, $B_{\mu\nu}$ say, which has a 1-form gauge symmetry. In other words in 4D we can write 
\begin{equation}
\frac{e}{16\pi G_N}\frac{1}{2}\epsilon^{\mu\nu\rho\sigma}B_{\mu\nu}\left(\partial_\rho\tau_\sigma-\partial_\sigma\tau_\rho\right)\,,
\end{equation}
which means that there is a gauge symmetry $\delta B_{\mu\nu}=\partial_\mu\Sigma_\nu-\partial_\nu\Sigma_\mu$.

\section{Coupling to matter}\label{sec:coupling_to_matter}
In this section we discuss the coupling of matter to the \regls{nrg} action obtained in the previous section. We consider this by expanding a generic matter Lagrangian using the same methods as used for the \regls{eh} Lagrangian. This will enable us to find the sourced equations of motion in the $1/c^2$ expansion,
and in particular those of NRG. We also discuss \reGlspl{wi} of the sources that follow from leading and subleading order diffeormphisms as well as the expansion of the relativistic conservation laws of the energy-momentum tensor. In order to make contact with the alternate (or primed) formulation of NRG, we will also discuss the boost-invariant currents that source the equations of motion in that case.
Finally, we show how the Poission equation of Newtonian gravity can be obtained from NRG coupled to matter.

\subsection{Expansion of the matter Lagrangian}\label{sec:matter_expansion_eoms}
\begin{figure}[ht!]
    \centering
\begin{tikzpicture}
\matrix (m) [matrix of math nodes,row sep=2em,column sep=2em,minimum width=2em]
{
        && \os{T}{-4}_{h}^{\mu\nu} &  \os{T}{-4}_{\tau}^{\mu}  & & \\
     & \mathcal T_{\Phi}^{\mu\nu} \equiv \os{T}{-2}_{\Phi}^{\mu\nu} & \mathcal T_{h}^{\mu\nu} \equiv \os{T}{-2}_{h}^{\mu\nu} & \mathcal{G}^{\mu}_{\tau}\equiv\os{T}{-2}_{\tau}^{\mu} & \mathcal T_{m}^{\mu} \equiv  \os{T}{-2}_{m}^{\mu} &\\};

\draw[-,white] (m-1-3) -- node [midway,sloped,black] {=} (m-2-2);
\draw[-,white] (m-1-4) -- node [midway,sloped,black] {=} (m-2-5);
\end{tikzpicture}
\caption{Structure of the currents in the $1/c^2$ expansion of the matter Lagrangian similar to Figure \ref{fig:structure_EOMs_expansion}, but with $\tau\wedge\d\tau=0$ imposed off shell. For this to be consistent, the leading order currents and those related to them by variational calculus identities must be zero.}
\label{fig:structure_matter_expansion}
\end{figure}
Consider any matter Lagrangian $\mathcal{L}_{\text{mat}}=\mathcal{L}_{\text{mat}}(c^2,\phi,\partial_\mu\phi)$ where $\phi$ is a generic matter field with the spacetime indices suppressed.
Let us suppose that the most leading term in the $1/c^2$ expansion is of order $c^N$. Since the Einstein--Hilbert Lagrangian is at most of order $c^6$ we assume $N\leq6$.

The expansion of the matter Lagrangian is performed using the general methods of Section \ref{sec:Lagrangians} as:
\begin{eqnarray}\label{eq:expLagrangian_Matter}
\mathcal{L}_{\text{mat}}(c^2,\phi,\partial_\mu\phi)&=&c^{N}\tilde{\mathcal{L}}_{\text{mat}}(\sigma)\nonumber\\
&=&c^{N}\;\oss{\mathcal{L}}{-N}{\text{mat},\,LO}+c^{N-2}\;\;\oss{\mathcal{L}}{2-N}{\text{mat},\,NLO}+c^{N-4}\;\;\oss{\mathcal{L}}{4-N}{\text{mat},\,NNLO}+\order{c^{N-6}}\,.
\end{eqnarray}
At each order $n\in\mathbb{N}$ including zero, we define matter currents as responses to varying the geometric fields as
\begin{eqnarray}
\os{T}{2n-N}^{\alpha\beta}_{h} &\equiv& 2e^{-1}\frac{\delta\;\;{\os{\mathcal{L}}{2n-N}}_{\text{mat}\,,\mathrm{N}^n\mathrm{LO}}}{\delta h_{\alpha\beta}}\,,\label{eq:Matter_couplings_expansion}\\
\os{T}{2n-N}^\alpha_{\tau} &\equiv& e^{-1}\frac{\delta\;\;{\os{\mathcal{L}}{2n-N}}_{\text{mat}\,,\mathrm{N}^n\mathrm{LO}}}{\delta \tau_{\alpha}}\,.
\end{eqnarray}
and similarly for the subleading fields mutatis mutandis as summarised in Figure \ref{fig:structure_matter_expansion}.
For $n=0$ we have $\os{T}{-N}^{\alpha\beta}_{\Phi}=\os{T}{-N}^\alpha_{m}=0$ since the \regls{nlo} fields do not appear at \regls{lo}.
Furthermore because of \eqref{eq:NNLO_Nvar_NLO_var} we have the relations
\begin{eqnarray}
\os{T}{2-N}^{\alpha\beta}_{\Phi} &=& \os{T}{-N}^{\alpha\beta}_{h}\,,\label{eq:Matter_couplings_currentsB1}\\
\os{T}{2-N}^\alpha_{m} &=& \os{T}{-N}^\alpha_{\tau}\label{eq:Matter_couplings_currentsB2}\,,
\end{eqnarray}
and similarly for the more subleading fields. 

These currents are natural to work with as they allow us to write the expansion of matter coupled general relativity to any desired order. The \reglspl{eom} with matter couplings at a given order $2m\geq-6$ then becomes
\begin{eqnarray}
\os{G}{2m}^{\alpha\beta}_{h} &=& 8\pi G_N\os{T}{2m}^{\alpha\beta}_{h} \,,\\
\os{G}{2m}^{\alpha}_{\tau} &=& 8\pi G_N\os{T}{2m}^{\alpha}_{\tau} \,,\\
\os{G}{2m}^{\alpha\beta}_{\Phi} &=& 8\pi G_N\os{T}{2m}^{\alpha\beta}_{\Phi} \,,\\
\os{G}{2m}^{\alpha}_{m} &=& 8\pi G_N\os{T}{2m}^{\alpha}_{m} \,,
\end{eqnarray}
and so on for even more subleading fields that we will not consider in this paper.

Notice that $\os{G}{-6}^{\alpha\beta}_{h},\,\os{G}{-6}^{\alpha}_{\tau}\propto\tau\wedge \d\tau$ so it is necessary to have matter with $N\leq 4$ such that $\os{T}{-6}^{\alpha\beta}_{h}=\os{T}{-6}^{\alpha}_{\tau}=0$ in order to avoid acausal non-relativistic \regls{nc} spacetimes. We will see in the next section that $N\leq 4$ in all the examples we have encountered.

Because of the above analysis we will for the rest of this section restrict to matter with $N\leq 4$ and study how they can source the NRG sector of the $1/c^2$ expansion of full \regls{gr}.
Recall that the NRG Lagrangian is defined to be the \regls{nnlo} Lagrangian with the off shell \regls{ttnc} condition $\tau\wedge\d\tau=0$.
This term appears at order $\order{c^2}$ and is therefore sourced by $\os{\mathcal{L}}{-2}_{\text{mat}}$.
We will likewise impose TTNC and define the currents as follows 
\begin{equation}\label{eq:Matter_couplings_1}
\delta\;\os{\mathcal{L}}{-2}_{\text{mat}}\vert_{\tau\wedge \d\tau=0} = e\left(\mathcal{T}_{\tau}^\mu \delta \tau_\mu + \mathcal{T}_{m}^\mu \delta m_\mu
+\frac{1}{2}\mathcal{T}_{h}^{\mu\nu}\delta h_{\mu\nu}+\frac{1}{2}\mathcal{T}_{\Phi}^{\mu\nu}\delta{\Phi}_{\mu\nu}\right)\,.
\end{equation}
We again use the notation that when we are dealing with TTNC off shell, the variations of the matter Lagrangian at order $\order{c^2}$ are denoted by (suppressing spacetime indices) $\os{T}{-2}_{\phi}\vert_{\tau\wedge \d\tau=0}=\mathcal{T}_\phi$ for the field $\phi$.
If $N=2$ the LO matter Lagrangian is of order $\order{c^2}$.
Therefore in that case $\mathcal{T}_{\Phi}^{\mu\nu}=\mathcal{T}_{m}^\mu=0$.
When $N=4$ the LO matter Lagrangian is of order $\order{c^4}$ and the NLO matter Lagrangian is the one that couples to NRG.
In the latter case the responses $\mathcal{T}_{\Phi}^{\mu\nu}$ and $\mathcal{T}_{m}^\mu$ are generically nonzero. The equations of motion of matter coupled NRG are
\begin{eqnarray}
\mathcal{G}^{\alpha\beta}_{h} &=& 8\pi G_N\mathcal{T}^{\alpha\beta}_{h} \,,\label{eq:EOMNRG1}\\
\tau_\alpha\mathcal{G}^{\alpha}_{\tau} &=& 8\pi G_N\tau_\alpha\mathcal{T}^{\alpha}_{\tau} \,,\\
\mathcal{G}^{\alpha\beta}_{\Phi} &=& 8\pi G_N\mathcal{T}^{\alpha\beta}_{\Phi} \,,\\
\mathcal{G}^{\alpha}_{m} &=& 8\pi G_N\mathcal{T}^{\alpha}_{m} \,,\label{eq:EOMNRG4}
\end{eqnarray}
where the left hand sides are given by \eqref{eq:NRG_EOM1}-\eqref{eq:NRG_EOM4}.

We can also relate the $1/c^2$ expansion of the relativistic energy-momentum tensors $E_\mathrm{mat}^\mu$ and $E_\mathrm{mat}^{\mu\nu}$ defined in \eqref{eq:rel_EM_tensors_matter_Lagrangian} to the expansion \eqref{eq:Matter_couplings_currentsB1}-\eqref{eq:Matter_couplings_currentsB2} and related equations.
Analogously to Section \ref{sec:NRG_EOMs} the relations are
\begin{eqnarray}
\mathcal T^{\mu\nu}_{h}&=& \os{E}{-2}^{\mu\nu}_{\text{mat}}+\left(\Phi+\frac{1}{2}h^{\alpha\beta}\Phi_{\alpha\beta}\right)\os{E}{-4}^{\mu\nu}_{\text{mat}}\,,\label{eq:EM_NRG_to_GR_NNLO1}\\
\mathcal T^\mu_{\tau} &=& \os{E}{-2}^\mu_{\text{mat}}+\left(\Phi+\frac{1}{2}h^{\alpha\beta}\Phi_{\alpha\beta}\right)\os{E}{-4}^\mu_{\text{mat}}\,,\label{eq:EM_NRG_to_GR_NNLO2}\\
\mathcal T^{\mu\nu}_{\Phi}&=& \os{E}{-4}^{\mu\nu}_{\text{mat}}\,,\label{eq:EM_NRG_to_GR_NNLO3}\\
\mathcal T^\mu_{m} &=& \os{E}{-4}^\mu_{\text{mat}}\label{eq:EM_NRG_to_GR_NNLO4}\,.
\end{eqnarray}
In deriving this result we expanded $E^\mu_m$ as 
\begin{equation}\label{eq:expEmat}
E^\mu_{\text{mat}}=\os{E}{-4}_{\text{mat}}^\mu+\sigma\;\os{E}{-2}_{\text{mat}}^\mu+\order{\sigma^2}\,,
\end{equation}
and similarly for $E^{\mu\nu}_{\text{mat}}$. To explain the notation we refer to \eqref{eq:rel_EM_tensors_matter_Lagrangian} (where we factored out $c^N$ with $N=4$).

\subsection{Ward identities}\label{sec:Ward_identities}
\begin{figure}[ht]
    \centering
\begin{tikzpicture}
\matrix (m) [matrix of math nodes,row sep=2em,column sep=0.5em,minimum width=2em]
{
     \text{LO currents} &\,&\,\\
     \text{NLO currents} &\,&\,\\
     \text{NNLO currents} &\,&\,\\};

\draw[->] (m-1-1) edge [out=-10,in=10,looseness=3] node[right] {$\mathcal{L}_\xi$} (m-1-1);
\draw[->] (m-2-1) edge [out=-7,in=7,looseness=4] node[right] {$\mathcal{L}_\xi$} (m-2-1);
\draw[->] (m-3-1) edge [out=-6,in=6,looseness=4] node[right] {$\mathcal{L}_\xi$} (m-3-1);

\draw[->] (m-2-2) edge [out=20,in=-20,looseness=2] node[right] {$\mathcal{L}_\zeta$} (m-1-2);
\draw[->] (m-3-2) edge [out=20,in=-20,looseness=2] node[right] {$\mathcal{L}_\zeta$} (m-2-2);

\draw[->] (m-3-2) edge [out=-5,in=5,looseness=2] node[right] {$\mathcal{L}_{\Xi_{(4)}}$} (m-1-2);

\end{tikzpicture}
\caption{Structure of the \reGlspl{wi}: At each order there are LO WIs generated by the LO vector field $\xi^\mu$ through the Lie derivative $\mathcal L_\xi$.
The subleading vector field $\zeta^\mu$ generates a WI through $\mathcal L_\zeta$, which is equivalent to the LO WI at LO.
Similarly $\mathcal L_\zeta$ at NNLO generates a WI which is equivalent to the LO WI at NLO.
This works similarly for subsubleading vector field $\Xi_{(4)}^\mu$ and is systematically extended to higher orders in the expansion.
Hence, when working at a particular order, energy--momentum conservation of the previous orders in the expansion is always included at that given order.
}
\label{fig:structure_Bianchi_expansion}
\end{figure}
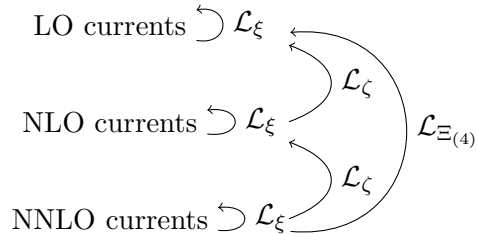

The matter sector must be invariant under the gauge transformations that act simultaneously on the geometric objects as well as on the matter fields. Since the latter variations are proportional to the matter equations of motion we can ignore these terms at the expense of being able to only derive on shell Ward identities. We can derive these on shell Ward identities from applying the transformation laws \eqref{eq:trafo5a}, \eqref{eq:trafo9}-\eqref{eq:torsional_u(1)_2} acting on the geometric fields in \eqref{eq:Matter_couplings_1} and requiring invariance of the matter action up to the matter equations of motion.
The structure of the tower of WIs that follows from this is summarised in Figure \ref{fig:structure_Bianchi_expansion}.

Diffeomorphism invariance of \eqref{eq:Matter_couplings_1} implies that we have the following conservation law\footnote{Diffeomorphisms commute with the TTNC condition.}:
\begin{eqnarray}\label{eq:EM_conservation_NNLO}
0&=&\tau_\rho\left(\check \nabla_\mu+2a_\mu\right) \mathcal{T}_{\tau}^\mu +m_\rho\left(\check \nabla_\mu+2a_\mu\right) \mathcal{T}_{m}^\mu +  h_{\rho\nu}\left(\check \nabla_\mu+a_\mu\right)\mathcal{T}_{h}^{\mu\nu} + \Phi_{\rho\nu}\left(\check \nabla_\mu+a_\mu\right)\mathcal{T}_{\Phi}^{\mu\nu}  \nonumber\\
&&-a_\rho \tau_\mu \mathcal{T}_{\tau}^\mu + F_{\mu\rho}\mathcal{T}_{m}^\mu-a_\rho m_\mu \mathcal{T}_{m}^\mu
- \tau_\rho K_{\mu\nu} \mathcal{T}_{h}^{\mu\nu} + \left(\check\nabla_\mu \Phi_{\rho\nu} - \frac{1}{2}\check\nabla_\rho \Phi_{\mu\nu} \right)\mathcal{T}_{\Phi}^{\mu\nu}\nonumber\\
&&-\left(a_\mu \tau_\rho - \tau_\mu a_\rho\right)v^\lambda\Phi_{\lambda\nu}\mathcal{T}_{\Phi}^{\mu\nu}\,.
\end{eqnarray}
The $v^\rho$ and $h^{\rho\sigma}$ projections are:
\begin{eqnarray}
0&=&-\left(\check \nabla_\mu+2a_\mu\right) \left(\mathcal{T}_{\tau}^\mu-v^\rho\Phi_{\rho\nu}\mathcal{T}_{\Phi}^{\mu\nu}\right) -\Phi\left(\check \nabla_\mu+2a_\mu\right) \mathcal{T}_{m}^\mu  
+v^\rho F_{\mu\rho}\mathcal{T}_{m}^\mu + K_{\mu\nu} \mathcal{T}_{h}^{\mu\nu}\nonumber\\
&&-\frac{1}{2}\mathcal{L}_v\left(P^\rho_\mu P^\sigma_\nu\Phi_{\rho\sigma}\right)\mathcal{T}_\Phi^{\mu\nu}\,, \label{eq:vprojectiondiffeoWI}\\
0&=&h^{\rho\sigma}m_\rho\left(\check \nabla_\mu+2a_\mu\right) \mathcal{T}_{m}^\mu  + h^{\rho\sigma}h_{\rho\nu}\left(\check \nabla_\mu+a_\mu\right)\mathcal{T}_{h}^{\mu\nu} + h^{\rho\sigma}\Phi_{\rho\nu}\left(\check \nabla_\mu+a_\mu\right)\mathcal{T}_{\Phi}^{\mu\nu}  \nonumber\\
&&- h^{\rho\sigma} a_\rho \tau_\mu\mathcal{T}_{\tau}^\mu + h^{\rho\sigma}F_{\mu\rho}\mathcal{T}_{m}^\mu- h^{\rho\sigma} a_\rho m_\mu\mathcal{T}_{m}^\mu
+h^{\rho\sigma}\left(\check\nabla_\mu \Phi_{\rho\nu} - \frac{1}{2}\check\nabla_\rho \Phi_{\mu\nu} \right)\mathcal{T}_{\Phi}^{\mu\nu}\,.\label{eq:hprojectiondiffeoWI}
\end{eqnarray}
In the first equation we used
\begin{eqnarray}
    -\frac{1}{2}\mathcal{T}_\Phi^{\mu\nu}\mathcal{L}_v\Phi_{\mu\nu}&=&-\frac{1}{2}\mathcal{T}_\Phi^{\mu\nu}\mathcal{L}_v\left(P^\rho_\mu P^\sigma_\nu\Phi_{\rho\sigma}-2\tau_\mu v^\rho\Phi_{\rho\nu}\right)\nonumber\\
&=&-\frac{1}{2}\mathcal{T}_\Phi^{\mu\nu}\mathcal{L}_v\left(P^\rho_\mu P^\sigma_\nu\Phi_{\rho\sigma}\right)+\mathcal{T}_\Phi^{\mu\nu}a_\mu v^\rho\Phi_{\rho\nu}\,,
\end{eqnarray}
and in the second equation we used $\tau_\mu\mathcal{T}_\Phi^{\mu\nu}=0$, a property that will be derived further below.

In addition the subleading diffeomorphism gives another conservation equation between the currents.
This is the same as the conservation equation at NLO because of \eqref{eq:Matter_couplings_currentsB1}-\eqref{eq:Matter_couplings_currentsB2}.
Explicitly we have
\begin{equation}\label{eq:EM_conservation_subleadingdiffeo_NNLO}
0=\tau_\rho\left(\check \nabla_\mu+2a_\mu\right) \mathcal{T}_{m}^\mu  +h_{\rho\nu} \left(\check \nabla_\mu+a_\mu\right)\mathcal{T}_{\Phi}^{\mu\nu} 
- a_\rho \tau_\mu\mathcal{T}_{m}^\mu 
- \tau_\rho K_{\mu\nu} \mathcal{T}_{\Phi}^{\mu\nu}\,.
\end{equation}
The two projections along $v^\rho$ and $h^{\rho\sigma}$, i.e.
\begin{eqnarray}
    \left(\check\nabla_\mu+2a_\mu\right)\mathcal{T}_m^\mu-K_{\mu\nu}\mathcal{T}_\Phi^{\mu\nu} & = & 0\,,\label{eq:WIm}\\
    \left(\check\nabla_\nu+a_\nu\right)\mathcal{T}_\Phi^{\mu\nu}-\mathcal{T}_m^\nu\tau_\nu h^{\mu\rho}a_\rho & = & 0\,,\label{eq:WIPhi}
\end{eqnarray}
give the matter counterparts of the equations \eqref{eq:LambdaBianchi} and \eqref{eq:zetaBianchi}. The $v^\rho$ projection agrees with the leading order term in the expansion of \eqref{eq:conservation1}. To show this we used \eqref{eq:expEmat} and \eqref{eq:EM_NRG_to_GR_NNLO1}-\eqref{eq:EM_NRG_to_GR_NNLO4}. Likewise the $h^{\rho\sigma}$ projection agrees with the leading order term in the expansion of \eqref{eq:conservation2}.

Finally the Galilean boost Ward identity is
\begin{equation}\label{eq:GalboostWImatter}
    \mathcal{T}_{m}^\mu e_\mu^a + \mathcal{T}_{h}^{\mu\nu} e_\mu^a \tau_\nu + \mathcal{T}_{\Phi}^{\mu\nu} \left(\tau_\mu \pi^a_\nu+m_\mu e^a_\nu\right)=0\,.
\end{equation}
This can be read as saying that the spatial components of $\mathcal{T}_{m}^\mu$ are completely determined in terms of the other currents. This is the matter counterpart of \eqref{eq:WIGalboosts} provided that $\mathcal{T}_{\Phi}^{\mu\nu}\tau_\mu =0$.
We can show that we always have that $\mathcal{T}_{\Phi}^{\mu\nu}\tau_\mu e_\nu^a=0$:
This follows from as a Ward identity for the subleading Galilean boosts with parameter $\eta_a$ (see equation \eqref{eq:torsional_u(1)_2}).
Later, in equation \eqref{eq:tauTPhi}, we will see that when we assume that the order $\order{c^2}$ matter Lagrangian does not depend on the NNLO field $B_\mu$ for the case with off shell TTNC that this in fact implies that $\mathcal{T}_{\Phi}^{\mu\nu}\tau_\mu =0$. This will be assumed to hold throughout.

Note that the $h^{\rho\sigma}$ projection of the diffeomorphism Ward identity, equation \eqref{eq:hprojectiondiffeoWI}, only contains $h^{\mu\rho}h^{\nu\sigma}\Phi_{\mu\nu}$ (to see this we need to use that $\mathcal{T}_{\Phi}^{\mu\nu}\tau_\mu =0$) which is the part of $\Phi_{\mu\nu}$ that appears in the NRG Lagrangian. On the other hand the $v^\rho$ projection \eqref{eq:vprojectiondiffeoWI} contains $v^\mu h^{\nu\rho}\Phi_{\mu\nu}$ which does not appear in the NRG Lagrangian. The terms involving $v^\mu\Phi_{\mu\nu}$ will therefore drop out from the Ward identity. They must cancel against $v^\mu\Phi_{\mu\nu}$ contributions to the currents. More specifically the combination $\mathcal{T}_{\tau}^\mu-v^\rho\Phi_{\rho\nu}\mathcal{T}_{\Phi}^{\mu\nu}$ does not depend on $v^\mu\Phi_{\mu\nu}$. This can be seen by using that the matter Lagrangian is at most of order $c^4$ so that at NLO, which is $c^2$, the field $\Phi_{\mu\nu}$ appears linearly in a term of the form $h^{\mu\rho}h^{\nu\sigma}\Phi_{\mu\nu}X_{\rho\sigma}$ where $X_{\rho\sigma}$ depends on the matter fields.
It is straightforward to see that then $\mathcal{T}_{\tau}^\mu-v^\rho\Phi_{\rho\nu}\mathcal{T}_{\Phi}^{\mu\nu}$ is independent of $v^\mu\Phi_{\mu\nu}$.

\subsubsection{Expansion of the Hilbert energy-momentum tensor}
We collect here a few general remarks about the $1/c^2$ expansion of the Hilbert energy-momentum tensor $T^{\mu\nu}$. From equations \eqref{eq:Matter_EH_tensor1} and \eqref{eq:Matter_EH_tensor2} we can see that for the case $N=4$ (which is the highest value $N$ can have for a causal spacetime, referring back to our earlier discussion in the beginning of the section) 
\begin{eqnarray}
T_\mu T^{\mu\nu} & = & \order{c^2}\,,\\
\Pi_{\mu\sigma}\Pi^{\sigma\rho} T^{\mu\nu} & = & \order{c^4}\,.
\end{eqnarray}
Hence we expand $T^{\mu\nu}$ as 
\begin{equation}\label{eq:EM_indicesup_expansion}
T^{\mu\nu}=c^4\; \os{T}{-4}^{\mu\nu}+c^2\os{T}{-2}^{\mu\nu}+\os{T}{0}^{\mu\nu}+\order{c^{-4}}\,,
\end{equation}
where $\tau_\mu\os{T}{-4}^{\mu\nu}=0$. More specifically equations \eqref{eq:Matter_EH_tensor1} and \eqref{eq:Matter_EH_tensor2} tell us that
\begin{eqnarray}
\mathcal{T}^\mu_m & = & -\os{T}{-2}^{\mu\nu}\tau_\nu-\os{T}{-4}^{\mu\nu}m_\nu\,,\\
\mathcal{T}^{\mu\nu}_\Phi & = & \os{T}{-4}^{\mu\nu}\,,\label{eq:calTm}\\
\mathcal{T}^\mu_\tau & = & -\os{T}{-4}^{\mu\nu}B_\nu-\os{T}{-2}^{\mu\nu}m_\nu-\os{T}{0}^{\mu\nu}\tau_\nu-\left(\Phi+\frac{1}{2}h^{\alpha\beta}\Phi_{\alpha\beta}\right)\left(\os{T}{-2}^{\mu\nu}\tau_\nu+\os{T}{-4}^{\mu\nu}m_\nu\right)\,,\label{eq:calTtau}\\
h_{\mu\rho}\mathcal{T}_h^{\mu\nu} & = & h_{\mu\rho}\os{T}{-2}^{\mu\nu}+\left(\Phi+\frac{1}{2}h^{\alpha\beta}\Phi_{\alpha\beta}\right)h_{\mu\rho}\os{T}{-4}^{\mu\nu}\,,\label{eq:calTh}
\end{eqnarray}
where we used \eqref{eq:expEmat} as well as \eqref{eq:EM_NRG_to_GR_NNLO1}-\eqref{eq:EM_NRG_to_GR_NNLO4}.

At the leading order the Ward identity derived from the conservation law \eqref{eq:EM_conservation_Hilbert} is given by
\begin{equation}\label{eq:WIA}
\left(\bar\nabla_\mu+\hat a_\mu\right) \os{T}{-4}^{\mu\nu}+h^{\nu\sigma}\hat a_\sigma\tau_\mu\tau_\rho \os{T}{-2}^{\mu\rho}=0\,.
\end{equation}
This equation is identical to \eqref{eq:WIPhi}. At NLO we find
\begin{equation}\label{eq:WIA_no_tau_contraction}
e^{-1}\partial_\mu\left(e \os{T}{-2}^{\mu\nu}\right)
+\Gamma^{\nu}_{(0)\mu\rho}\os{T}{-2}^{\mu\rho}
+\Gamma^\mu_{{(2)}\,\mu\rho}\os{T}{-4}^{\rho\nu}
+\Gamma^\nu_{{(2)}\,\mu\rho}\os{T}{-4}^{\mu\rho}
+\Gamma^\nu_{{(-2)}\,\mu\rho}\os{T}{0}^{\mu\rho}=0\,.
\end{equation}
Contracting this with $\tau_\nu$ we find
\begin{equation}\label{eq:WIB}
\left(\bar\nabla_\mu+2\hat a_\mu\right) \tau_\nu\os{T}{-2}^{\mu\nu}+\bar K_{\mu\nu} \os{T}{-4}^{\mu\nu}=0\,,
\end{equation}
which is the same as \eqref{eq:WIm}. We used the expansion of the Christoffel connection $\Gamma^\rho_{\mu\nu}$ discussed in appendix \ref{subapp:NRexpansionmetricGamma}. To show equality with \eqref{eq:WIm} we used equation \eqref{eq:relbarKtoK}.

\subsection{Boost invariant currents}
The matter currents that naturally couple to the boost invariant NRG formulation of the NNLO Lagrangian \eqref{eq:action_NRG_prime} are defined as
\begin{equation}\label{eq:Matter_couplings_2}
\delta\os{\mathcal{L_{\mathrm{M}}}}{-2} \equiv -e\left(\mathcal T_{\hat\Phi}\delta\hat\Phi-\mathcal T^{\hat v}_\mu\delta\hat v^\mu+\frac{1}{2}\mathcal T^h_{\mu\nu}\delta h^{\mu\nu}-\frac{1}{2}\mathcal T^{\mu\nu}_{\bar \Phi}\delta\bar{\Phi}_{\mu\nu}\right)\,,
\end{equation}
so that the equations of motion in the presence of matter become
\begin{eqnarray}
\mathcal G_{\hat\Phi} &=& 8\pi G_N\mathcal T_{\hat\Phi} \,,\\
\mathcal G^{\hat v}_\mu &=& 8\pi G_N\mathcal T^{\hat v}_\mu \,,\\
\mathcal G^h_{\mu\nu} &=& 8\pi G_N\mathcal T^h_{\mu\nu} \,,\\
\mathcal G^{\mu\nu}_{\bar \Phi} &=& 8\pi G_N\mathcal T^{\mu\nu}_{\bar \Phi} \,.
\end{eqnarray}
Like for the geometry part, we can use the variational relations \eqref{eq:variation_boostinv_var1}-\eqref{eq:variation_boostinv_var4} to express the boost invariant currents in terms of the currents defined from varying the set of fields $\tau_\mu, m_\mu, h_{\mu\nu}, \Phi_{\mu\nu}$:
\begin{eqnarray}
\mathcal{T}_{\tau}^\rho &=& -\mathcal T_{\hat\Phi} \Phi v^\rho+ \mathcal T_\mu^{\hat v}\left(\Phi h^{\mu\rho}+v^\mu \hat v^\rho\right) - \mathcal T^h_{\mu\nu}v^\mu h^{\nu\rho} - \mathcal T_{\bar\Phi}^{\rho\sigma}B_\sigma\,,\label{eq:tauvarmatcurrent}\\
\mathcal{T}_{m}^\rho &=& \mathcal T_{\hat\Phi} \hat v^\rho - \mathcal T_\mu^{\hat v} h^{\mu\rho} - \mathcal T_{\bar\Phi}^{\rho\sigma}m_\sigma\,,\\
\mathcal{T}_{h}^{\rho\sigma} &=& -2\mathcal T_{\hat\Phi} m_\mu h^{\mu(\rho} \left(v^{\sigma)}-\frac{1}{2}h^{\sigma)\nu}m_\nu\right) - 2\mathcal T_\mu^{\hat v} \hat v^{(\rho} h^{\sigma)\mu} + \mathcal T^{h}_{\rho\sigma}h^{\mu\rho}h^{\nu\sigma}\,,\\
\mathcal{T}_{\Phi}^{\rho\sigma} &=& \mathcal T_{\bar\Phi}^{\rho\sigma}\,.
\end{eqnarray}
The presence of the NNLO field $B_\sigma$ in \eqref{eq:tauvarmatcurrent} does not need to concern us as this is in agreement with the fact that only the $\tau_\mu$ variation in the direction of $\tau_\mu$ does not couple to NNLO fields (see the discussion of Section \ref{sec:NRG_EOMs}). The $\tau_\rho$ projection of \eqref{eq:tauvarmatcurrent} vanishes provided $\tau_\rho\mathcal{T}_{\Phi}^{\rho\sigma}=0$.

A set of currents that have all indices up and that is boost invariant is defined by using as independent variables $\tau_\mu,\,\bar h_{\mu\nu},\,\bar \Phi_{\mu\nu}$, i.e.
\begin{equation}\label{eq:Matter_couplings_3}
\delta\;\os{\mathcal{L}}{-2}_{\text{mat}} \equiv e\left(\bar{\mathcal{T}}_{\tau}^\mu\delta \tau_\mu+\frac{1}{2}\mathcal T_{\bar h}^{\mu\nu}\delta \bar h_{\mu\nu}+\frac{1}{2}\mathcal T^{\mu\nu}_{\bar \Phi}\delta\bar{\Phi}_{\mu\nu}\right)\,.
\end{equation}
Using \eqref{eq:variation_boostinv_var4} and
\begin{equation}
\delta \bar h_{\mu\nu} = \delta h_{\mu\nu}-2\tau_{(\mu}\delta m_{\nu)}-2m_{(\mu}\delta \tau_{\nu)}\,,
\end{equation}
we can relate them to the other currents via
\begin{eqnarray}
\mathcal{T}_{\tau}^\rho &=& \bar{\mathcal{T}}^\rho_\tau - \mathcal T_{\bar h}^{\rho\sigma}m_\sigma - \mathcal T_{\bar \Phi}^{\rho\sigma}B_\sigma\,,\\
\mathcal{T}_{m}^\rho &=& - \mathcal T_{\bar h}^{\rho\sigma}\tau_\sigma - \mathcal T_{\bar \Phi}^{\rho\sigma}m_\sigma\,,\\
\mathcal{T}_{h}^{\rho\sigma}e_\rho^a &=& \mathcal T_{\bar h}^{\rho\sigma}e_\rho^a\,,\\
\mathcal{T}_{\Phi}^{\rho\sigma} &=& \mathcal T_{\bar \Phi}^{\rho\sigma}\,.
\end{eqnarray}
The variation with respect to the NNLO field $B_\nu$ is $\tau_\mu\mathcal{T}_{\bar \Phi}^{\mu\nu}$. Since we will assume that there will be no $B_\mu$ dependence in the matter Lagrangian with off shell TTNC it follows that
\begin{equation}\label{eq:tauTPhi}
    \tau_\mu\mathcal{T}_{\bar \Phi}^{\mu\nu}=0\,.
\end{equation}
One could also formulate the equations of motions by varying the gravity Lagrangian with respect to $\tau_\mu,\,\bar h_{\mu\nu},\,\bar \Phi_{\mu\nu}$, but we shall refrain from doing that.

\subsection{Newtonian gravity}\label{sec:torsionless_Newtonian_gravity}
In Section \ref{sec:3D_CS_TTNC} it was shown that we cannot add a Lagrange multiplier to enforce $\d\tau=0$. We will see in the next section that the situation $\tau\wedge \d\tau=0$ versus $\d\tau=0$ is decided on by the nature of the $1/c^2$ expansion of the matter Lagrangian. This happens via equations of motion imposed by the matter fields. The matter equations of motion may sometimes force $\d\tau=0$. We will see this happening for a massive point particle, certain approximations of perfect fluids and for the Schr\"odinger field approximation to a massive complex scalar field.

In this section we will study the necessary conditions on the matter sector in order that $\d\tau=0$ is compatible with the equations of motion of gravity coupled to matter.

The special properties that the matter currents must satisfy can be understood completely from the scaling equation \eqref{eq:TTNCsourcing} which after using the Einstein equations becomes,
\begin{equation}\label{eq:scaling_eqn_Newtonian}
    8\pi G_N\left(-(d-2)\tau_\mu\mathcal{T}_m^\mu+h_{\mu\nu}\mathcal{T}_\Phi^{\mu\nu}\right)=(d-1)h^{\mu\nu}\left(\check\nabla_\mu+a_\mu\right)a_\nu=(d-1)e^{-1}\partial_\mu\left(eh^{\mu\nu}a_\nu\right)\,.
\end{equation}
This gives an equation for the non-relativistic lapse function $N$ defined as $\tau=N \d T$ since $h^{\mu\nu}a_\nu=h^{\mu\nu}N^{-1}\partial_\nu N$. It then follows that if there is to be no torsion the matter must satisfy the necessary condition
\begin{equation}\label{eq:EM_tensor_constraint_for_no_torsion}
(d-2)\tau_\mu\mathcal{T}_m^\mu=h_{\mu\nu}\mathcal{T}_\Phi^{\mu\nu}\,.
\end{equation}
Existence of non-trivial solutions to $e^{-1}\partial_\mu\left(eh^{\mu\nu}a_\nu\right)=0$ depends on boundary conditions and topology of the manifold.
This is crucial in establishing a twistless torsionful Schwarzschild-type vacuum solution to the EOMs in Section \ref{sec:strong_Schwarzschild}.

When $a_\mu=0$ the trace equation \eqref{eq:scaleeq} simplifies to
\begin{eqnarray}
&&8\pi G_N\left(-(d-2)\tau_\mu \mathcal{T}_{\tau}^\mu-(d-2)m_\mu \mathcal{T}_{m}^\mu+h_{\mu\nu}\mathcal{T}_{h}^{\mu\nu}+\Phi_{\mu\nu}\mathcal{T}_{\Phi}^{\mu\nu}\right)\nonumber\\
&&=(d-1)\Bigg[v^\mu v^\nu\check R_{\mu\nu}+h^{\mu\nu}\check\nabla_\rho\left(v^\rho\check\nabla_\mu m_\nu\right)+\frac{1}{4}h^{\mu\rho}h^{\nu\sigma}F_{\mu\nu}F_{\rho\sigma}+h^{\mu\nu}\check\nabla_\mu \partial_\nu\Phi\nonumber\\
&&-\left(2h^{\mu\rho}h^{\nu\sigma}-h^{\mu\nu}h^{\rho\sigma}\right)\check\nabla_\mu\left(m_\nu K_{\rho\sigma}\right)\Bigg]=(d-1)\hat v^\mu\hat v^\nu\bar R_{\mu\nu}\,.\label{eq:scaleeq_torsionless}
\end{eqnarray}
By changing the connection to \eqref{eq:Special TNC connection} all terms on the right hand side combine to form the scalar $\hat v^\mu\hat v^\nu\bar R_{\mu\nu}$.

For matter coupled Newtonian gravity what the examples in the next section will show is that in those cases the most leading order in the expansion of the matter Lagrangian is of order $c^2$ which guarantees that the currents $\mathcal{T}_m^\mu$ and $\mathcal{T}_\Phi^{\mu\nu}$ are both zero. In that case the equations \eqref{eq:scaleeq_torsionless} together with the $m_\mu$ and $\Phi_{\mu\nu}$ equations of motion are all independent of $\Phi_{\mu\nu}$. Hence for matter coupled Newtonian gravity the $m_\mu$ and $\Phi_{\mu\nu}$ equations of motion together with \eqref{eq:scaleeq_torsionless} can be used to solve for the fields $\tau_\mu$, $h_{\mu\nu}$ and $m_\mu$. When that happens $\Phi_{\mu\nu}$ decouples from the other fields $\tau_\mu$, $h_{\mu\nu}$ and $m_\mu$. In the next section we will see examples of this. The left hand side of \eqref{eq:scaleeq_torsionless} contains
$\tau_\mu \mathcal{T}_{\tau}^\mu$ which, as we will later see, is minus the mass density $-\rho$ that enters in the geometric Poisson equation \eqref{eq:Newton}.
In particular it is not a Bargmann mass as we elaborate on in appendix \ref{sec:newtonian_gravity}. When $\d\tau=0$ and the currents $\mathcal{T}_m^\mu$ and $\mathcal{T}_\Phi^{\mu\nu}$ are both zero the equations of motion of matter coupled Newtonian gravity are simply (after some rewriting)
\begin{eqnarray}
&& h^{\rho\nu}v^\sigma\check R_{\rho\sigma}
-\frac{1}{2}h^{\mu\rho}h^{\nu\sigma} \check{\nabla}_\mu F_{\rho\sigma}=0\,,\\
 && h^{\mu\rho}h^{\nu\sigma}\check R_{\mu\nu}=0
\,,\\
&  & v^\mu v^\nu\check R_{\mu\nu}+\frac{1}{4}h^{\mu\rho}h^{\nu\sigma}F_{\mu\nu}F_{\rho\sigma}+\check\nabla_\mu\left(h^{\mu\nu}v^\rho F_{\rho\nu}\right)\nonumber\\
&=&\frac{8\pi G_N}{d-1}\left(-(d-2)\tau_\mu \mathcal{T}_{\tau}^\mu+h_{\mu\nu}\mathcal{T}_{h}^{\mu\nu}\right) \,.\label{eq:scaleeq_torsionless2}
\end{eqnarray}
This can equivalently be written as
\begin{equation}\label{eq:mattercoupledNCgravity}
    \bar R_{\mu\nu}=\frac{8\pi G_N}{d-1}\left(-(d-2)\tau_\rho \mathcal{T}_{\tau}^\rho+h_{\rho\sigma}\mathcal{T}_{h}^{\rho\sigma}\right)\tau_\mu\tau_\nu\,.
\end{equation}
We will see explicit realisations of this equation in the next section.

\section{Examples of matter couplings}\label{sec:matter_examples}
In this section we study some canonical examples of matter that can be put on type II \regls{nc} geometries and their coupling to \regls{nrg}: Point particles, fluids, scalar fields and electrodynamics.
We always start with a $1/c^2$ expansion of their relativistic parent and derive the \reglspl{eom}.
Their currents and \reglspl{wi} are studied and we comment on what kind of NRG the theories can source.
For the point particles we see that there is basically a branching into the usual \regls{nr} particle and a novel type of particle motion that lives on a \regls{ttnc} geometry and couples to torsion.
To get a better conceptual understanding of the latter we study the motion on Rindler spacetime. 
Perfect fluids, which will play a distinguished role later on in Section \ref{sec:solutions}, are then studied with and without an extra $U(1)$ current.
We then turn to both real and complex scalar fields and see, among other results, that we can derive Schr\"odinger--Newton theory as a special case.
The $1/c^2$ expansion of Maxwell electrodynamics yields novel magnetic and electric theories on (type II) TTNC geometry that we study.
Finally we see how we can obtain \gls{ged} on torsionless NC geometry.

\subsection{Point particles}\label{sec:case1_particle}
\subsubsection{Lagrangian}
The proper time particle Lagrangian is 
\begin{equation}
\mathcal{L}=-mc\left(-g_{\mu\nu}\dot X^\mu\dot X^\nu\right)^{1/2}\,.
\end{equation}
In here $X^\mu(\lambda)$ are the embedding scalars and $\lambda$ is the geodesic parameter.
Expanding the metric according to \eqref{eq:metric_expansion1} we obtain the $1/c^2$ expansion of the Lagrangian
\begin{equation}
\label{eq:ppLag}
\mathcal{L}=-mc^2\tau_\mu\dot X^\mu+\frac{m}{2}\frac{\bar h_{\mu\nu}\dot X^\mu\dot X^\nu}{\tau_\rho\dot X^\rho}+\order{c^{-2}}\,.
\end{equation}
We still need to expand the embedding scalars according to \eqref{eq:expansion_field_general}:
\begin{equation}\label{eq:1c2_exp_embedding_scalar}
X^\mu=x^\mu+\frac{1}{c^2}y^\mu+\order{c^{-4}}\,.
\end{equation}
This is necessary for otherwise we would overconstrain the equations of motion for $X^\mu$. Note that $\tau_\mu$ is a function of $X$ and so we need to expand
\begin{equation}
\tau_\mu(X)=\tau_\mu(x)+\frac{1}{c^2}y^\sigma\partial_\sigma\tau_\rho(x)+\order{c^{-4}}\,.
\end{equation}
We obtain
\begin{equation}
\mathcal{L}=-mc^2\tau_\mu\dot x^\mu+m\left(-\tau_\mu\dot y^\mu-\dot x^\nu y^\mu\partial_\mu\tau_\nu+\frac{1}{2}\frac{\bar h_{\mu\nu}\dot x^\mu\dot x^\nu}{\tau_\rho\dot x^\rho}\right)+\order{c^{-2}}\,,
\end{equation}
where all functions $\tau_\mu$ and $\bar h_{\mu\nu}$ depend on $x^\mu(\lambda)$.
The \regls{lo} Lagrangian is
\begin{equation}\label{eq:pointparticle_LO_lagrangian}
\oss{\mathcal{L}}{-2}{\text{LO}}=-m\tau_\mu\dot x^\mu\,.
\end{equation}
After a partial integration the \regls{nlo} Lagrangian becomes
\begin{equation}\label{eq:NRparticle}
\oss{\mathcal{L}}{0}{\text{NLO}}=m\left(\left(\partial_\nu\tau_\mu-\partial_\mu\tau_\nu\right)\dot x^\nu y^\mu+\frac{1}{2}\frac{\bar h_{\mu\nu}\dot x^\mu\dot x^\nu}{\tau_\rho\dot x^\rho}\right)\,.
\end{equation}
This is the Lagrangian of a particle on type II TNC geometry.

The \reglspl{eom} of the LO Lagrangian are
\begin{equation}\label{eq:LOeom}
\dot x^\mu\left(\partial_\mu\tau_\nu-\partial_\nu\tau_\mu\right)=0\,,
\end{equation}
which is correctly reproduced by the EOMs of $y^\mu$ in the subleading Lagrangian.
On a TTNC background this becomes
\begin{equation}
    \dot x^\mu a_\mu=0\,,\qquad \dot x^\mu \tau_\mu a_\rho=0\,.
\end{equation}
Since we assumed that $\tau_\mu\dot x^\mu\neq 0$ the equations of motion force $a_\mu=0$. On a fixed NC background the action is the same as the standard point particle action on type I TNC geometry \cite{Kuenzle:1972uk, Andringa:2010it, Bergshoeff:2015uaa}.

Further notice that the LO action is order $c^2$ and that this couples to the \regls{nnlo} gravity action.
The NLO particle action therefore only backreacts to the N$^3$LO gravity action where it will source NNLO fields.
This means that we can solve the geodesic equation before we backreact the solution.

If we restrict ourselves to TTNC backgrounds we can rewrite \eqref{eq:NRparticle} to
\begin{equation}
\mathcal{L}_{\text{NR}}=m\left(\frac{1}{2}\frac{h_{\mu\nu}\dot x^\mu\dot x^\nu}{\tau_\rho\dot x^\rho}-m_\mu(x,y)\dot x^\mu\right)\,,
\end{equation}
where 
\begin{equation}
m_\mu(x,y)=m_\mu+\tau_\mu y^\nu a_\nu-a_\mu\tau_\nu y^\nu\,.
\end{equation}
This observation is useful for computing the $x^\mu$ equation of motion. This turns out to be identical to the type I geodesic equation on a background with $\d\tau=0$ (which is forced upon us by the $y^\mu$ equation of motion). In a gauge in which $\tau_\mu\dot x^\mu=1$ this equation is given by
\begin{equation}\label{eq:NLOeom}
\ddot x^\mu+\bar \Gamma^\mu_{\nu\rho}\dot x^\nu\dot x^\rho=0\,,
\end{equation}
where the connection depends on $m_\mu(x,y) $ which on shell is identical to $m_\mu$ since $a_\mu=0$. In other words the $y^\mu$ field has decoupled from the leading and subleading equations of motion, \eqref{eq:LOeom} and \eqref{eq:NLOeom} respectively. One could thus say that $y^\mu$ is a Lagrange multiplier for the condition $a_\mu=0$.

\subsubsection{Newtonian gravity coupled to point particles}\label{subsubsec:NCgravparticle}
Consider the NRG Lagrangian coupled to the LO point particle Lagrangian. They couple to each other because they both appear at order $c^2$. The combined system is what we call \gls{ncg} coupled to a point particle,
\begin{eqnarray}\label{eq:action_NRG_Newtonian}
\mathcal{L}_\mathrm{NCG} & = & \frac{e}{16\pi G_N}\Bigg[h^{\mu\rho}h^{\nu\sigma}K_{\mu\nu}K_{\rho\sigma}-\left(h^{\mu\nu}K_{\mu\nu}\right)^2
-2m_\nu\left(h^{\mu\rho}h^{\nu\sigma}-h^{\mu\nu}h^{\rho\sigma}\right)\check{\nabla}_\mu K_{\rho\sigma}\nonumber\\
&&+\Phi h^{\mu\nu}\check R_{\mu\nu}
+\frac{1}{4}h^{\mu\rho}h^{\nu\sigma}F_{\mu\nu}F_{\rho\sigma}
+\frac{1}{2}\zeta_{\rho\sigma}h^{\mu\rho}h^{\nu\sigma}(\partial_\mu\tau_\nu-\partial_\nu\tau_\mu)\nonumber\\
&&
-\Phi_{\rho\sigma}h^{\mu\rho}h^{\nu\sigma}\left(\check R_{\mu\nu}-\check{\nabla}_\mu a_\nu - a_\mu a_\nu
-\frac{1}{2}h_{\mu\nu}h^{\kappa\lambda}\check R_{\kappa\lambda}
+h_{\mu\nu} e^{-1} \partial_{\kappa}\left(e h^{\kappa\lambda} a_\lambda\right)\right)\Bigg]\nonumber\\
&&-m\int \d\lambda\, \delta(x-x(\lambda))\tau_\mu\dot x^\mu\,.
\end{eqnarray}
The equations of motion consist of \eqref{eq:mattercoupledNCgravity} where 
\begin{equation}
    \mathcal{T}_\tau^\mu=-m\int \d\lambda\,\frac{\delta(x-x(\lambda))}{e}\dot x^\mu\,,
\end{equation}
and $\mathcal{T}_h^{\mu\nu}=0$ as is easily obtained from the variations of \eqref{eq:pointparticle_LO_lagrangian}.
In a worldline gauge for which $\tau_\mu\dot x^\mu=1$ we thus have
\begin{equation}\label{eq:rho}
    \rho=-\tau_\mu\mathcal{T}_\tau^\mu=m\int \d\lambda\, \frac{\delta(x-x(\lambda))}{e}\,.
\end{equation}
Besides the NC equations of motion there are additional decoupled equations of motion for the field $\Phi_{\mu\nu}$ as well as the Lagrange multiplier. The latter can be replaced by the NNLO fields by replacing the gravity part of the above Lagrangian by the NNLO Lagrangian of the expansion of the EH Lagrangian. Finally there is the $x^\mu$ equation of motion which enforces $\d\tau=0$. This thus gives a complete off shell description of NC gravity with a point particle source.

\subsubsection{On shell expansion}
Equations \eqref{eq:LOeom} and \eqref{eq:NLOeom} can also be obtained from an on shell expansion by starting with the relativistic geodesic equation
\begin{equation}\label{eq:Releom}
\ddot X^\mu+\Gamma^\mu_{\nu\rho}\dot X^\nu\dot X^\rho=0\,,
\end{equation}
where $g_{\mu\nu}\dot X^\mu\dot X^\nu=-c^2$ and by expanding $X^\mu=x^\mu+\ldots$. The gauge choice $g_{\mu\nu}\dot X^\mu\dot X^\nu=-c^2$ tells us that $\tau_\mu\dot x^\mu=1$. The leading order term in the expansion of \eqref{eq:Releom} tells us that $a_\mu=0$. Using this result the new leading order expansion of \eqref{eq:Releom} gives us \eqref{eq:NLOeom}, so in this manner we never needed to work with $y^\mu$ at the level of the EOM.

The $\tau_\mu$ projection of \eqref{eq:NLOeom} is trivially satisfied using that $\tau_\mu\dot x^\mu=1$. This suggests that we should expand \eqref{eq:Releom} to one further subleading order and project that equation with $\tau_\mu$ to find something nontrivial. Indeed doing so leads to 
\begin{equation}
\frac{\d^2}{\d\lambda^2}\left(\tau_\mu y^\mu\right)+\bar K_{\mu\nu}\dot x^\mu\dot x^\nu+2\dot x^\mu\partial_\mu\hat\Phi=0\,.
\end{equation}
Furthermore from $g_{\mu\nu}\dot X^\mu\dot X^\nu=-c^2$ we also learn that
\begin{equation}
\frac{\d}{\d\lambda}\left(\tau_\mu y^\mu\right)=\frac{1}{2}\bar h_{\mu\nu}\dot x^\mu\dot x^\nu\,.
\end{equation}
We thus find
\begin{equation}\label{eq:conservationparticle}
\frac{\d}{\d\lambda}\left(\frac{1}{2}\bar h_{\mu\nu}\dot x^\mu\dot x^\nu+2\hat\Phi\right)+\bar K_{\mu\nu}\dot x^\mu\dot x^\nu=0\,,
\end{equation}
which is the expression for energy conservation:
Consider for example the geometry $\tau=\d t$, $h_{\mu\nu}\d x^\mu \d x^\nu=\d\vec x^2$ and $m=\Phi(x) \d t$ with $\Phi$ time independent.
In this case \eqref{eq:conservationparticle} becomes
\begin{equation}
\frac{\d}{\d\lambda}\left(\frac{1}{2}\dot{\vec x}\cdot\dot{\vec x}+\Phi\right)=0\,,
\end{equation}
which is the classical expression for energy conservation for a particle moving in a time-independent Newtonian potential.

\subsubsection{Fluid description}
Instead of using embedding coordinates we can also say that the geodesics correspond to the integral curves of a parallel transported  unit normalised vector field $U^\mu$, i.e.
\begin{equation}
U^\rho\nabla_\rho U^\mu=U^\rho\left(\partial_\rho U^\mu+\Gamma^\mu_{\rho\nu}U^\nu\right)=0\,,
\end{equation}
where everything is a function of the spacetime coordinates $x^\mu$ which are not expanded in $1/c$.
We can then expand $U^\mu$, which obeys $g_{\mu\nu}U^\mu U^\nu=-c^2$, according to \eqref{eq:expansion_field_general} as
\begin{equation}\label{eq:expU}
U^\mu=u^\mu+c^{-2}u_{(2)}^\mu+\order{c^{-4}}\,,
\end{equation}
where $\tau_\mu u^\mu=1$ and $\tau_\mu u_{(2)}^\mu=\frac{1}{2}\bar h_{\mu\nu}u^\mu u^\nu$. We obtain for $\d\tau=0$ the equations
\begin{eqnarray}
0 & = & u^\mu\bar\nabla_\mu u^\nu\,,\\
0 & = & u^\mu\partial_\mu\left(\frac{1}{2}\bar h_{\rho\sigma}u^\rho u^\sigma+2\hat\Phi\right)+\bar K_{\mu\nu}u^\mu u^\nu\,,
\end{eqnarray}
where the latter results from the $\tau_\mu$ projection of the subleading equation. If we multiply these equations with $\rho$ given in \eqref{eq:rho} then we obtain equations \eqref{eq:NLOeom} and \eqref{eq:conservationparticle}. Using that $\nabla_\mu (\rho U^\mu)=0$ at leading order implies  $\bar\nabla_\mu (\rho u^\mu)=0$ we can also write these as fluid-type conservation equations, i.e.
\begin{eqnarray}
0 & = & \bar\nabla_\mu\left(\rho u^\mu u^\nu\right)\,,\label{eq:massmomcons}\\
0 & = & \bar\nabla_\mu\left[\rho\left(\frac{1}{2}\bar h_{\rho\sigma}u^\rho u^\sigma+2\hat\Phi\right)u^\mu\right]+\bar K_{\mu\nu}\rho u^\mu u^\nu\,.\label{eq:energycons}
\end{eqnarray}
The latter equation is identically satisfied given the mass-momentum conservation equation \eqref{eq:massmomcons}.

\subsubsection{Coupling to electrodynamics}
We can easily generalise the action to couple the particle to the $1/c^2$ expansion of a background electromagnetic potential $A_\mu$,
whose dynamics we study further in Section \ref{sec:electrodynamics}.
The expansion of the electromagnetic potential is assumed to be a $1/c^2$ expansion to match the orders of the expansion of the point particle Lagrangian \eqref{eq:ppLag}, so that 
\begin{equation}
A_\mu\left(X\right) = c^2A^{(-2)}_\mu(x)+A^{(0)}_\mu(x)+y^\rho \partial_\rho A^{(-2)}_\mu(x)+\order{c^{-2}}\,.
\end{equation}
Expanding the usual electric coupling to electrodynamics with electric charge $q$ then yields 
\begin{equation}
\mathcal{L}_{\mathrm{EM}} = qc^2A^{(-2)}_\mu(x) \dot x^\mu+q\left[A^{(0)}_\mu(x) \dot x^\mu +F^{(-2)}_{\rho\mu}\dot x^\mu y^\rho\right]+\order{c^{-2}}\,.
\end{equation}
The resulting total Lagrangian for a massive point particle with mass $ m$ and charge $q$  is hence expanded as
\begin{eqnarray}
\mathcal{L}_{\mathrm{tot}}&=& \mathcal{L}_{\mathrm{NR,pp}} + \mathcal{L}_{\mathrm{EM}} = 
c^2\left(qA^{(-2)}_\mu-m\tau_\mu\right)\dot x^\mu  \nonumber\\
&&+\left(qF^{(-2)}_{\rho\mu}-m\tau_{\rho\mu}\right)\dot x^\mu y^\rho+\frac{m}{2}\frac{\bar h_{\mu\nu}\dot x^\mu\dot x^\nu}{\tau_\rho\dot x^\rho}+qA^{(0)}_\mu \dot x^\mu+\order{c^{-2}}\,.
\end{eqnarray}
In order to cancel the leading order term, we see that we need to take $A^{(-2)}_\mu=\frac{m}{q}\tau_\mu$.
In that case also the term that gives coupling to torsion at the next-to-leading order vanishes and we get
\begin{eqnarray}
\mathcal{L}_{\mathrm{tot},A^{(-2)}=\frac{m}{q}\tau}&=&\frac{m}{2}\frac{\bar h_{\mu\nu}\dot x^\mu\dot x^\nu}{\tau_\rho\dot x^\rho}+qA^{(0)}_\mu \dot x^\mu+\order{c^{-2}}\,,
\end{eqnarray}
This is exactly the Lagrangian of a (type I) Newton--Cartan particle coupled to electrodynamics \cite{Andringa:2010it, Andringa:2012uz}.
Notice that there is now no $y^\rho$ dependence enforcing torsionlessness so that these particles can propagate in torsionful geometry
\footnote{A similar feature is observed in the coupling of non-relativistic strings including a background $B$-field
\cite{Harmark:2019upf}.}.

We return to the  $1/c^2$  expansion of Maxwell electrodynamics  in section  \ref{sec:electrodynamics}.

\subsection{TTNC geodesics}\label{sec:Galilean_lightcones}
In the previous subsection we studied the $1/c^2$ expansion of the massive point particle Lagrangian and concluded that this is only consistent on a background with $\d\tau=0$. This begs the question what about point particles moving on a torsionful NC geometry.

Consider again the action
\begin{equation}\label{eq:actionparticle}
S=-mc\int \d\lambda\sqrt{-g_{\mu\nu}(X)\dot X^\mu\dot X^\nu}\,.
\end{equation}
This action is worldline reparameterisation invariant with respect to $\delta X^\mu=\xi\dot X^\mu$. The $X^\mu$ equation of motion is given by
\begin{equation}
\left(\delta^\sigma_\mu-\frac{g_{\mu\tau}\dot X^\tau\dot X^\sigma}{g_{\kappa\lambda}\dot X^\kappa\dot X^\lambda}\right)\left(\ddot X^\mu+\Gamma^\mu_{\rho\nu}\dot X^\rho\dot X^\nu\right)=0\,.
\end{equation}
If we fix the worldline reparameterisations by setting
\begin{equation}\label{eq:norm}
g_{\mu\nu}\dot X^\mu\dot X^\nu=-C^2\,,
\end{equation}
where $C$ is any constant, then the geodesic equation becomes
\begin{equation}\label{eq:geo}
\ddot X^\mu+\Gamma^\mu_{\nu\rho}(X)\dot X^\nu\dot X^\rho=0\,.
\end{equation}
Any solution to this equation obeys
\begin{equation}
\frac{\d}{\d\lambda}\left(g_{\mu\nu}\dot X^\mu\dot X^\nu\right)=0\,,
\end{equation}
so only the sign in \eqref{eq:norm} is not automatic. Since we are dealing with a massive point particle we will take it to be timelike.

The norm of the timelike tangent vector is
\begin{equation}
g_{\mu\nu}\dot X^\mu\dot X^\nu=-c^2\left(T_\mu(X)\dot X^\mu\right)^2+\Pi_{\mu\nu}(X)\dot X^\mu\dot X^\nu<0\,.
\end{equation}
In the previous subsection we took $T_\mu(X)\dot X^\mu=\order{c^0}$ and $\Pi_{\mu\nu}(X)\dot X^\mu=\order{c^0}$. Here we will instead take the following starting point
\begin{equation}
    T_\mu(X)\dot X^\mu=\order{c^{-1}}\,,\qquad\Pi_{\mu\nu}(X)\dot X^\mu=\order{c^0}\,,
\end{equation}
where both are expanded in a series of $1/c^2$.
We can achieve this by expanding the embedding scalar as
\begin{equation}\label{eq:expX}
X^\mu=x^\mu+\frac{1}{c}y^\mu+\order{c^{-2}}\,.
\end{equation}
The leading order equation obtained from \eqref{eq:geo} and the expansion of the Christoffel symbols \eqref{eq:Christoffel_1c2_expansion1} is
\begin{equation}
\left(\tau_\nu\dot x^\nu\right)^2h^{\mu\sigma}a_\sigma=0\,.
\end{equation}
For NC geometry this is automatic but for TTNC geometry this gives $\tau_\nu\dot x^\nu=0$. In the latter case the only way to keep the tangent vector $\dot X^\mu$ timelike is for there to be a term at order $1/c$.

Using
\begin{equation}
\tau_\mu(X)=\tau_\mu(x)+\frac{1}{c}y^\rho\partial_\rho\tau_\mu(x)+\order{c^{-2}}\,,
\end{equation}
we find
\begin{eqnarray}
\tau_\mu(X)\dot X^\mu & = & \frac{1}{c}\left(\tau_\mu\dot y^\mu+\dot x^\mu y^\nu\partial_\nu\tau_\mu\right)+\order{c^{-2}}\nonumber\\
& = & \frac{1}{c}\left(\frac{\d}{\d\lambda}\left(\tau_\mu y^\mu\right)-\dot x^\nu a_\nu \tau_\mu y^\mu\right)+\order{c^{-2}}\,.\label{eq:tauX}
\end{eqnarray}
This gives
\begin{equation}
g_{\mu\nu}\dot X^\mu\dot X^\nu=-\left(\frac{\d}{\d\lambda}\left(\tau_\mu y^\mu\right)-\dot x^\nu a_\nu \tau_\mu y^\mu\right)^2+h_{\mu\nu}\dot x^\mu\dot x^\nu+\order{c^{-1}}<0\,,
\end{equation}
where $\tau_\mu$ and $h_{\mu\nu}$ are functions of $x^\mu$. We conclude that in the large $c$ limit
\begin{equation}\label{eq:F}
F\equiv\left(\frac{\d}{\d\lambda}\left(\tau_\mu y^\mu\right)-\dot x^\nu a_\nu \tau_\mu y^\mu\right)^2-h_{\mu\nu}\dot x^\mu\dot x^\nu=C^2\,,
\end{equation}
where we took $C^2$ in \eqref{eq:norm} to be independent of $c$. Using \eqref{eq:geo}, \eqref{eq:Christoffel_1c2_expansion1}-\eqref{eq:Christoffel_1c2_expansion2}, \eqref{eq:expX} and \eqref{eq:tauX} we find that the leading order geodesic equation is
\begin{equation}\label{eq:LOgeo}
\ddot x^\mu+\check\Gamma^\mu_{\nu\rho}\dot x^\nu\dot x^\rho+h^{\mu\sigma}a_\sigma\left(\frac{\d}{\d\lambda}\left(\tau_\mu y^\mu\right)-\dot x^\nu a_\nu \tau_\mu y^\mu\right)^2=0\,,
\end{equation}
where we used that $\bar\Gamma^\mu_{\nu\rho}\dot x^\nu\dot x^\rho=\check\Gamma^\mu_{\nu\rho}\dot x^\nu\dot x^\rho$.
By contracting \eqref{eq:geo} with $\tau_\mu(X)$ and expanding up to order $\order{c^{-1}}$ we furthermore obtain
\begin{equation}\label{eq:y}
\frac{\d}{\d\lambda}\log\left(N\left(\tau_\mu\dot y^\mu+\dot x^\mu y^\nu\partial_\nu\tau_\mu\right)\right)=0\,,
\end{equation}
where we used $\tau=N \d T$. Here $T$ is a time function. This equation exists in any coordinate system, i.e. both $N$ and $T$ are scalar functions of the coordinates. 

Equations \eqref{eq:LOgeo} and \eqref{eq:y} can also be obtained from an action with \eqref{eq:F} appearing as a gauge fixing condition. To see this we will set $\tau_\nu\dot x^\nu=0$ off shell. The leading term in the expansion of \eqref{eq:actionparticle} is given by
\begin{equation}\label{eq:masslessGalparticle}
S=\int \d\lambda\mathcal{L}=-mc\int \d\lambda\left[\left(\tau_\mu\dot y^\mu+\dot x^\mu y^\nu\partial_\nu\tau_\mu\right)^2-h_{\mu\nu}\dot x^\mu\dot x^\nu\right]^{1/2}\,.
\end{equation}
We will define the Lagrangian as 
\begin{equation}
\mathcal{L}=-mc F^{1/2}\,.
\end{equation}
The Lagrangian is a function of $x^\mu$, $y^\mu$ and their derivatives. The EOMs can be written as
\begin{eqnarray}
0 & = & \frac{\d}{\d\lambda}\frac{\partial F}{\partial\dot x^\mu}-\frac{1}{2}\frac{1}{F}\frac{\d F}{\d\lambda}\frac{\partial F}{\partial\dot x^\mu}=\frac{\partial F}{\partial x^\mu}\,,\\
0 & = & \frac{\d}{\d\lambda}\frac{\partial F}{\partial\dot y^\mu}-\frac{1}{2}\frac{1}{F}\frac{\d F}{\d\lambda}\frac{\partial F}{\partial\dot y^\mu}=\frac{\partial F}{\partial y^\mu}\,.
\end{eqnarray} 
The $y^\mu$ EOM can be written as
\begin{equation}
\frac{\d}{\d\lambda}\log\frac{F}{N^2\left(\tau_\mu\dot y^\mu+\dot x^\mu y^\nu\partial_\nu\tau_\mu\right)^2}=0\,,
\end{equation}
where we used that for any TTNC geometry $h^{\mu\nu}a_\nu=h^{\mu\nu}N^{-1}\partial_\nu N$. This follows from $\tau_\mu=N\partial_\mu T$. This agrees with equation \eqref{eq:y} when $F=\text{constant}$ which is what was assumed in deriving \eqref{eq:y}. Let the integration constant be $a$, then we find
\begin{equation}
e^{-a}F=N^2\left(\tau_\mu\dot y^\mu+\dot x^\mu y^\nu\partial_\nu\tau_\mu\right)^2=N^4\left[\frac{\d}{\d\lambda}\left(y^\mu\partial_\mu T\right)\right]^2\,,
\end{equation}
which we note is manifestly positive as required in \eqref{eq:F}. A useful identity is
\begin{equation}
\tau_\mu\dot y^\mu+\dot x^\mu y^\nu\partial_\nu\tau_\mu=\frac{\d}{\d\lambda}\left(\tau_\mu y^\mu\right)-\dot x^\nu a_\nu \tau_\mu y^\mu=N\frac{\d}{\d\lambda}\left(y^\mu\partial_\mu T\right)\,.
\end{equation}
The $x^\mu$ equation of motion comes out to be
\begin{equation}
\ddot x^\mu+\check\Gamma^\mu_{\nu\rho}\dot x^\nu\dot x^\rho-\frac{1}{2F}\frac{\d F}{\d\lambda}\dot x^\mu+\left(\tau_\mu\dot y^\mu+\dot x^\mu y^\nu\partial_\nu\tau_\mu\right)^2h^{\mu\sigma}a_\sigma=0\,.
\end{equation}

We will now choose a gauge in which $F$ is constant with $F=e^a=C^2$. This implies that 
\begin{equation}\label{eq:yT}
\tau_\mu\dot y^\mu+\dot x^\mu y^\nu\partial_\nu\tau_\mu=N^{-1}\Longrightarrow \frac{\d}{\d\lambda}\left(y^\mu\partial_\mu T\right)=N^{-2}\,.
\end{equation}
Equation \eqref{eq:F} becomes
\begin{equation}\label{eq:geo1}
\frac{1}{2}h_{\nu\rho}\dot x^\nu\dot x^\rho-\frac{1}{2}N^{-2}=-\frac{1}{2}C^2\,.
\end{equation}
The $x^\mu$ equation of motion simplifies to 
\begin{equation}\label{eq:geo2}
\ddot x^\mu+\check\Gamma^\mu_{\nu\rho}\dot x^\nu\dot x^\rho=\frac{1}{2}h^{\mu\sigma}\partial_\sigma N^{-2}\,.
\end{equation}
The last two equations together with 
\begin{equation}\label{eq:geo3}
\tau_\mu\dot x^\mu=0\,,    
\end{equation}
determine the geodesics in a TTNC background. Note that $m_\mu$ and hence the Newtonian potential does not appear. Instead we now have a force that is dictated by minus the gradient of $-\frac{1}{2}N^{-2}$ which plays the role of potential energy. The fact that $C^2>0$ (for massive relativistic point particles moving below the speed of light) means that we only have bound states in this potential field. The last equation is automatically satisfied if one contracts it with $\tau_\mu$ or $h_{\mu\kappa}\dot x^\kappa$.

The fact that $\tau_\mu\dot x^\mu=0$ means that we cannot replace the $\lambda$ geodesic parameter with coordinate time. This makes a particle interpretation challenging. The objects probe only the LO fields $\tau_\mu$ and $h_{\mu\nu}$, which is dictated by local Galilean symmetries and perhaps one should think of these particles as (massless) Galilean particles.
In Section \ref{sec:geodesics_sphericalsym_background} we shall see how the above is realised explicitly for the case of a spherical symmetric Schwarzschild-type \regls{ttnc} background.
Regardless of conceptual difficulties, one nevertheless obtains the same orbits as from the relativistic geodesic equation in Schwarzschild spacetime. 
Finally we note that the LO Lagrangian \eqref{eq:masslessGalparticle} is $\order{c}$ and so backreactions of this object would require that we include odd powers of $1/c$ in the metric expansion \cite{Ergen:2020yop}.

\subsubsection{Rindler spacetime}\label{subsubsec:Rindler}
To illustrate the difference between Lorentzian and Newton--Cartan geometries and the role of geodesics we make
a slight digression and study here the simple case of 2-dimensional Rindler spacetime. In Section \ref{sec:solutions} we will consider many more examples of solutions of non-relativistic gravity.

Consider the 2D Lorentzian line element
\begin{equation}\label{eq:Mink}
\d s^2=-c^2\d t^2+\d x^2\,.
\end{equation}
Perform the following coordinate transformation
\begin{equation}\label{eq:diffeoRindler}
ct=R\sinh (cT)\,,\qquad x=R\cosh (cT)\,,
\end{equation}
where $T$ has dimensions of inverse velocity and $R$ has dimensions of length. We then find
\begin{equation}\label{eq:Rindler}
\d s^2=-c^2R^2\d T^2+\d R^2\,.
\end{equation}
This is 2D Rindler spacetime. It corresponds to the left and right wedges of a lightcone with centre at $(t,x)=(0,0)$. Lines of constant $T$ are straight lines through the origin since
\begin{equation}
\frac{ct}{x}=\tanh (cT)\,,
\end{equation}
and lines of constant $R$ are hyperbolae since
\begin{equation}
x^2-c^2t^2=R^2\,.
\end{equation}

In the sense of type II NC geometry the metrics \eqref{eq:Mink} and \eqref{eq:Rindler} give rise to
\begin{equation}\label{eq:Minklimit}
\tau=\d t\,,\qquad h=\d x^2\,,\qquad m=0\,,\qquad\Phi=0\,,
\end{equation}
and 
\begin{equation}\label{eq:Rindlerlimit}
\tau=R \d T\,,\qquad h= \d R^2\,,\qquad m=0\,,\qquad\Phi=0\,.
\end{equation}
Since the first of these has $\d\tau=0$ and the second has $\tau\wedge \d\tau=0$ but $\d\tau\neq 0$ they are clearly not diffeomorphic spacetimes. We thus learn that diffeomorphic spacetimes in Lorentzian geometry can correspond to non-diffeomorphic spacetimes in NC geometry. The reason in this case is because the diffeomorphism \eqref{eq:diffeoRindler} is not analytic in $1/c$.

We would like to understand the type II NC limit \eqref{eq:Rindlerlimit} of Rindler spacetime. Since the clock 1-form components vanish at $R=0$ we need to check if this is in fact a coordinate singularity. To this end we will perform the same coordinate transformation \eqref{eq:diffeoRindler}  as before. We will take $c=\hat c/\epsilon$ where $\hat c$ is numerically equal to the speed of light and $\epsilon$ is some dimensionless small quantity. The $1/c$ expansion then becomes an expansion around $\epsilon=0$. We will set $\hat c=1$ (so that $\tau$ has dimensions of length), and write 
\begin{equation}
t=R\sinh (T)\,,\qquad x=R\cosh (T)\,.
\end{equation}
We then have the time-like and space-like vielbeins
\begin{equation}
\label{eq:Rindlervielbeins}
\tau=R\d T=\frac{x}{\sqrt{x^2-t^2}}\d t-\frac{t}{\sqrt{x^2-t^2}}\d x\,,\qquad e=\d R=\frac{-t}{\sqrt{x^2-t^2}}\d t+\frac{x}{\sqrt{x^2-t^2}}\d x\,,
\end{equation}
where we write $h=ee$ for the metric on spatial slices.

The lines $t=\pm x$ correspond to $T=\pm\infty$. Since the total lapse of time (i.e. $\int_\gamma \tau$ along some curve $\gamma$)  is the same in all coordinate systems we can see that future and past infinity correspond to $T=\pm\infty$, i.e. $t=\pm x$, with the exception of the origin $R=0$ or what is the same $(t,x)=(0,0)$. The lapse of time along a curve with constant $R=R_0$ is  $R_0\int_{T_i}^{T_f} \d T$ from some initial to some final time. Clearly this goes to infinity for $T_i\rightarrow-\infty$ and $T_f\rightarrow\infty$. That means that in the $t,x$ coordinates we can consider $t=\pm x$ to represent the boundaries of spacetime except at the origin. 

To understand what happens at the origin consider a straight line $t=\alpha x$ where $\vert\alpha\vert<1$. It follows
from \eqref{eq:Rindlervielbeins} that along such a curve the components of $\tau$ become
\begin{equation}
\tau_t=\frac{\text{sign}(x)}{\sqrt{1-\alpha^2}}\,,\qquad\tau_x=-\frac{\alpha\text{sign}(x)}{\sqrt{1-\alpha^2}}\,, 
\end{equation}
while for those of the 1-form $e$ we get 
\begin{equation}
e_t=-\frac{\alpha\text{sign}(x)}{\sqrt{1-\alpha^2}}\,,\qquad e_x=\frac{\text{sign}(x)}{\sqrt{1-\alpha^2}}\,.
\end{equation}
Let us consider  first the time-like vielbein $\tau$. We see that the values of the components of $\tau$ depend on the direction with which one approaches the origin and secondly when passing through the origin the sign changes. Thus we observe a discontinuous change in $\tau_t$ and $\tau_x$ when passing through the origin. For example if we consider a curve along the $t=0$ axis then $\alpha=0$ so that we jump from $\tau_t=1$ and $\tau_x=0$ to $\tau_t=-1$ and $\tau_x=0$. Since we think of $\tau$ as the normal to constant time (here $T$) hypersurfaces we see that the direction of time is reversed. Going up in the right Rindler wedge is going to the future and in the left Rindler wedge going to the future means going down. This is simple to visualise by drawing a straight line $T=\text{constant}$ through the origin. When moving it forward in time on the right means that it is going down on the left. We do not observe such a feature with the $h$ `metric' because it is quadratic in $e$ and so the sign functions disappear. So as we go through the origin the metrics $\tau$ and $h$ depend on the direction through the origin and the sign of $\tau$ is flipped. To summarise the type II TNC version of Rindler spacetime can be visualised as the two Rindler wedges of Minkowski spacetime joined at the origin and with the Milne patches of Minkowski spacetime removed entirely.

We continue by considering the geodesics in this spacetime.  From the expansion of the timelike geodesics we find, using equations \eqref{eq:geo1}-\eqref{eq:geo3},
\begin{equation}
\dot T=0\,,\qquad \frac{\d}{\d\lambda} y^T=R^{-2}\,,\qquad \dot R^2=R^{-2}-C^2\,.
\end{equation}
In terms of the parameter $\lambda$ we have
\begin{equation}
R^2=C^{-2}-C^2(\lambda-\lambda_0)^2\,,\qquad y^T-y^T_0=\frac{1}{2}\log\vert\frac{1+C^2(\lambda-\lambda_0)}{1-C^2(\lambda-\lambda_0)}\vert\,.
\end{equation}
For finite values of the geodesic parameter $\lambda$ we reach $y^T=\pm\infty$. We can also write the solutions as
\begin{equation}
T=T_0\,,\qquad R\cosh(y^T-y^T_0)=\frac{1}{C}\,.
\end{equation}
We cannot view these solutions as curves that are entirely described in terms of the spacetime coordinates $T$ and $R$. The reason is that we can only replace the geodesic parameter by $y^T$ but the latter is a worldline scalar $y^T$ that is the subleading embedding function for the coordinate $T$.

If we were to define (note that due to the appearance of $y^T$ this is not a coordinate transformation)
\begin{equation}
\tilde x=R\cosh y^T\,,\qquad \tilde t=R\sinh y^T\,,
\end{equation}
then for $y_0^T=0$ we get $\tilde x=1/C$ and for $y_0^T\neq 0$ we get 
\begin{equation}
\tilde t=\tilde x\coth y^T_0-\frac{1}{C}\frac{1}{\sinh y^T_0}\,.
\end{equation}
These are straight lines with slopes larger than $+1$ or less than $-1$, i.e. the timelike geodesics of Minkowksi spacetime. However in the sense of NC geometry that interpretation is lost.

From the point of view of type II NC geometry the field $y^T$ is just a Lagrange multiplier in the action for a massless Galilean particle. Hence the only real type II geodesics are those for which $T=\text{constant}$. In the sense of an approximation of relativistic geodesics the field $y^T$ is of course important and simply captures the NLO effect, correctly reproducing the straight line geodesics of Minkowski spacetime.

\subsection{Perfect fluids}\label{sec:perfect_fluid}
We continue our study of the $1/c^2$ expansion of matter systems by presenting the case of a perfect relativistic fluid. 

We expand the normalised fluid velocity according to \eqref{eq:expansion_field_general} as
\begin{equation}
\label{eq:Umu} 
U^\mu=u^\mu+\frac{1}{c^2}u_{(2)}^\mu+\order{c^{-4}}\,.
\end{equation}
Then the normalisation condition 
\begin{equation}
g_{\mu\nu}U^\mu U^\nu=-c^2\,,
\end{equation}
together with the expansion \eqref{eq:metric_expansion1} of the metric implies the relations
\begin{eqnarray}
\tau_\mu u^\mu & = & 1\,,\\
\tau_\mu u_{(2)}^\mu & = & \frac{1}{2}\bar h_{\mu\nu}u^\mu u^\nu\,.
\end{eqnarray}
This means that the relativistic fluid velocity admits the expansion 
\begin{equation}
U_\mu=c^2\left(-\tau_\mu+\frac{1}{c^2}\left(-\tau_\mu\frac{1}{2}\bar h_{\rho\sigma}u^\rho u^\sigma+\bar h_{\mu\nu}u^\nu\right)+\order{c^{-4}}\right)\,.
\end{equation}

Next we turn to the relativistic energy-momentum tensor 
\begin{equation}
T^{\mu\nu}=\frac{E+P}{c^2}U^\mu U^\nu+Pg^{\mu\nu}\,,
\end{equation}
where $E$ and $P$ are the relativistic internal energy and pressure.
We will assume that these quantities have an expansion given by
\begin{eqnarray}
E & = & c^4E_{(-4)}+c^2E_{(-2)}+E_{(0)}+\order{c^{-2}}\,,\label{eq:perfect_fluid_sum_energy}\\
P & = & c^4P_{(-4)}+c^2P_{(-2)}+P_{(0)}+\order{c^{-2}}\,.\label{eq:perfect_fluid_sum_mom}
\end{eqnarray}
The energy-momentum tensor expands according to \eqref{eq:EM_indicesup_expansion}, which using 
\eqref{eq:Umu} and the expansion of the inverse metric in \eqref{eq:metric_expansion2} gives
\begin{eqnarray}
\os{T}{-4}^{\mu\nu} & = & P_{(-4)}h^{\mu\nu}\,,\\
\os{T}{-2}^{\mu\nu} & = & \left(E_{(-4)}+P_{(-4)}\right)u^\mu u^\nu-P_{(-4)}\left(\hat v^\mu\hat v^\nu+h^{\mu\rho}h^{\nu\sigma}\bar\Phi_{\rho\sigma}\right)+P_{(-2)}h^{\mu\nu}\,,\\
\os{T}{0}^{\mu\nu} & = & \left(E_{(-2)}+P_{(-2)}\right)u^\mu u^\nu+\left(E_{(-4)}+P_{(-4)}\right)\left(u^\mu u_{(2)}^\nu+u^\nu u_{(2)}^\mu\right)\nonumber\\
&&+P_{(0)}h^{\mu\nu}-P_{(-2)}\left(\hat v^\mu\hat v^\nu+h^{\mu\rho}h^{\nu\sigma}\bar\Phi_{\rho\sigma}\right)+P_{(-4)}\left(2\hat v^\mu \hat v^\nu \hat \Phi-\bar Y^{\mu\nu}\right)\,.
\end{eqnarray}

We can use the results of Section \ref{sec:Ward_identities} for the expansion of the relativistic energy-momentum conservation equation. The Ward identity \eqref{eq:WIA} becomes
\begin{equation}\label{eq:conservation_perfect_fluid_LO}
h^{\mu\nu}\partial_\nu P_{(-4)} + \left(E_{(-4)}+P_{(-4)}\right)h^{\mu\nu}a_\nu=0\,,
\end{equation}
while the subleading equation contracted with $\tau_\nu$ (see \eqref{eq:WIB})  gives
\begin{equation}\label{eq:LOenergycons}
u^\mu\partial_\mu E_{(-4)}+\left(E_{(-4)}+P_{(-4)}\right)\left(\bar\nabla_\mu+a_\mu\right)u^\mu=0\,.
\end{equation}
We observe that the quantity $P_{(-4)}$ is a `pressure' needed to balance the force due to torsion.

If $P_{(-4)}=E_{(-4)}=P_{(-2)}=0$ we find that we must have $a_\mu=0$ and hence $\d\tau=0$. In that case
\eqref{eq:WIB}  becomes a mass conservation equation with $E_{(-2)}$ the mass density. We set $E_{(-2)}=n$ for mass density and $P_{(0)}=\mathcal{P}$ for pressure. Then the conservation equation  \eqref{eq:WIA}  becomes
\begin{equation}\label{eq:massmomentumcon}
 \bar\nabla_\mu\mathcal{T}^{\mu\nu}=0\,,
\end{equation}
where
\begin{equation}
\os{T}{-2}^{\mu\nu}=\mathcal{T}^{\mu\nu} \equiv \mathcal{P}h^{\mu\nu}+n u^\mu u^\nu\,.
\end{equation}
This is the Cauchy stress-mass tensor. Equation \eqref{eq:massmomentumcon} describes mass-momentum conservation. Contracting \eqref{eq:massmomentumcon} with $\tau_\nu$ leads to the mass conservation equation $\bar\nabla_\mu(nu^\mu)=0$. The coupling to Newton--Cartan gravity \eqref{eq:mattercoupledNCgravity} can be found by considering \eqref{eq:calTtau} and \eqref{eq:calTh} which tells us that 
\begin{equation}
    \mathcal{T}_\tau^\mu=-n u^\mu\,,\qquad h_{\mu\rho}\mathcal{T}^{\mu\nu}_h=0\,.
\end{equation}

The subleading conservation equation \eqref{eq:WIA_no_tau_contraction} with $a_\mu=0$ can be written as
\begin{equation}
\bar\nabla_\mu \os{T}{0}^{\mu\nu}+\Gamma^\mu_{{(2)}\,\mu\rho}\mathcal{T}^{\rho\nu}+\Gamma^\nu_{{(2)}\,\mu\rho}\mathcal{T}^{\mu\rho}=0\,.
\end{equation}
If we contract this with $\tau_\nu$ we obtain
\begin{equation}\label{eq:NLOEMconservation}
\bar\nabla_\mu\left(\left[\frac{1}{2}\bar h_{\rho\sigma}u^\rho u^\sigma+2\hat\Phi\right] nu^\mu+\mathcal{P}\hat v^\mu\right)+\bar K_{\mu\nu}\mathcal{T}^{\mu\nu}+\bar\nabla_\mu\left(\left(E_{(0)}+\mathcal{P}\right)u^\mu+nu_{(2)}^\mu\right)+\Gamma^\mu_{{(2)}\,\mu\rho}n u^\rho=0\,.
\end{equation}
The relativistic conservation equations for a perfect fluid $\nabla_\mu T^{\mu\nu}=0$ when projected with $U_\nu$ give
\begin{equation}
\nabla_\mu\left[\left(E+P\right)U^\mu\right]=U^\mu\partial_\mu P\,,
\end{equation}
which at subleading order leads to 
\begin{equation}\label{eq:subcon}
\bar\nabla_\mu\left(\left(E_{(0)}+\mathcal{P}\right)u^\mu+nu_{(2)}^\mu\right)+\Gamma^\mu_{{(2)}\,\mu\rho}n u^\rho=u^\mu\partial_\mu \mathcal{P}\,.
\end{equation}
Combining this with \eqref{eq:NLOEMconservation} we obtain 
\begin{equation}\label{eq:Econ}
\bar\nabla_\mu\left(\left[\frac{1}{2}\bar h_{\rho\sigma}u^\rho u^\sigma+2\hat\Phi\right] nu^\mu+\mathcal{P}\hat v^\mu\right)+\bar K_{\mu\nu}\mathcal{T}^{\mu\nu}+u^\mu\partial_\mu \mathcal{P}=0\,.
\end{equation}
This equation is independent of $u_{(2)}^\mu$ and for zero pressure it reduces to the fluid description of the non-relativistic particle given in \eqref{eq:energycons}. It can be shown that upon using \eqref{eq:massmomentumcon} it is identically satisfied. In other words the energy conservation couples the LO fluid variables to the NLO fluid variable $u^\mu_{(2)}$.

In order to obtain the standard equations for a massive Galilean (Bargmann) fluid we need to include a relativistic conserved $U(1)$ current $J^\mu=QU^\mu$. We expand the charge density as
\begin{equation}
Q = c^2 n+Q_{(0)}+\order{c^{-2}}\,,
\end{equation}
then for $\d\tau=0$ the leading term in $\nabla_\mu J^\mu=0$ is $\bar\nabla_\mu(nu^\mu)=0$ while the subleading term reads
\begin{equation}
\bar\nabla_\mu\left(n u_{(2)}^\mu\right)+\overset{\scriptscriptstyle{(2)}}{\Gamma}{}^{\mu}_{\mu\nu}nu^\nu=-\bar\nabla_\mu\left(Q_{(0)}u^\mu\right)\,.
\end{equation}
Combining this with \eqref{eq:subcon} we obtain
\begin{equation}\label{eq:subcon2}
\bar\nabla_\mu\left(\left(E_{(0)}-Q_{(0)}+\mathcal{P}\right)u^\mu\right)=u^\mu\partial_\mu \mathcal{P}\,.
\end{equation}
If we use this to eliminate $u^\mu\partial_\mu \mathcal{P}$ from \eqref{eq:Econ} we obtain
\begin{equation}\label{eq:Econ2}
\bar\nabla_\mu\left(\left(\mathcal{E}+\mathcal{P}+\frac{1}{2}n\bar h_{\rho\sigma}u^\rho u^\sigma\right) u^\mu+2\hat\Phi nu^\mu+\mathcal{P}\hat v^\mu\right)+\bar K_{\mu\nu}\mathcal{T}^{\mu\nu}=0\,,
\end{equation}
where we defined $\mathcal{E}=E_{(0)}-Q_{(0)}$. Using \eqref{eq:Econ} this can be rewritten as 
\begin{equation}
\bar\nabla_\mu\left(\left(\mathcal{E}+\mathcal{P}\right) u^\mu\right)-u^\mu\partial_\mu\mathcal{P}=0\,.
\end{equation}

Defining the covariant energy-momentum tensor for a non-relativistic Bargmann fluid
\begin{equation}
\os{T}{\mathrm{NR}}^{\mu}{}_{\nu}= -\left(\mathcal{E}+\mathcal{P}+\frac{1}{2}n\bar h_{\kappa\lambda}u^\kappa u^\lambda\right)u^\mu\tau_\nu+n u^\mu\bar h_{\nu\rho}u^\rho+\mathcal{P}\delta^\mu_\nu\,,
\end{equation}
and extracting the energy current 
\begin{equation}
\mathcal{E}^\mu =  \hat v^\nu\os{T}{\mathrm{NR}}^{\mu}{}_{\nu}\,,
\end{equation}
we find the energy conservation equation
\begin{equation}
\bar\nabla_\mu\mathcal{E}^\mu+\mathcal{T}^{\mu\rho}\bar K_{\mu\rho}=0\,.
\end{equation}

What we thus see is that in the case of the $1/c^2$ expansion of the relativistic perfect fluid without the $U(1)$ current we find mass and momentum conservation at leading order but the energy conservation equation couples to the subleading field $u_{(2)}^\mu$. On the other hand, when there is a $U(1)$ current present one can find a limit in which all the usual non-relativistic fluid equations (on a type I NC geometry), i.e. mass, momentum and energy conservation, are obtained, forming a closed set of equations. In the case of the point particle this distinction did not arise because there is no internal energy.
We note that various non-relativistic fluids have been studied in the literature, see for example
\cite{Hassaine:1999hn, Horvathy:2009kz,Jensen:2014ama, Geracie:2015xfa, Hartong:2016nyx,deBoer:2017ing,deBoer:2017abi, Armas:2019gnb}.

\subsection{Scalar fields}
\subsubsection{Complex scalar field}
Consider the action of a complex scalar field
\begin{equation}
S=-c^{-1}\int \d^{d+1}x\sqrt{-g}\left(g^{\mu\nu}\partial_\mu\phi\partial_\nu\phi^\star+m^2 c^2\phi\phi^\star\right)\,.
\end{equation}
We split the field in terms of its modulus and phase according to $\phi=\frac{1}{\sqrt{2}}\varphi e^{ic^2\theta}$ so that
\begin{multline}\label{eq:complexscalar}
S=\int \d^{d+1}x E\Bigg[\frac{1}{2c^2}\left(T^{\mu}\partial_\mu\varphi\right)^2-\frac{1}{2}\Pi^{\mu\nu}\partial_\mu\varphi\partial_\nu\varphi\\+\varphi^2\left(\frac{c^2}{2}\left(T^{\mu}\partial_\mu\theta\right)^2-\frac{c^4}{2}\Pi^{\mu\nu}\partial_\mu\theta\partial_\nu\theta-\frac{1}{2}m^2c^2\right)\Bigg]\,,
\end{multline}
where we used \eqref{eq:metrics_explicit_c_factors_no_expansion} and \eqref{eq:defPi}.
Next we expand the modulus and phase of $\phi$ according to
\begin{eqnarray}
\varphi & = & \varphi_{(0)}+c^{-2}\varphi_{(2)}+c^{-4}\varphi_{(4)}+\order{c^{-6}}\,,\\
\theta & = & \theta_{(0)}+c^{-2}\theta_{(2)}+c^{-4}\theta_{(4)}+\order{c^{-4}}\,.
\end{eqnarray}
The expansion of the Lagrangian is
\begin{equation}
\mathcal{L}=c^{4}\;\oss{\mathcal{L}}{-4}{LO}+c^{2}\;\oss{\mathcal{L}}{-2}{NLO}+\order{c^0}\,.
\end{equation}

The leading order Lagrangian is given by 
\begin{equation}
\mathcal{L}_{\mathrm{LO}}=-\frac{1}{2}e\varphi_{(0)}^2h^{\mu\nu}\partial_\mu\theta_{(0)}\partial_\nu\theta_{(0)}\,,
\end{equation}
so that the $\varphi_{(0)}$ equation of motion tells us that 
\begin{equation}
h^{\mu\nu}\partial_\mu\theta_{(0)}\partial_\nu\theta_{(0)}=0\,.
\end{equation}
This condition is a sum of squares, so it implies  
\begin{equation}\label{eq:conditionscalarA}
h^{\mu\nu}\partial_\mu\theta_{(0)}=0\,.    
\end{equation}
As explained in Section \ref{sec:Lagrangians}, this condition will be repeated at any order of the Lagrangian through the equations of motion of the most subleading field in the expansion of $\phi$ .

With these comments, we can determine the NLO Lagrangian to be
\begin{eqnarray}
\oss{\mathcal{L}}{-2}{NLO} & = & e\left(-\varphi_{(0)}^2h^{\mu\nu}\partial_\mu\theta_{(0)}\partial_\nu\theta_{(2)}+\frac{1}{2}\varphi_{(0)}^2\left(\hat v^\mu\partial_\mu\theta_{(0)}\right)^2-\frac{1}{2}m^2\varphi_{(0)}^2-\varphi_{(0)}\varphi_{(2)}h^{\mu\nu}\partial_\mu\theta_{(0)}\partial_\nu\theta_{(0)}\right.\nonumber\\
&&\left.+\frac{1}{2}\varphi_{(0)}^2h^{\mu\rho}h^{\nu\sigma}\bar\Phi_{\rho\sigma}\partial_\mu\theta_{(0)}\partial_\nu\theta_{(0)}-\frac{1}{2}\varphi_{(0)}^2\left(\hat\Phi+\frac{1}{2}h^{\rho\sigma}\bar\Phi_{\rho\sigma}\right)h^{\mu\nu}\partial_\mu\theta_{(0)}\partial_\nu\theta_{(0)}\right)\,.\label{eq:NLOscalar}
\end{eqnarray}
Using that $h^{\mu\nu}\partial_\mu\theta_{(0)}=0$ (which now follows from the $\varphi_{(2)}$ EOM),
the EOM of $\varphi_{(0)}$ is,  
\begin{equation}\label{eq:conditionscalarB}
v^\mu\partial_\mu\theta_{(0)}=\pm m\,.
\end{equation}
Equations \eqref{eq:conditionscalarA} and \eqref{eq:conditionscalarB} then imply that
\begin{equation}
\tau_\mu=\mp \frac{1}{m}\partial_\mu\theta_{(0)}\,,
\end{equation}
so that $\tau$ is exact. The LO matter Lagrangian is $\order{c^4}$ but it does not source gravity at that order due to \eqref{eq:conditionscalarA}. The first sourcing of gravity appears at $\order{c^2}$. We thus couple \eqref{eq:action_NRG} to \eqref{eq:NLOscalar}. This means that the coupling of the matter Lagrangian \eqref{eq:complexscalar} to the EH Lagrangian will give rise to Newtonian gravity coupled to a scalar field in the large speed of light expansion.

The NNLO Lagrangian is 
\begin{eqnarray}
\oss{\mathcal{L}}{0}{N^2LO} & = & e\left(\hat\Phi+\frac{1}{2}h^{\rho\sigma}\bar\Phi_{\rho\sigma}\right)\left(\frac{1}{2}\varphi_{(0)}^2\left(\hat v^\mu\partial_\mu\theta_{(0)}\right)^2-\varphi_{(0)}^2h^{\mu\nu}\partial_\mu\theta_{(0)}\partial_\nu\theta_{(2)}-\frac{1}{2}m^2\varphi_{(0)}^2\right)\nonumber\\
&&+e\left(-\frac{1}{2}h^{\mu\nu}\partial_\mu\varphi_{(0)}\partial_\nu\varphi_{(0)}-\varphi_{(0)}^2 h^{\mu\nu}\partial_\mu\theta_{(0)}\partial_\nu\theta_{(4)}-\frac{1}{2}\varphi_{(0)}^2 h^{\mu\nu}\partial_\mu\theta_{(2)}\partial_\nu\theta_{(2)}\right.\nonumber\\
&&\left.-2\varphi_{(0)}\varphi_{(2)} h^{\mu\nu}\partial_\mu\theta_{(0)}\partial_\nu\theta_{(2)}+\varphi_{(0)}^2\hat v^\mu\partial_\mu\theta_{(0)}\hat v^\mu\partial_\mu\theta_{(2)}+\varphi_{(0)}\varphi_{(2)}\left(\hat v^\mu\partial_\mu\theta_{(0)}\right)^2\right.\nonumber\\
&&\left.+\varphi_{(0)}^2h^{\mu\rho}h^{\nu\sigma}\bar\Phi_{\rho\sigma}\partial_\mu\theta_{(0)}\partial_\nu\theta_{(2)}-m^2\varphi_{(0)}\varphi_{(2)}-\varphi_{(0)}^2\hat\Phi\left(\hat v^\mu\partial_\mu\theta_{(0)}\right)^2\right)\nonumber\\
&&+\frac{1}{2}e\varphi_{(0)}^2 Y^{\mu\nu}\partial_\mu\theta_{(0)}\partial_\nu\theta_{(0)}-\frac{1}{2}e\chi h^{\mu\nu}\partial_\mu\theta_{(0)}\partial_\nu\theta_{(0)}\,,\label{eq:NNLOScalar}
\end{eqnarray}
where $Y^{\mu\nu}$ is defined in \eqref{eq:metric_expansion2} and where we added a Lagrange multiplier $\chi$ to enforce $h^{\mu\nu}\partial_\mu\theta_{(0)}\partial_\nu\theta_{(0)}$ and where we ignored all terms quadratic in $h^{\mu\nu}\partial_\mu\theta_{(0)}=0$. The field $\chi$ is given by $\varphi_{(2)}^2+2\varphi_{(0)}\varphi_{(4)}$. The term $\varphi_{(0)}^2 Y^{\mu\nu}\partial_\mu\theta_{(0)}\partial_\nu\theta_{(0)}$ only contributes to the $\theta_{(0)}$ equation of motion because $Y^{\mu\nu}$ is of the form $h^{\mu\rho}Y_\rho{}^\nu+h^{\nu\rho}Y_\rho{}^\mu$ as it obeys $\tau_\mu\tau_\nu Y^{\mu\nu}=0$. Since we will not concern ourselves with NNLO fields we did not write out the $Y^{\mu\nu}$ term. The $\chi$ and $\varphi_{(2)}$ equations of motion set $\hat v^\mu\partial_\mu\theta_{(0)}=\pm m$. We will take the plus sign. The $\theta_{(4)}$ equation of motion is automatically satisfied. The $\theta_{(2)}$ equation of motion becomes
\begin{equation}\label{eq:Sch1}
\partial_\mu\left(-me\varphi_{(0)}^2\hat v^\mu+e\varphi_{(0)}^2 h^{\mu\nu}\partial_\nu\theta_{(2)}\right)=0\,.
\end{equation}
Finally the $\varphi_{(0)}$ equation of motion is
\begin{equation}\label{eq:Sch2}
e^{-1}\partial_\mu\left(eh^{\mu\nu}\partial_\nu\varphi_{(0)}\right)-\varphi_{(0)} h^{\mu\nu}\partial_\mu\theta_{(2)}\partial_\nu\theta_{(2)}+2m\varphi_{(0)}\hat v^\mu\partial_\mu\theta_{(2)}-2m^2\varphi_{(0)}\hat\Phi=0\,.
\end{equation}
If we restrict the NNLO Lagrangian \eqref{eq:NNLOScalar} to \eqref{eq:conditionscalarA} and \eqref{eq:conditionscalarB} we obtain the Schr\"odinger Lagrangian
\begin{equation}\label{eq:SchL}
    \mathcal{L}_{\text{Sch}}=e\left(m\varphi_{(0)}^2\hat v^\mu\partial_\mu\theta_{(2)}-\frac{1}{2}h^{\mu\nu}\partial_\mu\varphi_{(0)}\partial_\nu\varphi_{(0)}-\frac{1}{2}\varphi_{(0)}^2h^{\mu\nu}\partial_\mu\theta_{(2)}\partial_\nu\theta_{(2)}-m^2\varphi_{(0)}^2\hat\Phi\right)\,,
\end{equation}
whose equations of motion are \eqref{eq:Sch1} and \eqref{eq:Sch2}.

The diffeomorphisms generated by $\Xi^\mu=\xi^\mu+c^{-2}\zeta^\mu+\order{c^{-4}}$ (where $\zeta^\mu=-\Lambda\hat v^\mu+h^{\mu\nu}\zeta_\nu$) act on $\theta_{(2)}$ as 
\begin{equation}
\delta\theta_{(2)}=\xi^\mu\partial_\mu\theta_{(2)}-m\Lambda\,.
\end{equation}
Hence if we define the wavefunction $\psi=\sqrt{m}\varphi_{(0)}e^{i\theta_{(2)}}$ then this wavefunction satisfies the Schr\"odinger equation on a type I NC background with $\d\tau=0$. In other words equations \eqref{eq:Sch1} and \eqref{eq:Sch2} can be combined into the complex Schr\"odinger equation
\begin{equation}\label{eq:Scheq}
    -i\hat v^\mu\partial_\mu\psi-\frac{1}{2}i\psi e^{-1}\partial_\mu\left(e\hat v^\mu\right)+\frac{1}{2m}e^{-1}\partial_\mu\left(eh^{\mu\nu}\partial_\nu\psi\right)-m\hat\Phi\psi=0\,.
\end{equation}
For previous approaches to formulating the Schr\"odinger equation on Newton--Cartan spacetimes, see also \cite{Kuchar:1980tw, Duval:1983pb, Duval:1984cj,Jensen:2014aia,Hartong:2014pma}.

\subsubsection{Schr\"odinger--Newton theory}
Let us next consider the coupling of the complex scalar field to gravity. The variation of \eqref{eq:NLOscalar} with respect to $m_\mu$, $\Phi_{\mu\nu}$ and $h_{\mu\nu}$ all give zero upon using $h^{\mu\nu}\partial_\mu\theta_{(0)}=0$. On the other hand the $\tau$ variation in the direction of $\tau$ gives
\begin{equation}\label{eq:NSsource}
\frac{1}{e}\tau_\mu\frac{\delta\;\oss{\mathcal{L}}{-2}{NLO}}{\delta\tau_\mu}=\tau_\mu\mathcal{T}_\tau^\mu=-m^2\varphi_{(0)}^2=-m\psi\psi^*\,.
\end{equation}

Consider the NRG Lagrangian coupled to the NLO scalar Lagrangian. They couple to each other because they both appear at order $c^2$. The combined system is 
\begin{eqnarray}\label{eq:action_NRG_Sch}
\mathcal{L}_\mathrm{NS} & = & \frac{e}{16\pi G_N}\Bigg[h^{\mu\rho}h^{\nu\sigma}K_{\mu\nu}K_{\rho\sigma}-\left(h^{\mu\nu}K_{\mu\nu}\right)^2
-2m_\nu\left(h^{\mu\rho}h^{\nu\sigma}-h^{\mu\nu}h^{\rho\sigma}\right)\check{\nabla}_\mu K_{\rho\sigma}\nonumber\\
&&+\Phi h^{\mu\nu}\check R_{\mu\nu}
+\frac{1}{4}h^{\mu\rho}h^{\nu\sigma}F_{\mu\nu}F_{\rho\sigma}
+\frac{1}{2}\zeta_{\rho\sigma}h^{\mu\rho}h^{\nu\sigma}(\partial_\mu\tau_\nu-\partial_\nu\tau_\mu)\nonumber\\
&&
-\Phi_{\rho\sigma}h^{\mu\rho}h^{\nu\sigma}\left(\check R_{\mu\nu}-\check{\nabla}_\mu a_\nu - a_\mu a_\nu
-\frac{1}{2}h_{\mu\nu}h^{\kappa\lambda}\check R_{\kappa\lambda}
+h_{\mu\nu} e^{-1} \partial_{\kappa}\left(e h^{\kappa\lambda} a_\lambda\right)\right)\Bigg]\nonumber\\
&&+e\left(-\varphi_{(0)}^2h^{\mu\nu}\partial_\mu\theta_{(0)}\partial_\nu\theta_{(2)}+\frac{1}{2}\varphi_{(0)}^2\left(\hat v^\mu\partial_\mu\theta_{(0)}\right)^2-\frac{1}{2}m^2\varphi_{(0)}^2-\varphi_{(0)}\varphi_{(2)}h^{\mu\nu}\partial_\mu\theta_{(0)}\partial_\nu\theta_{(0)}\right.\nonumber\\
&&\left.+\frac{1}{2}\varphi_{(0)}^2h^{\mu\rho}h^{\nu\sigma}\bar\Phi_{\rho\sigma}\partial_\mu\theta_{(0)}\partial_\nu\theta_{(0)}-\frac{1}{2}\varphi_{(0)}^2\left(\hat\Phi+\frac{1}{2}h^{\rho\sigma}\bar\Phi_{\rho\sigma}\right)h^{\mu\nu}\partial_\mu\theta_{(0)}\partial_\nu\theta_{(0)}\right)\,.
\end{eqnarray}
If we include the next order in the expansion and restrict to EOM containing at most NLO fields we obtain the Schr\"odinger--Newton theory.

This theory is essentially the scalar field analogue of the massive point particle theory of Section \ref{subsubsec:NCgravparticle}. The Lagrangian for the actual Schr\"odinger equation appears at $\order{c^0}$ and so the backreaction problem is considerably simplified. One first solves the equations of motion of the Newtonian gravity \eqref{eq:mattercoupledNCgravity} with source given by \eqref{eq:NSsource}, i.e.
\begin{equation}\label{eq:NSgravity}
    \bar R_{\mu\nu}=8\pi G_N\frac{d-2}{d-1}m\psi\psi^*\tau_\mu\tau_\nu\,.
\end{equation}
This leads to a Newtonian potential that then appears at the next order in the Schr\"odinger equation giving rise to the well-known Schr\"odinger--Newton equation. 

To see this more explicitly, choose a background that solves \eqref{eq:NSgravity}, i.e. 
\begin{equation}
    \tau=\d t\,,\qquad h=\d\vec x\cdot \d\vec x\,,\qquad m=\Phi \d t\,,
\end{equation}
where 
\begin{equation}
    \partial_i\partial_i\Phi=8\pi G_N\frac{d-2}{d-1}m\psi\psi^*\,.
\end{equation}
If we take $d=3$ we can solve (using a Green's function) for $\Phi$ to give
\begin{equation}
    \Phi=-mG_N\int \d^3x'\frac{\psi(x')\psi^\star(x')}{\vert\vec x-\vec x'\vert}\,.
\end{equation}
The Schr\"odinger equation \eqref{eq:Scheq} becomes the Schr\"odinger--Newton equation:
\begin{equation}
    i\partial_t\psi(t,x)=\left(-\frac{1}{2m}\vec\partial^2-m^2G_N\int \d^3x'\frac{\psi(t,x')\psi^\star(t,x')}{\vert\vec x-\vec x'\vert}\right)\psi(t,x)\,.
\end{equation}

There is an extensive literature on the subject of the Schr\"odinger--Newton equation \cite{Karolyhazy:1966zz, BialynickiBirula:1976zp, Diosi:2014ura, Diosi:1986nu, Bahrami:2014gwa, Anastopoulos:2014yja}.
It appears in many different experimental, numerical and theoretical studies, including applications to quantum interference \cite{Colella:1975dq, Nesvizhevsky:2002ef} and laboratory tests of quantum gravity aspects \cite{Giulini:2011uw, Carney:2018ofe}.

\subsubsection{Real scalar field}
We briefly consider the case of a real Klein--Gordon scalar field with Lagrangian
\begin{equation}\label{eq:real_scalar_KG_lagrangian}
\mathcal{L}=-c^{-1}\sqrt{-g}\left(\frac{1}{2}g^{\mu\nu}\partial_\mu\varphi\partial_\nu\varphi+V(\varphi)\right)\,,
\end{equation}
where we assume that the potential $V$ does not depend on $c$ explicitly.
Let us expand the scalar field $\varphi$ as 
\begin{equation}
\varphi=\varphi_{(0)}+c^{-2}\varphi_{(2)}+\order{c^{-4}}\,.
\end{equation}
The $1/c^2$ expansion of the Lagrangian then becomes
\begin{equation}\label{eq:KG_Lagrangian_1c2_expansion}
\mathcal{L} = \oss{\mathcal{L}}{0}{LO}+c^{-2}\oss{\mathcal{L}}{2}{NLO}+\order{c^{-4}}\,,
\end{equation}
where we find
\begin{eqnarray}
\oss{\mathcal{L}}{0}{LO} & = & e\left( -\frac{1}{2}h^{\mu\nu}\partial_\mu\phi_{(0)}\partial_\nu\phi_{(0)}-V(\varphi_{(0)})\right)\,,\label{eq:KG_LO_action}\\
\oss{\mathcal{L}}{2}{NLO} & = & e\left(-h^{\mu\nu}\partial_\mu\phi_{(0)}\partial_\nu\phi_{(2)}-\frac{1}{2}\hat\Phi h^{\mu\nu}\partial_\mu\phi_{(0)}\partial_\nu\phi_{(0)}+\frac{1}{2}\left(\hat v^\mu\partial_\mu\phi_{(0)}\right)^2\right.\nonumber\\
&&\left.+\frac{1}{2}\left(h^{\mu\rho}h^{\nu\sigma}\bar\Phi_{\rho\sigma}-\frac{1}{2}h^{\mu\nu}h^{\rho\sigma}\bar\Phi_{\rho\sigma}\right)\partial_\mu\phi_{(0)}\partial_\nu\phi_{(0)}-V^\prime(\varphi_{(0)})\varphi_{(2)}\right)\,.\label{eq:KG_NLO_action}
\end{eqnarray}

In a Kaluza--Klein reduction it would be more natural to give the scalar field Lagrangian the same prefactor as the EH Lagrangian. In that case we have to  multiply \eqref{eq:real_scalar_KG_lagrangian} by $\frac{c^4}{16\pi G_N}$. When we do this the LO scalar field Lagrangian couples to Galilean gravity, i.e. the NLO Lagrangian in the expansion of the EH Lagrangian.

\subsection{Electrodynamics}\label{sec:electrodynamics}
Non-relativistic versions of electrodynamics have been investigated in various formulations in the literature \cite{LeBellac, Santos:2004pq, DeMontigny:2005oib, Duval:2014uoa, Bagchi:2014ysa, Bleeken:2015ykr}.
In this section we will consider the $1/c^2$ expansion of Maxwellian electrodynamics.

Consider the Maxwell Lagrangian
\begin{equation}\label{eq:Maxwell_lagrangian}
\mathcal{L}_{\text{Max}}=-\frac{1}{4c}\sqrt{-g}g^{\mu\rho}g^{\nu\sigma}F_{\mu\nu}F_{\rho\sigma}
\end{equation}
where $F_{\mu\nu}=\partial_\mu A_\nu-\partial_\nu A_\mu$.
We will expand $A_\mu$  as
\begin{equation}
A_\mu=c^2 A^{(-2)}_\mu+A^{(0)}_\mu+\order{c^{-2}}\,, 
\end{equation}
following our general expression \eqref{eq:expansion_field_general}. 
The expansion of the transformation of $A_\mu$ under diffeomorphisms and gauge transformations (expanded as $\Sigma =c^2\sigma^{(-2)}+ \sigma^{(0)}+\order{c^{-2}}$), yields the transformations of the LO and NLO vector fields 
\begin{eqnarray}
\delta A^{(-2)}_\mu &=&\mathcal{L}_\xi A_\mu^{(-2)} + \partial_\mu \sigma^{(-2)}\,,\\
\delta A^{(0)}_\mu &=&\mathcal{L}_\xi A_\mu^{(0)} + \mathcal{L}_\zeta A_\mu^{(-2)} + \partial_\mu \sigma^{(0)}\,.
\end{eqnarray}
The Maxwell Lagrangian thus starts at order $\order{c^4}$, so that
\begin{equation}
\mathcal{L}_{\text{Max}} = c^4\; \oss{\mathcal{L}}{-4}{\text{Max},\,LO} + c^2\; \oss{\mathcal{L}}{-2}{\text{Max},\,NLO} +\order{c^0}\,,
\end{equation}
 with the LO and NLO Lagrangians 
\begin{eqnarray}
\oss{\mathcal{L}}{-4}{\text{Max},\,LO} & = & -\frac{e}{4}h^{\mu\rho}h^{\nu\sigma}F^{(-2)}_{\mu\nu}F^{(-2)}_{\rho\sigma}\,,\\
\oss{\mathcal{L}}{-2}{\text{Max},\,NLO} & = & e\left[ -\frac{1}{2}h^{\mu\rho}h^{\nu\sigma}F^{(-2)}_{\mu\nu}F^{(0)}_{\rho\sigma}-\frac{1}{4}\left(\hat\Phi+\frac{1}{2}h^{\alpha\beta}\bar\Phi_{\alpha\beta}\right)h^{\mu\rho}h^{\nu\sigma}F^{(-2)}_{\mu\nu}F^{(-2)}_{\rho\sigma}\right.\nonumber\\
&&\left.+\frac{1}{2}h^{\mu\rho}\left(\hat v^\nu\hat v^\sigma+h^{\nu\alpha}h^{\sigma\beta}\bar\Phi_{\alpha\beta}\right)F^{(-2)}_{\mu\nu}F^{(-2)}_{\rho\sigma}\right]\,.\label{eq:ED_NLO_action}
\end{eqnarray}
The $A^{(-2)}_\mu$ equations of motion at LO and NLO are
\begin{eqnarray}
0 & = & \partial_\mu\left(e h^{\mu\rho}h^{\nu\sigma}F^{(-2)}_{\rho\sigma}\right)\,,\label{eq:ED_EOM_1}\\
0 & = & \partial_\mu\Bigg(e\Bigg[h^{\mu\rho}h^{\nu\sigma}F^{(0)}_{\rho\sigma}+\left(\hat\Phi+\frac{1}{2}h^{\alpha\beta}\bar\Phi_{\alpha\beta}\right)h^{\mu\rho}h^{\nu\sigma}F^{(-2)}_{\rho\sigma}\nonumber\\
&&-2h^{\rho[\mu}\left(\hat v^{\nu]}\hat v^\sigma+h^{\nu]\alpha}h^{\sigma\beta}\bar\Phi_{\alpha\beta}\right)F^{(-2)}_{\rho\sigma}\Bigg]\Bigg)\,.\label{eq:ED_EOM_2}
\end{eqnarray}
As usual, the $A^{(0)}_\mu$ equation of motion of the NLO Lagrangian is the same as the $A^{(-2)}_\mu$ equation of motion of the LO Lagrangian.

The NLO action \eqref{eq:ED_NLO_action} is similar but not the same as the covariantised version of Galilean electrodynamics (GED) presented in \cite{Festuccia:2016caf}.
We here have additional couplings to $\hat \Phi,\,\bar \Phi_{\mu\nu}$, that were not present in this previous work. The reason is that GED is naturally formulated on a (type I) Newton--Cartan background and obtained through a different non-relativistic limit:
To obtain GED one takes a strict $c\rightarrow\infty$ limit where a different scaling of the temporal and spatial components of the gauge field is allowed in addition to an extra coupling to a real scalar field. We shall see later that it is also possible to obtain GED via a $1/c^2$ expansion.

\subsubsection{Magnetic theory}
It is useful to decompose the leading order Maxwell field 
\begin{eqnarray}
A^{(-2)}_\mu & = & -v^\rho A^{(-2)}_\rho\tau_\mu+A^{(-2)}_\rho e^\rho_a e^a_\mu \equiv  -\tilde\varphi\tau_\mu+\check A_\mu\,, 
\end{eqnarray}
separating the temporal and spatial components since $v^\mu \check A_\mu=0$ . 
It follows that its field strength $\check F_{\rho\sigma}$ is basically the magnetic field strength tensor.
On the other hand the temporal component $\tilde\varphi$ is closely related to the electric potential.
The transformation of $\tilde\varphi$ and $\check A_\mu$ under diffeomorphisms and gauge transformations are:
\begin{eqnarray}
\delta \tilde\varphi &=&\mathcal{L}_\xi\tilde\varphi + v^\lambda\partial_\lambda \sigma^{(-2)}\,,\\
\delta \check A_\mu &=&\mathcal{L}_\xi \check A_\mu + e_\mu^a e_a^\lambda\partial_\lambda \sigma^{(-2)}\,.
\end{eqnarray}
They both transform under the $U(1)$ gauge transformation, but with temporal and spatial derivatives, respectively.

We find that the LO and NLO Lagrangians can be written in these variables as
\begin{eqnarray}
\oss{\mathcal{L}}{-4}{mag,\,LO} & = & -\frac{e}{4}h^{\mu\rho}h^{\nu\sigma}\check F_{\mu\nu}\check F_{\rho\sigma}\,,\\
\oss{\mathcal{L}}{-2}{mag,\,NLO} & = & e\left[-\frac{1}{2}h^{\mu\rho}h^{\nu\sigma}\check F_{\mu\nu} F^{(0)}_{\rho\sigma}-\frac{1}{4}\left(\hat\Phi+\frac{1}{2}h^{\alpha\beta}\bar\Phi_{\alpha\beta}\right)h^{\mu\rho}h^{\nu\sigma}\check F_{\mu\nu}\check F_{\rho\sigma}\right.\nonumber\\
&&+\frac{1}{2}h^{\mu\rho}h^{\nu\alpha}h^{\sigma\beta}\bar\Phi_{\alpha\beta}\check F_{\mu\nu}\check F_{\rho\sigma}+\left(\partial_\mu+a_\mu\right)\tilde\varphi \left(\partial_\rho+a_\rho\right)\tilde\varphi\nonumber\\
&&\left.
+\frac{1}{2}h^{\mu\rho}\left(\hat v^\nu\hat v^\sigma \check F_{\mu\nu}\check F_{\rho\sigma}
+2\hat v^\nu \check F_{\mu\nu} \left(\partial_\rho+a_\rho\right)\tilde\varphi\right)\right]\,.
\end{eqnarray}
As the $\order{c^{4}}$ term is non-zero, spacetime torsion will be sourced in the gravity EOMs \eqref{eq:EOMNRG1}-\eqref{eq:EOMNRG4}.
The LO and NLO equations of motion \eqref{eq:ED_EOM_1}-\eqref{eq:ED_EOM_2} when TTNC is imposed become
\begin{eqnarray}
0 & = & \partial_\mu\left(e h^{\mu\rho}h^{\nu\sigma}\check F_{\rho\sigma}\right)\,,\label{eq:ED_EOM_var_1}\\
0 & = & \partial_\mu\left(e\left[h^{\mu\rho}h^{\nu\sigma}F^{(0)}_{\rho\sigma}+\left(\hat\Phi+\frac{1}{2}h^{\alpha\beta}\bar\Phi_{\alpha\beta}\right)h^{\mu\rho}h^{\nu\sigma}\check F_{\rho\sigma}-2h^{\rho[\mu}h^{\nu]\alpha}h^{\sigma\beta}\bar\Phi_{\alpha\beta}\check F_{\rho\sigma}\right.\right.\nonumber\\
&&\left.\left.-2h^{\rho[\mu}\hat v^{\nu]}\left(\hat v^\sigma \check F_{\rho\sigma}+\left(\partial_\rho+a_\rho\right)\tilde\varphi\right)\right]\right)\,.\label{eq:ED_EOM_var_2}
\end{eqnarray}
Since the spatial component of the gauge field dominates the expansion we refer to this as the magnetic limit. The equation of motion for $\tilde\varphi$ is given by the $\tau_\nu$ projection of the latter, which gives
\begin{equation}\label{eq:ED_EOM_var_2_tau_proj}
\left(\partial_\mu-a_\mu\right)\left[e h^{\mu\rho}\left(\hat v^\sigma \check F_{\rho\sigma}+\left(\partial_\rho+a_\rho\right)\tilde\varphi\right)\right]=0\,.
\end{equation}

When $\tau \wedge \d\tau=0$ equations \eqref{eq:ED_EOM_var_1} and \eqref{eq:ED_EOM_var_2_tau_proj}
are the same as in the magnetic limit of Maxwell's equations coupled to TTNC geometry studied in \cite{Festuccia:2016caf}.
In addition we still have the equation from the spatial projection of the equation of motion \eqref{eq:ED_EOM_var_2},
which is not present in that work.
However, this is the only equation to involve the subleading gauge field $A^{(0)}_\mu$,
so we do not lose compatibility with the magnetic limit.
The reason for the extra equation is that the magnetic limit studied in \cite{Festuccia:2016caf} is not the same kind of non-relativistic expansion as done here.
In the previous work we scaled the temporal and spatial components of the gauge field differently and took $c\rightarrow\infty$ as a strict limit (on shell), which projects out the extra equation that appears here.

\subsubsection{Electric theory}
Let us now see how the electric limit of Maxwell's equations studied in  \cite{Festuccia:2016caf} fits into the non-relativistic expansions studied here.

First we impose off shell that
\begin{equation}
A^{(-2)}_\mu=-\varphi\tau_\mu\,.
\end{equation}
and let all gauge transformations start at order $\order{1}$ so that $\varphi$ is a scalar transforming as $\delta \varphi =\mathcal{L}_\xi\varphi$.
Using off shell TTNC and $\check F_{\mu\nu}=0$ we find that the Maxwell Lagrangian \eqref{eq:Maxwell_lagrangian} now starts at order $\order{c^2}$ and expands as
\begin{equation}
\mathcal{L}_\mathrm{elec} =  c^2\;\oss{\mathcal{L}}{-2}{elec,\,LO}  + \oss{\mathcal{L}}{0}{elec,\,NLO} +\order{c^{-2}}\,,
\end{equation}
where
\begin{eqnarray}
\oss{\mathcal{L}}{-2}{elec,\,LO} & = & \frac{e}{2}h^{\mu\nu}\left(\partial_\mu+a_\mu\right)\varphi\left(\partial_\nu+a_\nu\right)\varphi\,,\\
\oss{\mathcal{L}}{0}{elec,\,NLO} & = & e\left[-\frac{1}{2}\hat\Phi h^{\mu\nu}\left(\partial_\mu+a_\mu\right)\varphi\left(\partial_\nu+a_\nu\right)\varphi+ \frac{1}{4}h^{\alpha\beta}\bar\Phi_{\alpha\beta} h^{\mu\nu}\left(\partial_\mu+a_\mu\right)\varphi\left(\partial_\nu+a_\nu\right)\varphi\right.\nonumber\\
&&\left. -\frac{1}{2}h^{\mu\alpha}h^{\nu\beta}\bar\Phi_{\alpha\beta}\left(\partial_\mu+a_\mu\right)\varphi\left(\partial_\nu+a_\nu\right)\varphi-\frac{1}{4}h^{\mu\rho}h^{\nu\sigma}F^{(0)}_{\mu\nu}F^{(0)}_{\rho\sigma}\right.\nonumber\\
&&\left.-h^{\mu\rho}\hat v^\nu F^{(0)}_{\nu\rho}\left(\partial_\mu+a_\mu\right)\varphi \right]\,.\label{eq:ED_electric_Lagrangian_NLO}
\end{eqnarray}
The LO equation of motion is
\begin{equation}
\left(\partial_\mu-a_\mu\right)\left[e h^{\mu\rho}\left(\partial_\rho+a_\rho\right)\varphi\right]=0\,.
\end{equation}
The $A^{(0)}_\mu$ equation of motion is
\begin{equation}\label{eq:electric}
\partial_\rho\left(e\left[h^{\mu\rho}h^{\nu\sigma}F^{(0)}_{\mu\nu}-\left(h^{\mu\sigma}\hat v^\rho-h^{\mu\rho}\hat v^\sigma\right)\left(\partial_\mu+a_\mu\right)\varphi\right]\right)=0\,.
\end{equation}
Contracting \eqref{eq:electric} with $\tau_\sigma$ reproduces the LO equation of motion which is thus contained as a NLO equation of motion as expected.
For $\d\tau=0$ equation \eqref{eq:electric} agrees with the electric limit of Maxwell's equations coupled to NC geometry as studied in \cite{Festuccia:2016caf}.
In addition we have the variation of the NLO action wrt. $\varphi$ which gives
\begin{equation}
0 =  \left(\partial_\mu-a_\mu\right)\left[e\left(h^{\mu\nu}\hat\Phi-\frac{1}{2}h^{\mu\nu}h^{\alpha\beta}\bar\Phi_{\alpha\beta}
+h^{\mu\alpha}h^{\nu\beta}\bar\Phi_{\alpha\beta}\right) \left(\partial_\nu+a_\nu\right)\varphi
+eh^{\mu\rho}\hat v^\nu F^{(0)}_{\nu\rho}\right] \,.
\end{equation}
This latter equation does not appear in the on shell strict $c\rightarrow\infty$ limit of the electric limit of Maxwell's equations. 
Similar to the previous section the origin of this extra equation can be traced back to the fact that here we are performing an expansion as opposed to taking a limit as in \cite{Festuccia:2016caf}.

\subsubsection{Galilean electrodynamics}
We would like to obtain Galilean electrodynamics (GED) from a $1/c^2$ expansion of a relativistic theory.
To do this we couple relativistic electrodynamics with gauge field $A_\mu$ to a massless real free scalar $\Psi$ and study a particular expansion.
Consider then the decomposition
\begin{eqnarray}
A_\mu & = & -c^2 \varphi\tau_\mu+\check A_\mu- \tilde\varphi\tau_\mu+\chi \tau_\mu+\order{c^{-2}}\,,\\
\Psi & = & -c \varphi+c^{-1} \chi+\order{c^{-3}}\,.
\end{eqnarray}
We split the $\order{1}$ component of $A_\mu$ along $\tau_\mu$ into two scalars: $\tilde \varphi$ that transforms under gauge transformations and $\chi$ that does not transform under gauge transformations. The expansion of the gauge parameter starts at order $\order{1}$.
The resulting transformations of the fields are 
\begin{eqnarray}
\delta \varphi &=&\mathcal{L}_\xi\varphi\,,\\
\delta \tilde\varphi &=&\mathcal{L}_\xi\tilde\varphi - \varphi v^\lambda \mathcal{L}_\zeta \tau_\lambda + v^\lambda\partial_\lambda \sigma\,,\\
\delta \chi &=& \mathcal{L}_\xi\chi - \mathcal{L}_\zeta \varphi\,,\\
\delta \check A_\mu &=&\mathcal{L}_\xi \check A_\mu + \mathcal{L}_\zeta \check A_\mu - \varphi e_\mu^a e^\lambda_a \mathcal{L}_\zeta \tau_\lambda + e_\mu^a e_a^\lambda\partial_\lambda \sigma\,.
\end{eqnarray}
The sum of the relativistic Lagrangians \eqref{eq:real_scalar_KG_lagrangian} (with $\varphi$ replaced by $\Psi$ and no potential term) and \eqref{eq:Maxwell_lagrangian} starts at order $\order{c^2}$ and yields:
\begin{eqnarray}
\hspace{-1.5cm}\oss{\mathcal{L}}{-2}{LO} & = & eh^{\mu\nu}a_\mu\varphi\left(\partial_\nu +\frac{1}{2}a_\nu\right)\varphi\,,\\
\hspace{-1.5cm}\oss{\mathcal{L}}{0}{NLO} & = & e\left[-\frac{1}{4}h^{\mu\rho}h^{\nu\sigma}\bar F_{\mu\nu}\bar F_{\rho\sigma}+h^{\mu\rho}\hat v^\nu \bar F_{\mu\nu} \left(\partial_\rho+a_\rho\right)\varphi+\frac{1}{2}\left(\hat v^\mu \partial_\mu \varphi\right)^2\right.\nonumber\\
\hspace{-1.5cm}&&\left.-\frac{1}{2}\hat \Phi h^{\mu\nu}\left[\left(\partial_\mu+a_\mu\right)\varphi \left(\partial_\nu+a_\nu\right)\varphi+\partial_\mu \varphi\partial_\nu\varphi\right]-\varphi \chi h^{\mu\nu}\bar \nabla_\mu a_\nu \right.\nonumber\\
\hspace{-1.5cm}&&\left.+\frac{1}{2}\left(h^{\mu\alpha}h^{\nu\beta}\bar\Phi_{\alpha\beta}-\frac{1}{2}h^{\mu\nu}h^{\alpha\beta}\bar\Phi_{\alpha\beta}\right) \left[\partial_\mu\varphi \partial_\nu\varphi-\left(\partial_\mu+a_\mu\right)\varphi \left(\partial_\nu+a_\nu\right)\varphi\right]\right]\,,
\end{eqnarray}
where we defined
\begin{equation}
\bar F_{\mu\nu} \equiv \check F_{\mu\nu} - 2\partial_{[\mu}\left(\tilde \varphi \tau_{\nu]}\right)\,,
\end{equation}
and with TTNC off shell we have $h^{\mu\rho}h^{\nu\sigma}\bar F_{\mu\nu}=h^{\mu\rho}h^{\nu\sigma}\check F_{\mu\nu}$.
We see that in the leading order Lagrangian the term with two derivatives cancels out, leaving just a single spatial derivative and torsion vector terms.

The leading order equation of motion from the $\varphi$ variation is
\begin{equation}
\varphi h^{\mu\nu}\bar \nabla_\mu a_\nu = 0\,,
\end{equation}
consistent with what one gets at NLO by varying $\chi$ as it should be.
We must have either $\varphi=0$ or $h^{\mu\nu}\bar \nabla_\mu a_\nu=e^{-1}\partial_\mu\left(eh^{\mu\nu} a_\nu\right)-h^{\mu\nu}a_\mu a_\nu=0$.
In particular $a_\mu=0\Leftrightarrow\d\tau =0$ is a solution.

When $\d\tau =0$ we obtain GED exactly as it appears in \cite{Festuccia:2016caf} with $\oss{\mathcal{L}}{0}{NLO} \equiv\mathcal{L}_{\text{GED}}$ and
\begin{equation}
e^{-1}\mathcal{L}_{\text{GED}} = -\frac{1}{4}h^{\mu\rho}h^{\nu\sigma}\bar F_{\mu\nu}\bar F_{\rho\sigma}+h^{\mu\rho}\hat v^\nu \bar F_{\mu\nu} \partial_\rho\varphi+\frac{1}{2}\left(\hat v^\mu \partial_\mu \varphi\right)^2-\hat \Phi h^{\mu\nu}\partial_\mu \varphi\partial_\nu\varphi\,.
\end{equation}
and the subleading scalar $\chi$ and the field $\bar\Phi_{\alpha\beta}$ decouple.

\section{Solutions of non-relativistic gravity}\label{sec:solutions}
In this section we will consider solutions to the \regls{nrg} theory.
Here we will explicitly demonstrate that it is much richer than just Newtonian gravity.
We see that many of the canonical \regls{gr} solutions are also exact solutions to NRG.
The discussion is initiated by looking at isometries of \regls{ttnc} spacetimes and how to do gauge fixing.
The $1/c^2$ expansion of the Schwarzschild solution is studied first and done in two different ways: a) A weak field expansion related to the \gls{pn} expansion and b) a strong field expansion that will be an exact torsionful solution of NRG.
For the latter we also study the geodesics, which turn out to be the same as in GR, albeit conceptually different.
Next the \regls{tov} fluid star is studied and we show that the TOV equation can be derived entirely in our NR framework.
We then look at cosmological solutions and show that the \regls{flrw} spacetime is also an exact solution of NRG.
We conclude this section by discussing inequivalent spacetimes that arise from different $1/c^2$ expansions of the \regls{ads} spacetime.

\subsection{Isometries and gauge fixing}\label{sec:ansaetze_solutions}
The geometric fields $\tau_\mu$, $h_{\mu\nu}$, $m_\mu$ and $\Phi_{\rho\sigma}$  transform according to studied in Section \ref{sec:expansion_metric} by \eqref{eq:trafo1}, \eqref{eq:trafo9}, \eqref{eq:trafo5} and \eqref{eq:torsional_u(1)_1}.
An isometry is a transformation which leaves the fields unchanged, that is (with $\tau\wedge \d\tau = 0$):
\begin{eqnarray}
0=\delta\tau_\mu & = & \mathcal{L}_\xi\tau_\mu\,,\\
0=\delta h_{\mu\nu} & = & \mathcal{L}_\xi h_{\mu\nu}+\tau_\mu\lambda_\nu+\tau_\nu\lambda_\mu\,,\\
0=\delta m_\mu & = & \mathcal{L}_\xi m_\mu+\partial_\mu\Lambda-a_\mu\Lambda+\tau_\mu h^{\nu\rho}a_\nu\zeta_\rho+\lambda_\mu\,,\\
0=h^{\mu\rho}h^{\nu\sigma}\delta\Phi_{\mu\nu} & = & h^{\mu\rho}h^{\nu\sigma}\left(\mathcal{L}_\xi\Phi_{\mu\nu}+2\Lambda K_{\mu\nu}+\check\nabla_\mu\zeta_\nu+\check\nabla_\nu\zeta_\mu\right)\,.
\end{eqnarray}
We can also say that these are diffeomorphisms generated by $K^\mu$ for which there exist $\lambda_\mu$, $\Lambda$ and $\zeta_\mu$ such that
\begin{eqnarray}
 \mathcal{L}_K\tau_\mu & = & 0\,,\\
\mathcal{L}_K m_\mu & = & -\partial_\mu\Lambda+a_\mu\Lambda-\tau_\mu h^{\nu\rho}a_\nu\zeta_\rho-\lambda_\mu\,,\label{eq:KE2}\\
\mathcal{L}_K h_{\mu\nu} & = & -\tau_\mu\lambda_\nu-\tau_\nu\lambda_\mu\,,\\
h^{\mu\rho}h^{\nu\sigma}\mathcal{L}_K\Phi_{\mu\nu} & = & -h^{\mu\rho}h^{\nu\sigma}\left(2\Lambda K_{\mu\nu}+\check\nabla_\mu\zeta_\nu+\check\nabla_\nu\zeta_\mu\right)\,.\label{eq:KE4}
\end{eqnarray}

By fixing diffeomorphisms and Milne boosts we can always write
\begin{eqnarray}
\tau_\mu \d x^\mu & = & N\d t\,,\label{eq:gauge1}\\
h_{\mu\nu}\d x^\mu \d x^\nu & = & \gamma_{ij}\d x^i \d x^j\,,\label{eq:gauge2}
\end{eqnarray}
where $x^\mu=(t,x^i)$. This means that
\begin{eqnarray}
v^\mu & = & -N^{-1}\delta^\mu_t\,,\\
h^{\mu\nu} & = & \gamma^{ij}\delta^\mu_i\delta^\nu_j\,.
\end{eqnarray}
Demanding that $\delta\tau_i=0$, $\delta h_{tt}=\delta h_{ti}=0$ leads to the residual gauge transformations (using \eqref{eq:trafo1} and \eqref{eq:trafo5})
\begin{eqnarray}
\partial_i\xi^t & = & 0\,,\\
\lambda_t & = & 0\,,\\
\lambda_i & = & -N^{-1}\gamma_{ij}\partial_t\xi^j\,.
\end{eqnarray}
The nonzero components transform as
\begin{eqnarray}
\delta N & = & \xi^i\partial_i N+\partial_t\left(\xi^t N\right)\,,\\
\delta \gamma_{ij} & = & \xi^k\partial_k\gamma_{ij}+\gamma_{ik}\partial_j\xi^k+\gamma_{kj}\partial_i\xi^k+\xi^t\partial_t\gamma_{ij}\,.
\end{eqnarray}
The residual gauge transformations act on $m_\mu$ as
\begin{eqnarray}
\delta m_t & = &\xi^i\partial_i m_t+m_i\partial_t\xi^i+\partial_t(\Lambda+ \xi^t m_t)+\gamma^{ij}\partial_iN(\zeta_j+\Lambda m_j)\,,\\
\delta m_i & = & \xi^t\partial_t m_i+\xi^j\partial_j m_i+m_j\partial_i\xi^j+N\partial_i(N^{-1}\Lambda)-N^{-1}\gamma_{ij}\partial_t\xi^j\,.
\end{eqnarray}

We have defined the general notion of a Killing vector and discussed a convenient gauge for the \regls{lo} fields.
In principle one could next study ans\"atze that preserve certain symmetries, but instead we will simply discuss a number of solutions to the \reglspl{eom}.

For the special case of solutions of the LO theory that are also exact solutions of GR, i.e. for which $m_\mu=0$ and $\Phi_{\mu\nu}=0$ the equations of motion \eqref{eq:EOMNRG1}-\eqref{eq:EOMNRG4}, where the left hand side is given by \eqref{eq:NRG_EOM1}-\eqref{eq:NRG_EOM4}, reduce to
\begin{eqnarray}
&& -\frac{1}{2}\left(h^{\mu\nu}K_{\mu\nu}\right)^2+\frac{1}{2}h^{\mu\rho}h^{\nu\sigma}K_{\mu\nu}K_{\rho\sigma}=8\pi  G_N\tau_\mu\mathcal{T}_\tau^\mu\,,\label{eq:NRG_EOM1_S}\\
&& \left(h^{\mu\rho}h^{\nu\sigma}-h^{\mu\nu}h^{\rho\sigma}\right)\check{\nabla}_\mu K_{\rho\sigma}
+\frac{1}{2}v^\nu h^{\mu\nu}\check R_{\mu\nu}=8\pi G_N\mathcal{T}_m^\nu
\,,\label{eq:NRG_EOM2_S}\\
&& h^{\mu\rho}h^{\nu\sigma}\left(\check R_{\mu\nu}
-\frac{1}{2}h_{\mu\nu}h^{\kappa\lambda}\check R_{\kappa\lambda}-\left(\check{\nabla}_\mu +a_\mu\right)a_\nu 
+h_{\mu\nu}h^{\kappa\lambda} \left(\check\nabla_{\kappa}+a_\kappa\right)a_\lambda\right)=8\pi G_N\mathcal{T}_\Phi^{\rho\sigma}\,,\label{eq:NRG_EOM3_S}\\
&& -\frac{1}{2}h^{\alpha\beta}\left(h^{\mu\rho}h^{\nu\sigma}-h^{\mu\nu}h^{\rho\sigma}\right)K_{\mu\nu}K_{\rho\sigma}-\check\nabla_\lambda\left(v^\lambda\left(h^{\mu\alpha}h^{\nu\beta}-h^{\alpha\beta}h^{\mu\nu}\right)K_{\mu\nu}\right)=8\pi G_N\mathcal{T}_h^{\mu\nu}P_\mu^\alpha P_\nu^\beta\,,\label{eq:NRG_EOM4_S}\nonumber
\end{eqnarray}
where $K_{\mu\nu}=\frac{1}{2N}\partial_t h_{\mu\nu}$.
We will study solutions of these matter coupled equations in the last two subsections. We refer to \cite{VandenBleeken:2019gqa} for more details and comments about the structure of the equations of motion in the gauge \eqref{eq:gauge1} and \eqref{eq:gauge2}.

\subsection{Weak gravity expansion of the Schwarzschild metric}\label{sec:weak_Schwarzschild}
One way to generate solutions to non-relativistic gravity is by considering the $1/c^2$ expansion of GR. Therefore, let us consider the Schwarzschild metric with factors of $c$ restored:
\begin{equation}\label{eq:schwarzschild_metric}
\d s^2 = -c^2 \left(1-\frac{2G_N m}{c^2 r}\right)\d t^2+\left(1-\frac{2G_Nm}{c^2 r}\right)^{-1} \d r^2 + r^2 \d\Omega_{S^2}\,.
\end{equation}

We can perform two different physically relevant expansions depending on how we treat the mass parameter as a function of $c^2$.
The first option is to take $m$ constant as we expand.
In that case, by comparing to \eqref{eq:metric_expansion2}, \eqref{eq:metric_expansion2}, we can read off the fields in the expansion of the Lorentzian metric as
\begin{eqnarray}
\tau_\mu \d x^\mu &=& \d t\,,\\
m_\mu \d x^\mu &=& -\frac{G_Nm}{r}\d t=\Phi \d t\,,\\
h_{\mu\nu}\d x^\mu \d x^\nu &=& \d r^2 + r^2\d\Omega_{S^2} \,,\\
\Phi_{\mu\nu}\d x^\mu \d x^\nu &=& \frac{2G_Nm}{r} \d r^2 = -2\Phi \d r^2\,.
\end{eqnarray}
The result is a flat torsionless Newton--Cartan spacetime with non-zero subleading fields $m_\mu$ and $\Phi_{\mu\nu}$. One can verify that this is a vacuum solution of the EOMs \eqref{eq:NRG_EOM1}-\eqref{eq:NRG_EOM4}.
The solution is expressed in terms of the Newtonian potential $\Phi\equiv-v^\mu m_\mu=-G_Nm/r$.
In this case the $1/c^2$ expansion does not terminate. The expansion of the temporal vielbein is completely captured by $\tau_\mu$ and $m_\mu$ (with $B_\mu$ and further subleading fields equal to zero), while the $1/c^2$ expansion of the spatial part of the metric does not terminate.
In fact the higher order spatial fields $\Phi^{(n)}_{\mu\nu}$ encoding the remaining post-Newtonian effects all take the simple form
\begin{equation} 
\Phi^{(2n)}_{\mu\nu}\d x^\mu \d x^\nu = \left(\frac{2G_Nm}{r}\right)^n \d r^2=(-2\Phi)^n \d r^2\,,
\end{equation}
where $n\in \mathbb{N}$ and $\Phi^{(2)}_{\mu\nu}=\Phi_{\mu\nu}$.
When all the fields in the expansion are resummed, we obtain the Lorentzian metric again.
Since the torsion is zero, the expansion is really describing weak gravitational fields.
From the study of geodesics in Section \ref{sec:case1_particle} we see that the geodesic equation simply becomes \eqref{eq:conservationparticle}.
In particular at this order in the expansion we do not see any terms that would give rise to the deflection of light etc., i.e. everything agrees with the prediction of Newtonian gravity.

\subsection{Strong gravity expansion of the Schwarzschild metric}\label{sec:strong_Schwarzschild}
The situation is quite different if we perform an expansion of the Schwarzschild metric \eqref{eq:schwarzschild_metric} where we take the mass to be of order $c^2$ so that $M=m/c^2=\mathrm{constant}$ as was done in \cite{VandenBleeken:2017rij}.
This is a strong gravity expansion of the Schwarzschild metric, i.e. one not captured by Newtonian gravity, but still described as a Newton--Cartan geometry.

This provides us with a different approximation of GR as compared to the post-Newtonian expansion.
In this case the expansion terminates at NLO and the geometry is described the by the LO fields
\begin{eqnarray}
\tau_\mu \d x^\mu &=& \sqrt{1-\frac{2G_NM}{r}}\d t\,,\label{eq:TTNCsol1}\\
m_\mu \d x^\mu &=& 0\,,\\
h_{\mu\nu}\d x^\mu \d x^\nu &=& \left(1-\frac{2 G_NM}{r}\right)^{-1}\d r^2 + r^2\d\Omega_{S^2} \,,\\
\Phi_{\mu\nu}\d x^\mu \d x^\nu &=& 0\,.\label{eq:TTNCsol4}
\end{eqnarray}
This is a torsionful Newton--Cartan spacetime which is actually a solution of the equations of motion of the NLO Lagrangian \eqref{eq:NLOLag} in the expansion of the EH Lagrangian (Galilean gravity)  as it does not involve the subleading fields.

This is a vacuum solution with torsion. In Section \ref{subsec:TOV} we will show that this can be viewed as the exterior solution of a fluid star which can be interpreted as a source for the torsion.

\subsubsection{Geodesics in static and spherically symmetric backgrounds}\label{sec:geodesics_sphericalsym_background}
Let us consider the results of Section \ref{sec:Galilean_lightcones} and apply them to the case of geodesics in the torsionful geometry \eqref{eq:TTNCsol1}-\eqref{eq:TTNCsol4}. This will lead to the results reported in \cite{Hansen:2019vqf}.

We will start with a slightly more general case than
the one in \eqref{eq:TTNCsol1}-\eqref{eq:TTNCsol4} and consider a geometry with spherical symmetry which can be written as
\begin{equation}
\tau_\mu \d x^\mu=N(r)\d t\,,\qquad h_{\mu\nu}\d x^\mu \d x^\nu=\gamma_{ij}\d x^i \d x^j=R^2(r)\d r^2+r^2\left(\d\theta^2+\sin^2\theta \d\phi^2\right)\,.
\end{equation}
The relevant equations  for geodesic motion are given in  \eqref{eq:geo1}-\eqref{eq:geo3}. The time component of \eqref{eq:geo2} is automatically satisfied because of $\tau_\mu\dot x^\mu=0$ which implies $\dot t=0$. The spatial components obey
\begin{equation}\label{eq:geo3b}
\ddot x^i+\frac{1}{2}\gamma^{il}\left(\partial_j\gamma_{kl}+\partial_k\gamma_{jl}-\partial_l\gamma_{jk}\right)\dot x^j\dot x^k=\frac{1}{2}\gamma^{ij}\partial_j N^{-2}\,.
\end{equation}
We will consider motion in the equatorial plane only, i.e. $\dot\theta=0$ and $\theta=\pi/2$. In this case \eqref{eq:geo3b} reduces to
\begin{eqnarray}
0 & = & \ddot r+\frac{1}{2}R^{-2}\partial_r R^2\dot r^2-r^{-3}R^{-2}l^2-\frac{1}{2}R^{-2}\partial_r N^{-2}\,,\label{eq:X}\\
0 & = & \ddot \phi+2r^{-1}\dot\phi\dot r=r^{-2}\frac{\d}{\d\lambda}\left(r^2\dot\phi\right)\,.
\end{eqnarray}
The latter equation can be integrated to $r^2\dot\phi=l$. This has been used in the first equation. Equation \eqref{eq:geo1} becomes
\begin{equation}
N^{-2}=C^2+R^2(r)\dot r^2+r^2\dot\phi^2\,,
\end{equation}
which can be rewritten as
\begin{equation}\label{eq:radial}
\dot r^2+C^2R^{-2}-N^{-2}R^{-2}+R^{-2}r^{-2}l^2=0\,.
\end{equation}
The $\lambda$ derivative of this equation gives \eqref{eq:X}. Conversely, integrating \eqref{eq:X} gives \eqref{eq:radial} with integration constant $C^2$. The geodesic equations have an overall scale symmetry which involves rescaling the geodesic parameter $\lambda$ and thus the angular momentum $l$ as well as $N^{-2}$ (which in $\tau$ can be compensated by rescaling $t$). This means that the value of $C^2$ is not important. The only thing that matters is whether it is zero, positive or negative. For timelike geodesics it should be positive.

We now specialise to the geometry described by \eqref{eq:TTNCsol1}-\eqref{eq:TTNCsol4} where we will call $r_s=2G_N M$ the Schwarzschild radius (treated as independent of $c$). Let us restrict attention to geodesics for which $\dot\phi\neq 0$ then after rearranging we find
\begin{equation}
    \left(\frac{\d r}{\d\phi}\right)^2=\frac{r^4}{l^2}-\left(1-\frac{r_s}{r}\right)\left(\frac{C^2}{l^2}r^4+r^2\right)\,.
\end{equation}
This is a well-known equation describing planetary motion (for $C^2>0$) in the Schwarzschild geometry including the effects of the perihelion precession. It also captures the phenomenon of light deflection (for $C^2=0$). For more discussion we refer to \cite{Hansen:2019vqf}.

\subsection{Tolman--Oppenheimer--Volkoff equation}\label{subsec:TOV}
In this section we will show that the  Tolman--Oppenheimer--Volkoff (TOV) equation for the hydrostatic equilibrium of a spherically symmetric isotropic body (fluid star) can be derived entirely within the non-relativistic gravity framework.

The solution we are after is known to be static, and hence we need $K_{\mu\nu}=0$. From the equations of motion
\eqref{eq:NRG_EOM1_S}-\eqref{eq:NRG_EOM4_S}  we infer that the sources must obey
\begin{equation}
    \tau_\mu\mathcal{T}_\tau^\mu=0\,,\qquad \mathcal{T}_m^\mu\propto v^\mu\,,\qquad\mathcal{T}_h^{\mu\nu}P_\mu^\alpha P_\nu^\beta=0\,.
\end{equation}
From the boost Ward identity \eqref{eq:GalboostWImatter} we also learn $\mathcal{T}_h^{\mu\nu}P_\mu^\alpha\tau_\nu=0$. In order not to source any subleading orders we can fulfil these conditions if we take a perfect fluid as defined in  section \ref{sec:perfect_fluid} with only $E_{(-4)}$ and $P_{(-4)}$ nonzero. This can be seen to follow from equations \eqref{eq:calTm}-\eqref{eq:calTh}  We furthermore take for the fluid velocity
\begin{equation}
    U^\mu=-v^\mu\,.
\end{equation}
so that he fluid energy-momentum tensor reads 
\begin{equation}
    T^{\mu\nu}=c^2E_{(-4)}v^\mu v^\nu+c^4P_{(-4)}h^{\mu\nu}\,.
\end{equation}
The equations of motion \eqref{eq:NRG_EOM1_S}-\eqref{eq:NRG_EOM4_S} then
reduce to 
\begin{eqnarray}
&&e^{-1}\partial_\mu\left(eh^{\mu\nu}a_\nu\right)=8\pi G_N\frac{1}{d-1}\left(dP_{(-4)}+(d-2)E_{(-4)}\right)\,,\label{eq:TOV1}\\
&&h^{\mu\rho}h^{\nu\sigma}\left(\check R_{\mu\nu}-\left(\check\nabla_\mu+a_\mu\right)a_\nu\right)=8\pi G_N h^{\rho\sigma}\frac{1}{d-1}\left(E_{(-4)}-P_{(-4)}\right)\,.\label{eq:TOV2}
\end{eqnarray}
It can be shown that the $1/c^2$ expansion found in \cite{VandenBleeken:2017rij} agrees with equations \eqref{eq:TOV1} and \eqref{eq:TOV2}. The fluid equations of motion are \eqref{eq:conservation_perfect_fluid_LO} and \eqref{eq:LOenergycons}.

Let us now turn to the most general $d=3$ static spherically symmetric ansatz for the spacetime geometry that follows from using the results of Section \ref{sec:ansaetze_solutions} and requiring the relevant isometries, summarized by
\begin{eqnarray}
\tau_{\mu} & = &N(r)\delta_{\mu}^{t}= e^{\alpha\left(r\right)}\delta_{\mu}^{t}\,,\label{eq:spherical_ansatz1}\\
v^{\mu} & = & -e^{-\alpha\left(r\right)}\delta_{t}^{\mu}\,,\\
h^{\mu\nu} & = & \mathrm{diag}\left(0,e^{-2\beta\left(r\right)},1/r^{2},1/(r^{2}\sin^{2}\theta)\right)\,,\\
h_{\mu\nu} & = & \mathrm{diag}\left(0,e^{+2\beta\left(r\right)},r^{2},r^{2}\sin^{2}\theta\right)\,.
\end{eqnarray}
where $\alpha(r)$ and $\beta(r)$ are arbitrary functions.
The same ansatz can be obtained from the $1/c^2$ expansion of the 
corresponding analysis for a Lorentzian metric $g_{\mu\nu}$ using Birkhoff's theorem.

Inserting the static spherically symmetric ansatz into \eqref{eq:TOV1} and \eqref{eq:TOV2} we find that the equations of motion take the form 
\begin{eqnarray}
&&\frac{e^{-2 \beta} \left(r \alpha''+
   r \alpha'^2-r \alpha'\beta'+2\alpha'\right)}{r}=4\pi G_N\left(3P_{(-4)}+E_{(-4)}\right)\,,\label{eq:TOV1ansatz}\\
&& \frac{e^{-2 \beta} \left(-r
   \alpha ''+2 \beta '+r\alpha ' \beta '-r\alpha
   '^2\right)}{r}
   =4\pi G_N\left(E_{(-4)}-P_{(-4)}\right)\,,\label{eq:TOV2ansatz}\\
&& \frac{e^{-2 \beta} \left(-r \alpha '+r
   \beta '+e^{2 \beta}-1\right)}{r^2}
   =4\pi G_N\left(E_{(-4)}-P_{(-4)}\right)\,.\label{eq:TOV3ansatz}
\end{eqnarray}
From these we can solve for $P_{(-4)},\,E_{(-4)}$ to find
\begin{eqnarray}
&&\frac{e^{-2 \beta}}{r} \left[ 2\beta'+r^{-1} \left(e^{2\beta}-1\right) \right]=8\pi G_N E_{(-4)}=8\pi G_N c^{-4} E\,,\label{eq:E4_TOV}\\
&& \frac{e^{-2 \beta}}{r} \left[ 2\alpha'-r^{-1} \left(e^{2\beta}-1\right) \right]=8\pi G_N P_{(-4)} = 8\pi G_N c^{-4} P\,,\label{eq:P4_TOV}
\end{eqnarray}
where we restored the full energy and pressure according \eqref{eq:perfect_fluid_sum_energy} and \eqref{eq:perfect_fluid_sum_mom} in the last equalities.

It is convenient to define a function $M$ (with dimensions of mass over velocity squared) through
\begin{equation}
E_{\left(-4\right)} = \frac{1}{4\pi r^{2}}M^{\prime}\left(r\right)\,.
\end{equation}
The solution to the first equation can be written as
\begin{equation}
e^{-2\beta}=1-\frac{8\pi G_N}{r}\int_{r_{0}}^{r}s^{2}E_{\left(-4\right)}\left(s\right)\mathrm{d}s\,.
\end{equation}
The conservation equation \eqref{eq:conservation_perfect_fluid_LO} gives us
\begin{equation}
P_{\left(-4\right)}^{\prime}=-\alpha^{\prime}\left(P_{\left(-4\right)}+E_{\left(-4\right)}\right)\label{eq:TOV4}\,.
\end{equation}
This conservation equation is exactly the same as the one that appears in the relativistic case. With this one finds that the remaining equations can be rewritten as 
\begin{equation}
P^{\prime}=-\frac{G_N}{r^{2}}\left(P+E\right)\left(M(r)+4\pi r^{3}c^{-4}P\right)\left(1-\frac{2M(r)G_N}{r}\right)^{-1}\,,
\end{equation}
after reinstating factors of $c^2$. This is exactly the relativistic TOV equation for a stellar body of mass of the order of $c^2$, i.e. for which $M(r)=\frac{m(r)}{c^2}$ is order $c^0$, with $m$ the dimensions of mass. This is the same point of view as taken in \cite{VandenBleeken:2017rij} and in section \ref{sec:strong_Schwarzschild} of this paper. 

We thus conclude that  the physical structure of stellar bodies can be described completely by non-relativistic (strong) gravity.  Its description does not require the principle of relativity.

\subsection{Cosmological solutions and Friedmann equations} \label{sec:FRW_cosmology_solutions}
We will next show that \regls{flrw} solution solves the LO equations of motion \eqref{eq:NRG_EOM1_S}-\eqref{eq:NRG_EOM4_S}. Using again the form \eqref{eq:gauge1}, \eqref{eq:gauge2} for the LO fields, we can show
that in this case we must have 
\begin{equation}
    N=1\,,\qquad \gamma_{ij}=a^2(t)\sigma_{ij}\,,
\end{equation}
where $\sigma_{ij}$ is constant in time and describes a maximally symmetric space in $d$ dimensions. We assume that the scale factor $a$ is independent of $c$. It follows that the acceleration and extrinsic curvature satisfy
\begin{equation}
    a_\mu=0\,,\qquad K_{\mu\nu}=\frac{\dot a}{a}h_{\mu\nu}\,,
\end{equation}
where the dot denotes differentiation with respect to time. The equations of motion \eqref{eq:NRG_EOM1_S}-\eqref{eq:NRG_EOM4_S} then reduce to 
\begin{eqnarray}
-\frac{1}{2}d(d-1)\left(\frac{\dot a}{a}\right)^2 & = & 8\pi G_N\tau_\mu\mathcal{T}^\mu_\tau\,,\\
\frac{1}{2}v^\nu h^{\rho\sigma}\check R_{\rho\sigma} & = & 8\pi G_N\mathcal{T}_m^\nu\,,\\
h^{\mu\rho}h^{\nu\sigma}\check R_{\mu\nu}-\frac{1}{2}h^{\rho\sigma}h^{\mu\nu}\check R_{\mu\nu} & = & 8\pi G_N\mathcal{T}_\Phi^{\rho\sigma}\,,\\
-\frac{1}{2}d(d-1)\left(\frac{\dot a}{a}\right)^2h^{\alpha\beta}-(d-1)h^{\alpha\beta}\frac{\d}{\d t}\left(\frac{\dot a}{a}\right) & = & 8\pi G_N\mathcal{T}_h^{\mu\nu}P_\mu^\alpha P_\nu^\beta\,.
\end{eqnarray}

Just like in the previous subsection we will translate this set of equations into conditions on the $1/c^2$ expansion of a perfect fluid. Using equations \eqref{eq:calTm}-\eqref{eq:calTh} it follows that we need to take a perfect fluid as defined in section \ref{sec:perfect_fluid} with only $E_{(-4)}$, $P_{(-4)}$, $E_{(-2)}$ and $P_{(-2)}$ nonzero. For the fluid velocity we will take again
\begin{equation}
    U^\mu=-v^\mu\,.
\end{equation} 
This means that the fluid energy-momentum tensor takes the form 
\begin{equation}
    T^{\mu\nu}=c^4P_{(-4)}h^{\mu\nu}+c^2E_{(-4)}v^\mu v^\nu+c^2P_{(-2)}h^{\mu\nu}+E_{(-2)}v^\mu v^\nu\,.
\end{equation}
The above equations of motion then simplify further to the set of equations
\begin{eqnarray}
\frac{1}{2}d(d-1)\left(\frac{\dot a}{a}\right)^2 & = & 8\pi G_N E_{(-2)}\,,\\
\frac{2k}{a^2} &=& 8\pi G_N \left(P_{(-4)}+E_{(-4)}\right)\,,\\
0 & = & dP_{(-4)}+(d-2)E_{(-4)} \,,\\
-(d-1)\frac{\d}{\d t}\left(\frac{\dot a}{a}\right) & = & 8\pi G_N\left(P_{(-2)}+E_{(-2)}\right)\,,
\end{eqnarray}
where $k=-1,0,1$ depending whether the spatial metric $\sigma_{ij}$ in $\gamma_{ij}=a^2(t)\sigma_{ij}$ is a maximally symmetric space  of constant negative, zero or positive curvature. The third equation follows from the absence of a source for torsion. We see that the sources for the spatial derivatives are $P_{(-4)}$ and $E_{(-4)}$ while the sources for the time derivatives are $P_{(-2)}$ and $E_{(-2)}$.

After resumming the energy $E$ and pressure $P$ according to \eqref{eq:perfect_fluid_sum_energy} and \eqref{eq:perfect_fluid_sum_mom}, one obtains the Friedmann equations for a $d+1$-dimensional cosmological spacetime.
It is straightforward to see that for $d=3$ one can put these equations in the conventional form
\begin{eqnarray}
\left(\frac{\dot a}{a}\right)^2 & = & \frac{8\pi G_N}{3c^2}E-\frac{c^2k}{a(t)^{2}}\,,\label{eq:TNC_Friedmann1}\\
\frac{\d}{\d t}\left(\frac{\dot a}{a}\right)+\left(\frac{\dot a}{a}\right)^2 & = & -\frac{4\pi G_N}{3c^2}\left(E+3P\right)\,.\label{eq:TNC_Friedmann2}
\end{eqnarray}
The cosmology one obtains from non-relativistic gravity thus agrees with the (relativistic) Friedmann equations obtained from GR. If the spatial curvature $k$ vanishes, then the spacetime can be sourced by a perfect fluid with $E_{(-4)}=P_{(-4)}=0$.

These equations were derived from an Einstein equation written as $G_{\mu\nu}=\frac{8\pi G_N}{c^4} T_{\mu\nu}$. In the presence of a cosmological constant $\Lambda$ this is also written as $G_{\mu\nu}+\Lambda g_{\mu\nu}=\frac{8\pi G_N}{c^4}T'_{\mu\nu}$. If we define $T'_{\mu\nu}=\frac{E'+P'}{c^2}U_\mu U_\nu+P'g_{\mu\nu}$ then we have the relations $P'=P+\frac{c^4}{8\pi G_N}\Lambda$ and $E'=E-\frac{c^4}{8\pi G_N}\Lambda$. In the case of de Sitter spacetime we have $a=\exp{(Ht)}$ where $H$ is the constant Hubble parameter and $k=0$. This leads to $E_{(-4)}=P_{(-4)}=0$ and $E_{(-2)}=-P_{(-2)}=\frac{3H^2}{8\pi G_N}$, so that $E=-P=\frac{3c^2H^2}{8\pi G_N}=\frac{c^4}{8\pi G_N}\Lambda$ with $\Lambda=3\frac{H^2}{c^2}$. The de Sitter radius is $c/H$. This implies that $E'=0=P'$ as it should.

The leading order fluid conservation equations are given by \eqref{eq:conservation_perfect_fluid_LO} and \eqref{eq:LOenergycons}, reflecting that the quantities $E_{(-4)},\,P_{(-4)}$ (which are nonzero for $k\neq 0$) are homogeneous and that energy is conserved.
The subleading conservation equations similarly become:
\begin{eqnarray}
0 &=& h^{\mu\nu}\partial_\nu P_{(-2)}\,,\\
0 &=&\check\nabla_\mu\left[ E_{(-2)}u^\mu u^\nu\right]+ P_{(-2)} u^\nu h^{\mu\rho} K_{\mu\rho} \qquad \Rightarrow\\
0 &=& u^\mu \partial_\mu E_{(-2)} + d\frac{\dot a}{a}\left(E_{(-2)} + P_{(-2)}\right)\,,
\end{eqnarray}
and when we rewrite these in terms of $E$ and $P$, they are equivalent to the conservation equations appearing in GR.
For different and more canonical approaches to Newtonian cosmology, see references \cite{Milgrom:1983zz, Buchert:1995fz, Ellis:1998ct, Aldrovandi:1998im, Chisari:2011iq, Benisty:2019wpm}.

\subsection{\texorpdfstring{$1/c^2$}{1/c2} expansion of \texorpdfstring{$\mathrm{AdS}$}{AdS} spacetimes}\label{sec:AdS_spacetime_1c}
As a final example of general interest  we consider the $1/c^2$ expansion of $\mathrm{AdS}_{d+1}$ and illustrate the dependence
on the coordinates that are chosen before taking the limit. 

The $\mathrm{AdS}_{d+1}$ metric in global coordinates (with factors of $c$ restored) is
\begin{equation}\label{eq:AdS_global}
\d s^{2}=-c^{2}\cosh^{2}\rho \d t^{2}+l^2\left(\d\rho^{2}+\sinh^{2}\rho \d\Omega_{d-1}^{2}\right)\,,
\end{equation}
where $l$ is the AdS radius, $\rho>0$ is dimensionless and $t$ has dimensions of time (if we keep $\hat c$) or length (if we set $\hat c=1$). In this coordinate system we can use again 
\eqref{eq:metric_expansion2}, \eqref{eq:metric_expansion2} to read off the corresponding type II NC geometry 
\begin{eqnarray}
\tau_{\mu}d x^{\mu} & = & \cosh\rho \d t\,,\\
h_{\mu\nu}d x^{\mu}d x^{\nu} & = & l^2\left(\d\rho^{2}+\sinh^{2}\rho \d\Omega_{d-1}^{2}\right)\,,\\
m_\mu & = & 0\,,\\
\Phi_{\mu\nu} & = & 0\,.
\end{eqnarray}

Obviously the $1/c^{2}$ expansion terminates immediately.
This NC spacetime is torsionful with torsion vector given by
\begin{equation}
a_{\mu}\d x^{\mu}=\tanh\rho \d\rho\,.    
\end{equation}
On the other hand because this coordinate system is static the extrinsic curvature vanishes, i.e. $K_{\mu\nu}=0$.
The same can be done starting with Poincaré coordinates, also leading to the result that the spacetime obtained from the $1/c^2$ expansion is torsionful.

However, the situation is different in FLRW coordinates. In this case the metric takes the form
\begin{equation}\label{eq:AdS_RW}
\d s^{2}=-c^{2}\d T^{2} + l^2\cos^2\left(\frac{cT}{l}\right) \d s^2_{\mathbb H^d}\,,
\end{equation}
where $\d s^2_{\mathbb H^d}$ is the metric of hyperbolic $d$-space which has $k=-1$ with $k$ defined in the previous subsection. In order to obey the $1/c^2$ expansion ansatz of the metric $l$ cannot depend on $c$ but then the argument of the scale factor $a=l\cos\left(\frac{cT}{l}\right)$ depends on $c$. The $l$ prefactor in $a$ is necessary in order that \eqref{eq:TNC_Friedmann1} gives $E=-P=\frac{c^4}{8\pi G_N}\Lambda$ with $\Lambda=-\frac{3}{l^2}$ so that $E'=0=P'$, as it should, with $E'$ and $P'$ defined at the end of the previous subsection. This is different from what we found in the de Sitter case with $\Lambda=3\frac{H^2}{c^2}$ where the de Sitter radius $c/H$ was chosen to be of order $c$. We conclude that we cannot expand \eqref{eq:AdS_RW} analytically in $1/c^2$.

Let us consider a slightly different coordinate system for AdS by defining $r=l\sinh\rho$, leading to the metric 
\begin{equation}
    \d s^2=-c^2\left(1+\frac{r^2}{l^2}\right)\d t^2+\frac{\d r^2}{1+\frac{r^2}{l^2}}+r^2\d\Omega_{d-1}^2\,.
\end{equation}
If we replace $l^2$ by $-l^2$ this gives us to the static patch of de Sitter spacetime. If we now define $l=\frac{c}{H}$ with $H$ independent of $c$ and we treat both signs of $l^2$ at the same time we find
\begin{equation}
    \d s^2=-c^2\d t^2\mp H^2r^2\d t^2+\frac{\d r^2}{1\pm\frac{H^2r^2}{c^2}}+r^2\d\Omega_{d-1}^2\,,
\end{equation}
where the upper sign is for AdS and the lower sign is for dS spaces. Expanding  this to NLO the resulting NC geometry can be read off as
\begin{equation}\label{eq:NHspaces}
    \tau=\d t\,,\qquad h_{\mu\nu}\d x^\mu \d x^\nu=\d\vec x\cdot \d\vec x\,,\qquad m=\pm \frac{1}{2}H^2 \vec x^2\d t\,,
\end{equation}
where we left out $\Phi_{\mu\nu}$ and where we transformed to Cartesian coordinates. Such a NC geometry  is known as the Newton--Hooke spacetime. In \cite{Grosvenor:2017dfs} we showed that such a spacetime can be written in the form of a non-relativistic FLRW geometry with flat spatial slices by a sequence of NC gauge transformations. For the AdS case, i.e. the upper sign in \eqref{eq:NHspaces}, one can show using the techniques of \cite{Grosvenor:2017dfs} that this can be written as
\begin{equation}
    \tau=\d t'\,,\qquad h'_{\mu\nu}\d x^\mu \d x^\nu=\cos^2(Ht)\d\vec x'\cdot \d\vec x'\,,\qquad m'=0\,,
\end{equation}
where we transformed $h_{\mu\nu}$ and $m_\mu$ using a Galilean boost and abelian gauge transformation and where we furthermore transformed the coordinates. For the dS case we similarly find 
\begin{equation}
    \tau=\d t'\,,\qquad h'_{\mu\nu}\d x^\mu \d x^\nu=e^{2Ht}\d\vec x'\cdot \d\vec x'\,,\qquad m'=0\,.
\end{equation}
These should however not be confused with the FLRW spacetimes discussed above as the latter result from a different $1/c^2$ expansion.

We conclude that, starting with the Lorentzian $\mathrm{AdS}$ spacetime 
one encounters a situation similar to the two different $1/c^2$ expansions of the Schwarzschild geometry discussed at the beginning of this section. In that case,  the difference depends on how  the mass as a function of $c^2$ is treated. In analogy, we see here that 
the expansion depends on how we treat the cosmological constant as a function of $c^2$.

\section{Discussion and outlook}\label{sec:discussion}
The main purpose of this  paper has been the development of \regls{nrg} as it appears from a large speed of light expansion of \regls{gr}. We have given a detailed introduction to the underlying geometry, which we dubbed type II \regls{nc} geometry. We have presented the gauge transformations of the fields and how they can be thought of as arising from the gauging of an algebra that in turn can be obtained from an algebra expansion of the Poincar\'e algebra. We defined the Lagrangian of NRG to be given by the \regls{nnlo} Lagrangian in the $1/c^2$ expansion of the Einstein--Hilbert Lagrangian in which we impose the \regls{ttnc} condition for a global foliation in terms of constant time slices with the help of a Lagrange multiplier. We derived this Lagrangian using two different methods: by direct $1/c^2$ expansion and by using gauge invariance under type II gauge transformations. We have subsequently discussed  the resulting equations of motion and the coupling to matter. We have furthermore described some of the main examples of matter couplings, i.e. point particles, perfect fluids, real and complex scalar fields and electrodynamics. Finally, we have presented some of the simplest solutions of non-relativistic gravity (coupled to a perfect fluid) and commented on their physical
relevance. 

\subsubsection*{Open problems and future directions}
As a first avenue of further analysis, understanding NRG from a Hamiltonian perspective would tell us more about the number of degrees of freedom. This can be achieved by the usual counting of the phase space dimension and constraints per spacetime point. The Hamiltonian perspective would furthermore provide us with natural candidates for the definition of asymptotic charges such as mass, energy, momentum, angular momentum etc. In this light it would for example be interesting to see what would happen with asymptotic symmetry groups in the non-relativistic regime.
This might help us in understanding if NRG has the potential to admit a holographic dual description.
For example in the case of the AdS/CFT correspondence one could wonder about what happens with the Brown--Henneaux analysis in 3 dimensions \cite{Brown:1986nw, Henneaux:2010xg} when we expand in $1/c^2$ or what happens to the fluid/gravity correspondence \cite{Hubeny:2011hd, Davison:2016auk} in the \regls{nr} expansion.
One could also examine how to implement the $1/c^2$ expansion in the bulk at the level of the boundary theory of known dualities.
More generally, it is interesting to speculate whether there is a relation with the entropic and emergent gravity ideas of \cite{Verlinde:2010hp,Verlinde:2016toy} which are also connected to Newtonian gravity and modifications thereof.

Another perspective on the theory would be provided by performing a detailed analysis of the linearised spectrum (for example around flat NC space).
We do not expect that the theory has propagating degrees of freedom, and hence we expect that the gravitational interactions are instantaneous as in Newtonian gravity. Nevertheless, it would be interesting to understand the structure of the propagators and how the theory would behave from a perturbative QFT point of view.
Obviously, it would be important to study further $1/c^2$ expansions of relativistic solutions in detail.
This will teach us more about the conceptual nature of non-relativistic gravity.
In particular it would be interesting to see how the $1/c^2$ expansion of the Kerr geometry fits into this framework.
This is a sufficiently general spacetime to study in order to understand if there is a notion of a non-relativistic black hole.
We can then also hopefully shine some more light on the correct interpretation of the geodesics studied in Section \ref{sec:Galilean_lightcones}.

It is clear from the analysis presented here that if one were to continue the expansion of the \regls{eh} Lagrangian beyond NNLO it would quickly become very challenging.
We expect that performing the same analysis in first order formalism should be more suited to a higher order expansion.
As we have stressed in this paper, we know the underlying symmetry principle at any given order along with the systematics of the expansion of the EH action, but one needs to develop an efficient way to extract results.
To this end we plan to pursue the analysis of the $1/c$ expansion in first order formalism in \cite{Hansen:2020a}.
In this connection see also the references \cite{Cariglia:2018hyr, Bergshoeff:2019ctr, Gomis:2019fdh}. 
A related point that needs to be addressed is the question about the status of the odd powers of $1/c$. These have been discarded in this work as a simplifying assumption, but ultimately we need to understand their physical significance.
In this light we refer to \cite{Ergen:2020yop}.

One of the possible applications of a higher order analysis would be to make contact with the \regls{pn} approximation.
In particular it would be of interest to construct a map relating non-relativistic gravity in our formalism to more conventional PN parameterisations since it goes beyond the approximation where the torsion is zero and encodes strong field effects more naturally.
Importantly, there may be relevant domains of validity in physical processes, such as the early phase of inspirals of compact objects, where non-relativistic gravity can either give new results or alternative methods to check known results.
Examining the two-body problem in non-relativistic gravity will thus be important as well.
In another direction, it would be worthwhile to obtain the action and equations of motion at higher order in the $1/c^2$ expansion. 

It would of course also be important to examine how the non-relativistic action is related to string theory.
The current state-of-the-art includes non-relativistic strings that probe type I NC geometry, as well as the closely related \regls{snc} geometry.
It would thus be very interesting to uncover how strings couple to the type II NC geometry and in particular whether the beta-functions of this putative theory reproduce our NRG theory.
More generally, it would be interesting to see how branes couple to type II TNC geometry differently from type I TNC geometry as studied in \cite{Pereniguez:2019eoq, Romano:2019ulw}.

Finally there are of course various other open issues one could consider, for example the coupling of a non-relativistic spinning particle to type II TNC, fermionic matter actions and adding spacetime supersymmetry\footnote{Supersymmetric actions for other types of non-relativistic gravity theories, e.g. with Bargmann symmetry, have been considered in \cite{Andringa:2013mma, Bergshoeff:2015ija, Bergshoeff:2015uaa, Bergshoeff:2016lwr, deAzcarraga:2019mdn}.}.
With the richness of non-relativistic physics demonstrated so far there are certainly still numerous other interesting studies to be done.

\section*{Acknowledgements}
We thank Eric Bergshoeff,  Jan de Boer, Shira Chapman, Jos\'e Figueroa-O'Farrill, Troels Harmark, Emil Have, Donal O'Connell, Gerben Oling, Bernd Schroers, Haopeng Yan,  Dieter Van den Bleeken, Stefan Vandoren, Erik Verlinde and Manus Visser  for useful discussions.
The work of DH is supported by the Swiss National Science Foundation through the NCCR SwissMAP.
The work of JH is supported by the Royal Society University Research Fellowship ``Non-Lorentzian Geometry in Holography'' (grant number UF160197).
The work of NO is supported in part by the project ``Towards a deeper understanding of  black holes with non-relativistic holography'' of the Independent Research Fund Denmark (grant number DFF-6108-00340) and the Villum Foundation Experiment project 00023086.

\appendix

\section{Notation and conventions}\label{sec:notation_conventions}

The number of spatial dimensions is denoted by $d$. For indices we use the following:
\begin{itemize}
\item $a,b,\ldots$ are spatial (tangent space) indices, $a=1,\ldots,d$.
\item $A,B,\ldots$ (beginning of alphabet) are Lorentzian (tangent space) indices, $A=0,1,\ldots,d$.
\item $\mu,\nu,\ldots$ are coordinate indices, $\mu=0,\ldots,d$.
\item $M,N,\ldots$ (middle of alphabet) are coordinate indices of Lorentzian metrics used in null reductions, $N=0,1,\ldots,d,d+1$.
\end{itemize}
A superscript of the type $\os{X}{n}$ indicates the order of some coefficient of a Laurent/Taylor expansion in $1/c$ for some object $X\left(c\right)$. There is one exception to this rule. When expanding a field $\phi$ whose $1/c^2$ expansion starts at order $c^0$ we write instead
\begin{equation}
    \phi=\phi_{(0)}+c^{-2}\phi_{(2)}+c^{-4}\phi_{(4)}+O(c^{-6})\,.
\end{equation}

\subsection{Acronyms}
\printglossary[type=\acronymtype]

\subsection{Curvature}\label{subapp:curv}
For any torsionful connection $\Gamma^\rho_{\mu\nu}$ with covariant derivative $\nabla_\mu$ the Riemann tensor ${R}_{\mu\nu\sigma}{}^{\rho}$  and torsion tensor $T^\rho{}_{\mu\nu}$ are universally defined through
\begin{eqnarray}
\left[\nabla_\mu,\nabla_\nu\right]X_\sigma & = & R_{\mu\nu\sigma}{}^\rho X_\rho-T^\rho{}_{\mu\nu}\nabla_\rho X_\sigma\,,\\
\left[\nabla_\mu,\nabla_\nu\right]X^\rho & = & -R_{\mu\nu\sigma}{}^\rho X^\sigma-T^\sigma{}_{\mu\nu}\nabla_\sigma X^\rho\,,
\end{eqnarray}
so explicitly
\begin{eqnarray}
{R}_{\mu\nu\sigma}{}^{\rho}&\equiv&-\partial_{\mu}{\Gamma}_{\nu\sigma}^{\rho}+\partial_{\nu}{\Gamma}_{\mu\sigma}^{\rho}-{\Gamma}_{\mu\lambda}^{\rho}{\Gamma}_{\nu\sigma}^{\lambda}+{\Gamma}_{\nu\lambda}^{\rho}{\Gamma}_{\mu\sigma}^{\lambda}\label{eq:Riemann_tensor}\,,\\
T^\rho{}_{\mu\nu} &\equiv& 2\Gamma^\rho_{[\mu\nu]}\,.\label{eq:torsion_tensor}
\end{eqnarray}
The Bianchi identities are
\begin{eqnarray}
    R_{[\mu\nu\sigma]}{}^\rho & = & T^\lambda{}_{[\mu\nu}T^\rho{}_{\sigma]\lambda}-\nabla_{[\mu}T^\rho{}_{\nu\sigma]}\,,\\
    \nabla_{[\lambda}R_{\mu\nu]\sigma}{}^\kappa & = & T^\rho{}_{[\lambda\mu}R_{\nu]\rho\sigma}{}^\kappa\,.
\end{eqnarray}

The Ricci tensor is also universally defined as
\begin{equation}\label{eq:Ricci_tensor_LC}
R_{\mu\nu} \equiv {R}_{\mu\rho\nu}{}^{\rho}\,.
\end{equation}
We will always work with a connection such that 
\begin{equation}
    \Gamma^{\rho}_{\mu\rho}=\partial_\mu\log M\,,
\end{equation}
where $M$ is the measure. This implies that
\begin{equation}
R_{\mu\nu\rho}{}^\rho=0\,,
\end{equation}
and hence that the antisymmetric part of the Ricci tensor is 
\begin{equation}
    2R_{[\mu\nu]}=-2T^\lambda{}_{\rho[\mu}T^\rho{}_{\nu]\lambda}+T^\lambda{}_{\mu\nu}T^\rho{}_{\lambda\rho}+
    \nabla_\mu T^\rho{}_{\nu\rho}-\nabla_\nu T^\rho_{\mu\rho}+\nabla_\rho T^\rho{}_{\mu\nu}\,.
\end{equation}

In this paper we use three different choices of affine connections. The formulae of this appendix apply to all of these three choices.

\subsection{Comparison of notations}\label{sec:conventions_change}
\begin{table}[ht]
\begin{center}
\begin{tabular}{  l  l  l  p{5cm} }
\toprule
& This work & Hansen et. al. 2018 \cite{Hansen:2018ofj} & Van den Bleeken 2017 \cite{VandenBleeken:2017rij} \\ \toprule
Clock 1-form & $\tau_\mu$ & $\tau_\mu$ & $\tau_\mu$ \\ 
& $v^\mu$ & $v^\mu$ & $-\tau^\mu$ \\
& $\hat v^\mu$ & $\hat v^\mu$ & $-\hat \tau^\mu$ \\ \midrule
Spatial metrics & $h^{\mu\nu}$ & $h^{\mu\nu}$ & $h^{\mu\nu}$ \\ 
 & $h_{\mu\nu}$ & $h_{\mu\nu}$ & $h_{\mu\nu}$ \\ 
 & $\bar h_{\mu\nu}$ & $\bar h_{\mu\nu}$ & $\hat h_{\mu\nu}-2\hat \Phi \tau_\mu\tau_\nu$ \\ \midrule
Subleading fields& $m_\mu$ & $m_\mu$ & $ -C_\mu$ \\ 
& $B_\mu$ & - & $-B_\mu$ \\ 
& $\Phi_{\mu\nu}$ & - & $-h_{\mu\rho}h_{\nu\sigma}\beta^{\rho\sigma}$ \\ 
& $\bar\Phi_{\mu\nu}$ & $2\tau_{(\mu}\hat B_{\nu)}-\bar h_{\mu\rho}\bar h_{\nu\sigma} \hat \beta^{\rho\sigma}$ & - \\ 
& $Y^{\mu\nu}$ & - & $2\hat \tau^{(\mu}h^{\nu)\rho}\hat B_\rho + \hat \gamma^{\mu\nu}$ \\ \midrule
Torsion vectors& $\hat a_\mu$ & $a_\mu$ & $ \hat a_\mu$ \\ 
& $a_\mu$ & - & -  \\ \midrule
Extrinsic curvatures& $K_{\mu\nu}$ & - & - \\ 
& $\bar K_{\mu\nu}$ & $\bar K_{\mu\nu}$ & $\hat K_{\mu\nu}-2\hat\Phi\tau_{(\mu}\hat a_{\nu)}-\tau_\mu\tau_\nu \hat\tau^\rho \partial_\rho \hat\Phi$ \\ 
&$\hat \Phi$ &$\tilde \Phi$& $\hat \Phi$ \\ \midrule
Connections & $\bar \Gamma^\rho_{\mu\nu}$ & $\bar \Gamma^\rho_{\mu\nu}$ & $\os{\Gamma}{\mathrm{nc}}^\rho_{\mu\nu}$ \\ 
& $\check \Gamma^\rho_{\mu\nu}$ & - & -  \\ \bottomrule
\end{tabular}
\caption{Comparison of notation used in three different papers including the present one. A `-' denotes that the corresponding object has not been defined in the corresponding paper.}\label{table:notation}
\end{center}
\end{table}

The notation in this paper have been streamlined and differs from some of the choices in previous works.
To make comparison easier we present in this appendix Table \ref{table:notation} with notations used in two other papers \cite{Hansen:2018ofj,VandenBleeken:2017rij}.

\section{Review of Torsional Newton--Cartan geometry}\label{sec:NC review}
\reGls{tnc} geometry has been reviewed extensively in the literature.
We repeat here the most fundamental aspects, see also references \cite{Duval:1984cj, Andringa:2010it, Christensen:2013lma, Christensen:2013rfa, Geracie:2015dea, Hartong:2015wxa, Hartong:2015zia}.

TNC geometry is characterised by three tensors $\tau_\mu,\,h_{\mu\nu}, m_\mu$ with $h_{\mu\nu}$ symmetric and of signature $(0,1,\ldots,1)$,  subject to the following gauge redundancies
\begin{eqnarray}
\delta\tau_\mu & = & \mathcal{L}_\xi\tau_\mu\,,\\
\delta h_{\mu\nu} & = & \mathcal{L}_\xi h_{\mu\nu}+\tau_\mu\lambda_\nu+\tau_\nu\lambda_\mu\,,\\
\delta m_\mu & = & \mathcal{L}_\xi m_\mu+\partial_\mu\sigma+\lambda_\mu\,,\label{eq:typeItrafom}
\end{eqnarray}
where $\lambda_\mu$ obeys $v^\mu\lambda_\mu=0$ with $v^\mu$ defined as follows.
The inverse of $-\tau_\mu\tau_\nu+h_{\mu\nu}$ is given by $-v^\mu v^\nu+h^{\mu\nu}$ with $v^\mu\tau_\mu=-1$, $v^\mu h_{\mu\nu}=0$, $\tau_\mu h^{\mu\nu}=0$ and $h_{\mu\rho}h^{\rho\nu}=\delta^\nu_\mu+v^\nu\tau_\mu$. The inverse objects transform as
\begin{eqnarray}
\delta v^\mu & = & \mathcal{L}_\xi v^\mu+h^{\mu\nu}\lambda_\nu\,,\\
\delta h^{\mu\nu} & = & \mathcal{L}_\xi h^{\mu\nu}\,.
\end{eqnarray}
The parameter $\lambda_\mu$ corresponds to local Galilean (or Milne) boosts and the parameter $\sigma$ to Abelian gauge transformations associated with particle number conservation.

TNC geometries only admit degenerate metric structuresgiven by $\tau_\mu\tau_\nu$ and $h^{\mu\nu}$ respectively. For example, lower indices can   no longer raise be raised at will because contravariant and covariant tensors of the same rank can not be mapped to each other in a one-to-one way.
The non-uniqueness in $v^\mu,\,h_{\mu\nu}$ can be interpreted as the ambiguity due to frames related by local Galliean boost transformations (also sometimes called Milne boosts in the literature).

In addition to $\tau_\mu$ and $h^{\mu \nu}$, one  can define 
the following Galilean boost-invariant spacetime tensors 
\begin{equation}
\hat{v}^\mu\equiv v^\mu-h^{\mu\nu}m_\nu\,\quad\,,\quad\bar{h}_{\mu\nu}\equiv h_{\mu\nu}-2\tau_{(\mu}m_{\nu)}\,,\qquad \hat{\Phi} \equiv -v^{\mu}m_{\mu}+\frac{1}{2}h^{\mu\nu}m_{\mu}m_{\nu}\,.
\end{equation}
These form a convenient set of variables to describe TNC geometry and they satisfy the completeness relations
\begin{equation}
\hat{v}^{\mu}\tau_{\mu} = -1\,\,\,\,\,,\,\,\,\,\,\hat{v}^{\mu}\tau_{\nu} = -\delta_{\nu}^{\mu}+h^{\mu\lambda}\bar{h}_{\lambda\nu}\,.
\end{equation}
It should be noted that $\hat v^\mu$, $\bar h_{\mu\nu}$ and $\hat\Phi$ are not invariant under the Abelian gauge transformation with parameter $\sigma$.

One can also define an affine connection $\Gamma_{\mu\nu}^{\lambda}$ so that we may take covariant derivatives. It is natural to require the TNC equivalent of metric compatibility $\nabla_\rho \tau_\mu=0,\,\nabla_\rho h^{\mu\nu}=0$. The first of these conditions implies that any metric compatible connection must have the same temporal projection of the torsion tensor $2\tau_\rho\Gamma_{[\mu\nu]}^{\rho}=2\partial_{[\mu}\tau_{\nu]}$.
Thus constraints on torsion imply restrictions on the geometric data in contradistinction the case of Riemannian geometry.

We distinguish between three possible classes of Newton--Cartan geometry:
\begin{enumerate}
    \item (Torsionless) \regls{nc} geometry, $\d\tau=0$.
    \item \reGls{ttnc} geometry, $\tau\wedge\d\tau=0$.
    \item \reGls{tnc} geometry, $\tau\wedge\d\tau\neq0$.
\end{enumerate}
The full TNC case is acausal as has been argued in \cite{Geracie:2015dea}, but is still interesting in applications to field theory and holography because the energy current is the response to varying an unconstrained $\tau_\mu$. On the other hand in the torsionless Newton--Cartan geometry there is a notion of absolute time as $\oint \tau =0$ implies that all observers agree on the time interval between events.
For most purposes we will restrict ourselves to the twistless torsional Newton--Cartan geometry which defines a spacetime foliation whose normal 1-form is $\tau$.

In \cite{Festuccia:2016awg,Festuccia:2016caf} it was shown that any boost invariant TNC connection may be written as
\begin{equation}
\Gamma_{\mu\nu}^{\lambda}=\bar{\Gamma}_{\mu\nu}^{\lambda}+C_{\mu\nu}^{\lambda}\,,
\end{equation}
where we define a distinguished TNC connection as
\begin{equation}
\bar{\Gamma}_{\mu\nu}^{\lambda}\equiv-\hat{v}^{\lambda}\partial_{\mu}\tau_{\nu}+\frac{1}{2}h^{\lambda\sigma}\left(\partial_{\mu}\bar{h}_{\nu\sigma}+\partial_{\nu}\bar{h}_{\mu\sigma}-\partial_{\sigma}\bar{h}_{\mu\nu}\right)\label{eq:Special TNC connection}
\end{equation}
and $C_{\mu\nu}^{\lambda}$ is a spacetime tensor; a TNC analogue of the ``contortion'' tensor. For $\bar{\Gamma}_{\mu\nu}^{\lambda}$ the torsion tensor is given by $\bar{T}_{\mu\nu}^{\lambda}=-2\hat{v}^{\lambda}\partial_{[\mu}\tau_{\nu]}$. The connection $\bar\Gamma^\rho_{\mu\nu}$ is manifestly boost invariant while, unless $\d\tau=0$, it  is in general not invariant under the Abelian gauge transformation with parameter $\sigma$. 

We will also make use of the non-boost invariant connection
\begin{equation}
    \check\Gamma^\rho_{\mu\nu} \equiv -v^\rho\partial_\mu\tau_\nu+\frac{1}{2}h^{\rho\sigma}\left(\partial_\mu h_{\nu\sigma}+\partial_\nu h_{\mu\sigma}-\partial_\sigma h_{\mu\nu}\right)\,,
\end{equation}
which has the nice property that it does not contain $m_\mu$ and is therefore invariant under the $\sigma$ gauge transformation. It has non-zero torsion given by $\check{T}_{\mu\nu}^{\lambda}=-2{v}^{\lambda}\partial_{[\mu}\tau_{\nu]}$.

\section{Null reduction of Einstein gravity}\label{sec:null-reduction}
\subsection{General properties of null reductions}\label{subsec:null-redGR}
It is well-known that what we refer in this paper to as type I NC geometry can be obtained from null reduction of a Lorentzian metric with a null isometry, see for example \cite{Julia:1994bs,Hartong:2015xda}.
We will denote the null Killing vector by $\frac{\partial}{\partial u}$. If the $d+1$ dimensional NC spacetime has coordinates $x^\mu$ then the null uplifted Lorentzian geometry has coordinates $x^M=(u, x^\mu)$. 

Any Lorentzian metric with a null isometry and its inverse can be written as
\begin{equation}
g_{MN}	=	\left(\begin{array}{cc}
g_{\mu\nu} & g_{\mu u}\\
g_{u\nu} & g_{uu}
\end{array}\right)=\left(\begin{array}{cc}
\bar{h}_{\mu\nu} & \tau_{\mu}\\
\tau_{\nu} & 0
\end{array}\right)\,,
\end{equation}
\begin{equation}
g^{MN}	=	\left(\begin{array}{cc}
g^{\mu\nu} & g^{\mu u}\\
g^{u\nu} & g^{uu}
\end{array}\right)=\left(\begin{array}{cc}
h^{\mu\nu} & -\hat{v}^{\mu}\\
-\hat{v}^{\nu} & 2\hat{\Phi}
\end{array}\right)\,,
\end{equation}
where $g_{uu}=0$ due to the existence of the null Killing vector $\partial_u$.
The null reduction of the components of the $(d+2)$-dimensional Levi--Civita connection  $\hat{\Gamma}_{MN}^L$ (without any constraints on $\tau_\mu$) is
\begin{eqnarray}
\hat{\Gamma}_{\mu\nu}^{\rho}	&=&	-\hat{v}^{\rho}\partial_{(\mu}\tau_{\nu)}+\frac{1}{2}h^{\rho\sigma}\left(\partial_{\mu}\bar{h}_{\nu\sigma}+\partial_{\nu}\bar{h}_{\mu\sigma}-\partial_{\sigma}\bar{h}_{\mu\nu}\right)=\bar\Gamma^\lambda_{(\mu\nu)}\,,\label{eq:LC_connection_nullred_1}\\
\hat{\Gamma}_{\mu\nu}^{u}	&=&	-\bar{K}_{\mu\nu}-2\tau_{(\mu}\partial_{\nu)}\hat{\Phi}\,,\\
\hat{\Gamma}_{\mu u}^{\rho}=\hat{\Gamma}_{u\mu}^{\rho}	&=&	\frac{1}{2}h^{\rho\sigma}\left(\partial_{\mu}\tau_{\sigma}-\partial_{\sigma}\tau_{\mu}\right)=\frac{1}{2}h^{\rho\sigma}\tau_{\mu\sigma}\,,\\
\hat{\Gamma}_{\mu u}^{u}=\hat{\Gamma}_{u\mu}^{u}	&=&	\frac{1}{2}\hat{v}^{\sigma}\left(\partial_{\sigma}\tau_{\mu}-\partial_{\mu}\tau_{\sigma}\right)=\frac{1}{2}\hat a_\mu\,,\\
\hat{\Gamma}_{uu}^{\rho} = \hat{\Gamma}_{uu}^{u}	&=&	0\,.\label{eq:LC_connection_nullred_5}
\end{eqnarray}
We denote the higher-dimensional Levi--Civita connection and associated curvatures with a hat. Note that the null-reduced Levi--Civita connection is equal to the symmetric part of the boost invariant NC connection \eqref{eq:Special TNC connection}. The definitions of extrinsic curvatures $\bar K_{\mu\nu},\,\hat a_\mu$ can be found in Table \ref{table:defsfields}.

A very useful object is the null reduction of the Ricci tensor, which has the following components
\begin{eqnarray}
\hat R_{\mu\nu}&=&\bar{R}_{\mu\nu}-\bar{\nabla}_{\rho}\bar{\Gamma}_{[\mu\nu]}^{\rho}-\bar{\nabla}_{\mu}\hat a_{\nu}-\frac{1}{2}\hat a_{\mu}\hat a_{\nu}+h^{\rho\sigma}\left(\bar{K}_{\rho(\mu|}+\tau_{(\mu|}\partial_{\rho}\hat{\Phi}\right)\tau_{|\nu)\sigma}\,,\label{eq:nullredmunu_fullTNC}\\
\hat R_{\mu u} & = &-\frac{1}{2}\tau_\mu e^{-1}\partial_\nu \left(eh^{\nu\rho}\hat a_\rho\right)
\,,\label{eq:hatRmuu_fullTNC}\\
\hat R_{uu} & = & \frac{1}{4}h^{\mu\rho}h^{\nu\sigma}\tau_{\mu\nu}\tau_{\rho\sigma}\label{eq:hatRuu_fullTNC}\,,
\end{eqnarray}
where $\bar{R}_{\mu\nu}$ is the Ricci tensor corresponding to the
connection \eqref{eq:Special TNC connection}.

The extremely useful property of these expressions is that they transform nicely under the Bargmann $U(1)$ gauge transformation of type I NC geometry. These transformations are easy to derive using the fact that the $U(1)$ corresponds to the following higher dimensional diffeomorphism
\begin{equation}
u'=u+\sigma(x)\,,\qquad x'^\mu=x^\mu\,,
\end{equation}
with
\begin{equation}\label{eq:U1_trafo_null_red_m}
m'_\mu=m_\mu+\partial_\mu\sigma\,.
\end{equation}
Infinitesimally this reads $\delta u=-\xi^u$ where $\xi^u$ is the $u$-component of $\xi^M$, the generator of 
$(d+2)$-dimensional diffeomorphisms. Using the tensorial transformation rule of $\hat R_{MN}$ under a diffeomorphism generated by $\xi^M=-\sigma\delta^M_u$ one shows that
\begin{eqnarray}
\delta \hat R_{\mu\nu} & = & -\hat R_{\mu u}\partial_\nu\sigma-\hat R_{\nu u}\partial_\mu\sigma\,,\qquad\delta\hat R_{\mu u}=-\hat R_{uu}\partial_\mu\sigma\,,\qquad\delta\hat R_{uu}=0\,,\nonumber\\
\delta\hat R^{\mu\nu} & = & 0\,,\qquad\delta \hat R^{\mu u}=\hat R^{\mu\nu}\partial_\nu\sigma\,,\qquad\delta\hat R^{uu}=2\hat R^{\mu u}\partial_\mu\sigma\,.\label{eq:U1_trafo_null_red_Ricci}
\end{eqnarray}
These transformation rules are fully general and thus true for any TNC geometry.

From the higher-dimensional Bianchi identities for the Einstein tensor, i.e. $\hat\nabla_M\hat G^{MN}=0$, we can derive two very important results
\begin{eqnarray}
e^{-1}\partial_\mu\left(e\hat G^\mu{}_u\right) & = & 0\,,\label{eq:nullredBI1}\\
e^{-1}\partial_\mu\left(e\hat G^{\mu u}\right) & = & -\hat a_\mu\hat G^{\mu u}+\bar K_{\mu\nu}\hat G^{\mu\nu}+2\hat G^\mu{}_u\partial_\mu\hat\Phi\,.\label{eq:nullredBI2}
\end{eqnarray}
They are true for any TNC geometry. In fact, since these are identities, they are true regardless of which $U(1)$ transformation we assign the fields to have! To derive these results one needs to use the null reduction formulae for the higher dimensional Christoffel connection \eqref{eq:LC_connection_nullred_1}-\eqref{eq:LC_connection_nullred_5}.

The first of these Bianchi identities, \eqref{eq:nullredBI1}, is the geometrical counterpart of Bargmann mass conservation. This follows by using the $(d+2)$-dimensional Einstein equation and recognising $\hat T^\mu{}_u$ as the mass current of the lower-dimensional theory, see e.g. appendix A of \cite{Hartong:2016nyx}. The second identity \eqref{eq:nullredBI2} is the geometrical counterpart of energy conservation. The difference between the two is just a raising or lowering of the $u$ index. Furthermore, using an argument similar to the one leading up to \eqref{eq:U1_trafo_null_red_Ricci}, it can be shown that $\hat G^\mu{}_u$ is Bargmann $U(1)$ invariant for any TNC geometry whereas $\hat G^{\mu u}$ is not, not even for TTNC geometry.

If we specialise to TTNC geometries the null reduction of the Ricci tensor simplifies to
\begin{eqnarray}
\hat R_{\mu\nu}&=&\bar R_{(\mu\nu)}-\frac{1}{2}\hat a_\mu \hat a_\nu-\bar\nabla_{(\mu}\hat a_{\nu)}+\tau_\mu\tau_\nu\hat\Phi h^{\rho\sigma}\hat a_\rho \hat a_\sigma+\hat a_\sigma\tau_{(\mu}\bar\nabla_{\nu)}\hat v^\sigma\,,\label{eq:nullredmunu}\\
\hat R_{\mu u} & = & -\frac{1}{2}\tau_\mu e^{-1}\partial_\nu \left(eh^{\nu\rho}\hat a_\rho\right)\,,\label{eq:hatRmuu}\\
\hat R_{uu} & = & 0\,.
\end{eqnarray}
The higher-dimensional Ricci scalar $\hat R$ for TTNC geometries is given by
\begin{equation}\label{eq:hatRicci}
\hat R=h^{\mu\nu}\bar R_{\mu\nu}-2e^{-1}\partial_\mu\left(eh^{\mu\nu}\hat a_\nu\right)+\frac{1}{2}h^{\mu\nu}\hat a_\mu \hat a_\nu\,.
\end{equation}
Furthermore, for TTNC geometries we have that $\hat G^\mu{}_u=0$ and $\hat G^{uu}$ is given by
\begin{equation}\label{eq:Guu}
\hat G^{uu}=\hat v^\mu\hat v^\nu\bar R_{\mu\nu}-\hat\Phi h^{\mu\nu}\bar R_{\mu\nu}+\frac{1}{2}\hat\Phi h^{\mu\nu}\hat a_\mu \hat a_\nu\,.
\end{equation}
Using standard manipulations with commutators and the definition of the Riemann tensor it can be shown that (up to a total derivative)
\begin{equation}\label{eq:kinetic}
\hat v^\mu\hat v^\nu\bar R_{\mu\nu}=\left(h^{\mu\nu}\bar K_{\mu\nu}\right)^2-h^{\mu\rho}h^{\nu\sigma}\bar K_{\mu\nu}\bar K_{\rho\sigma}\,.
\end{equation}
We will thus consider $\hat G^{uu}$ as a nicely transforming completion of the kinetic term.

\subsection{Newtonian gravity vs. null reduced general relativity}\label{sec:newtonian_gravity}
In this section we review how to obtain standard Newtonian gravity in the torsionless Newton--Cartan framework. We will also compare this to the null reduction of \regls{gr}.

When $\d\tau=0$ we can write the connection \eqref{eq:Special TNC connection} as 
\begin{equation}
\bar\Gamma^\rho_{\mu\nu}=\check\Gamma^\rho_{\mu\nu}+K^\rho_{\mu\nu}\,,
\end{equation}
where 
\begin{eqnarray}
\check\Gamma^\rho_{\mu\nu} & = & -v^\rho\partial_\mu\tau_\nu+\frac{1}{2}h^{\rho\sigma}\left(\partial_\mu h_{\nu\sigma}+\partial_\nu h_{\mu\sigma}-\partial_\sigma h_{\mu\nu}\right)\,,\\
K^\rho_{\mu\nu} & = & -\frac{1}{2}h^{\rho\sigma}\left(\tau_\mu F_{\nu\sigma}+\tau_\nu F_{\mu\sigma}\right)\,,
\end{eqnarray}
where $F_{\mu\nu}=\partial_\mu m_\nu-\partial_\nu m_\mu$ is the Bargmann $U(1)$ curvature.
The connection $\check\Gamma^\rho_{\mu\nu}$ only depends on the $U(1)$ invariant fields $\tau_\mu$ and $h_{\mu\nu}$.
This connection is not Galilean boost invariant. Note that the dependence of $\bar\Gamma^\rho_{\mu\nu}$ on $m_\mu$ is via the addition of a tensor $K^\rho_{\mu\nu}$. The sourceless NC equations of motion $\bar R_{\mu\nu}=0$ can then be written as
\begin{eqnarray}
h^{\mu\rho}h^{\nu\sigma}\check R_{\rho\sigma} & = & 0\,,\label{eq:Ricciflat}\\
h^{\mu\rho}v^{\sigma}\check R_{\rho\sigma} & = & -\frac{1}{2}e^{-1}\partial_\lambda\left(eh^{\mu\rho}h^{\lambda\sigma}F_{\rho\sigma}\right)\,,\\
v^{\rho}v^{\sigma}\check R_{\rho\sigma} & = & -e^{-1}\partial_\rho\left(e v^\nu h^{\rho\sigma}F_{\nu\sigma}\right)-\frac{1}{4}h^{\mu\nu}h^{\rho\sigma}F_{\mu\rho}F_{\nu\sigma}\,.
\end{eqnarray}
In NC gravity this should be supplemented with the condition $\d\tau=0$. 
The left hand side is pure geometric data and the right hand side depends entirely on the ``electric" and ``magnetic" field strength components of $F_{\mu\nu}$. The divergence of the electric field strength in the third equation, i.e. $e^{-1}\partial_\rho\left(e v^\nu h^{\rho\sigma}F_{\nu\sigma}\right)$, is what gives rise to Newton's law of gravity when appropriately sourced by a mass density.

The null reduction of the Einstein equations of motion, in the absence of sources, also leads to $\bar R_{\mu\nu}=0$. It does not lead to $\d\tau=0$ on the nose. However one can make the argument that the sourceless null reduced Einstein equations of motion force $\d\tau=0$.
This happens in two steps. First of all the equation of motion \eqref{eq:hatRuu_fullTNC}
leads to the TTNC condition and secondly the EOM $\hat v^\mu\hat R_{\mu u}=0$ implies that 
\begin{equation}
\partial_\mu\left(e h^{\mu\nu}a_\nu\right)=0\,.
\end{equation}
Since furthermore \eqref{eq:spatialcurva} holds, which means that $h^{\mu\nu}a_\nu=N^{-1}h^{\mu\nu}\partial_\nu N$ for some function $N$ defined via $\tau=N \d T$, this equation is a Laplacian acting on $N$. This follows from the fact that $e=N\sqrt{\gamma}$ where $\gamma$ is the metric on the $T=\text{constant}$ slices. These spatial slices are described by $d$-dimensional Riemannian geometry. In particular they are Ricci flat as follows from the equation $\bar R_{\mu\nu}=0$. In the absence of sources the clock 1-form $\tau$, which defines the foliation of the spacetime, must be everywhere regular.
That means that $N$ is a bounded harmonic function on a Ricci flat $d$-dimensional geometry.
Such functions must be constant and so $a_\mu=0$ implying that $\d\tau=0$.

Hence for $\d\tau=0$ we observe a Bargmann invariance of the sourceless equations of NC geometry. It is therefore tempting to suggest that the sourceful generalisation should also obey Bargmann invariance. The sourceful NC equations that correspond to Newtonian gravity are given by\footnote{We ignore here the second term on the right hand side of \eqref{eq:mattercoupledNCgravity} and focus only on the coupling to mass.}
\begin{equation}\label{eq:Newton}
\bar R_{\mu\nu} = 8\pi G_N \frac{d-2}{d-1}\rho\tau_\mu\tau_\nu\,.
\end{equation}
We will now argue that this equation is not compatible with a Bargmann invariant coupling of NC geometry to matter.

The mass current $J^\mu$ in a Bargmann invariant theory is $U(1)$ invariant and conserved. The only candidate geometrical quantity that obeys the same properties is the $\hat G^\mu{}_u$ component 
 of the Einstein tensor on a background with a null isometry
 (see \eqref{eq:nullredBI1}). Hence the coupling must be
\begin{equation}
\hat G^\mu{}_u\propto J^\mu\,.
\end{equation}
From a null reduction point of view we have of course that $J^\mu=\hat T^\mu{}_u$, where $\hat T_{MN}$ is the null uplifted energy-momentum tensor. This equation implies that upon contraction with $\tau_\mu$ we obtain
\begin{equation}
\hat R_{uu}\propto\hat T_{uu}=\rho\,.
\end{equation}
From the form of $\hat R_{uu}$ in \eqref{eq:hatRuu_fullTNC} we thus see that mass sources $\tau\wedge\d\tau\neq 0$. This is in direct conflict with Newtonian gravity described in the previous section because in that theory the notion of mass is compatible with $\d\tau=0$.
\emph{Hence $\rho$ in \eqref{eq:Newton} is not a Bargmann mass density}.
We thus conclude that Newtonian gravity cannot originate from a Bargmann invariant theory\footnote{Indeed, as shown in \cite{Hansen:2018ofj} and elaborated on in section \ref{sec:Poincare_algebra_expansion} the underlying algebra that follows from the $1/c^2$ expansion of the Poincar\'e algebra is different.}.

We can make this a bit more explicit by performing a null reduction of the Einstein--Hilbert Lagrangian, which up to total derivatives yields
\begin{equation}\label{eq:nullredEH}
\mathcal{L}=\frac{e}{16\pi G_N}\left(h^{\mu\nu}\bar R_{\mu\nu}+\frac{1}{2}h^{\mu\nu}\hat a_\mu \hat a_\nu-\frac{1}{2}\hat\Phi h^{\mu\rho}h^{\nu\sigma}\tau_{\mu\nu}\tau_{\rho\sigma}\right)\,.
\end{equation}
This is not a consistent reduction but the inconsistency is extremely mild in that all of its equations of motion agree with the null reduction of Einstein's equation. It only fails to reproduce the $\hat G^{uu}\propto\hat T^{uu}$ equation of motion. The reason for this is simply that the null reduction sets $\hat g_{uu}=0$ off shell and so we cannot vary this component. Furthermore, the $\hat G^{uu}\propto\hat T^{uu}$ equation of motion does not impose any constraints on any of the other equations of motion that follow from the null reduced action. The reason is that for any given $\hat\Phi$, the equations of motion $\hat G^{\mu\nu}\propto\hat T^{\mu\nu}$ together with $\hat G^{\mu u}\propto\hat T^{\mu u}$ form a closed set. The remaining equation $\hat G^{uu}\propto\hat T^{uu}$, instead of imposing a constraint merely completes the other equations of motion by suppling an equation of motion from which $\hat\Phi$ can be determined. 

Finally we remark that the $\hat T^{uu}$ component of the null reduced energy-momentum is not a new independent source but a composite object that is formed from various other sources. All of the above implies that we can use the null reduced \regls{eh} action to study the coupling between geometry and matter for Bargmann invariant theories. On shell one then simply complements the equations of motion with $\hat G^{uu}\propto\hat T^{uu}$ to provide an equation for $\hat\Phi$. The point is the when adding matter in a Bargmann invariant manner the field $\hat\Phi$ couples to the mass density. It then follows that the $\hat\Phi$ equation of motion leads to the same conclusion as before, namely that Bargmann mass density sources TNC torsion for which $\tau\wedge\d\tau \neq 0$. We conclude that in order to couple matter to NC gravity we cannot use type I NC geometry. In particular, in Section \ref{sec:coupling_to_matter} we show that type II NC geometry will lead to a consistent coupling between gravity and matter.

\section{Torsional Newton--Cartan identities}\label{sec:NC_identities}
In this appendix we collect useful identities that hold for type I and type II NC geometries. Throughout this appendix we will not impose any restrictions on the clock 1-form $\tau$.

\subsection{Summary of definitions\label{subsec:Completeness-relations-and}}
\begin{table}[ht]
\begin{center}
\begin{tabular}{  l   l  l }
\toprule
Boost invariant metrics & $\hat v^\mu\equiv v^\mu-h^{\mu\nu}m_\nu$ \\ 
& $\bar{h}_{\mu\nu} \equiv  h_{\mu\nu}-2\tau_{(\mu}m_{\nu)}$\\ 
& $\hat{\Phi}  \equiv -v^{\mu}m_{\mu}+\frac{1}{2}h^{\mu\nu}m_{\mu}m_{\nu}$\\ 
& $\bar \Phi_{\mu\nu}  \equiv \Phi_{\mu\nu}-m_\mu m_\nu-2B_{(\mu}\tau_{\nu)}$\\ \midrule
Torsion vectors & $a_\mu \equiv \mathcal{L}_v \tau_\mu$ \\ 
& $\hat a_\mu \equiv \mathcal{L}_{\hat v} \tau_\mu$ \\ \midrule
Extrinsic curvatures & $K_{\mu\nu} \equiv -\frac{1}{2}\mathcal{L}_v h_{\mu\nu}$ \\ 
&$\bar K_{\mu\nu} \equiv -\frac{1}{2}\mathcal{L}_{\hat v} \bar h_{\mu\nu}$ \\ \midrule
Other curvatures & $\tau_{\mu\nu} \equiv \partial_\mu \tau_\nu - \partial_\nu\tau_\mu$ \\ 
& $F_{\mu\nu} \equiv \partial_\mu m_\nu - \partial_\nu m_\mu - a_\mu m_\nu + a_\nu m_\mu$ \\ \midrule
Connections & $\check\Gamma^\rho_{\mu\nu} \equiv -v^\rho\partial_\mu\tau_\nu+\frac{1}{2}h^{\rho\sigma}\left(\partial_\mu h_{\nu\sigma}+\partial_\nu h_{\mu\sigma}-\partial_\sigma h_{\mu\nu}\right)$ \\ 
& $\bar\Gamma^\rho_{\mu\nu} \equiv -\hat v^\rho\partial_\mu\tau_\nu+\frac{1}{2}h^{\rho\sigma}\left(\partial_\mu \bar h_{\nu\sigma}+\partial_\nu \bar h_{\mu\sigma}-\partial_\sigma \bar h_{\mu\nu}\right)$ \\
\bottomrule
\end{tabular}
\caption{Definitions of fields and derived objects such as torsion vectors, extrinsic curvatures and affine connections used in the main text.}\label{table:defsfields}
\end{center}
\end{table}
We present in in Table \ref{table:defsfields} a summary of the definitions of fields used throughout the paper.

\subsection{Identities for covariant derivatives, Riemann and Ricci tensors}\label{sec:Riemann_tensor}
The torsion tensor of $\check{\Gamma}_{\mu\nu}^{\lambda}$ is given by 
\begin{equation}
{\check{T}^{\lambda}}_{\mu\nu}\equiv 2\check{\Gamma}_{[\mu\nu]}^{\lambda} =-v^{\lambda}\tau_{\mu\nu}\,,
\end{equation}
and we may calculate the covariant derivatives
\begin{eqnarray}
\check{\nabla}_\rho \tau_\mu &=& 0\,,\label{eq:Cov_Der_check_metrics_1}\\
\check{\nabla}_\rho h^{\mu\nu} &=& 0\,,\\
\check{\nabla}_\rho v^\mu &=& -h^{\mu\nu}K_{\rho\nu}\,,\\
\check{\nabla}_\rho h_{\mu\nu} &=& -\tau_\mu K_{\nu\rho} -\tau_\nu K_{\mu\rho}\label{eq:Cov_Der_check_metrics_4}\,.
\end{eqnarray}
Likewise the torsion tensor of the boost invariant connection $\bar{\Gamma}_{\mu\nu}^{\lambda}$ is given by 
\begin{equation}
{\bar{T}^{\lambda}}_{\mu\nu}\equiv 2\bar{\Gamma}_{[\mu\nu]}^{\lambda} =-\hat{v}^{\lambda}\tau_{\mu\nu}\,,
\end{equation}
and we can derive the following covariant derivatives
\begin{eqnarray}
\bar{\nabla}_\rho \tau_\mu &=& 0\,,\\
\bar{\nabla}_\rho h^{\mu\nu} &=& 0\,,\\
\bar{\nabla}_{\rho}\hat{v}^{\mu} & = & -h^{\mu\lambda}\left(\bar{K}_{\rho\lambda}+\tau_{\rho}\partial_{\lambda}\hat{\Phi}-\hat{\Phi}\tau_{\rho\lambda}\right)\label{eq:extrinsic_cov_vhat}\\
\bar{\nabla}_{\rho}\bar{h}_{\mu\nu} & = & 2\hat{\Phi}\tau_{\rho(\mu}\tau_{\nu)}-2\tau_{\mu}\tau_{\nu}\partial_{\rho}\hat{\Phi}-2\tau_{\rho}\tau_{(\mu}\partial_{\nu)}\hat{\Phi}-2\tau_{(\mu}\bar{K}_{\nu)\rho}\,.\label{eq:extrinsic_cov_hbar}
\end{eqnarray}
From the above we may derive the following useful contractions and projections:
\begin{eqnarray}
\hat{v}^{\sigma}\bar{\nabla}_{\sigma}\hat{v}^{\rho} & = & +2\hat{\Phi}h^{\rho\lambda}\hat a_{\lambda}+h^{\rho\lambda}\partial_{\lambda}\hat{\Phi}\,,\\
h^{\kappa\sigma}\bar{\nabla}_{\sigma}\hat{v}^{\rho} & = & -h^{\kappa\sigma}h^{\rho\lambda}\left(\bar{K}_{\sigma\lambda}-\hat{\Phi}\tau_{\sigma\lambda}\right)\,,\\
h^{\lambda\nu}\bar{\nabla}_{\rho}\bar{h}_{\mu\nu} & = & \tau_{\mu}\bar{\nabla}_{\rho}\hat{v}^{\lambda}\\
\hat{v}^{\rho}\bar{\nabla}_{\rho}\bar{h}_{\mu\nu} & = & 2\tau_{(\mu}\left(\partial_{\nu)}\hat{\Phi}+2\hat{\Phi}\hat a_{\nu)}\right)\\
\hat{v}^{\nu}\bar{\nabla}_{\rho}\bar{h}_{\mu\nu} & = & 2\tau_{\mu}\partial_{\rho}\hat{\Phi}+\tau_{\rho}\partial_{\mu}\hat{\Phi}+\bar{K}_{\rho\mu}-\hat{\Phi}\tau_{\rho\mu}\\
\hat{v}^{\rho}\hat{v}^{\nu}\bar{\nabla}_{\rho}\bar{h}_{\mu\nu} & = & \tau_{\mu}\hat{v}^{\rho}\partial_{\rho}\hat{\Phi}-\partial_{\mu}\hat{\Phi}-2\hat{\Phi}\hat a_{\mu}\\
\hat{v}^{\rho}\hat{v}^{\mu}\hat{v}^{\nu}\bar{\nabla}_{\rho}\bar{h}_{\mu\nu} & = & -2\hat{v}^{\rho}\partial_{\rho}\hat{\Phi}
\end{eqnarray}
The Riemann tensor $\bar R_{\mu\nu\sigma}{}^\rho$ as defined in \eqref{eq:Riemann_tensor} and Ricci tensor $\bar R_{\mu\nu} = \bar R_{\mu\rho \nu}{}^\rho$ enjoy a number of useful properties and identities:
\begin{eqnarray}
\bar R_{\mu\nu\rho}{}^\rho &=&0\,,\\
\bar R_{[\mu\nu\sigma]}{}^{\rho}&=&-\nabla_{[\mu}{\bar T}_{\nu\sigma]}^{\rho}+{\bar T}_{[\mu\nu}^{\lambda}{\bar T}_{\sigma]\lambda}^{\rho}\,,\\
\bar R_{[\mu\nu]} &=& -\frac{1}{2}\bar{\nabla}_{\rho}\bar{T}_{\mu\nu}^{\rho}-\bar{\nabla}_{[\mu}\hat a_{\nu]}\,,\\
\bar R_{\mu\nu\lambda}{}^\rho h^{\lambda\sigma} &=& -\bar R_{\mu\nu\lambda}{}^\sigma h^{\rho\lambda}\,,\\
3\bar \nabla_{[\lambda} \bar R_{\mu\nu]\sigma}{}^\kappa &=& \bar T^\rho_{\mu\nu} \bar R_{\lambda\rho\sigma}{}^\kappa+\bar T^\rho_{\lambda\mu} \bar R_{\nu\rho\sigma}{}^\kappa+\bar T^\rho_{\nu\lambda} \bar R_{\mu\rho\sigma}{}^\kappa\,, \label{eq:Bianchi2}
\end{eqnarray}
with analogous identities for $\check R_{\mu\nu\sigma}{}^\rho$.
If we now contract $\lambda$ and $\kappa$ in \eqref{eq:Bianchi2} and furthermore contract that equation with $v^\nu h^{\mu\sigma}$ we can derive the following contracted Bianchi identity
\begin{equation}
e^{-1}\partial_\rho\left(e\left[h^{\rho\nu}\hat v^\mu\bar R_{\mu\nu}-\frac{1}{2}\hat v^\rho h^{\mu\nu}\bar R_{\mu\nu}\right]\right)=-h^{\mu\rho}h^{\nu\sigma}\bar K_{\rho\sigma}\bar R_{\mu\nu}+\frac{1}{2}h^{\rho\sigma}\bar K_{\rho\sigma}h^{\mu\nu}\bar R_{\mu\nu}\,.
\end{equation}
This is a bit similar to the divergence of the Einstein tensor in GR except that here we are considering the $h^{\rho\nu}\hat v^\mu$ component of the Ricci tensor.
There are similar identities for the other components.

\subsection{Identities for extrinsic curvatures\label{subsec:Extrinsic-curvatures}}
We have the following useful contractions
\begin{eqnarray}
\hat{v}^{\mu}\hat a_{\mu} & = & 0\,,\\
\hat a_{\lambda}\bar{T}_{\,\,\,\mu\nu}^{\lambda} & = & 0\,,\\
h^{\mu\rho}\bar\nabla_\rho\hat v^\nu &=&-h^{\mu\rho}h^{\nu\sigma}\bar K_{\rho\sigma}\,.
\end{eqnarray}
Using that the Lie derivative satisfies the Leibniz rule one may derive

\begin{eqnarray}
h^{\mu\lambda}\hat a_{\lambda} & = & 2K^{\mu\lambda}\tau_{\lambda}=h^{\mu\lambda}\bar{T}_{\,\,\,\rho\lambda}^{\rho}\,,\\
\bar{K}_{\mu\nu}\hat{v}^{\nu} & = & -\hat a_{\mu}\hat{\Phi}-\tau_{\mu}\mathcal{L}_{\hat{v}}\hat{\Phi}\,,\\
\bar{K}_{\mu\nu}\hat{v}^{\mu}\hat{v}^{\nu} & = & \mathcal{L}_{\hat{v}}\hat{\Phi}\,,\\
\hat a_{\mu}\hat{v}^{\sigma} &=&\mathcal{L}_{\hat v} h^{\sigma\lambda}\bar{h}_{\lambda\mu}-2h^{\sigma\lambda}\bar{K}_{\lambda\mu}\label{eq:Lie_derivative_CR}\,.
\end{eqnarray}

\subsection{Variational calculus}\label{sec:variational_calculus_identities}
\subsubsection{Basic relations}
A complete set of type I NC data can be formed from the set of fields $h^{\mu\nu},\,\hat{v}^{\rho},\,\hat{\Phi}$. Hence when considering variations we can consider the variations of $h^{\mu\nu},\,\hat{v}^{\rho},\,\hat{\Phi}$. This set is convenient when working with manifestly boost invariant objects. Alternatively we can work with the independent set of fields $\tau_\mu$ and $\bar h_{\mu\nu}$. They are related via
\begin{eqnarray}
\delta\bar{h}_{\mu\nu} & = & -2\tau_{\mu}\tau_{\nu}\delta\hat{\Phi}+\left(\tau_{\mu}\bar{h}_{\nu\rho}+\tau_{\nu}\bar{h}_{\mu\rho}\right)\delta\hat{v}^{\rho}-\bar{h}_{\mu\rho}\bar{h}_{\nu\sigma}\delta h^{\rho\sigma}\label{eq:TNC_variations0}\\
\delta\tau_{\mu} & = & \tau_{\mu}\tau_{\nu}\delta\hat{v}^{\nu}-\bar{h}_{\mu\rho}\tau_{\nu}\delta h^{\nu\rho}\label{eq:TNC_variations1}\\
\delta e & = & -e\left(\frac{1}{2}\bar{h}_{\mu\nu}\delta h^{\mu\nu}-\tau_{\mu}\delta\hat{v}^{\mu}\right)\label{eq:TNC_variations2}\,,
\end{eqnarray}
where
\begin{equation}
\tau_{\mu}\tau_{\nu}\delta h^{\mu\nu} =0\,.
\end{equation}
Conversely we also have the relations \eqref{eq:variation_boostinv_var1}-\eqref{eq:variation_boostinv_var4} derived in the main part of the paper.

\subsubsection{Connection and torsion}\label{sec:variation_connection}
We find for the variation of $\check{\Gamma}_{\mu\nu}^{\rho} $:
\begin{eqnarray}
\delta \check \Gamma^\rho_{\mu\nu} & = & - v^\rho\check\nabla_\mu\delta \tau_\nu   +  h^{\rho\sigma}K_{\mu\nu}\delta \tau_\sigma -\frac{1}{2}v^\lambda h^{\rho\sigma}\tau_{\nu\sigma}\delta h_{\mu\lambda}-\frac{1}{2}v^\lambda h^{\rho\sigma}\tau_{\mu\sigma}\delta h_{\nu\lambda}  \nonumber\\
&& + \frac{1}{2}\tau_{\mu\nu} h^{\rho\sigma} v^\lambda \delta h_{\sigma\lambda} + \frac{1}{2} h^{\rho\sigma}\left(\check\nabla_\mu \delta h_{\nu\sigma} + \check\nabla_\nu\delta  h_{\mu\sigma} - \check\nabla_\sigma\delta h_{\mu\nu} \right)\,,\label{eqn:variation_connection_check}
\end{eqnarray}
while the variation of $\bar{\Gamma}_{\mu\nu}^{\rho} $ is given by: 
\begin{eqnarray}
\delta \bar \Gamma^\rho_{\mu\nu} & = &{- \frac{1}{2}\tau_{\mu\nu} h^{\rho\sigma}\bar h_{\sigma\lambda} \delta \hat v^\lambda} {- \tau_\sigma(\tau_\mu \partial_\nu\hat{\Phi} + \tau_\nu\partial_\mu \hat{\Phi})\delta h^{\rho\sigma}} - \bar K_{\mu\nu}\tau_\sigma \delta h^{\rho\sigma} + 2h^{\rho\sigma} \hat a_\sigma \tau_\mu \tau_\nu\delta \hat{\Phi}\nonumber\\
&& +\hat{\Phi} h^{\rho\sigma} \hat a_\sigma (\tau_\mu\delta \tau_\nu +\tau_\nu\delta\tau_\mu) - \frac{1}{2}h^{\rho\sigma} \hat a_\sigma(\tau_\mu \bar h_{\nu\lambda} + \tau_\nu \bar h_{\mu\lambda})\delta \hat v^\lambda\nonumber\\
&& {-\hat v^\rho\nabla_\mu\bar\delta \tau_\nu + \frac{1}{2} h^{\rho\sigma}\left(\bar\nabla_\mu \delta \bar h_{\nu\sigma} + \bar\nabla_\nu\delta \bar h_{\mu\sigma} - \bar\nabla_\sigma\delta\bar h_{\mu\nu} \right)}\,.\label{eq:special TNC connection_variation}
\end{eqnarray}
Useful contractions and projections are:
\begin{eqnarray}
h^{\mu\nu}\delta\bar{\Gamma}_{\mu\nu}^{\rho} & = & +\left[+h^{\rho\sigma}\hat a_{\sigma}\tau_{\eta}+h^{\rho\sigma}\tau_{\eta\sigma}\right]\delta\hat{v}^{\eta}
 +\left[+\frac{1}{2}h^{\rho\sigma}\bar{h}_{\mu\nu}-h^{\sigma\lambda}\bar{h}_{\lambda\mu}\delta_{\nu}^{\rho}\right]\bar{\nabla}_{\sigma}\delta h^{\mu\nu}\,,\\
\delta\bar{\Gamma}_{\rho\nu}^{\rho} & = & +\left[- \tau_{\rho}\hat a_{\nu}-\tau_{\rho\nu}\right]\delta\hat{v}^{\rho}
+\left[-\hat{v}^{\rho}\right]\bar{\nabla}_{\rho}\delta\tau_{\nu}
+\left[-\frac{1}{2}\bar{h}_{\rho\sigma}\right]\bar{\nabla}_{\nu}\delta h^{\rho\sigma}\,,\\
h^{\rho\nu}\delta\bar{\Gamma}_{\mu\nu}^{\mu} & = & +\left[-h^{\rho\sigma}\tau_{\eta}\hat a_{\sigma}-h^{\rho\sigma}\tau_{\eta\sigma}\right]\delta\hat{v}^{\eta}
 +\left[-\frac{1}{2}h^{\rho\sigma}\bar{h}_{\mu\nu}+\hat{v}^{\sigma}\tau_{\nu}\delta_{\mu}^{\rho}\right]\bar{\nabla}_{\sigma}\delta h^{\mu\nu}\,.
\end{eqnarray}
From this we can derive the variation of the torsion vector:
\begin{eqnarray}
\delta \hat a_{\mu} & = & \left[\hat v^\lambda \tau_{\mu\rho}\right]\delta\left(\tau_{\lambda}\hat{v}^{\rho}\right)+\left[-2\hat{v}^{\rho}\right]\bar{\nabla}_{[\mu}\delta\tau_{\rho]}\,.\label{eq:a_var_Lie_derivative}
\end{eqnarray}

\subsubsection{Ricci tensor}\label{sec:curvature_variations}
In the conventions introduced in Section \ref{subapp:curv} the variation of the Ricci tensor is given in terms of variations of the connection as:
\begin{equation}\label{eq:deltabarR}
\delta R_{\mu\nu}=\nabla_\rho\delta\Gamma^\rho_{\mu\nu}-\nabla_\mu\delta\Gamma^\rho_{\rho\nu}-2\Gamma^\lambda_{[\mu\rho]}\delta\Gamma^\rho_{\lambda\nu}\,.
\end{equation}
The (spatial) trace is easier to calculate because of metric compatibility and gives:
\begin{equation}\label{eq:deltabarRtrace}
h^{\mu\nu}\delta R_{\mu\nu}=\nabla_\rho\left(h^{\mu\nu}\delta\Gamma^\rho_{\mu\nu}\right)-\nabla_\mu\left(h^{\mu\nu}\delta\Gamma^\rho_{\rho\nu}\right)-2\Gamma^\lambda_{[\mu\rho]}\left(h^{\mu\nu}\delta\Gamma^\rho_{\lambda\nu}\right)\,.
\end{equation}
Assuming that it does not multiply anything except the measure we can calculate $h^{\mu\nu}\delta \check R_{\mu\nu}$ as found in \eqref{eq:varRic_tau}-\eqref{eq:varRic_h}.
For $h^{\mu\nu}\delta \bar R_{\mu\nu}$ we find

\begin{eqnarray}\label{eq:varRic_bar}
h^{\mu\nu}\delta\bar{R}_{\mu\nu} & = & +\Big[h^{\rho\sigma}h^{\kappa\lambda}\tau_{\lambda\sigma}\tau_{\rho\kappa}\Big]\delta\hat{\Phi}\nonumber \\
 &  & +\Big[2h^{\mu\nu}a_{\mu}a_{\nu}\tau_{\rho}+2h^{\mu\nu}a_{\mu}\tau_{\rho\nu}-\hat{\Phi}h^{\kappa\sigma}h^{\mu\nu}\tau_{\rho}\tau_{\mu\kappa}\tau_{\nu\sigma}
 +h^{\mu\nu}\bar{\nabla}_{\mu}a_{\nu}\tau_{\rho}+h^{\mu\nu}\bar{\nabla}_{\mu}\tau_{\rho\nu}\Big]\delta\hat{v}^{\rho}\nonumber \\
 &  & +\Big[-h^{\rho\sigma}a_{\rho}a_{\sigma}\bar{h}_{\mu\nu}-h^{\rho\sigma}\bar{\nabla}_{\rho}a_{\sigma}\bar{h}_{\mu\nu}+a_{\mu}a_{\nu}+\bar{\nabla}_{\mu}a_{\nu}+\tau_{\mu}\hat{v}^{\sigma}\bar{\nabla}_{\sigma}a_{\nu}\nonumber \\
 &  & -h^{\rho\sigma}\bar{K}_{\mu\rho}\tau_{\nu\sigma}-h^{\rho\sigma}\bar{K}_{\rho\sigma}a_{\mu}\tau_{\nu}+h^{\rho\sigma}a_{\rho}\tau_{\mu}\bar{K}_{\nu\sigma}-\hat{\Phi}h^{\rho\sigma}\tau_{\mu\rho}\tau_{\nu\sigma}\nonumber \\
  &  & +3\hat{\Phi}h^{\rho\sigma}a_{\rho}\tau_{\mu}\tau_{\nu\sigma}
 +2\hat{\Phi}h^{\rho\sigma}\tau_{\mu}\tau_{\nu\sigma}
 +h^{\rho\sigma}\tau_{\mu}\tau_{\nu\rho}\partial_{\sigma}\hat{\Phi}\Big]\delta h^{\mu\nu}\,.
\end{eqnarray}

\subsubsection{Extrinsic curvatures}
We find for the variation of $\bar K_{\mu\nu}$:
\begin{eqnarray}
\delta\bar{K}_{\mu\nu} & = & -\frac{1}{2}\mathcal{L}_{\delta\hat{v}}\bar{h}_{\mu\nu}-\frac{1}{2}\mathcal{L}_{\hat{v}}\delta\bar{h}_{\mu\nu}\nonumber \\
 & = & +\left[\hat a_{\mu}\tau_{\nu}+\hat a_{\nu}\tau_{\mu}\right]\delta\hat{\Phi}+\left[\tau_{\mu}\tau_{\nu}\hat{v}^{\rho}\right]\partial_{\rho}\delta\hat{\Phi}\nonumber \\
 &  &
+\left[-\frac{1}{2}\hat a_{\mu}\bar{h}_{\nu\rho}-\frac{1}{2}\hat a_{\nu}\bar{h}_{\mu\rho}+\tau_{\mu}\bar{K}_{\nu\rho}+\tau_{\nu}\bar{K}_{\mu\rho}\right]\delta\hat{v}^{\rho}+\left[-\frac{1}{2}\tau_{\mu}\bar{h}_{\nu\rho}-\frac{1}{2}\tau_{\nu}\bar{h}_{\mu\rho}\right]\mathcal{L}_{\hat{v}}\delta\hat{v}^{\rho}\nonumber \\
 &  & 
+\left[-\bar{K}_{\mu\rho}\bar{h}_{\nu\sigma}-\bar{h}_{\mu\rho}\bar{K}_{\nu\sigma}\right]\delta h^{\rho\sigma}
 +\left[+\frac{1}{2}\bar{h}_{\mu\rho}\bar{h}_{\nu\sigma}\right]\mathcal{L}_{\hat{v}}\delta h^{\rho\sigma}+\left[-\frac{1}{2}\bar{\nabla}_{\rho}\bar{h}_{\mu\nu}\right]\delta\hat{v}^{\rho}\nonumber \\
 &  &  +\left[-\frac{1}{2}\bar{h}_{\rho\nu}\right]\bar{\nabla}_{\mu}\delta\hat{v}^{\rho}+\left[-\frac{1}{2}\bar{h}_{\mu\rho}\right]\bar{\nabla}_{\nu}\delta\hat{v}^{\rho} +\left[\hat{\Phi}\tau_{\rho\mu}\tau_{\nu}+\hat{\Phi}\tau_{\rho\nu}\tau_{\mu}\right]\delta\hat{v}^{\rho}\,.
\end{eqnarray}
Useful projections are:
\begin{eqnarray}
h^{\mu\nu}\delta\bar{K}_{\mu\nu} & = & +\left[-\hat a_{\rho}\right]\delta\hat{v}^{\rho} +\left[-\delta_{\rho}^{\mu}-\tau_{\rho}\hat{v}^{\mu}\right]\bar{\nabla}_{\mu}\delta\hat{v}^{\rho} \nonumber \\
 &  & +\left[-\bar{K}_{\rho\sigma}+2\hat{\Phi}\tau_{\rho}a_{\sigma}+\tau_{\rho}\partial_{\sigma}\hat{\Phi}\right]\delta h^{\rho\sigma} +\left[+\frac{1}{2}\hat{v}^{\lambda}\bar{h}_{\rho\sigma}\right]\bar{\nabla}_{\lambda}\delta h^{\rho\sigma}\,,\label{eq: h=00005CdeltaK_contract_var}\\
h^{\rho\mu}h^{\sigma\nu}\delta\bar{K}_{\mu\nu} & = & +\left[-h^{\rho\mu}\hat a_{\mu}\left(\delta_{\lambda}^{\sigma}+\tau_{\lambda}\hat{v}^{\sigma}\right)\right]\delta\hat{v}^{\lambda} +\left[-h^{\rho\mu}\left(\delta_{\lambda}^{\sigma}+\tau_{\lambda}\hat{v}^{\sigma}\right)\right]\bar{\nabla}_{\mu}\delta\hat{v}^{\lambda}\nonumber \\
 &  & +\left[-2h^{\rho\mu}\bar{K}_{\mu\lambda}\left(\delta_{\kappa}^{\sigma}+\tau_{\kappa}\hat{v}^{\sigma}\right)-\hat{v}^{\rho}\hat{v}^{\sigma}\hat a_{\lambda}\tau_{\kappa}\right]\delta h^{\lambda\kappa}\nonumber \\
 &  &  +\left[+\frac{1}{2}\delta_{\lambda}^{\rho}\delta_{\kappa}^{\sigma}+\tau_{\lambda}\delta_{\kappa}^{\rho}\hat{v}^{\sigma}\right]\mathcal{L}_{\hat{v}}\delta h^{\lambda\kappa}\,. \label{eq: hh=00005CdeltaKvar}
\end{eqnarray}

\section{Twistless Torsional Newton–Cartan identities}\label{sec:TTNC_identities_special}

This appendix is very similar to the previous one except that we now assume that $\tau$ obeys the hypersurface orthogonality (or TTNC) condition $\tau\wedge\d\tau=0$. All of the identities of appendix \ref{sec:NC_identities} of course all apply here as well but there are many simplifications when the TTNC condition is imposed.
\subsection{Special TTNC identities}
The most fundamental identity for TTNC geometry is
\begin{equation}
\partial_\mu\tau_\nu-\partial_\nu\tau_\mu=a_\mu\tau_\nu-a_\nu\tau_\mu\,,
\end{equation}
where $a_\mu=\mathcal{L}_v\tau_\mu$ and $\hat a_\mu=\mathcal{L}_{\hat v}\tau_\mu$.
A second useful TTNC identity is the result that
\begin{equation}\label{eq:spatialcurva}
h^{\mu\rho}h^{\nu\sigma}\left(\partial_\mu a_\nu-\partial_\nu a_\mu\right)=0\,.
\end{equation}

Due to the presence of torsion one can show that, using the Bianchi identity for $\bar R_{[\mu\nu\sigma]}{}^\rho$, 
\begin{eqnarray}
3\bar{R}_{[\mu\nu\sigma]}{}^\rho & = &\left(\nabla_\mu\hat v^\rho\right)\left(\partial_\nu\tau_\sigma-\partial_\sigma\tau_\nu\right)+\left(\nabla_\sigma\hat v^\rho\right)\left(\partial_\mu\tau_\nu-\partial_\nu\tau_\mu\right)\nonumber\\
&&+\left(\nabla_\nu\hat v^\rho\right)\left(\partial_\sigma\tau_\mu-\partial_\mu\tau_\sigma\right)\,,\label{eq:totallyASpartRiem}
\end{eqnarray}
The antisymmetric part of the Ricci tensor is nonzero and equal to
\begin{equation}\label{eq:ASpartRicci}
2\bar R_{[\mu\nu]}=(\tau_\mu a_\nu-\tau_\nu a_\mu)\bar\nabla_\rho\hat v^\rho+\hat v^\rho(\tau_\mu\bar\nabla_\nu \hat a_\rho-\tau_\nu\bar\nabla_\mu \hat a_\rho)\,.
\end{equation}
If instead of contracting the second Bianchi identity (for $\kappa=\lambda$) with $v^\nu h^{\mu\sigma}$ we contract it with two inverse spatial metrics we obtain (for TTNC),
\begin{equation}
\bar\nabla_\mu\left(h^{\mu\nu}h^{\rho\sigma}\bar R_{\nu\rho}-\frac{1}{2}h^{\mu\sigma}h^{\nu\rho}\bar R_{\nu\rho}\right)=0\,.
\end{equation}

\subsection{Variational calculus}\label{sec:TTNC_variational_calculus}
\subsubsection{Basic identities}
For TTNC geometry we only need to calculate some projective variations.
The $\tau_\mu$ variation along $\tau_\mu$ is best computed by setting
\begin{equation}
\delta_\Omega\tau_\mu \equiv\Omega\tau_\mu\,.
\end{equation}
for arbitrary $\Omega$.
In order to compute the $h$ variation it is sufficient to consider
\begin{equation}
\delta_P h_{\mu\nu}\equiv P^\rho_\mu P^\sigma_\nu\delta h_{\rho\sigma}   
\end{equation}
where $P^\mu_\nu\equiv h^{\mu\lambda} h_{\lambda\nu}$. The $P$ variations of the other geometric objects are given by:
\begin{eqnarray}
\delta_P v^\mu &=&0\,, \\
\delta_P\tau_\mu&=&0\,,\\
\delta_P h^{\mu\nu}&=&-h^{\mu\rho}h^{\nu\sigma}\delta_P h_{\rho\sigma}\,,\\
\delta_P\Phi&=&0\,.
\end{eqnarray}

\subsubsection{Connections, Ricci tensor \texorpdfstring{$\check{R}_{\mu\nu}$}{check R} and extrinsic curvatures}
It can be shown  that for the connection $\check\Gamma^\rho_{\mu\nu}$ and its associated Ricci tensor we have simplifications compared to the general variations \eqref{eqn:variation_connection_check}, \eqref{eq:special TNC connection_variation}, \eqref{eq:varRic_tau}-\eqref{eq:varRic_h} and \eqref{eq:varRic_bar}:
\begin{eqnarray}
\delta_P\check\Gamma^{\rho}_{\mu\nu} & = & \frac{1}{2}h^{\rho\sigma}\left(\check\nabla_\mu\delta_P h_{\nu\sigma}+\check\nabla_\nu\delta_P\delta h_{\mu\sigma}-\check\nabla_\sigma\delta_P h_{\mu\nu}\right)\,,\\
h^{\mu\rho}h^{\nu\sigma}\delta_P\check R_{\mu\nu} & = & h^{\mu\rho}h^{\nu\sigma}\left(\check\nabla_\lambda\delta_P\check\Gamma^\lambda_{\mu\nu}-\check\nabla_\mu\delta_P\check\Gamma^\lambda_{\lambda\nu}\right)\,.
\end{eqnarray}

We have some relevant variations that are needed in Section \ref{sec:NRG_EOMs}.
\begin{eqnarray}
\delta_P\left(h^{\rho\nu} K_{\mu\nu}\right) & = & -v^\sigma\delta_P\check\Gamma^\rho_{\mu\sigma}\,,\\
\delta_P a_\mu &=&0\,.
\end{eqnarray}
Furthermore, it can be shown that  varying $\hat\Phi$ in the connection $\bar\Gamma^\rho_{\mu\nu}$ gives for twistless torsion
\begin{equation}
\left.\delta_{\hat\Phi}\bar\Gamma^\rho_{\mu\nu}\right|_{\tau\wedge \d \tau =0}=\tau_\mu\tau_\nu h^{\rho\sigma}\partial_\sigma\delta\hat\Phi+2\tau_\mu\tau_\nu h^{\rho\sigma}\hat a_\sigma\delta\hat\Phi\,.
\end{equation}
With this result it follows that the variation of $\hat v^\mu\hat v^\nu\bar R_{\mu\nu}$ with respect to $\hat\Phi$ results in a total derivative.

\subsubsection{Ricci tensor \texorpdfstring{$\bar{R}_{\mu\nu}$}{bar R}}\label{sec:Variation_hR_contraction}
We will now study various projections of the variations of $\bar{R}_{\mu\nu}$ that are needed in Section \ref{sec:NRGprime_EOMs}.
First of all considering the spatial trace multiplied by a scalar function $X$ under the assumption of hypersurface orthogonality of $\tau_\mu$ leads to
\begin{eqnarray}
X\left(h^{\mu\nu}\delta\bar{R}_{\mu\nu}\right)
 & = & \Biggl[ h^{\mu\nu}\left(\hat a_{\mu}+\bar{\nabla}_{\mu}\right)\left[\left(\hat a_{\nu}X+\bar{\nabla}_{\mu}X\right)\bar{h}_{\rho\sigma}\right]\nonumber \\
 &  & -\tau_{\rho}\bar{\nabla}_{\lambda}\left[\hat{v}^{\lambda}\left(\hat a_{\sigma}X+2\bar{\nabla}_{\sigma}X\right)\right]\nonumber \\
 &  & -\left[\hat a_{\rho}+\bar{\nabla}_{\rho}\right]\left[\hat a_{\sigma}X+\bar{\nabla}_{\sigma}X\right]\nonumber \\
 &  & +Xh^{\mu\nu}\hat a_{\mu}\bar{K}_{\nu\rho}\tau_{\sigma}\Biggr]\delta h^{\rho\sigma}\,.
\end{eqnarray}

Secondly, we also need the projection $h^{\mu\kappa}h^{\nu\sigma}\delta\bar{R}_{\mu\nu}$, which contracted with a symmetric tensor $X_{\mu\nu}$ after a bit of work gives
\begin{eqnarray}
X_{\rho\sigma}\left(h^{\mu\rho}h^{\nu\sigma}\delta\bar{R}_{\mu\nu}\right) & = & \Biggl[h^{\mu\nu}\left(\hat a_{\mu}+\bar{\nabla}_{\mu}\right)\left[\left(\hat a_{\nu}X_{\eta\rho}+\bar{\nabla}_{\nu}X_{\eta\rho}-\hat a_{\rho}X_{\nu\eta}-\bar{\nabla}_{\rho}X_{\nu\eta}\right)\tau_{\sigma}\hat{v}^{\eta}\right]\nonumber \\
 &  & -h^{\mu\nu}\left(\hat a_{\mu}+\bar{\nabla}_{\mu}\right)\left[\hat a_{\rho}X_{\nu\sigma}+\bar{\nabla}_{\rho}X_{\nu\sigma}\right]\nonumber \\
 &  & +\frac{1}{2}h^{\mu\nu}\left(\hat a_{\mu}+\bar{\nabla}_{\mu}\right)\left[\hat a_{\nu}X_{\rho\sigma}+\bar{\nabla}_{\nu}X_{\rho\sigma}\right]\nonumber \\
 &  & +\frac{1}{2}h^{\mu\nu}\left(\hat a_{\mu}+\bar{\nabla}_{\mu}\right)\left[\left(\hat a_{\lambda}X_{\kappa\nu}+\bar{\nabla}_{\lambda}X_{\kappa\nu}\right)h^{\lambda\kappa}\bar{h}_{\rho\sigma}\right]\nonumber \\
 &  & -h^{\mu\nu}\bar{\nabla}_{\lambda}\left[\hat{v}^{\lambda}\tau_{\sigma}\bar{\nabla}_{\mu}X_{\nu\rho}\right]\Biggr]\delta h^{\rho\sigma}\,.
\end{eqnarray}

\subsection{\texorpdfstring{$1/c^2$}{1c2} expansion formulae}\label{subapp:NRexpansionmetricGamma}
Expanding the measure gives
\begin{eqnarray}
\frac{1}{c}\sqrt{-g} & = &  e\left(1+\frac{1}{c^2}\hat\Phi+\frac{1}{2c^2}h^{\mu\nu}\bar\Phi_{\mu\nu}+\order{c^{-4}}\right) \nonumber\\
& = & e\left(1+\frac{1}{c^2}\Phi+\frac{1}{2c^2}h^{\mu\nu}\Phi_{\mu\nu}+\order{c^{-4}}\right) \label{eq:measure_1c2_exo}\,,
\end{eqnarray}
The Levi--Civita connection \eqref{eq:LC_connection_explicit_factors1c2} is expanded as
\begin{equation}
\Gamma^\rho_{\mu\nu}=c^2\Gamma^\rho_{{(-2)}\mu\nu}+\Gamma^\rho_{{(0)}\mu\nu}+c^{-2}\Gamma^\rho_{{(2)}\mu\nu}+\order{c^{-4}}\,.
\end{equation}
which can be expressed in terms of Galilean boost invariant objects according to 
\begin{eqnarray}
\Gamma^\rho_{{(-2)}\,\mu\nu} & = & h^{\rho\lambda}a_{\lambda}\tau_{\mu}\tau_{\nu}\,,\label{eq:Christoffel_1c2_expansion1}\\
\Gamma^\rho_{{(0)}\,\mu\nu} & = & \bar{\Gamma}_{\mu\nu}^{\rho}-\hat{v}^{\rho}\tau_{\mu}a_{\nu}-h^{\rho\kappa}h^{\lambda\eta}\bar\Phi_{\kappa\eta}a_{\lambda}\tau_{\mu}\tau_{\nu}\,,\label{eq:Christoffel_1c2_expansion2}\\
h^{\mu\alpha}h^{\nu\beta}\tau_\rho\Gamma^\rho_{(2)\mu\nu} & = & h^{\mu\alpha}h^{\nu\beta}\bar K_{\mu\nu}\,.
\end{eqnarray}
We placed the power of $c^{-1}$ symbol as a subscript to distinguish it from the expansion in equation \eqref{eq:LC_connection_explicit_factors1c2}.

\addcontentsline{toc}{section}{References}
\bibliography{NRG_bibliography}
\bibliographystyle{newutphys}

\end{document}